\documentclass[10pt, final, journal, letterpaper, oneside, twocolumn]{IEEEtran}

\usepackage{cite}

\usepackage{graphicx}
\usepackage{mathrsfs}
\usepackage{stfloats}
\usepackage{dsfont}
\usepackage{setspace}
\usepackage{array}
\usepackage{bbm}
\usepackage{hyperref}
\usepackage{here}
\usepackage{amsmath,amsthm,amsfonts,amssymb}
\usepackage{color}
\usepackage{accents}

\usepackage{epstopdf}
\usepackage[centerlast,small]{caption}
\usepackage{subcaption}
\usepackage{bbm}
\usepackage{easytable}
\usepackage{booktabs}

\usepackage{fancyvrb} 
\VerbatimFootnotes    

\newcommand{\norm}[1]{\left\lVert#1\right\rVert}

\def\SS{\mathcal{S}}

\newcommand{\vect}[1]{\mathbf{#1}}

\begin{document}

\title{Communication Models for Reconfigurable Intelligent Surfaces: From Surface Electromagnetics to Wireless Networks Optimization}	
	
\author{Marco~Di~Renzo,~\IEEEmembership{Fellow,~IEEE}, Fadil~H.~Danufane, and Sergei~Tretyakov,~\IEEEmembership{Fellow,~IEEE}

\thanks{Manuscript received October 2, 2021; {revised July 5, 2022}. (Corresponding author: \textit{Marco Di Renzo}.)}
\thanks{M. Di Renzo and F. H. Danufame are with Universit\'e Paris-Saclay, CNRS, CentraleSup\'elec, Laboratoire des Signaux et Syst\`emes, 3 Rue Joliot-Curie, 91192 Gif-sur-Yvette, France. (marco.di-renzo@universite-paris-saclay.fr).}
\thanks{S. Tretyakov is with the Department of Electronics and Nanoengineering, School of Electrical Engineering, Aalto University, Maarintie 8, 02150 Espoo, Finland.}
}

\maketitle

\begin{abstract}
A reconfigurable intelligent surface (RIS) is a planar structure that is engineered to dynamically control the electromagnetic waves. In wireless communications, RISs have recently emerged as a promising technology for realizing programmable and reconfigurable wireless propagation environments through nearly passive signal transformations. With the aid of RISs, a wireless environment becomes part of the network design parameters that are subject to optimization.

In this tutorial paper, we focus our attention on communication models for RISs. First, we review the communication models that are most often employed in wireless communications and networks for analyzing and optimizing RISs, and elaborate on their advantages and limitations. Then, we concentrate on models for RISs that are based on inhomogeneous sheets of surface impedance, and offer a step-by-step tutorial on formulating electromagnetically-consistent analytical models for optimizing the surface impedance. The differences between local and global designs are discussed and analytically formulated in terms of surface power efficiency and reradiated power flux through the Poynting vector. Finally, with the aid of numerical results, we discuss how approximate global designs can be realized by using locally passive RISs with zero electrical resistance (i.e., inhomogeneous reactance boundaries with no local power amplification), even for large angles of reflection and at high power efficiency.
\end{abstract}

\section{Introduction}
The history of wireless communications started with understanding fundamental electric and magnetic phenomena, as well as with related experiments and inventions that were carried out during the last half of the eighteenth century and the first decades of the nineteenth century \cite{History}. Wireless communications (often, just wireless) are defined as and are characterized by the transfer of information between two or more points without the need of using an electrical conductor as the medium to perform the transfer. The most common wireless technologies use electromagnetic waves. Thanks to the development and wide adoption of five wireless telecommunication standards and the recently started activities on the sixth generation of wireless systems and networks, we do live in a world of electromagnetic waves.

In our daily life, we observe plenty of concrete examples of electromagnetic phenomena, especially in the visible spectrum. For example, the visible light that is specularly reflected when it hits a smooth surface, so that we can see ourselves in a mirror; the visible light that changes its route when traveling from one medium to another, which causes, e.g., the virtual distortion of objects in water; or the visible light that creates complicated rainbow effects formed as a combination of reflection, refraction, and dispersion phenomena. These electromagnetic effects are governed by fundamental laws of physics and are, therefore, ultimately dictated by nature. More precisely, these examples of electromagnetic effects in the visible spectrum are determined by the interactions between the electromagnetic waves and the materials that are hit by them. When an arbitrary electromagnetic wave illuminates a material object, more precisely, it excites oscillations of the charged particles that constitute the material. These oscillating particles act, in turn, as secondary sources that radiate electromagnetic waves into the space, thus producing different wave phenomena. During hundreds of years of research in the field of electromagnetics, today we can not only understand these phenomena, but we can control the electromagnetic waves, and we can even create new wave effects that go beyond those governed solely by nature \cite{Capasso_Science}.

The development of electromagnetics, which is often defined as the theory of electromagnetic fields and waves, has greatly helped us to qualitatively and quantitatively comprehend how the waves propagate and how they interact with material objects \cite{SEM_Book}. This understanding has inspired researchers to engineer and manufacture artificial electromagnetic materials with controllable material parameters, which, when illuminated by appropriate electromagnetic waves, are capable of realizing wave effects (or transformations) that do not exist in nature. Engineered materials of this kind are referred to as metamaterials, which are often broadly defined as an effective homogeneous material formed by an arrangement of engineered structural elements that are designed to achieve specified and unusual electromagnetic properties \cite{Metamaterial_Book}. A typical example is constituted by a material that does not reflect the light in agreement with the law of reflection, i.e., the angle of reflection coincides with the angle at which the light illuminates the material, but according to the generalized law of reflection, i.e., the engineered material is capable of bending the light towards specified directions of reradiation that are different from the angle of incidence \cite{Capasso_Science}.

Metamaterials are three-dimensional artificial (engineered) materials, which are usually bulky, heavy, and often difficult to be fabricated. Due to the inevitable material losses, metamaterials may strongly attenuate the electromagnetic waves that penetrate through them. One possible alternative to overcome the inherent limitations of metamaterials is the use of metasurfaces, which are electrically thin artificial layers with sub-wavelength inclusions \cite{SEM_Book}. Metasurfaces are often referred to as the bi-dimensional version of metamaterials, which, by virtue of the surface equivalence theorem, have the same capabilities of shaping the propagation of the electromagnetic waves that interact with them, while being less bulky, lossier, and easier to be fabricated and to be deployed than metamaterials.

\subsection{Programmable Wireless Environments}
In current wireless telecommunication standards, different kinds of electromagnetic waves constitute the vehicle for enabling the transmission of information and for allowing users and devices to communicate. Therefore, equipping current wireless telecommunication standards or even designing a new wireless telecommunication standard with the inherent capability of controlling and shaping how the electromagnetic waves propagate in a complex wireless environment and how they interact with material objects (walls, buildings, etc.) would be beneficial \cite{Richie_Industrial}. Indeed, the potential application of metasurfaces in the context of wireless communication systems and networks has recently attracted the interest of wireless researchers and engineers. Examples of papers include \cite{Subrt-Pechac__2012}, \cite{Fink}, \cite{Kyle_2017}, \cite{Liaskos_ACM}, \cite{Liaskos_COMMAG}, \cite{MDR_EURASIP}, \cite{MDR_Reflections}, \cite{MDR_Access}, \cite{Kyle_2019}, \cite{MDR_Holos}, \cite{MDR_Relays}, \cite{MDR_JSAC}, \cite{MDR_Wireless2.0}, \cite{Rui_COMMAG}, \cite{Rui_Tutorial}, \cite{Cunhua_Magazine}. A short technology note that summarizes recent developments and ongoing pre-standardization activities is available in \cite{MDR_CTN}.

Current wireless systems utilize a variety of transmission technologies, communication protocols, and network deployment strategies. They include millimeter-wave communications, massive multi-input multi-output systems (MIMO), and ultra-dense heterogeneous networks. Currently available solutions are often based on the deployment, design, and optimization of transmitters, receivers, and  network infrastructure elements with power amplification and digital signal processing capabilities, as well as backhaul and power grid availability. Communication engineers usually design transmitters, receivers, network elements, and transmission protocols by assuming not to be able to control how the electromagnetic waves propagate through a wireless environment and how they interact with the material objects that exist in the considered environment. When an electromagnetic wave impinges, for example, upon a metallic wall or upon a glass window, the reflected and refracted waves are not directly controlled by the network operator but are determined by the properties of the electromagnetic waves and the constitutive elements of the material objects that interact with the electromagnetic waves. If the material objects in the wireless environment were coated with or were even made of metamaterials (engineered materials), we could control their interactions with the impinging electromagnetic waves and we could appropriately shape them as desired. This would enable us to co-design and jointly optimize the electromagnetic waves emitted by the transmitters, how they interact with the surrounding material objects, and how they are decoded by the receivers.

\begin{figure}[!t]
{\includegraphics[width=\columnwidth]{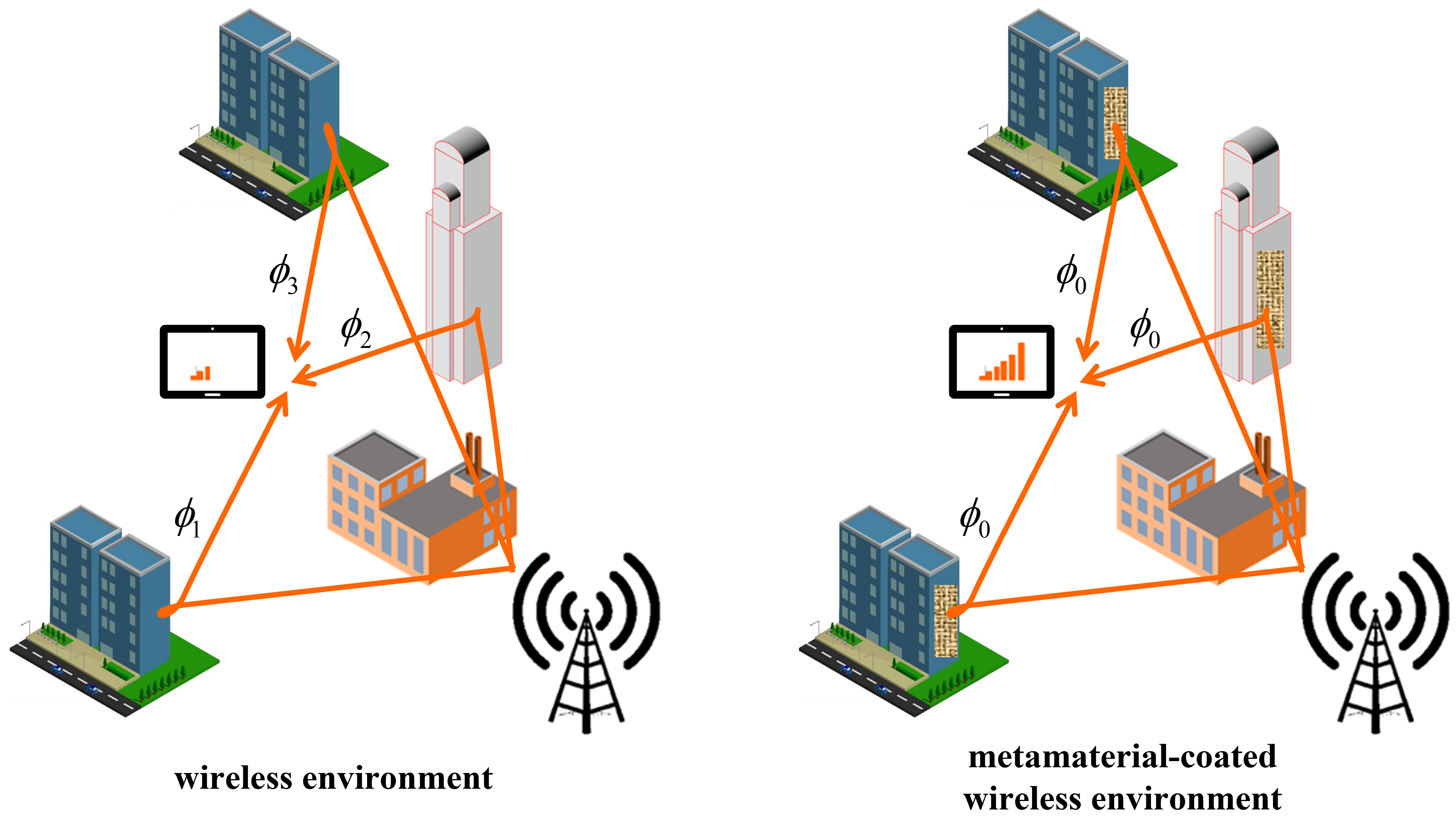}}
\caption{Illustration of a wireless environment and a smart (programmable) radio environment.}\label{fig:SREs}
\end{figure}
Metamaterial-coated wireless networks are an emerging design paradigm that is often referred to as programmable wireless environment or smart radio environment (SRE) \cite{MDR_EURASIP}. An example of SRE is illustrated in Fig. \ref{fig:SREs}. In a conventional wireless environment, the electromagnetic waves that are reflected or refracted by material objects are out of control of the system designer. As shown in Fig. \ref{fig:SREs}, the reflected electromagnetic waves reach the intended receiver with different phases that may partially cancel out. This phenomenon can be alleviated by equipping, whenever possible, the transmitters and receivers with multiple antennas or by deploying additional infrastructure elements with signal processing units, power amplifiers, and multiple radio frequency chains. In an SRE, on the other hand, the same material objects are coated with metamaterial sheets (i.e., metasurfaces) that shape the reradiated electromagnetic waves so that they reach the intended receiver with approximately the same phase. By co-designing the metamaterial sheets, the transmitters, and the receivers, the performance of wireless networks may be further improved.

\subsection{Reconfigurable Intelligent Surfaces}
In the context of wireless communication systems and networks, as exemplified in Fig. \ref{fig:SREs}, the use of planar metamaterial structures or metasurfaces is receiving major attention from the wireless community, see, e.g., \cite{MDR_JSAC}, \cite{Cunhua_Magazine}. The reason lies, as mentioned, in the reduced losses and less complex design of two-dimensional (either planar or conformal) metasurfaces as compared with three-dimensional metamaterials. Broadly speaking, a metasurface is a metamaterial sheet of sub-wavelength thickness. Despite their negligible thickness compared with the wavelength of the electromagnetic waves, metasurfaces can be as powerful as metamaterials in terms of wave manipulations while avoiding some of their drawbacks. This is ensured by the surface equivalence theorem, which states that the electromagnetic fields excited by arbitrary sources located in a volumetric material sample can be equivalently created by surface currents enclosing the volume. Therefore, any metamaterial sample can be replaced by electrically thin metasurfaces that are engineered to produce the same scattered electromagnetic waves \cite{SEM_Book}.

In wireless communications, the metasurfaces need to be reconfigurable so as to ensure that they can shape the electromagnetic waves based on the network conditions. Wireless researchers have adopted different names to refer to a reconfigurable metasurface \cite{MDR_CTN}. In the present tutorial paper, we adopt the term reconfigurable intelligent surface (RIS), since it is adopted by a recently established industry specification group (ISG) within the European telecommunications standards institute (ETSI) \cite{ETSI_ISG-RIS}. Broadly speaking, an RIS is an engineered surface that is intelligent (or smart) because it is capable of (i) applying wave transformations that go beyond those governed solely by nature and (ii) being configured any time that the propagation and network conditions require it.

Compared with other technologies, RISs have advantages and limitations, as recently summarized in \cite[Table 1]{MDR_CTN}. Within the recently established ETSI-ISG on RISs, an RIS is usually defined as a nearly-passive reconfigurable engineered surface that (i) is implemented by using passive scattering elements, (ii) does not require high-cost active components, such as power amplifiers, (iii) does not possess sophisticated signal processing capabilities, but only the necessary low-power electronic circuits for enabling its reconfigurability, and (iv) is not equipped with multiple radio frequency chains for data transmission, but requires a simple front-end to receive and send control signals. These characteristics suggest that an RIS may be considered as a sustainable and environmentally friendly technology solution. The absence of power amplifiers and digital signal processing capabilities naturally pose, however, important design and deployment challenges to be solved. This includes the impossibility of on-board channel estimation, signal regeneration, and amplification, which are currently being tackled by wireless researchers and engineers \cite{MDR_JSAC}, \cite{Cunhua_Magazine}. {Let us consider the problem of channel estimation as an example. The absence of power amplifiers and radio frequency chains makes not possible for an RIS to transmit pilot signals for channel estimation. Likewise, the absence of signal processing units makes not possible for an RIS to detect the pilot signals emitted by other devices. Therefore, efficient methods need to be developed for channel estimation and for optimizing RISs based, for example, on partial channel state information, so as to reduce the excessive overhead for channel estimation \cite{pan2021overview}. Another recently proposed approach is based on developing hybrid RISs, which are endowed with integrated communication and sensing capabilities \cite{NEC_Infocom-2022}.}

In wireless communications, an RIS has many potential applications, which go beyond its use to turn the environmental objects into digitally controllable smart scatterers, as shown in Fig. \ref{fig:SREs}. Other applications include the design of multi-stream multi-antenna transmitters with a single radio frequency chain (often called holographic surfaces and holographic MIMO) \cite{Wankai_Transmitter} and reconfigurable ambient backscatterers \cite{Fara_Backscattering}. In general terms, an RIS is a candidate future wireless technology for controlling and shaping the electromagnetic waves in a dynamic and goal-oriented manner, possibly turning the wireless environment into a service \cite{RIS_EuCNC}, and for realizing new transceiver designs and network elements at a lower complexity and power consumption. {System-level simulations for evaluating the performance gains offered by the deployment of RISs in a typical urban city served by a fifth-generation cellular network have recently been reported in \cite{RIS_SLS}}.

\begin{figure}[!t]
{\includegraphics[width=\columnwidth]{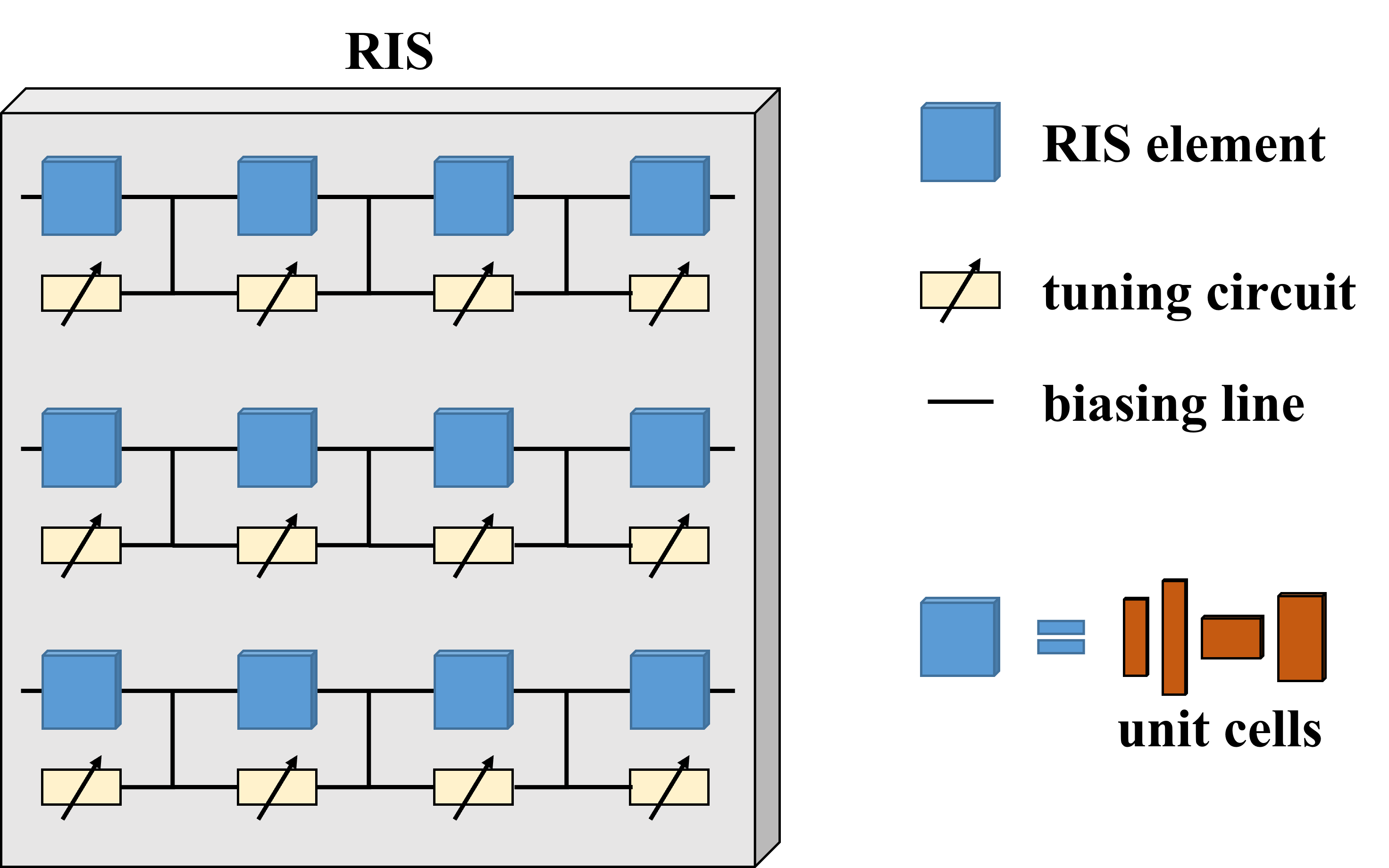}}
\caption{Conceptual architecture of a reconfigurable intelligent surface.}\label{fig:RIS}
\end{figure}
\begin{figure}[!t]
{\includegraphics[width=\columnwidth]{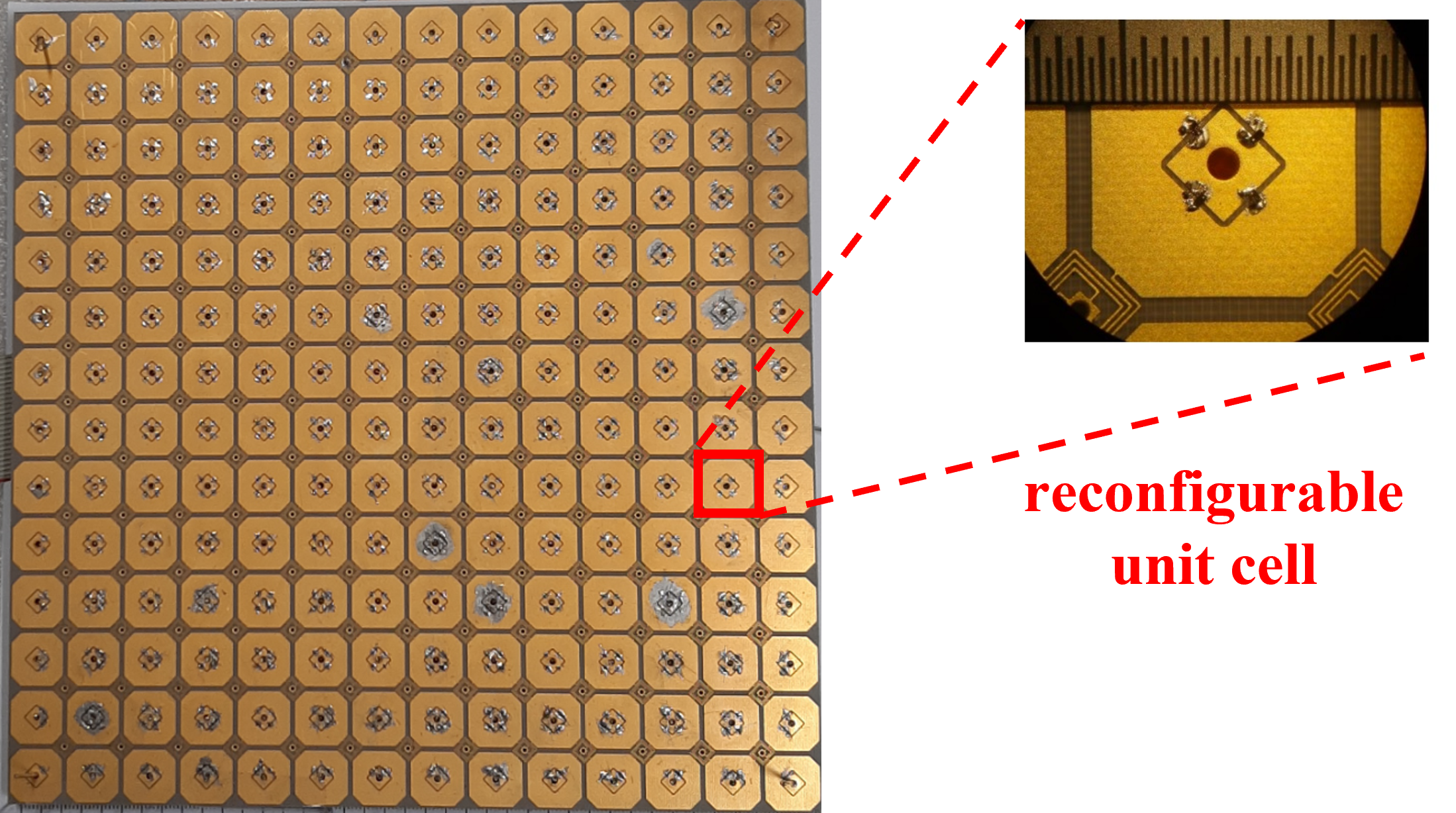}}
\caption{Example of manufactured reconfigurable intelligent surface made of 196 identical elements (unit cells) and 4 voltage-controlled varactors for each cell \cite{Romain_RIS-Prototype}.}\label{fig:OrangeRIS}
\end{figure}
\begin{figure}[!t]
{\includegraphics[width=\columnwidth]{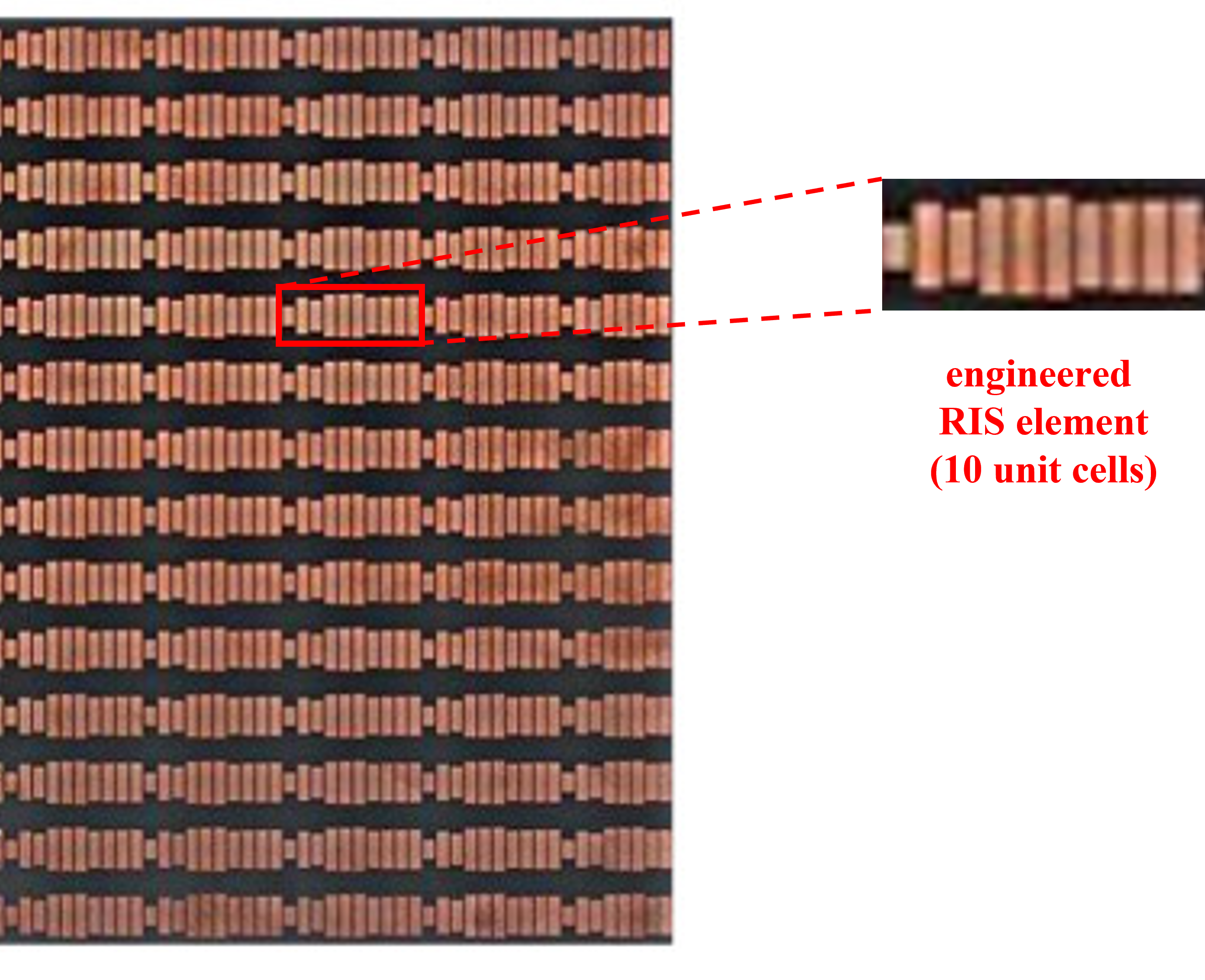}}
\caption{Example of manufactured engineered surface whose elements are made of 10 appropriately engineered unit cells \cite{Sergei_RealizedSurface}.}\label{fig:AaltoRIS}
\end{figure}
The conceptual structure of an RIS is sketched in Fig. \ref{fig:RIS}. As illustrated, an RIS is a planar surface that consists of an array of scattering elements, each of which can independently impose the required phase shift, and possibly an amplitude gain, on the incident electromagnetic waves. By carefully adjusting the phase shifts (and the amplitudes) of all the scattering elements, the reradiated electromagnetic waves can be shaped to propagate towards specified directions. Each RIS element may consist of multiple constitutive elements, which are usually referred to as unit cells. The unit cells that constitute each single RIS element have, in general, different shapes and sizes. If the RIS elements are made of the same unit cells and if they are arranged on a spatially periodic array, the resulting RIS is a quasi-periodic structure and the inter-distance between the RIS elements is usually referred to as the period of the metasurface. Once the unit cells of each RIS element are designed, the wave transformation that the RIS applies to the incident signals is fixed. The reconfigurability of the RIS is ensured by a network of tuning circuits and a biasing line that control the unit cells. For example, the tuning circuits in Fig. \ref{fig:RIS} may be positive-intrinsic negative (PIN) diodes or voltage-controlled varactors. Depending on the control voltage applied throughout the biasing line, the scattering properties of the RIS are adapted to the channel and network conditions, making it a digitally controllable scatterer. The tuning circuit and the biasing line may either control each individual unit cell, or each RIS element individually, or even multiple RIS elements together. Making each unit cell reconfigurable through an independent tuning circuit offers a finer control of the electromagnetic waves at the cost of a higher implementation complexity and power consumption. Two examples of manufactured engineered metasurfaces are illustrated in Figs. \ref{fig:OrangeRIS} \cite{Romain_RIS-Prototype} and \ref{fig:AaltoRIS} \cite{Sergei_RealizedSurface}. The metasurface in Fig. \ref{fig:OrangeRIS} is an RIS made of 196 identical unit cells. Each unit cell is digitally controlled by four varactors, which determine the reflection properties of the unit cell. The metasurface in Fig. \ref{fig:AaltoRIS} is a non-reconfigurable engineered surface, whose elements comprise ten different unit cells. The sizes and arrangements of the ten unit cells are jointly designed to realize a perfect anomalous reflector towards a fixed angle of reradiation with high power efficiency. {Interested readers may refer to \cite[Sec. 7]{mmWall} for information about the implementation cost and power consumption of deploying RISs instead of multiple access points in millimeter-wave networks. In \cite[Fig. 2]{Wankai_PathLoss-mmWave}, in addition, a general discussion on the fundamental tradeoffs offered by an RIS in terms of scattering performance, power consumption, and area of each unit cell is presented. In general, the implementation cost and power consumption of an RIS are greatly determined by the operating frequency and by the complexity of the circuits that enable the reconfigurability of the RIS elements. Simple unit cell designs are often preferred to reduce the cost and the power consumption, at the cost of some inherent performance tradeoffs. A comprehensive study has recently been conducted in \cite{DigitalRIS_Elsevier}.}

There exist multiple methods for designing an RIS. Interested readers may consult \cite{Capasso_Science}, \cite{Eleftheriades_Tutorial}, \cite{Epstein_Tutorial}, \cite{Eleftheriades_Circuit}, \cite{Cui_InfoMeta}, \cite{Cui_CodingMeta}, \cite{Giovampaola}, \cite{Caloz_Nanophotonics}, \cite{Caloz_Multiphysics}, \cite{Caloz_TAP}, \cite{Vincenzo}, \cite{Sergei_PerfectReflector}, \cite{Monticone}, \cite{Alu_PerfectReflector} for further information. Two typical design methods are the following.
\begin{itemize}
    \item The first method is a one-step approach, which departs from the design of an individual RIS element, which may or may not be made of multiple unit cells. In this method, the constitutive RIS element is designed in order to realize some predefined phase shifts (and possibly amplitude gains and losses) when a given electromagnetic wave impinges upon it. The RIS element may realize a discrete set of phase shifts or the phase shift may be continuously controlled (as for the RIS in Fig. \ref{fig:OrangeRIS}). The scattering properties, e.g., the phase and the amplitude of the reflection coefficient for reflecting surfaces, are usually characterized with the aid of full-wave numerical simulations. The outcome of this phase consists of defining the size, geometry, thickness, composite material, and the control circuitry to realize the desired set of phases and gains. In wireless communications, this is referred to as the RIS alphabet \cite{DigitalRIS_Elsevier}. This characterization is usually performed by applying locally (at the level of the RIS element) periodic boundary conditions, which mimic an infinite homogeneous surface whose constitutive elements are all identical. If the RIS element is made of a single unit cell, the periodic boundary conditions are applied at unit cell level. If the RIS element is made of multiple unit cells, the scattering response of all the unit cells is jointly characterized. Further information on using periodic boundary conditions for designing an RIS and their inherent advantages and limitations are elaborated in Section II (see Fig. \ref{fig:PBCs}). Once the electromagnetic characterization of the RIS element is complete, the RIS operates by joint optimizing the scattering response of the RIS elements in order to realize the desired wave transformations.
    \item The second method is a two-step approach, whose first step consists of engineering the entire RIS surface as a whole. This first phase is a macroscopic design in which the surface position-dependent properties of the RIS are formulated in terms of the specific functionality (e.g., reflection, refraction, beam splitting) or set of functionalities (e.g., joint reflection and refraction or dynamic switching between reflection and refraction) that the RIS needs to realize. The RIS is usually modeled as a location-dependent continuous sheet of electric surface impedance and magnetic surface admittance. This design method is further elaborated in Section II and it is embraced in Section III to illustrate the design and analysis of RISs in a step-by-step and tutorial-like manner. Once the electric surface impedance and magnetic surface admittance are determined, the second step consists of identifying the physical microscopic implementation of the unit cells for the entire RIS and the associated tuning circuits for realizing the electric surface impedance and magnetic surface admittance in practice. During this phase, typically, one departs from a unit cell design of a given shape and optimizes the sizes, inter-distances, material, and control circuits of the entire RIS to obtain the target surface impedance and admittance (surface modulation). If the surface impedance and admittance are periodic functions in space, one can jointly optimize only the unit cells that constitute a single period.
\end{itemize}

The first design approach is inherently local (at the granularity of either the RIS element or the unit cell), while the second design is inherently global (the entire RIS is optimized). {Usually, the second method has a higher complexity but it typically results in superior performance. If correctly implemented, the second method accounts, in a more accurate manner, for the interactions (mutual coupling) among unit cells whose size and inter-distance are smaller than half of the wavelength. In the next sections, the advantages and limitations of these methods are discussed. Specifically, further details on the design and optimization complexity of local and global designs are given}. In rest of the present paper, for the avoidance of doubt, we utilize the term \textit{local design} to refer to designs of RISs in which each unit cell is optimized individually. On the other hand, we utilize the term \textit{global design} to refer to designs of RISs in which groups of unit cells are jointly optimized. With reference to Fig. \ref{fig:RIS}, the local design may correspond to an RIS in which each RIS element comprises a single unit cell. The global design may correspond, on the other hand, to an RIS in which each RIS element comprises several unit cells that are jointly optimized. More in general, several RIS elements may be jointly optimized in a global design.

\begin{figure*}[!t]
		\begin{center}
			\begin{subfigure}{0.66\columnwidth}
				{\includegraphics[width=\linewidth]{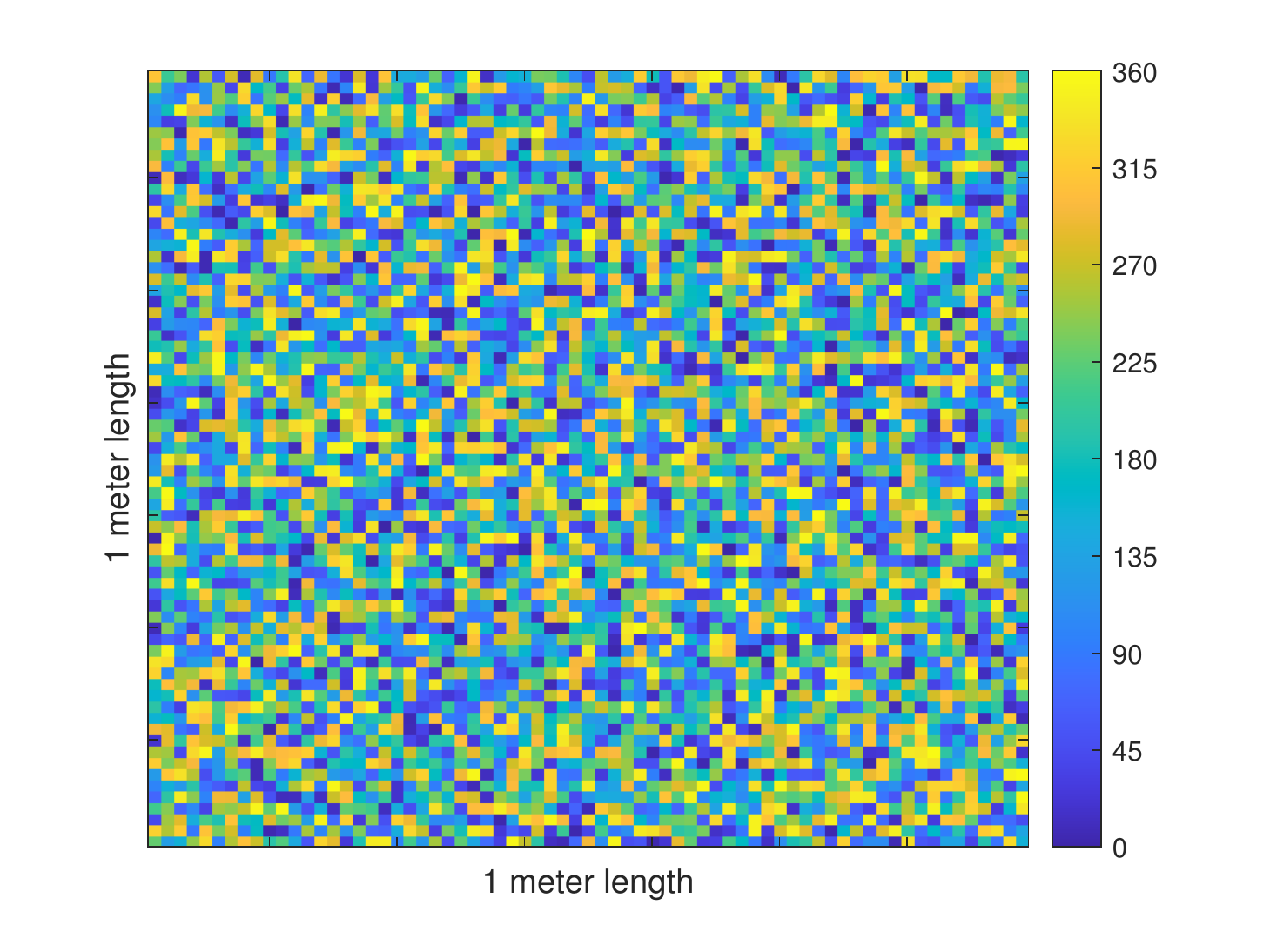}}
				\caption{{Example of aperiodic RIS (virtually continuous spatial sampling).}}\label{fig:RIS_PeriodicAperiodic__Aperiodic}
			\end{subfigure}
			\begin{subfigure}{0.66\columnwidth}
				{\includegraphics[width=\linewidth]{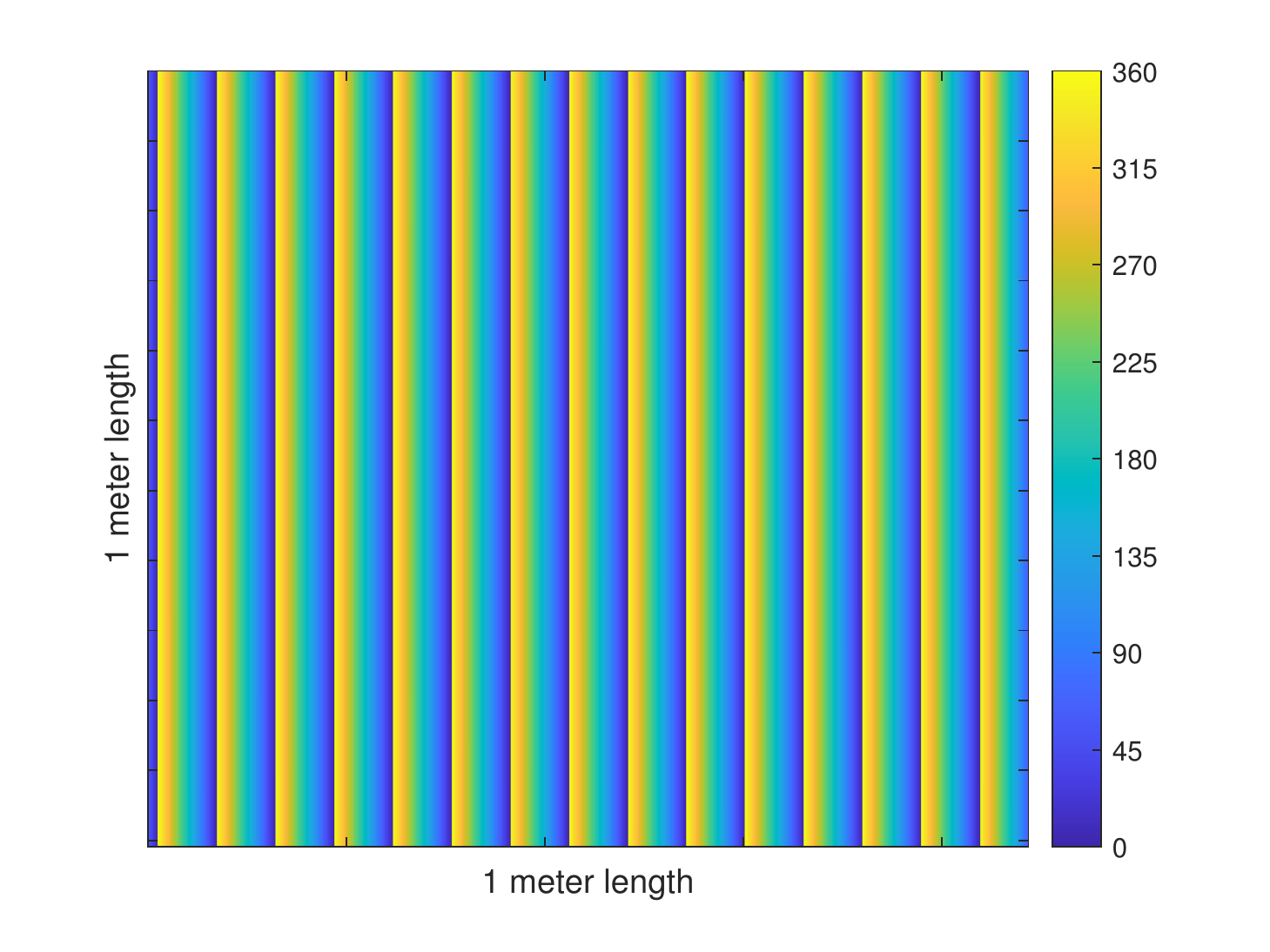}}
				\caption{{Example of periodic RIS (virtually continuous spatial sampling).}}\label{fig:RIS_PeriodicAperiodic__Periodic}
			\end{subfigure}
			\begin{subfigure}{0.66\columnwidth}
				{\includegraphics[width=\linewidth]{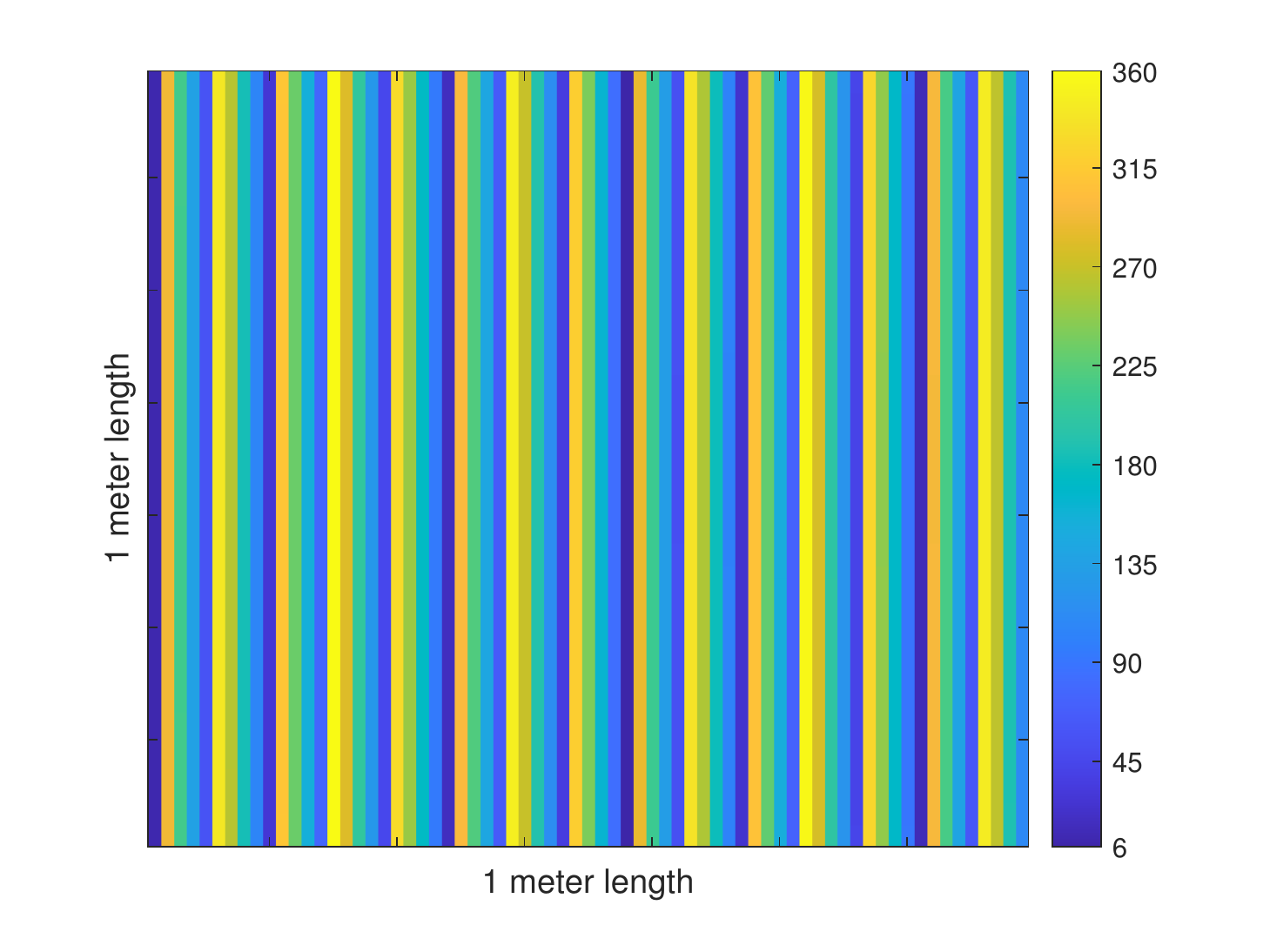}}
				\caption{{The periodic RIS in (b) with spatial sampling equal to $(1/4)$th of the wavelength.}}\label{fig:RIS_PeriodicAperiodic__AperiodicSampled}
			\end{subfigure}
		\end{center}
		\caption{Comparison between periodic and aperiodic RISs. The color scale represents the phase modulation applied by each unit cell.}
		\label{fig:RIS_PeriodicAperiodic}
	\end{figure*}
{Before proceeding, it is instructive to discuss the classification of RISs in terms of their power expenditure, i.e., the difference between passive, nearly-passive, and hybrid RISs, and the difference between periodic and aperiodic RISs.}

{\textbf{Passive, nearly-passive, and hybrid RISs}. In previous text, we have defined an RIS as a nearly-passive surface. In this context, the term \textit{nearly-passive} is referred to an RIS whose unit cells cannot amplify the incident electromagnetic waves, but some power is needed for operating the electronic circuits that make an RIS reconfigurable. As a consequence, an RIS cannot be a passive device, because, otherwise, it is not reconfigurable. Therefore, the RISs illustrated in Figs. \ref{fig:RIS} and \ref{fig:OrangeRIS} are not passive surfaces. Engineered surfaces without electronic circuits, and hence not reconfigurable, can, on the other hand, be passive surfaces. This definition of passive and nearly-passive RISs is usually referred to each individual unit cell. Stated differently, each unit cell cannot amplify the incident electromagnetic waves and the only power it needs is that necessary for ensuring its reconfigurability. Therefore, this definition usually applies without ambiguity to a local design. If several unit cells of an RIS are grouped and jointly optimized (e.g., the unit cells of one RIS element in Fig. \ref{fig:RIS}), the notion of passive and nearly-passive RISs needs to be refined. In this case, an RIS is defined as nearly-passive, in a global sense, if the group of unit cells that are jointly optimized do not amplify the incident electromagnetic waves. This definition does not necessarily imply that each unit cell of the group does not locally amplify the incident electromagnetic waves. In fact, some unit cells may accept power and send it to some other unit cells, which may in turn radiate more power than that received locally from the incident electromagnetic wave, so that the total reradiated power is not greater than the total incidence power. From an implementation standpoint, a global nearly-passive RIS can be realized either by explicitly deploying small amplifiers on the surface or by realizing passive mechanisms that enable the transfer of power from the unit cells that attenuate the incident electromagnetic waves towards the unit cells that amplify the incident electromagnetic waves. An example of this implementation is the RIS illustrated in Fig. \ref{fig:AaltoRIS}, and further information can be found in \cite{Sergei_RealizedSurface}. Recently, finally, researchers have been investigating the design of hybrid RISs in which some (usually a relatively small fraction of all the) unit cells amplify the incident electromagnetic waves so that the total reradiated power may be greater than the total incident power. These solutions require active power amplifiers, and incur in a higher cost and power consumption. However, they may facilitate critical tasks for the deployment of RISs in wireless networks, e.g., channel estimation, and they usually increase the transmission distance (coverage) thanks to the additional available power \cite{pan2021overview}.}

{\textbf{Periodic vs. aperiodic RISs: Why unit cells with sizes and inter-distances smaller than half of the wavelength?} To better understand the design challenges and tradeoffs for RISs, the notion of periodicity deserves to be elaborated in detail. For ease of understanding, let us start by comparing the two RISs illustrated in Figs. \ref{fig:OrangeRIS} and \ref{fig:AaltoRIS}. In Fig. \ref{fig:OrangeRIS}, we see that the surface is characterized by a \textit{geometric period} that coincides with the size of each unit cell (including the associated electronic circuits) on the surface. The entire RIS can hence be viewed as a periodic surface whose geometric period is the size of the unit cell. In \ref{fig:AaltoRIS}, on the other hand, the geometric period of the surface does not coincide with the size of one unit cell. In fact, the unit cells within one RIS element (i.e., 10 unit cells in Fig. \ref{fig:AaltoRIS}) have different sizes and inter-distances. The geometric period of the RIS in Fig. \ref{fig:AaltoRIS} is given by a group of 10 unit cells, which are repeated along the rows and columns of the surface. We observe, in addition, that no phase modulation is applied along the columns of the surface, since the same unit cell is repeated along the columns. In an engineered surface, in fact, the shape and size of each unit cell determine the applied phase modulation. The comparison between Figs. \ref{fig:OrangeRIS} and \ref{fig:AaltoRIS} allows us to understand that the geometric period of an RIS may not necessarily coincide with the size of one unit cell, but it may encompass several unit cells. Also, the geometric period may be different along the two sides of an RIS: In Fig. \ref{fig:AaltoRIS}, the geometric period along the rows is 1 unit cell and the geometric period along the columns is 10 unit cells.}

{Besides the geometric period just introduced, the design of an RIS depends on the specific function that it needs to realize, e.g., the reflection of an electromagnetic wave towards an angle different from the angle of incidence. Some wave transformations may result in configuring the unit cells of an RIS according to a repetitive pattern, which leads to a spatially-periodic configuration of the unit cells. This concept is illustrated in Fig. \ref{fig:RIS_PeriodicAperiodic}. More precisely, Figs. \ref{fig:RIS_PeriodicAperiodic__Aperiodic} and \ref{fig:RIS_PeriodicAperiodic__Periodic} show an example of aperiodic and periodic RIS, respectively. 
The figure shows an RIS whose geometric period is equal to $p$ along the rows and the columns. Also, the unit cells in Fig. \ref{fig:RIS_PeriodicAperiodic__Periodic} are configured according to a period $P$ that comprises several unit cell along the rows, similar to the RIS in Fig. \ref{fig:AaltoRIS}. We refer to the period $P$ as the \textit{RIS period}, which primarily depends on the function that an RIS needs to realize. RISs that realize wave transformations of this kind are referred to as periodic surfaces with period $P$. Otherwise, they are referred to as aperiodic surfaces. Sometimes, the same function (like anomalous reflection) can be realized by both periodic and aperiodic surfaces.}

\begin{table*}[!t]
		\centering
		\caption{{Steering angles ${{\theta _r}}$ that can be realized by an anomalous reflector with RIS period $P = \lambda /\sin \left( {{\theta _r}} \right)$ and geometric period $p=\lambda/N$ with $N \ge 2$, under the assumption that the RIS is optimized as a periodic surface.}}
		\label{Table_PeriodicVsPeriodic}
		\footnotesize
		{\begin{tabular}{c||c|c|c|c|c} \hline
			$N$ &  $n=N+1$ & $n=N+2$ & $n=N+3$ & $n=N+4$ & $n=N+10$ \\ \hline \hline
			$N=2$ &  $\theta _r \approx 41.81^\circ$ & $\theta _r \approx 30.00^\circ$ & $\theta _r \approx 23.58^\circ$ & $\theta _r \approx 19.47^\circ$ & $\theta _r \approx 9.59^\circ$ \\ \hline
			$N=3$ &  $\theta _r \approx 48.59^\circ$ & $\theta _r \approx 36.87^\circ$ & $\theta _r \approx 30.00^\circ$ & $\theta _r \approx 25.37^\circ$ & $\theta _r \approx 13.34^\circ$ \\ \hline
			$N=4$ &  $\theta _r \approx 53.13^\circ$ & $\theta _r \approx 41.81^\circ$ & $\theta _r \approx 34.85^\circ$ & $\theta _r \approx 30.00^\circ$ & $\theta _r \approx 16.60^\circ$ \\ \hline
			$N=5$ &  $\theta _r \approx 56.44^\circ$ & $\theta _r \approx 45.58^\circ$ & $\theta _r \approx 38.68^\circ$ & $\theta _r \approx 33.75^\circ$ & $\theta _r \approx 19.47^\circ$ \\ \hline
			$N=8$ &  $\theta _r \approx 62.73^\circ$ & $\theta _r \approx 53.13^\circ$ & $\theta _r \approx 46.66^\circ$ & $\theta _r \approx 41.81^\circ$ & $\theta _r \approx 26.39^\circ$ \\ \hline
			$N=16$ &  $\theta _r \approx 70.25^\circ$ & $\theta _r \approx 62.73^\circ$ & $\theta _r \approx 57.36^\circ$ & $\theta _r \approx 53.13^\circ$ & $\theta _r \approx 37.98^\circ$ \\ \hline
			$N=32$ &  $\theta _r \approx 75.86^\circ$ & $\theta _r \approx 70.25^\circ$ & $\theta _r \approx 66.10^\circ$ & $\theta _r \approx 62.73^\circ$ & $\theta _r \approx 49.63^\circ$ \\ \hline
			\end{tabular}}
	\end{table*}
{The optimization of a periodic RIS may be easier than for an aperiodic RIS. The reason is that the optimization complexity of a periodic RIS scales with the number of unit cells in the period $P$, rather than with the total number of unit cells in the RIS. To employ optimization methods for periodic surfaces is, however, necessary that the ratio $P/p$ is a positive integer number. This implies that RISs that are realized as a periodic arrangement of unit cells with geometric period $p$ cannot, in general, be designed as periodic surfaces, since it is not possible to obtain, for every wave transformation, a perfect periodic pattern on the surface. In Fig. \ref{fig:AaltoRIS}, this issue is avoided, since the authors have first identified the correct RIS period $P$ based on the function to realize, and they have then optimized the size and the number of unit cells within the RIS period $P$ to ease the design of the entire RIS (since only 10 unit cells need to be tuned). The downside of this approach is that it is difficult to realize reconfigurable structures, since the RIS period $P$ needs to be changed according to the specified wave transformation. This is typically not possible or it is very difficult to realize. On the other hand, the reconfigurable surface in Fig. \ref{fig:OrangeRIS} may be optimized under the assumption of being periodic. To do this, however, the RIS can be utilized to realize only the subset of wave transformations for which the ratio $P/p$ is a positive integer number. To better understand, let us consider an RIS that realizes the wave transformation of anomalous reflection, e.g., an electromagnetic wave is steered from an angle of incidence equal to $0$ degrees towards an angle of reflection ($\theta_r$) equal to $75$ degrees. Assuming that the electromagnetic waves are plane waves, it is known from \cite[Eq. 16]{MDR_JSAC} that the RIS period of this wave transformation is $P = {\lambda  \mathord{\left/ {\vphantom {\lambda  {\sin \left( {{\theta _r}} \right)}}} \right. \kern-\nulldelimiterspace} {\sin \left( {{\theta _r}} \right)}} = 1.0353 \lambda$, where $\lambda$ is the wavelength. Let us assume $p=\lambda/4$ as for the RIS in Fig. \ref{fig:OrangeRIS}. We obtain $P/p = 4.1411$, which is not an integer number, and, therefore, the RIS needs to be optimized as an aperiodic surface even though it realizes a periodic wave transformation. This is due to the finite value of the geometric period $p$: The smaller $p$ is, the finer the spatial resolution is and the more likely an RIS can be treated as a periodic surface. If an RIS is (virtually) a continuous surface with $p \to 0$, then any RIS period $P$ can be, in theory, realized. This possibility of avoiding the need of global optimization can be viewed as a simple justification of the advantages of designing RISs with a geometric period smaller than $\lambda/2$. More generally, the use of sub-wavelength unit cells allows the control of the near-field distribution that is necessary for highly efficient implementations of RISs. Additional discussions can be found in \cite{DigitalRIS_Elsevier} and \cite{Sergei_Tutorial}. If $p$ is finite, in general, an RIS that operates as an anomalous reflector can be designed as a periodic surface only to reflect the incident electromagnetic waves towards a discrete number of directions. If $p=\lambda/N$ with $N$ being an integer number with $N \ge 2$, the admissible angles of reflection in agreement with the RIS period $P = {\lambda  \mathord{\left/ {\vphantom {\lambda  {\sin \left( {{\theta _r}} \right)}}} \right. \kern-\nulldelimiterspace} {\sin \left( {{\theta _r}} \right)}}$ are those for which the identity $\sin \left( {{\theta _r}} \right) = {N \mathord{\left/ {\vphantom {N n}} \right.\kern-\nulldelimiterspace} n}$ for $n=N+1, N+2, \ldots$ is fulfilled. To better elucidate this concept, some examples of angles $\theta _r$ that fulfill this condition are given in Table \ref{Table_PeriodicVsPeriodic} for different values of $N$. From the table, we see that the geometric period $p$ needs to be very small, i.e., $N$ needs to be large, with respect to the wavelength in order to realize large steering angles. Also, the corresponding number of unit cells per RIS period $P$, i.e. $n \ge N+1$, needs to be large as well.} 

{In conclusion, the design of RISs based on the RIS period $P$ is a well consolidated approach in the field of engineered surfaces \cite{Sergei_RealizedSurface}. This approach may, however, not be readily applicable to the design of reconfigurable surfaces, since the sizes and the inter-distances of the unit cells need to be adapted according to $P$, which is in turn determined by the specified wave transformation to realize. In wireless communications, it is customary to treat a reconfigurable surface (i.e., an RIS) as a periodic arrangement, with geometric period $p$, of unit cells and to treat it as an aperiodic surface, regardless of whether the RIS needs to realize wave transformations that result in periodic or aperiodic configurations of the unit cells along the entire surface. This leads to the need of developing efficient and scalable optimization algorithms \cite{Rui_Tutorial}, \cite{pan2021overview}, \cite{DigitalRIS_Elsevier}. The impact of the geometric period $p$, i.e., the spatial discretization of an RIS, has recently been analyzed in \cite{DigitalRIS_Elsevier} with the aid of numerical simulations.}

\subsection{Research Opportunities and Challenges}
RIS-empowered SREs are an emerging field of research in wireless communications with several open research issues to be tackled, in order to quantify the gains that can be expected in realistic wireless network deployments. The major open research challenges have been addressed in many recent papers, e.g., \cite{MDR_JSAC}, \cite{Rui_Tutorial}, \cite{Cunhua_Magazine}, and they encompass how to efficiently perform channel estimation, how to enable the control of an RIS, where to best deploy RISs, how to efficiently integrate RISs in system-level and ray tracing simulators, etc. A major open research issue, in addition, consists of developing models for RISs that are electromagnetically consistent and sufficiently tractable for evaluating the performance and for optimizing RIS-assisted wireless networks from a signal-level and system-level perspective. A summary and comparison of currently available research efforts can be found in \cite[Table 1]{Vittorio_RayTracing}. The focus of the present tutorial paper is on this latter open research issue. More precisely, we aim to overview, in a tutorial manner, electromagnetically consistent communication models for RISs that are represented as thin sheets of electromagnetic material. Specifically, the focus of the present tutorial paper is on the differences and similarities between local and global design criteria for realizing anomalous reflectors, and how, departing from Maxwell's equations, optimization problems for designing RISs with unitary power efficiency can be formulated and numerically solved.

\subsection{Paper Organization}
The remainder of the present tutorial paper is organized as follows. In Section II, we overview the most widely used communication models for RISs. In Section III, we depart from models for RISs that fulfill Maxwell's equations, and discuss and compare local and global designs for RISs. Also, we formulate optimization problems for designing RISs that are globally optimal and can be realized with purely reactive impedance sheets. In Section IV, numerical results are illustrated in order to quantitatively compare the different designs for RISs that are presented in the previous sections. Finally, Section V concludes this tutorial paper.

\textit{Disclaimer}: Since the present paper is a tutorial and not a survey paper, we limit ourselves to report only examples of research works that can guide the readers to retrieve further information on modeling, analyzing, and optimizing metasurfaces in general and RISs for wireless applications in particular. A more comprehensive reference list can be found in, e.g., \cite{MDR_JSAC}, \cite{Rui_Tutorial}.

\section{Models for RISs Widely Used in Wireless Communications}
In this section, we overview three communication models for RISs that have recently been proposed in the literature. The considered communication models are given as examples, in order to clarify the modeling assumptions and the conditions under which they can be applied. The third model introduced in this section is further elaborated in Section III with the aid of step-by-step examples and is utilized to formulate optimization problems for designing RISs. In order to keep the focus on the key aspects of the communication models for ensuring their electromagnetic consistency and validity, we consider an RIS that is deployed in a free-space propagation environment. Multipath propagation can be added to the considered channel model as described in \cite{MDR_StatisticalLocation}, \cite{MDR_MutualImpedancesMIMO}, \cite{Trinh}. {Readers who are interested in comparing the electromagnetically consistent models summarized in the present paper against conventional models utilized in the vast majority of research works in communications can consult \cite{Rui_Tutorial} and \cite[Sec. 4]{DigitalRIS_Elsevier}.}

\subsection{Locally Periodic Discrete Model}
As mentioned in the previous section, a widely used model for RISs is based on a locally periodic design, in which periodic boundary conditions are applied at the unit cell level, see, e.g., \cite{Cui_InfoMeta}, \cite{Cui_CodingMeta}, \cite{Vincenzo}. In general, each RIS element is assumed to be comprised of several identical unit cells for reasons that are elaborated next. To illustrate this communication model, which is widely utilized in wireless communications, we consider the analytical formulation in \cite{Wankai_PathLoss-mmWave}, which has been experimentally validated by the authors with the aid of measurements in an indoor environment. An early version of the same communication model is available in \cite{Wankai_PathLoss}. {An in-depth evaluation of this model is presented in \cite{DigitalRIS_Elsevier}.}

\begin{table*}[!t]
		\centering
		\caption{Examples of reflection and transmission coefficients for RISs with discrete-valued phase shifts (two-state and four-state control).}
		\label{Table_RelectionCoefficientDiscrete}
		\footnotesize
		\newcommand{\tabincell}[2]{\begin{tabular}{@{}#1@{}}#2\end{tabular}}
		\begin{tabular}{c||c|c} \hline
			
			Reference &  Reflection Coefficient & Transmission Coefficient \\ \hline \hline
			
			\cite{Wankai_PathLoss-mmWave} ($f= 27$ GHz) & $\begin{array}{l} \left| {{\Gamma _1}} \right| = 0.9,\quad \angle {\Gamma _1} = {165^\circ }\\ \left| {{\Gamma _2}} \right| = 0.7,\quad \angle {\Gamma _2} = {0^\circ }\end{array}$ & -- \\ \hline
			
			\cite{Wankai_PathLoss-mmWave} ($f= 33$ GHz) & $\begin{array}{l} \left| {{\Gamma _1}} \right| = 0.8,\quad \angle {\Gamma _1} = {150^\circ }\\ \left| {{\Gamma _2}} \right| = 0.8,\quad \angle {\Gamma _2} = {0^\circ }\end{array}$ & -- \\ \hline
			
			\cite{Hongliang_OmniSurface} ($f= 3.6$ GHz) & $\begin{array}{l} \left| {{\Gamma _1}} \right| = 0.46,\quad \angle {\Gamma _1} = {20^\circ }\\ \left| {{\Gamma _2}} \right| = 0.55,\quad \angle {\Gamma _2} = {215^\circ }\end{array}$ & $\begin{array}{l} \left| {{T_1}} \right| = 0.58,\quad \angle {T _1} = {300^\circ }\\ \left| {{T_2}} \right| = 0.81,\quad \angle {T_2} = {123^\circ }\end{array}$ \\ \hline
			
			\cite{Linglong_Testbed} ($f= 2.3$ GHz) & $\begin{array}{l} \left| {{\Gamma _1}} \right| = -1.2  \, {\rm{dB}},\quad \angle {\Gamma _1} = {-205.5^\circ }\\ \left| {{\Gamma _2}} \right| = -1.2 \, {\rm{dB}},\quad \angle {\Gamma _2} = {-383.2^\circ } \\ \left| {{\Gamma _3}} \right| = -0.8  \, {\rm{dB}},\quad \angle {\Gamma _3} = {-290.2^\circ } \\ \left| {{\Gamma _4}} \right| = -0.7  \, {\rm{dB}},\quad \angle {\Gamma _4} = {-110.3^\circ }\end{array}$ & -- \\ \hline
			
		\end{tabular}
	\end{table*}
\begin{table}[!t]
		\centering
		\caption{Example of reflection coefficient for an RIS with continuous-valued phase shifts \cite{Romain_RIS-Prototype}.}
		\label{Table_RelectionCoefficientContinuous}
		\footnotesize
		\newcommand{\tabincell}[2]{\begin{tabular}{@{}#1@{}}#2\end{tabular}}
		\begin{tabular}{c||c|c} \hline
			
			Voltage &  $\begin{array}{c} {\rm{Reflection \; coefficient}}\\ {\rm{amplitude \; (}}\left| \Gamma  \right|{\rm{)}} \end{array}$ & $\begin{array}{c} {\rm{Reflection \; coefficient}}\\ {\rm{phase \; (}}\angle \Gamma  {\rm{)}} \end{array}$ \\ \hline \hline
			
			0 $V$ & -1.517 dB & 32.798$^\circ$ \\ \hline
            0.25 $V$ & -1.807 dB & 40.854$^\circ$ \\  \hline
            0.5 $V$ & -3.156 dB & 46.807$^\circ$ \\   \hline
            0.75 $V$ & -5.59 dB & 53.543$^\circ$ \\   \hline
            1 $V$ & -9.576 dB & 70.32$^\circ$ \\  \hline
            1.25 $V$ & -20.563 dB & -167.158$^\circ$ \\   \hline
            1.5 $V$ & -6.615 dB & -73.171$^\circ$ \\  \hline
            1.75 $V$ & -3.029 dB & -49.627$^\circ$ \\ \hline
            2 $V$ & -1.959 dB & -35.908$^\circ$ \\    \hline
            2.5 $V$ & -0.874 dB & -23.263$^\circ$ \\  \hline
            3 $V$ & -0.749 dB & -16.087$^\circ$ \\    \hline
            3.5 $V$ & -0.469 dB & -12.663$^\circ$ \\  \hline
            4 $V$ & -0.528 dB & -9.925$^\circ$ \\ \hline
            5 $V$ & -0.439 dB & -6.906$^\circ$ \\ \hline
			
		\end{tabular}
	\end{table}
The RIS is modeled as illustrated in Fig. \ref{fig:RIS}. For ease of description, we assume that (i) each RIS element is constituted by a single unit cell, (ii) all the unit cells have the same size and shape, and (iii) the inter-distance between adjacent unit cells is the same. Therefore, the RIS is modeled as a periodic arrangement of identical unit cells. The scattering response of each unit cell is configured thanks to the tuning circuit and the biasing line, as illustrated in Fig. \ref{fig:RIS}. We assume that there exist $M$ unit cells in each row and $N$ unit cell in each column of the surface. Therefore, the total number of reconfigurable unit cells is $MN$. The surface area of each unit cell is $d_x d_y$, with $d_x$ and $d_x$ being the horizontal and vertical sizes of each unit cell, respectively.

The RISs considered in \cite{Wankai_PathLoss-mmWave} operate as reflecting surfaces and, therefore, each unit cell is characterized by a complex reflection coefficient, which is defined as the ratio between the reflected electric field and the incident electric field. We denote the reflection coefficient of the $(m,n)$th unit cell as $\Gamma_{m,n}$. Specifically, the RISs in \cite{Wankai_PathLoss-mmWave} comprise unit cells that can apply two phase shifts (binary cells) depending on the configuration of the tuning circuit. For illustrative purposes, the values of the reflection coefficients are reported in Table \ref{Table_RelectionCoefficientDiscrete}. Further details on how these reflections coefficients are computed are given next. In Table \ref{Table_RelectionCoefficientDiscrete} and Table \ref{Table_RelectionCoefficientContinuous}, for completeness, we report two other examples of RISs that are modeled based on the same principle as the RISs considered in \cite{Wankai_PathLoss-mmWave}. One of the examples reported in Table \ref{Table_RelectionCoefficientDiscrete} considers the RIS introduced in \cite{Hongliang_OmniSurface}, which can simultaneously reflect and refract the incident electromagnetic waves. For this reason, it is characterized by a reflection coefficient and by a transmission coefficient, $T_{m,n}$, which is defined as the ratio between the refracted electric field and the incident electric field. Similar to the RISs in \cite{Wankai_PathLoss-mmWave}, the unit cells of the RIS in \cite{Hongliang_OmniSurface} can be configured in two different states that are characterized by the pairs $(\Gamma_1, T_1)$ and $(\Gamma_2, T_2)$. The other example reported in Table \ref{Table_RelectionCoefficientDiscrete} is the RIS introduced in \cite{Linglong_Testbed}, which operates as a reflecting surface but its unit cells can be configured in four different states. The RIS in Table \ref{Table_RelectionCoefficientContinuous} is modeled as a periodic array of unit cells, similar to the RISs in \cite{Wankai_PathLoss-mmWave}, \cite{Hongliang_OmniSurface} and \cite{Linglong_Testbed}. Similar to \cite{Wankai_PathLoss-mmWave} and \cite{Linglong_Testbed}, in addition, it operates only in reflection mode and is characterized by the reflection coefficient $\Gamma_{m,n}$. However, the reflection coefficient of each unit cell can be varied continuously as a function of a control voltage. Therefore, the phase shift applied by each unit cell can be tuned more finely. In the five examples of RISs reported in Table \ref{Table_RelectionCoefficientDiscrete} and Table \ref{Table_RelectionCoefficientContinuous}, we note that the amplitude and the phase of the reflection (and transmission) coefficient are not independent of each other. Also, the amplitude of the reflection coefficient is not unitary and it is not independent of the phase shift. In general, in addition, the reflection and transmission coefficients reported in Table \ref{Table_RelectionCoefficientDiscrete} and Table \ref{Table_RelectionCoefficientContinuous} depend on the angle of incidence of the electromagnetic waves, as shown in \cite[Fig. 2]{Romain_RIS-Prototype} and \cite[Fig. 4]{Epstein_Tutorial}. The examples reported in the two tables are referred to the canonical case of normal incidence. {Interested readers are invited to consult  \cite{DigitalRIS_Elsevier} for an in-depth numerical evaluation and comparison of the reflection properties of the RISs in Table \ref{Table_RelectionCoefficientDiscrete} and Table \ref{Table_RelectionCoefficientContinuous}, with focus on the impact of non-ideal implementation constraints, i.e., the non-unitary amplitude of the reflection coefficients, the dependency between the amplitude and the phase of the reflection coefficients, and the non-constant phase differences between the reflection coefficients of the RIS alphabet.}

Assuming that the set of possible reflection coefficients (the RIS alphabet), as a function of the tuning circuit, of a single unit cell of the RIS is given, the authors of \cite{Wankai_PathLoss-mmWave} have introduced an analytical model for computing the power observed at a given location of an RIS-assisted communication link. The RIS is assumed to be centered at the origin and to lie in the $xy$ plane (i.e., $z=0$). The received power can be formulated as follows:
\begin{align} \label{Eq_WankaiModel}
\frac{{{P^{\left( {{\rm{Rx}}} \right)}}}}{{{P^{\left( {{\rm{Tx}}} \right)}}}} &= \frac{{{G^{\left( {{\rm{Tx}}} \right)}}{G^{\left( {{\rm{Rx}}} \right)}}}{{\left( {{d_x}{d_y}} \right)}^2}}{{16{\pi ^2}}} \nonumber \\ &* {\left| {\sum\limits_{m = 1}^M {\sum\limits_{n = 1}^N {\frac{{\sqrt {{F_{m,n}}} {\Gamma _{m,n}}}}{{r_{m,n}^{\left( {{\rm{Tx}}} \right)}r_{m,n}^{\left( {{\rm{Rx}}} \right)}}}{e^{ - j\frac{{2\pi }}{\lambda }\left( {r_{m,n}^{\left( {{\rm{Tx}}} \right)} + r_{m,n}^{\left( {{\rm{Rx}}} \right)}} \right)}}} } } \right|^2}
\end{align}
where
\begin{align} \label{Eq_WankaiModel--1}
{F_{m,n}} &= {\left( {\frac{{{{\left( {d_0^{\left( {{\rm{Tx}}} \right)}} \right)}^2} + {{\left( {r_{m,n}^{\left( {{\rm{Tx}}} \right)}} \right)}^2} - {{\left( {{d_{m,n}}} \right)}^2}}}{{2d_0^{\left( {{\rm{Tx}}} \right)}r_{m,n}^{\left( {{\rm{Tx}}} \right)}}}} \right)^{ - 1 + {{{G^{\left( {{\rm{Tx}}} \right)}}} \mathord{\left/
 {\vphantom {{{G^{\left( {{\rm{Tx}}} \right)}}} 2}} \right.
 \kern-\nulldelimiterspace} 2}}} \nonumber \\
& *\left( {\frac{{{z^{\left( {{\rm{Tx}}} \right)}}}}{{r_{m,n}^{\left( {{\rm{Tx}}} \right)}}}} \right)\left( {\frac{{{z^{\left( {{\rm{Rx}}} \right)}}}}{{r_{m,n}^{\left( {{\rm{Rx}}} \right)}}}} \right)\\
& *{\left( {\frac{{{{\left( {d_0^{\left( {{\rm{Rx}}} \right)}} \right)}^2} + {{\left( {r_{m,n}^{\left( {{\rm{Rx}}} \right)}} \right)}^2} - {{\left( {{d_{m,n}}} \right)}^2}}}{{2d_0^{\left( {{\rm{Rx}}} \right)}r_{m,n}^{\left( {{\rm{Rx}}} \right)}}}} \right)^{ - 1 + {{{G^{\left( {{\rm{Rx}}} \right)}}} \mathord{\left/
 {\vphantom {{{G^{\left( {{\rm{Rx}}} \right)}}} 2}} \right.
 \kern-\nulldelimiterspace} 2}}} \nonumber
\end{align}
and the following notation is used:
\begin{itemize}
    \item ${{P^{\left( {{\rm{Tx}}} \right)}}}$ and ${{P^{\left( {{\rm{Rx}}} \right)}}}$ are the transmitted and received powers, respectively;
    \item ${{G^{\left( {{\rm{Tx}}} \right)}}}$ and ${{G^{\left( {{\rm{Rx}}} \right)}}}$ are the antenna gains of the transmitter and receiver, respectively;
    \item $\lambda$ is the wavelength of the electromagnetic wave and $j=\sqrt{-1}$ is the imaginary unit;
    \item ${r_{m,n}^{\left( {{\rm{Tx}}} \right)}}$ is the distance between the transmitter and the center point of the $(m,n)$th unit cell, and ${r_{m,n}^{\left( {{\rm{Rx}}} \right)}}$ is the distance between the center point of the $(m,n)$th unit cell and the receiver;
    \item ${{d_{m,n}}}$ is the distance between the center point of the $(m,n)$th unit cell and the center point of the RIS (i.e., the origin);
    \item ${d_0^{\left( {{\rm{Tx}}} \right)}}$ is the distance between the transmitter and the center point of the RIS, and ${d_0^{\left( {{\rm{Rx}}} \right)}}$ is the distance between the center point of the RIS and the receiver;
    \item ${{z^{\left( {{\rm{Tx}}} \right)}}}$ and ${{z^{\left( {{\rm{Rx}}} \right)}}}$ are the Cartesian coordinates of the transmitter and receiver on the $z$-axis, respectively.
\end{itemize}

{The analytical model in \eqref{Eq_WankaiModel} has recently been generalized in \cite{DigitalRIS_Elsevier} by capitalizing on scattering theory rather than utilizing antenna theory as in \cite{Wankai_PathLoss-mmWave}. Interested readers are referred to \cite{DigitalRIS_Elsevier} for further information.}

By using \eqref{Eq_WankaiModel}, it is possible to formulate the received power at any locations of the transmitter and receiver as a function of the location of the RIS and of the configuration of the unit cells. Therefore, the optimal configuration of the $MN$ unit cells of the RIS can be identified in order to, e.g., maximize the received power depending on the location of the receiver. More precisely, let ${\boldsymbol{\Gamma }}$ denote the $M \times N$ matrix of the reflection coefficients $\Gamma_{m,n}$ and let ${\Gamma _{m,n}} \in \left\{ {{\Gamma _1},{\Gamma _2}, \ldots ,{\Gamma _\Sigma }} \right\}$ be the $\Sigma$ possible reflection coefficients of each unit cell of the RIS, i.e., the RIS alphabet. In Table \ref{Table_RelectionCoefficientDiscrete} and Table \ref{Table_RelectionCoefficientContinuous}, we have $\Sigma=2$ or $\Sigma=4$ and $\Sigma=14$. With this notation, a typical problem formulation reads as follows:
\begin{subequations} \label{eq:OPT-RIS-Discrete:main}
\begin{align}
& \mathop {\max }\limits_{\boldsymbol{\Gamma }} {P^{\left( {{\rm{Rx}}} \right)}}\left( {\boldsymbol{\Gamma }} \right) \tag{\ref{eq:OPT-RIS-Discrete:main}}\\
{\rm{s}}{\rm{.t}}{\rm{.}}\quad &{\Gamma _{m,n}} \in \left\{ {{\Gamma _1},{\Gamma _2}, \ldots ,{\Gamma _\Sigma }} \right\}\quad \forall m,n \label{eq:OPT-RIS-Discrete:a}
\end{align}
\end{subequations}

{A simple algorithm for solving \eqref{eq:OPT-RIS-Discrete:main}, which is based on the alternating optimization method, can be found in \cite{DigitalRIS_Elsevier}. The main feature of the algorithm in \cite{DigitalRIS_Elsevier} lies in its applicability to any RIS alphabet without imposing any prior assumptions on the structure of the reflection coefficients $\Gamma_{m,n}$, e.g., the amplitude and phase of $\Gamma_{m,n}$ are independent of one another or the amplitude of $\Gamma_{m,n}$ is unitary.}

\begin{figure*}[!t]
{\includegraphics[width=\linewidth]{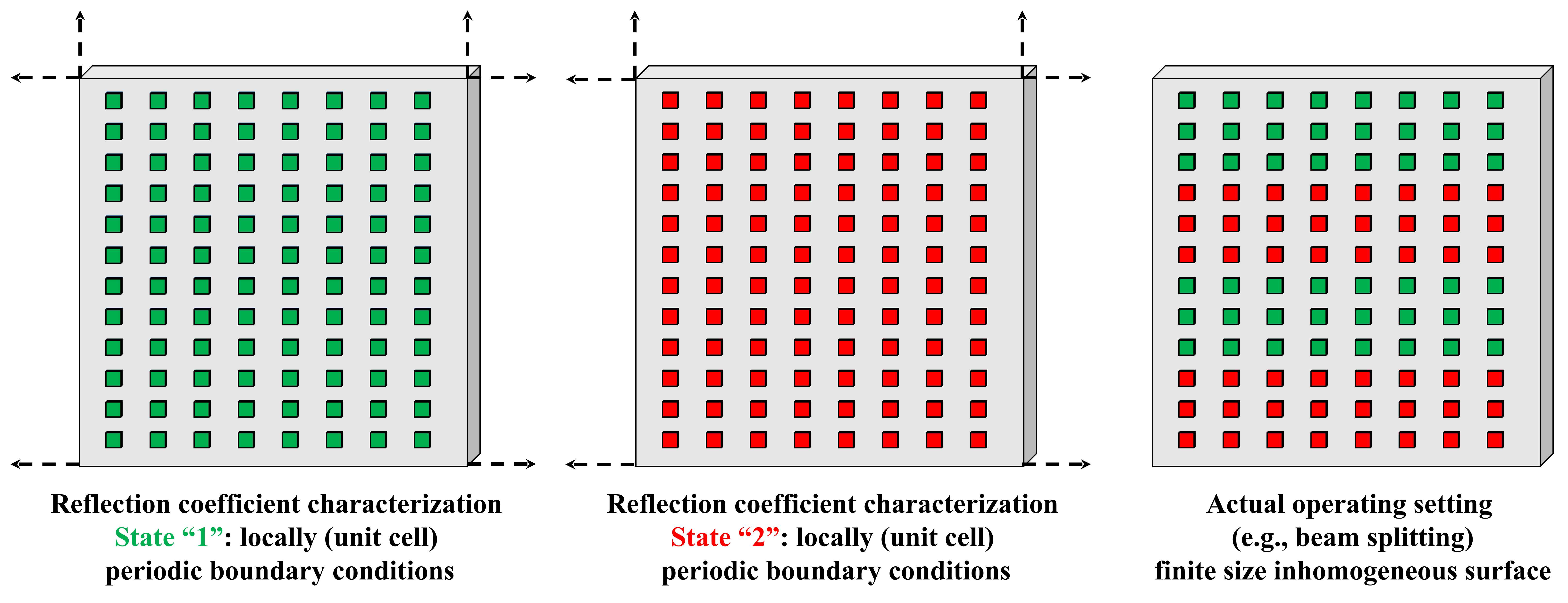}}
\caption{Illustration of the concept of (locally) periodic boundary conditions.}\label{fig:PBCs}
\end{figure*}
As mentioned, the set of $\Sigma$ states in \eqref{eq:OPT-RIS-Discrete:main}, i.e., the RIS alphabet, is determined by characterizing the electromagnetic response of the constituent unit cell of the RIS by employing a local design. In order to understand the applicability and accuracy of the received power model in \eqref{Eq_WankaiModel}, based on the solution of the optimization problem in \eqref{eq:OPT-RIS-Discrete:main}, it is instructive to analyze in detail the meaning of local design at the unit cell level and the concept of periodic boundary conditions mentioned in Section II. To this end, we consider, as an example, a binary unit cell that can take only two states, i.e., $\Sigma=2$ and ${\Gamma _{m,n}} \in \left\{ {{\Gamma _1},{\Gamma _2}} \right\}$.

The reflection coefficients $\Gamma_1$ and $\Gamma_2$ are obtained by utilizing the procedure sketched in Fig. \ref{fig:PBCs}. The reflection coefficient $\Gamma_1$ is estimated by considering an infinite-size RIS whose unit cells are all identical and the tuning circuits are set to the same configuration. Therefore, the RIS is effectively turned into an infinite and spatially homogeneous sheet with no phase variation along the entire surface. In this configuration, the surface reflection coefficient is well defined as the ratio between the tangential component of the reflected electric field and the tangential component of the incident electric field \cite[Chapter 7]{Orfanidis_Book}. Since the surface is spatially homogeneous and of infinite extent, only specular reflection is allowed. The obtained structure is usually analyzed with the aid of full-wave electromagnetic simulators, which model the infinite size of the surface and the periodic repetition of the elementary unit cell by applying the so-called periodic boundary conditions (see, e.g., \cite[Fig. 4]{Epstein_Tutorial}). Thanks to this procedure, the reradiation characteristics of the unit cell in the first possible state are characterized by taking assuming that it is surrounded by a neighborhood of identical unit cells. This implies that the mutual coupling and the interactions among all the identical unit cells in the considered homogeneous sheet are inherently taken into account when characterizing $\Gamma_1$. The same procedure is repeated for estimating $\Gamma_2$, with the only difference being that the tuning circuits are set to the configuration that results in the reflection coefficient $\Gamma_2$.

Having characterized the reflection coefficients $\Gamma_1$ and $\Gamma_2$, if necessary as a function of the angle of incidence of the electromagnetic wave, the communication model in \eqref{Eq_WankaiModel} stipulates that we may configure the state (either $\Gamma_1$ and $\Gamma_2$ in the example of Fig. \ref{fig:PBCs}) of each unit cell independently of the others and regardless of the states of the neighboring cells. An example is given in the right-hand side illustration of Fig. \ref{fig:PBCs}, in which the unit cells are configured to realize beam splitting, \cite{DigitalRIS_Elsevier}, \cite[Fig. 4]{Cui_InfoMeta}. However, caution needs to be paid when \eqref{Eq_WankaiModel} is utilized and the reflection coefficients $\Gamma_1$ or $\Gamma_2$ are obtained by applying locally periodic boundary conditions at the unit cell level. The reflection coefficients $\Gamma_1$ and $\Gamma_2$ are, in fact, determined by assuming that a unit cell configured in a given state is surrounded by an infinite (homogeneous) repetition of identical unit cells. When an RIS is configured to operate in practice, as sketched in the right-hand side illustration of Fig. \ref{fig:PBCs}, each unit cell is, however, immersed in a spatially inhomogeneous array whose neighboring unit cells can be all different from one another. This implies that the spatial symmetry imposed by the periodic boundary conditions does not hold anymore and the interactions (mutual coupling) among nearby unit cells are taken into account only in an approximate manner. In addition, a practical RIS is not of infinite extent but it has a finite size, even though it can be electrically (very) large. This implies that the notion of reflection coefficient is only an approximation and it holds only under the limit of physical optics \cite{Osipov_Book}. For these reasons, the unit cells are not typically optimized individually and independently of one another, but they are optimized in groups (often called macrocells), so as to ensure that the periodic boundary conditions utilized when characterizing each unit cell individually are approximately fulfilled during the normal operation of the RIS \cite{Vincenzo}, \cite{Grbic_Homogenization}. The right-hand side illustration of Fig. \ref{fig:PBCs} is a typical example in which the unit cells are split in groups, each containing 24 unit cells, and the groups are optimized such that the states ($\Gamma_1$ or $\Gamma_2$) of all the unit cells in a group are the same. The minimum required size of the group of unit cells for ensuring that \eqref{Eq_WankaiModel} is accurate enough for wireless applications is usually characterized with the aid of full-wave simulations.

\subsection{Mutually Coupled Antenna Elements}
The communication model for RISs introduced in the previous sub-section is widely employed in wireless communications and several optimization frameworks, under some simplifying assumptions, have been proposed based on it \cite{MDR_JSAC}, \cite{Rui_Tutorial}, \cite{DigitalRIS_Elsevier}. The local design at the unit cell level is a widely used method for characterizing the reflection and transmission characteristics of an RIS. As mentioned, however, the mutual coupling among the unit cells is only approximately taken into account, since the reflection coefficients of each unit cell (i.e., the RIS alphabet) are typically characterized by applying periodic boundary conditions at the unit cell level \cite{Grbic_Homogenization}. The accuracy of the model in \eqref{Eq_WankaiModel} can be improved by, e.g., not characterizing the reradiation properties of each unit cell individually but by analyzing, with full-wave simulations, the reradiation of groups of unit cells as a function of all the possible combinations of their states \cite{Caloz_Sergei}. In this case, periodic boundary conditions may be applied at the granularity of a group of unit cells in lieu of a single unit cell. The accuracy of this enhanced model is usually improved, but at the expenses of increasing the modeling and optimization complexity.

In \cite{MDR_MutualImpedances}, the authors have recently introduced a communication model for RISs that explicitly accounts for the mutual coupling among the RIS elements and for the control circuit of the unit cells. The communication model in \cite{MDR_MutualImpedances} is based on the theory of mutually coupled antennas and is directly applicable in multiple-antenna communication systems, since it resembles a MIMO communication channel. In \cite{MDR_MutualImpedancesOpt} and \cite{MDR_MutualImpedancesMIMO}, it has recently been shown that the model is suitable for formulating optimization problems in general wireless networks, such as the MIMO interference channel, and that it can be utilized to optimize an RIS by explicitly taking into account the mutual coupling among the RIS elements. In this sub-section, we first introduce the communication model in \cite{MDR_MutualImpedances} and we then elaborate on the assumptions under which it is developed and hence the conditions under which it can be utilized.

\begin{figure}[!t]
{\includegraphics[width=\linewidth]{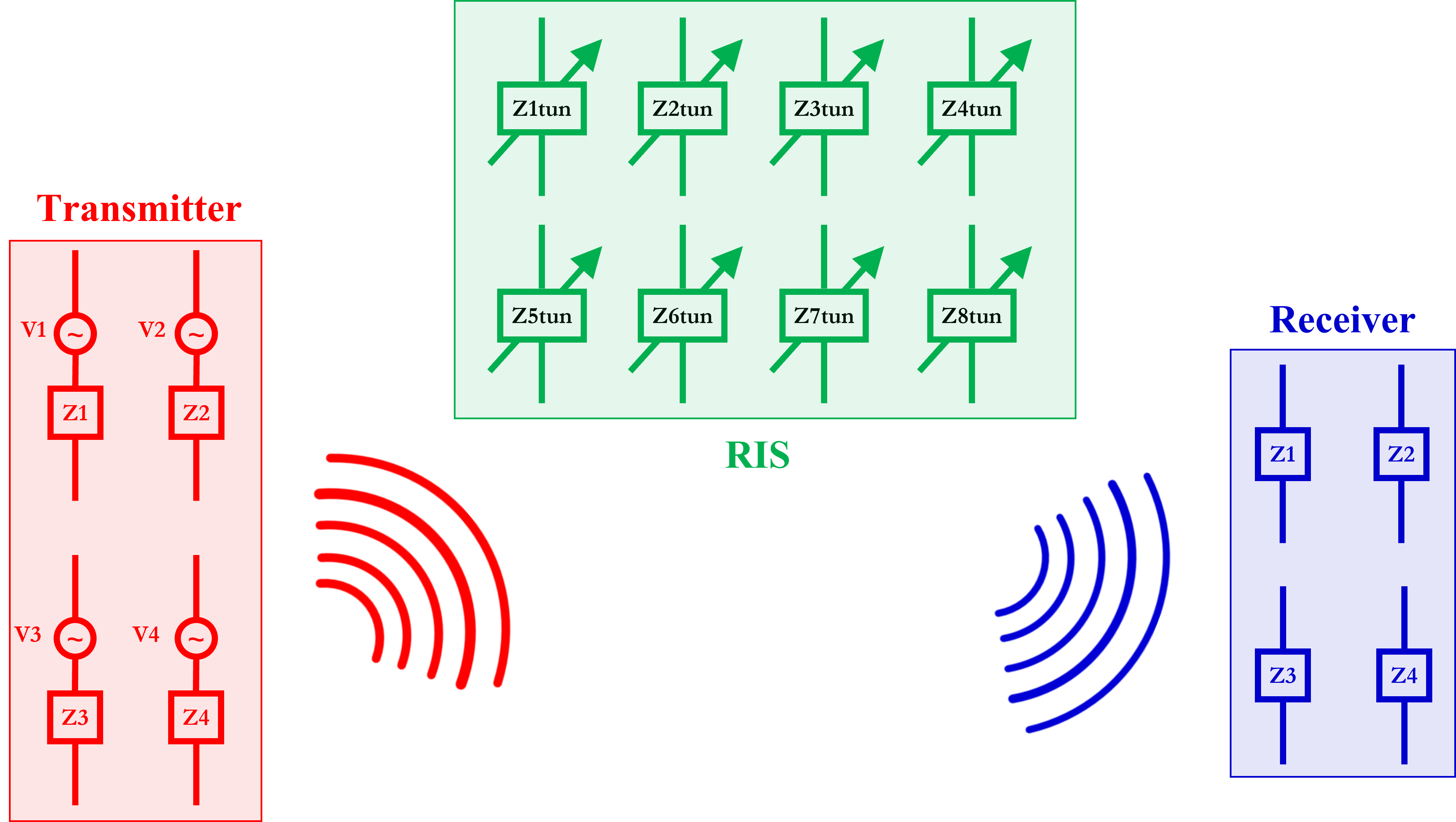}}
\caption{Communication model of a reconfigurable intelligent surface based on mutually coupled impedances.}\label{fig:ImpedanceModel}
\end{figure}
The RIS-assisted communication model introduced in \cite{MDR_MutualImpedances} is illustrated in Fig. \ref{fig:ImpedanceModel}. The model resembles a conventional single transmitter-receiver pair MIMO communication link in the presence of an RIS. The transmitter and the receiver are equipped with $M_0$ and $L_0 \le M_0$ antenna elements, respectively. The antenna elements are assumed to be thin wire dipoles of perfectly conducting material. The model can be generalized for application to radiating elements different from thin wire dipoles, which are considered in \cite{MDR_MutualImpedances} for analytical tractability. Each thin wire dipole at the transmitter is driven by a voltage generator that models the transmit feed line, and each thin wire dipole at the receiver is connected to a load impedance that mimics the receive electric circuit. For simplicity, we assume that the number of symbols (streams) sent by the transmitter is equal to the number of receive antennas. The transmission between the transmitter and the receiver is assisted by an RIS, which comprises $P$ nearly passive thin wire dipoles that are independently configurable (by an external controller) through tunable impedances. Compared with the illustration of the RIS in Fig. \ref{fig:RIS}, a thin wire dipole in Fig. \ref{fig:ImpedanceModel} can be viewed as an approximation for a unit cell. The model can be generalized to different physical structures for the unit cells, e.g., patch antennas. The physical model based on dipoles is considered in \cite{MDR_MutualImpedances} because relatively simple analytical or integral expressions for the current distribution of closely spaced thin wire dipoles are available in the literature \cite[Chapter 25]{Orfanidis_Book}.

Based on the system model in Fig. \ref{fig:ImpedanceModel}, an RIS-assisted channel is optimized by appropriately setting the tunable impedances connected to the thin wire dipoles of the RIS. More specifically, the authors of \cite{MDR_MutualImpedances}  have introduced an $L_0 \times M_0$ end-to-end channel matrix that formulates the voltage measured at the ports of the receive antennas as a function of the voltage generators connected to the ports of the transmit antennas, i.e., ${{\bf{v}}_{{\rm{Rx}}}} = {\bf{H}}{{\bf{v}}_{{\rm{Tx}}}}$, where ${\bf{v}}_{\rm{Tx}}$ is the $M_0 \times 1$ vector that collects the driving voltages at the transmitter, ${\bf{v}}_{\rm{Rx}}$ is the $L_0 \times 1$ vector that collects the voltages measured at the ports of the antennas at the receiver, and ${\bf{H}}$ is the $L_0 \times M_0$ channel matrix that accounts for the radiating elements (the thin wire dipoles) and the propagation of the electromagnetic waves. From \cite[Theorem 1]{MDR_MutualImpedances}, ${\bf{H}}$ can be formulated as follows:
\begin{align}
{\bf{H}} &=  {\left( {{{\bf{I}}_{L_0}} + {\bf{\Psi }}_{r,r}{\bf{Z}}_r^{ - 1} - {\bf{\Psi }}_{r,t}{{\left( {{\bf{\Psi }}_{t,t} + {{\bf{Z}}_t}} \right)}^{ - 1}}{\bf{\Psi }}_{t,r}{\bf{Z}}_r^{ - 1}} \right)^{ - 1}} \nonumber \\
& * {\bf{\Psi }}_{r,t}{\left( {{\bf{\Psi }}_{t,t} + {{\bf{Z}}_t}} \right)^{ - 1}} \label{HE2E}
\end{align}
\noindent where ${{{\bf{I}}_{L_0}}}$ is the $L_0 \times L_0$ identity matrix, and ${{{\bf{Z}}_t}}$ and ${{{\bf{Z}}_r}}$ are the $M_0 \times M_0$ and $L_0 \times L_0$ diagonal matrices that comprise the internal impedances of the transmit generators and the load impedances of the receive antennas, respectively. Furthermore, the following shorthand notation is introduced:
\begin{align}
 & {\bf{\Psi }}_{t,t} = {{\bf{Z}}_{t,t}} - {{\bf{Z}}_{t,s}}{\left( {{{\bf{Z}}_{s,s}} + {{\bf{Z}}_{{\rm{tun}}}}} \right)^{ - 1}}{{\bf{Z}}_{s,t}} \label{HE2E_1a} \\
&  {\bf{\Psi }}_{t,r} = {{\bf{Z}}_{t,r}} - {{\bf{Z}}_{t,s}}{\left( {{{\bf{Z}}_{s,s}} + {{\bf{Z}}_{{\rm{tun}}}}} \right)^{ - 1}}{{\bf{Z}}_{s,r}} \label{HE2E_1b} \\
&  {\bf{\Psi }}_{r,t} = {{\bf{Z}}_{r,t}} - {{\bf{Z}}_{r,s}}{\left( {{{\bf{Z}}_{s,s}} + {{\bf{Z}}_{{\rm{tun}}}}} \right)^{ - 1}}{{\bf{Z}}_{s,t}} \label{HE2E_1c} \\
& {\bf{\Psi }}_{r,r} = {{\bf{Z}}_{r,r}} - {{\bf{Z}}_{r,s}}{\left( {{{\bf{Z}}_{s,s}} + {{\bf{Z}}_{{\rm{tun}}}}} \right)^{ - 1}}{{\bf{Z}}_{s,r}} \label{HE2E_1d}
\end{align}
\noindent where ${{\bf{Z}}_{x,y}}$, for $x,y \in \left\{ {t,s,r} \right\}$ with $t$, $s$, and $r$ identifying the transmitter, the RIS, and the receiver, respectively, is the matrix of mutual (or self if $x=y$) impedances between the thin dipoles of $y$ and the thin dipoles of $x$, which characterizes the signal propagation and the mutual coupling between $x$ and $y$, and ${{\bf{Z}}_{{\rm{tun}}}^{(k)}}$ is the $P \times P$ diagonal matrix of tunable impedances of the RIS. The matrices ${{\bf{Z}}_{x,y}}$ for $x,y \in \left\{ {t,s,r} \right\}$ account for the microscopic structure of the RIS and the locations of the transmitter, RIS, and receiver. They can be either computed with the aid of full-wave simulators or can be computed analytically by relying on some approximations. For example, the authors of \cite{MDR_MutualImpedances} have used the induced electromagnetic field method for computing the mutual and self impedances, as well as a sinusoidal approximation for the current distribution on the thin wire dipoles. Under these modeling assumptions, given the RIS microstructure and the system topology, the impedances in ${{\bf{Z}}_{x,y}}$ for $x,y \in \left\{ {t,s,r} \right\}$ need to be computed only once, and are not usually considered optimization variables in the context of wireless communication systems.

On the other hand, the matrix ${{\bf{Z}}_{{\rm{tun}}}}$, which ensures the reconfigurability of the RIS, is the matrix to be optimized for steering the electromagnetic wave that is emitted by the transmitter and impinges upon the RIS towards the location of the receiver. Let us consider, for example, that the transmitter and the receiver are equipped with a single antenna, i.e., $M_0=L_0=1$, and that the objective is to maximize the power at the location of the receiver. Then, the matrix $\bf{H}$ is a scalar, i.e., ${v_{{\rm{Rx}}}} = H{v_{{\rm{Tx}}}}$, and the optimization problem as a function of ${{\bf{Z}}_{{\rm{tun}}}}$ can be formulated as follows:
\begin{subequations} \label{eq:OPT-RIS-Impedance:main}
\begin{align}
& \mathop {\max }\limits_{{{\bf{Z}}_{{\rm{tun}}}}} \left| {H\left( {{{\bf{Z}}_{{\rm{tun}}}}} \right)} \right| \tag{\ref{eq:OPT-RIS-Impedance:main}}\\
&{\rm{s}}{\rm{.t}}{\rm{.}}\quad \;{Z_{{\rm{tun}},p}} \in \left\{ {{{\mathcal{Z}}_1},{{\mathcal{Z}}_2}, \ldots ,{{\mathcal{Z}}_\Xi }} \right\}\quad \forall p = 1,2, \ldots ,P \label{eq:OPT-RIS-Impedance:a}
\end{align}
\end{subequations}
where ${Z_{{\rm{tun}},p}}$ is the $p$th element of ${{{\bf{Z}}_{{\rm{tun}}}}}$ and $\left\{ {{{\rm{Z}}_1},{{\rm{Z}}_2}, \ldots ,{{\rm{Z}}_\Xi }} \right\}$ is the set of $\Xi$ possible discrete values of the tuning impedances that can be implemented.

In \cite{MDR_MutualImpedancesOpt}, the authors have recently solved the optimization problem in \eqref{eq:OPT-RIS-Impedance:main} under the assumption that the real part of the tunable impedances is fixed and greater than zero, and that its imaginary part can be any real value. The constraint that the real part of the impedances is greater than zero ensures that the RIS does not amplify the incident electromagnetic wave and, therefore, no power amplifiers are needed. in fact, a negative resistance is equivalent to the need of using a power amplifier. In \cite{MDR_MutualImpedancesMIMO}, the authors have formulated a more complex optimization problem that maximizes the rate of a MIMO interference channel.

The matrices defined in \eqref{HE2E_1a}-\eqref{HE2E_1d} have a physical meaning and interpretation. For example, ${\bf{\Psi }}_{r,t}$ represents the transfer matrix (the channel) between the transmitter and the receiver, which accounts for the direct link (${{\bf{Z}}_{r,t}}$) and the RIS-reradiated link ${{\bf{Z}}_{r,s}}{\left( {{{\bf{Z}}_{s,s}} + {{\bf{Z}}_{{\rm{tun}}}}} \right)^{ - 1}}{{\bf{Z}}_{s,t}}$. This latter term is the product of three factors: ${{\bf{Z}}_{s,t}}$ represents the transfer function from the transmitter to the RIS; ${{\bf{Z}}_{r,s}}$ represents the transfer function from the RIS to the receiver; and ${\left( {{{\bf{Z}}_{s,s}} + {{\bf{Z}}_{{\rm{tun}}}}} \right)^{ - 1}}$ models the reradiation from the RIS. The matrix ${\bf{Z}}_{s,s}$ is, in general, a full matrix, which reduces to an almost diagonal matrix, i.e., the amplitudes of the elements in the main diagonal are much larger than the amplitudes of the off-diagonal elements, if the mutual coupling between the thin wire dipoles is negligible.

The channel matrix in \eqref{HE2E} can be simplified in several scenarios of practical relevance. If, for example, the transmitter, the receiver, and the RIS are in the far-field of each other, \eqref{HE2E} can be simplified without ignoring the mutual coupling among the thin dipoles. The self impedances ${{\bf{Z}}_{x,x}}$ are, in fact, independent of the transmission distances of the transmitter-receiver, transmitter-RIS, and RIS-receiver links, and they depend only on the inter-distances between the thin wire dipoles that comprise the transmitter, the RIS, and the receiver. In the far-field region, thus, the following simplifications can be applied:
\begin{align} \label{HE2E_2}
& {\bf{\Psi }}_{t,t} \approx {{\bf{Z}}_{t,t}} \\
& {\bf{\Psi }}_{r,r} \approx {{\bf{Z}}_{r,r}} \\
& {\bf{\Psi }}_{r,r}{\bf{Z}}_r^{ - 1} - {\bf{\Psi }}_{r,t}{\left( {{\bf{\Psi }}_{t,t} + {{\bf{Z}}_t}} \right)^{ - 1}}{\bf{\Psi }}_{t,r}{\bf{Z}}_r^{ - 1} \approx {\bf{\Psi }}_{r,r}{\bf{Z}}_r^{ - 1}
\end{align}

In the far-field region, therefore, $\mathbf{H}$ in \eqref{HE2E} can be approximated as follows:
\begin{align} \label{HE2E_3}
{\bf{H}}_{r,t} &\approx {\left( {{{\bf{I}}_{L_0}} + {{\bf{Z}}_{r,r}}{\bf{Z}}_r^{ - 1}} \right)^{ - 1}}{{\bf{Z}}_{r,t}}{\left( {{{\bf{Z}}_{t,t}} + {{\bf{Z}}_t}} \right)^{ - 1}} \\
& \quad- {\left( {{{\bf{I}}_{L_0}} + {{\bf{Z}}_{r,r}}{\bf{Z}}_r^{ - 1}} \right)^{ - 1}} \nonumber \\ & \quad * {{\bf{Z}}_{r,s}}{\left( {{{\bf{Z}}_{s,s}} + {\bf{Z}}_{{\rm{tun}}}} \right)^{ - 1}}  {{\bf{Z}}_{s,t}} \nonumber \\ & \quad*{\left( {{{\bf{Z}}_{t,t}} + {{\bf{Z}}_t}} \right)^{ - 1}} \nonumber
\end{align}

In \eqref{HE2E_3}, it is not difficult to recognize that the first addend on the right-hand side corresponds to the direct link between the transmitter and the receiver, and that the second addend on the right-hand side corresponds to the RIS-reradiated link that accounts for the internal impedances of the voltage generator at the transmitter, the load impedances at the receiver, the transfer matrices ${\bf{Z}}_{s,t}$ and ${\bf{Z}}_{r,s}$ that characterize the propagation of the electromagnetic wave from the transmitter to the RIS and from the RIS to the receiver, respectively, and the term ${\left( {{{\bf{Z}}_{s,s}} + {\bf{Z}}_{{\rm{tun}}}} \right)^{ - 1}}$ that accounts for the mutual coupling and the tuning circuits of the RIS.

The communication model in \eqref{HE2E}, and especially the simplified version in \eqref{HE2E_3}, is relatively simple to use in wireless communication systems thanks to the resemblance of $\bf{H}$ with a typical MIMO channel model. It is necessary to understand, however, the assumptions and the conditions under which \eqref{HE2E} can be applied. {Besides the computation of the matrix ${\bf{Z}}_{s,s}$ that, if done analytically, usually requires some approximations, the main assumption made to obtain $\bf{H}$ in \eqref{HE2E} lies in the considered expressions of the surface currents along the thin wire dipoles. Regardless of whether the wire dipoles operate in transmission, scattering, or reception modes, specifically, it is assumed that their surface current is a sinusoidal function \cite[Eq. (2)]{MDR_MutualImpedances}, which fulfills the property to be equal to zero along the entire antenna element if it is open-circuited, i.e., the load impedance connected to the port of the antenna element is equal to infinity. Specifically, the model introduced in \cite{MDR_MutualImpedancesMIMO} assumes that the surface currents in transmission, scattering, and reception modes have a sinusoidal shape whose amplitude evaluated at the port of the antenna element is an unknown variable that is determined by the parameters of the transmitter, RIS, and receiver, e.g., the tunable load impedances of the RIS, upon imposing appropriate boundary conditions. In the case of thin wire dipoles, the tangential component of the electric fields is imposed to be equal to zero on the surface of the dipole. In \cite{MDR_MutualImpedancesMIMO}, stated differently, the antenna elements are assumed to be canonical minimum scattering antennas. In simple terms, this assumption implies that a radiating element (a thin wire dipole in Fig. \ref{fig:ImpedanceModel}) does not radiate if it is open-circuited, and, therefore, it is like if it is not present (it is ``invisible'') in the network \cite{MinimumScatteringAntennas}. Concretely, this implies that the $p$th thin wire dipole that constitutes the RIS in Fig. \ref{fig:ImpedanceModel} does not reradiate in the presence of an electromagnetic wave if ${Z_{{\rm{tun}},p}} \to \infty$ (i.e., the current flowing in the antenna port of the thin wire dipole is zero) and, therefore, it can be removed from the system model. This statement can be readily verified by direct inspection of the simplified channel model in \eqref{HE2E_3}: We see, in fact, that the second added on the right-hand side of \eqref{HE2E_3}, i.e., the contribution of the RIS, tends to zero if ${Z_{{\rm{tun}},p}} \to \infty$ for every antenna element. This behavior is, of course, an approximation, since the presence of a thin wire dipole in the environment always perturbs (even if it is open-circuited) the electromagnetic field. In general, in fact, the current at the port of a thin wire dipole is equal to zero if ${Z_{{\rm{tun}},p}} \to \infty$, but it is not necessarily identically equal to zero along the entire thin dipole when it operates in scattering or reception modes. Therefore, caution needs to be paid when applying the communication model in \eqref{HE2E} to arbitrary antenna elements, since the approximation of canonical minimum scattering antennas needs to be evaluated. In addition, the communication model in \eqref{HE2E} is derived under the assumption that the shape (a sinusoidal function) of the surface current density of each thin wire dipole is not influenced by the proximity of other radiating elements.}

\subsection{Inhomogeneous Sheets of Surface Impedance}
In this sub-section, we consider models for RISs that abstract their microscopic structure and that are focused on the specific wave transformations that the metasurface, as a whole, is intended to realize. More precisely, a metamaterial-based RIS whose unit cells have sizes and inter-distances much smaller than the wavelength is homogenizable and can be modeled as a continuous surface sheet through appropriate surface functions, e.g., surface impedances \cite{MDR_JSAC}, \cite{Epstein_Tutorial}, \cite{Caloz_TAP}, \cite{Sergei_PerfectReflector}, \cite{Monticone}, \cite{Alu_PerfectReflector}, \cite{Vittorio_RayTracing}, \cite{Grbic_Homogenization}, \cite{Caloz_Sergei}, \cite{Sergei_MacroscopicModel}, \cite{Fadil_TCOM}, \cite{MDR_SPAWC}. This modeling approach is not dissimilar from the characterization of bulk (three-dimensional) metamaterials, which are usually represented through effective permittivity and permeability functions that determine the wave phenomena based on Maxwell's equations. The only difference is that a metasurface is better modeled by effective surface parameters, which manifest themselves in electromagnetic problems that are formulated as effective boundary conditions. These boundary conditions can be expressed in terms of surface polarizabilities, surface susceptibilities, or surface impedances (or admittances) \cite{SEM_Book}. In the present tutorial paper, we focus our attention on modeling an RIS through surface impedances.

\begin{figure}[!t]
		\begin{center}
			\includegraphics[width=\linewidth]{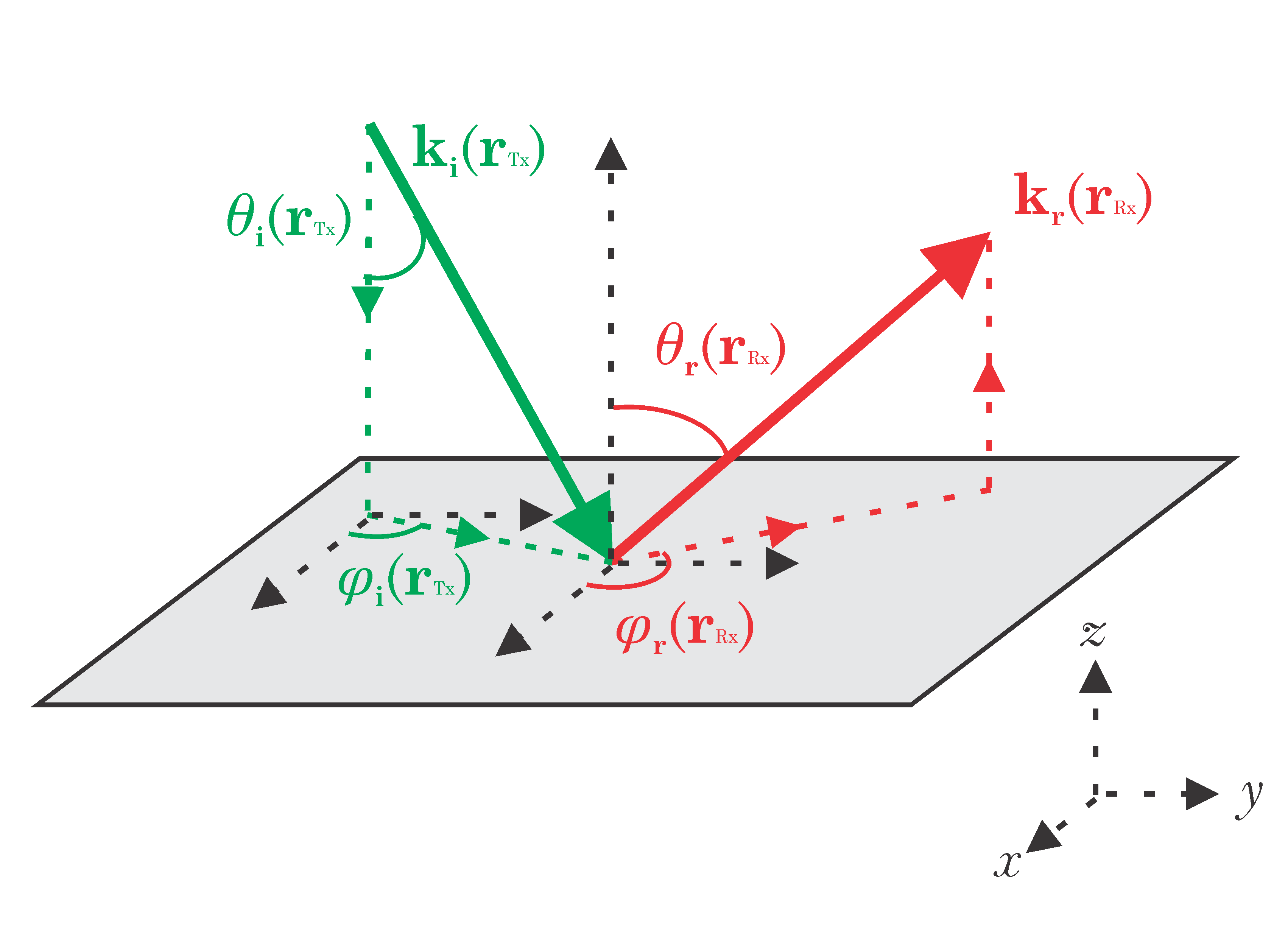}
			\caption{System model: An RIS as an inhomogeneous sheet (inhomogeneous boundary if the RIS is impenetrable) of surface impedance}
			\label{fig:SystemModel}
		\end{center}
	\end{figure}

The adopted modeling approach is, specifically, referred to as macroscopic \cite{MDR_JSAC}, \cite{Vittorio_RayTracing}, \cite{Sergei_MacroscopicModel}. Classical wave phenomena in materials or metamaterials are determined by the collective effects of a very large number of atoms that interact with the incident electromagnetic waves. The electromagnetic fields around individual atoms can be described by microscopic Maxwell’s equations. If the sizes of the atoms that constitute the material and the distances between them are much smaller than the wavelength, the electromagnetic fields and the sources in the material can be spatially averaged, thus effectively transforming microscopic Maxwell’s equations into macroscopic Maxwell’s equations. For a metasurface-based RIS, the same principle applies: If the RIS is electrically large and is made of sub-wavelength reconfigurable scattering elements (unit cells) whose inter-distances are much smaller than the wavelength, it is homogenizable and can be modeled through continuous surface averaged (macroscopic) surface impedances. Specifically, two conditions need to be fulfilled to make an RIS homogenizable \cite[Section 2.1]{Monticone}: (i) the first homogenization condition requires that the incident field varies little over one geometric period (the largest inter-distance among the unit cells) of the RIS, i.e., $\max \left\{ {{d_x},{d_y}} \right\} \ll \lambda$; and (ii) the second homogenization condition requires that the evanescent field scattered by the RIS is negligible at the observation point, i.e., $\left| z \right| > \max \left\{ {{d_x},{d_y}} \right\}$ for the RIS in Fig. \ref{fig:SystemModel}, where $\lambda$ is the wavelength of the electromagnetic wave, and $d_x$ and $d_y$ are the horizontal and vertical sizes of its unit cells (assuming that all the unit cells are identical in size). The second condition is typically fulfilled in the far-field of the RIS microstructure \cite[Fig. 29]{MDR_JSAC}, i.e., for observation distances of interest in wireless communications.

Under these assumptions, an RIS can be modeled as an inhomogeneous sheet of polarizable particles (the unit cells) that is characterized by an electric surface impedance and a magnetic surface admittance, which, for general wave transformations, are dyadic tensors. These two dyadic tensors constitute the macroscopic homogenized model of an RIS. The average total electric and magnetic fields that illuminate an RIS induce electric and magnetic currents that introduce a discontinuity between the electromagnetic fields on the two sides of an RIS (below and above the surface), which provides the means for manipulating the wavefront of the incident electromagnetic waves. Once the homogenized and continuous electric surface impedance and magnetic surface admittance are obtained based on the desired wave transformations, the microscopic structure and physical implementation of the RIS in terms of unit cells are obtained, by using, e.g., the method described in \cite{MDR_JSAC}, \cite{Caloz_TAP}. Generally speaking, once the macroscopic surface impedance and admittance are determined, appropriate geometric arrangements of sub-wavelength unit cells and the associated tuning circuits that exhibit the corresponding electric and magnetic response are characterized by, typically, using full-wave electromagnetic simulations \cite{Epstein_Tutorial}.

Based on this modeling approach, an RIS is characterized by a set of algebraic equations that result in boundary conditions for the electromagnetic fields at the two sides of the surface. This set of equations is referred to as generalized sheet transition conditions \cite{Holloway_2003}, \cite{Holloway_2005}. Under the assumption that only the tangential components of the electric and magnetic polarization densities are induced in the metasurface and that the RIS lies in the $xy$-plane (i.e., $z=0$) as illustrated in Fig. \ref{fig:SystemModel}, the generalized sheet transition conditions can be formulated as follows \cite{Epstein_Tutorial}:
\begin{align} \label{GSTC_E}
{\bf{E}}_{{\rm{tot}}}^t\left( {x,y,z = {0^ + }} \right) &+ {\bf{E}}_{{\rm{tot}}}^t\left( {x,y,z = {0^ - }} \right)\\
 & \hspace{-1.5cm} = 2{\overline{\overline {\bf{Z}}} _{se}}\left( {x,y} \right)\left( {{\bf{\hat z}} \times {{\bf{H}}_{{\rm{tot}}}}\left( {x,y,z = {0^ + }} \right)} \right) \nonumber \\
 & \hspace{-1.5cm} - 2{\overline{\overline {\bf{Z}}} _{se}}\left( {x,y} \right)\left( {{\bf{\hat z}} \times {{\bf{H}}_{{\rm{tot}}}}\left( {x,y,z = {0^ - }} \right)} \right) \nonumber
\end{align}
\begin{align} \label{GSTC_H}
{\bf{H}}_{{\rm{tot}}}^t\left( {x,y,z = {0^ + }} \right) &+ {\bf{H}}_{{\rm{tot}}}^t\left( {x,y,z = {0^ - }} \right)\\
 & \hspace{-1.5cm} =  - 2{\overline{\overline {\bf{Y}}} _{sm}}\left( {x,y} \right)\left( {{\bf{\hat z}} \times {{\bf{E}}_{{\rm{tot}}}}\left( {x,y,z = {0^ + }} \right)} \right) \nonumber \\
 & \hspace{-1.5cm} + 2{\overline{\overline {\bf{Y}}} _{sm}}\left( {x,y} \right)\left( {{\bf{\hat z}} \times {{\bf{E}}_{{\rm{tot}}}}\left( {x,y,z = {0^ - }} \right)} \right) \nonumber
\end{align}
where ${\overline{\overline {\bf{Z}}} _{se}}\left( {x,y} \right)$ and ${\overline{\overline {\bf{Y}}} _{sm}}\left( {x,y} \right)$ are the electric surface impedance and the magnetic surface admittance dyadic tensors that constitute the homogenized macroscopic model of an RIS. In addition, the following definitions for the electric and magnetic fields in \eqref{GSTC_E} and \eqref{GSTC_H} hold:
\begin{align}
{{\bf{F}}_{{\rm{tot}}}}\left( {x,y,z = {0^ + }} \right)& = {{\bf{F}}_{{\rm{inc}}}}\left( {x,y,z = {0^ + }} \right) \nonumber \\ & + {{\bf{F}}_{{\rm{ref}}}}\left( {x,y,z = {0^ + }} \right)
\end{align}
\begin{align} \label{eq:temp}
{{\bf{F}}_{{\rm{tot}}}}\left( {x,y,z = {0^ - }} \right) = {{\bf{F}}_{{\rm{tra}}}}\left( {x,y,z = {0^ - }} \right)
\end{align}
\begin{align}
{\bf{F}}_{{\rm{tot}}}^t\left( {x,y,z = {0^ \pm }} \right) = \left( {{\bf{\hat z}} \times {{\bf{F}}_{{\rm{tot}}}}\left( {x,y,z = {0^ \pm }} \right)} \right) \times {\bf{\hat z}}
\end{align}
where ${{\bf{\hat z}}}$ is the unit norm vector that is normal to the RIS as illustrated in Fig. \ref{fig:SystemModel}, and ${{\bf{F}}_{{\rm{inc}}}}\left( {x,y,z = {0^ + }} \right)$, ${{\bf{F}}_{{\rm{ref}}}}\left( {x,y,z = {0^ + }} \right)$ and ${{\bf{F}}_{{\rm{tra}}}}\left( {x,y,z = {0^ - }} \right)$ with ${\bf{F}} = \left\{ {{\bf{E}},{\bf{H}}} \right\}$ are the incident, reflected, and transmitted (refracted) electric and magnetic fields evaluated on the two sides of the RIS, respectively.

The equations in \eqref{GSTC_E} and \eqref{GSTC_H} completely characterize an RIS in terms of wave transformations and they can be utilized for the analysis and synthesis of an RIS. As far as the analysis is concerned, it is usually assumed that ${\overline{\overline {\bf{Z}}} _{se}}\left( {x,y} \right)$ and ${\overline{\overline {\bf{Y}}} _{sm}}\left( {x,y} \right)$ are known, and one is interested in solving \eqref{GSTC_E} and \eqref{GSTC_H} for obtaining the surface electric and magnetic fields in the close vicinity of the RIS, but at distances at which the homogenized model can be applied (as detailed in further text). As far as the synthesis is concerned, it is usually assumed that either the surface electric and magnetic fields in the close vicinity of the RIS are explicitly known or that an objective function that depend on them is known, and one is interested in identifying the corresponding functions ${\overline{\overline {\bf{Z}}} _{se}}\left( {x,y} \right)$ and ${\overline{\overline {\bf{Y}}} _{sm}}\left( {x,y} \right)$ that provide the desired electromagnetic fields or that maximize the objective function of interest. These techniques are referred to as direct and inverse source problems, respectively \cite{Brown_InverseSource}. In Section III, we present some examples to understand the optimization of RISs as a function of the surface impedance and for different design criteria.

\begin{table*}[!t]
		\centering
		\caption{{Comparison of the three RIS models.}}
		\label{table:ComparisionModels}
		\footnotesize
		{\begin{tabular}{l|l} \hline
			\hspace{1.75cm} \textbf{RIS model}  &  \hspace{4.5cm} \textbf{Main features and assumptions} \\ \hline
			Locally periodic discrete model				& $\begin{array}{l}
{\bullet \; \text{The incident electromagnetic waves are assumed to be reflected, at any point of the RIS, specularly}}\\
{\bullet \; \text{The RIS alphabet can be characterized for an individual unit cell}}
\end{array}$ \\ \hline
			Mutually coupled antenna elements				& $\begin{array}{l}
{\bullet \; \text{The mutual coupling among the RIS elements is explicitly modeled}}\\
{\bullet \; \text{The shape of the surface currents of each radiating element is assumed not to be influenced by the}}\\ 
{\text{\; \, proximity of the other radiating elements}}\\
{\bullet \; \text{In \cite{MDR_MutualImpedances}, the shape of the surface currents is assumed to be sinusoidal for transmitting and receiving}}\\ {\text{\; antennas, and for the RIS elements}}
\end{array}$ \\ \hline
			Inhomogeneous sheets of surface impedance			& $\begin{array}{l}
{\bullet \; \text{The RIS is modeled ``as a whole'' surface}}\\
{\bullet \; \text{Impedance matching between the incident and reflected electromagnetic waves is ensured by design}}\\
{\bullet \; \text{The incident electromagnetic waves are not assumed to be reflected, at any point of the RIS, specularly}}
\end{array}$ \\ \hline
		\end{tabular}}
	\end{table*}
One of the main advantages of modeling an RIS through inhomogeneous sheets of impedance and admittance dyadic tensors lies in the possibility of incorporating them into Maxwell's equations by leveraging the equivalence principle and the radiation integrals, which allow us to express the electric and magnetic fields anywhere in the volume of interest directly as a function of ${\overline{\overline {\bf{Z}}} _{se}}\left( {x,y} \right)$ and ${\overline{\overline {\bf{Y}}} _{sm}}\left( {x,y} \right)$ \cite[Chapter 18] {Orfanidis_Book}. The main assumption for using this approach consists of resorting to the physical optics approximation \cite[Chatper 8]{Osipov_Book}. Even though some approximations are usually needed to obtain the reradiated electromagnetic field, the resulting analytical framework is electromagnetically consistent and accounts for the physical implementation of the RIS. This procedure is elaborated in Section III through some specific examples, and the resulting analytical formulation is utilized for analyzing the radiated electromagnetic field. Another main advantage of the model for RISs based on inhomogeneous sheets of impedance and admittance dyadic tensors is that the mutual coupling among all the constitutive elements of the RIS (the unit cells) is inherently taken into account, since the model inherently abstracts, at the design stage, the physical implementation of the surface. The subsequent discretization of the RIS in unit cells based on the functions ${\overline{\overline {\bf{Z}}} _{se}}\left( {x,y} \right)$ and ${\overline{\overline {\bf{Y}}} _{sm}}\left( {x,y} \right)$ implicitly accounts for the local interactions and for the mutual coupling among the unit cells \cite{MDR_JSAC}, \cite{Caloz_Sergei}.

When modeling an RIS as an inhomogeneous sheet of impedance and admittance dyadic tensors, caution needs to be paid, however, to some implicit assumptions that are made. One of these assumptions is that an RIS is modeled as a device with zero thickness. In practice, however, an RIS has a finite thickness and is made of discrete and finite-size unit cells. In order for the homogenized (continuous) version of an RIS, which is utilized at the design stage, to accurately represent the reradiation properties of the manufactured metasurface, it is necessary that the thickness of the surface and the cross section of the unit cells are much smaller than the wavelength of the electromagnetic waves. If these conditions are met, the components of the surface fields that are related to the discretization of the surface in unit cells can be ignored, provided that the observation point is not too close to the surface of the RIS. As a rule of thumb, the observation point should be at least at a distance $\left| z \right| > {t \mathord{\left/ {\vphantom {t 2}} \right. \kern-\nulldelimiterspace} 2} + \max \left\{ {d_x,d_y} \right\}$, where $t$ is the thickness of the surface and $d_x$ and $d_y$ are the horizontal and vertical sizes of its unit cells (assuming that all the unit cells are identical in size) \cite{Epstein_Tutorial}. In practice, this condition does not pose any constraints in the context of wireless communication systems, since we are not usually interested in observation points that are so close to an RIS.

Due to these positive features, representing an RIS as an inhomogeneous sheet of surface impedance constitutes a suitable abstraction model for understanding the achievable performance limits of RISs in wireless networks and for their optimization as a function of different design criteria. In the next section, we utilize this modeling approach and elaborate how it can be leveraged for obtaining electromagnetically consistent analytical frameworks for RISs that are suitable for performance evaluation and for wireless networks optimization. In the next section, more specifically, we focus our attention on an RIS that operates in reflection mode and that is impenetrable, i.e., the electric and magnetic fields at $z=0^-$ are equal to zero in \eqref{eq:temp}. In this case, an RIS is referred to as an inhomogeneous boundary of surface impedance.

\begin{figure*}[!t]
		\begin{center}
			\begin{subfigure}{0.85\columnwidth}
			\centering
				{\includegraphics[width=0.85\columnwidth]{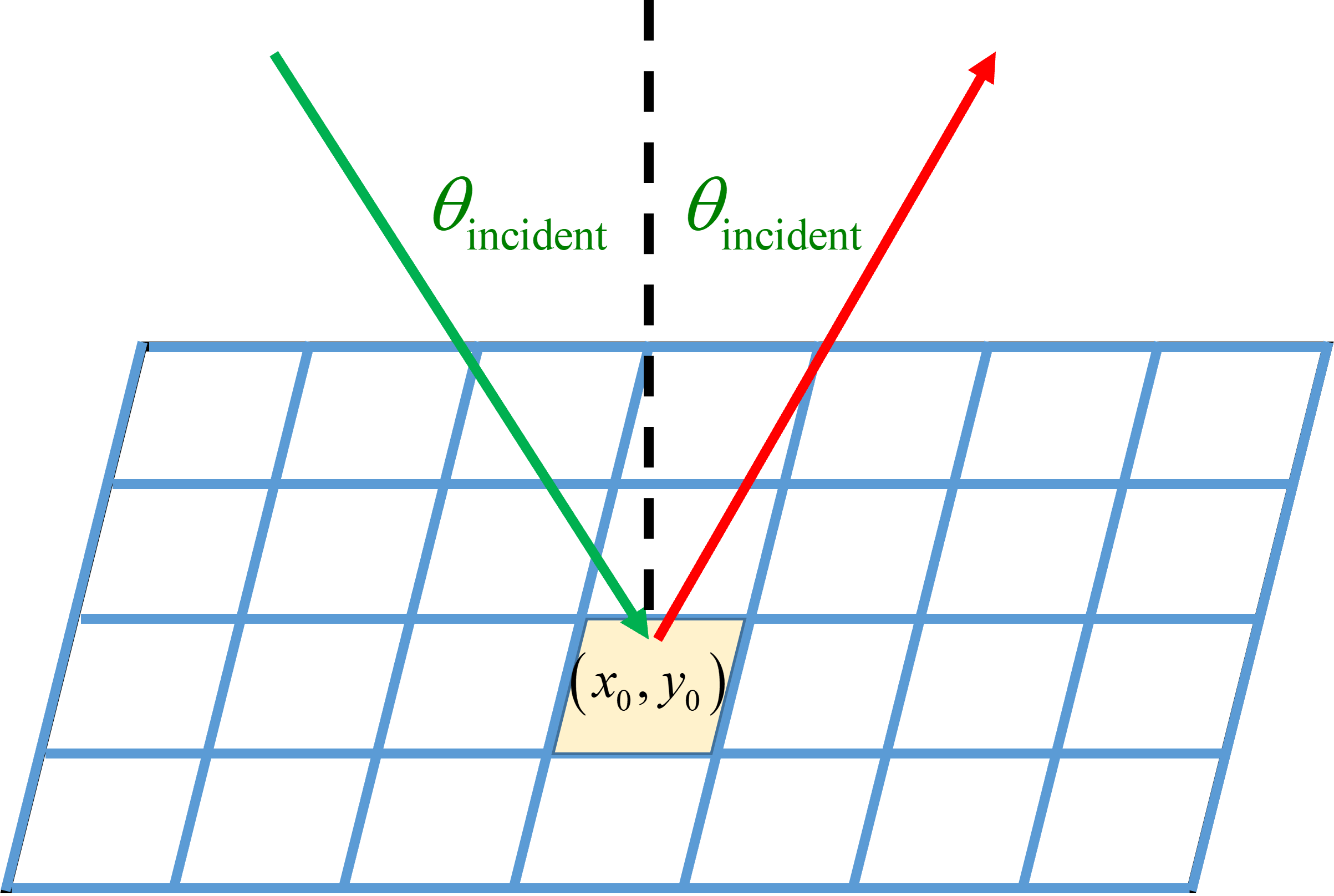}}
				\caption{Local specular reflection inherently assumed by the locally periodic discrete model. Anomalous reflection is obtained by applying non-uniform phase shifts.}\label{fig:LocalModel}
			\end{subfigure}\hfill%
			\begin{subfigure}{0.85\columnwidth}
			\centering
				{\includegraphics[width=0.85\columnwidth]{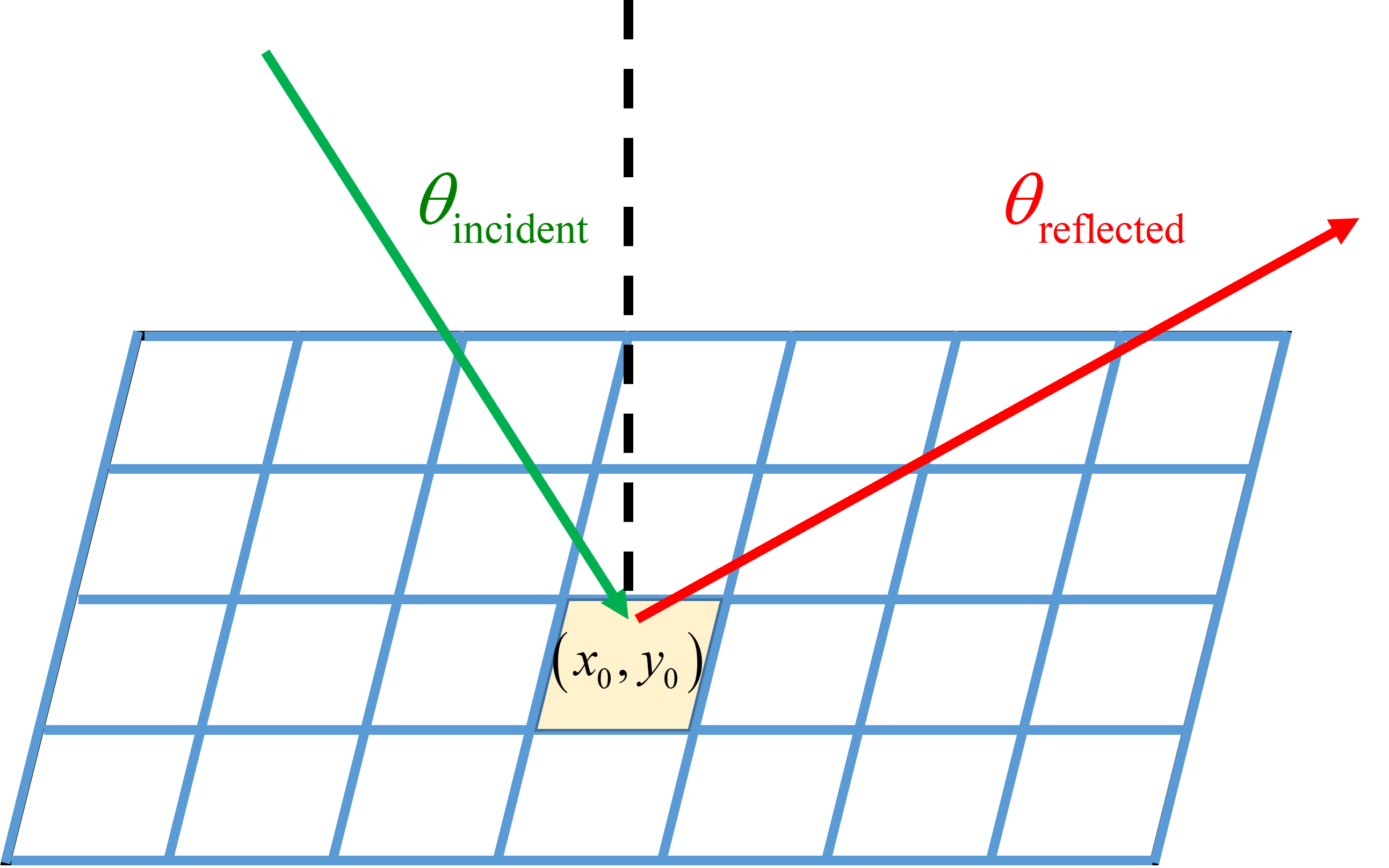}}
				\caption{Local non-specular reflection allowed by the inhomogeneous sheet of surface impedance model thanks to the impedance matching in \eqref{GSTC_E} and \eqref{GSTC_H} of the RIS viewed ``as a whole''.}\label{fig:SheetModel}
			\end{subfigure}
			\end{center} \caption{Illustrative comparison between the locally periodic discrete model and the inhomogeneous sheet of surface impedance model.} \label{fig:ComparisonModels}
	\end{figure*}
{\subsection{Comparison of the Three RIS Models}}
{The three models for RISs described in the previous sub-sections are developed under different modeling assumptions, and, therefore, their accuracy against full-wave electromagnetic simulations is, in general, different. A summary of the main features and modeling assumptions for each model is available in Table \ref{table:ComparisionModels}, by considering the canonical example of reflecting surfaces for ease of description. The \textit{locally periodic discrete model} relies on the assumption that the incident electromagnetic waves are reflected, at any point $(x_0, y_0)$ of the RIS, specularly, even though a different reflection amplitude and phase is applied at each point. In other words, the reflection coefficient is defined as if the RIS were an infinite and locally homogeneous surface. For this reason, this model for RISs may not always be accurate. The main advantage of this model is that the RIS alphabet can be characterized for an individual unit cell. The model based on \textit{mutually coupled antenna elements} has the advantage of explicitly modeling the mutual coupling among the unit cells of the RIS and of being formulated in terms of voltages at the input and output ports of the transmitter and receiver, respectively. This ease the utilization of the model for wireless networks optimization. The model proposed in \cite{MDR_MutualImpedances} relies on surface currents that are assumed to have a simple sinusoidal shape when the RIS elements operate in transmission, scattering, and reception modes. The surface currents depend, in general, on the specific radiating/scattering elements, on the electronic circuits that control the radiating/scattering elements, and whether the devices operate in transmission, scattering, or reception modes. In addition, it is assumed that the shape of the surface currents of each radiating element is not influenced by the proximity of the other radiating elements. The RIS model based on \textit{inhomogeneous sheets of surface impedance} has the advantage of modeling an RIS ``as a whole'', and of not relying on the assumption that the incident electromagnetic waves are reflected, at any point $(x_0, y_0)$ of the RIS, specularly. Since the RIS is modeled as a whole surface, the reflection at any point of the surface can be non-specular. This is illustrated in Fig. \ref{fig:ComparisonModels}, and it is further elaborated in the next section. This RIS model ensures that the impedances of the incident and reflected electromagnetic waves are matched, since the fulfillment of the boundary conditions in \eqref{GSTC_E} and \eqref{GSTC_H} is imposed by design. In addition, this RIS model can be easily integrated into Maxwell's equations, under the assumptions of the physical optics approximation, i.e., the currents induced on a finite-size metasurface are the same as those on an infinite-size metasurface (the perturbations of the induced currents near the edges of the metasurface are ignored), and by resorting to vector diffraction theory. This is detailed in the next section. In general terms, the RIS models based on the locally periodic discrete condition and the inhomogeneous sheet of surface impedance characterize the electromagnetic fields on the surface of an RIS, i.e., at the spatial discontinuity created by the metasurface. On the other hand, the RIS model based on mutually coupled antenna elements characterizes the voltages at the input and output ports of the transmitter and receiver, respectively. Different physical quantities under different assumptions are therefore modeled.}

\begin{figure}[!t]
{\includegraphics[width=\columnwidth]{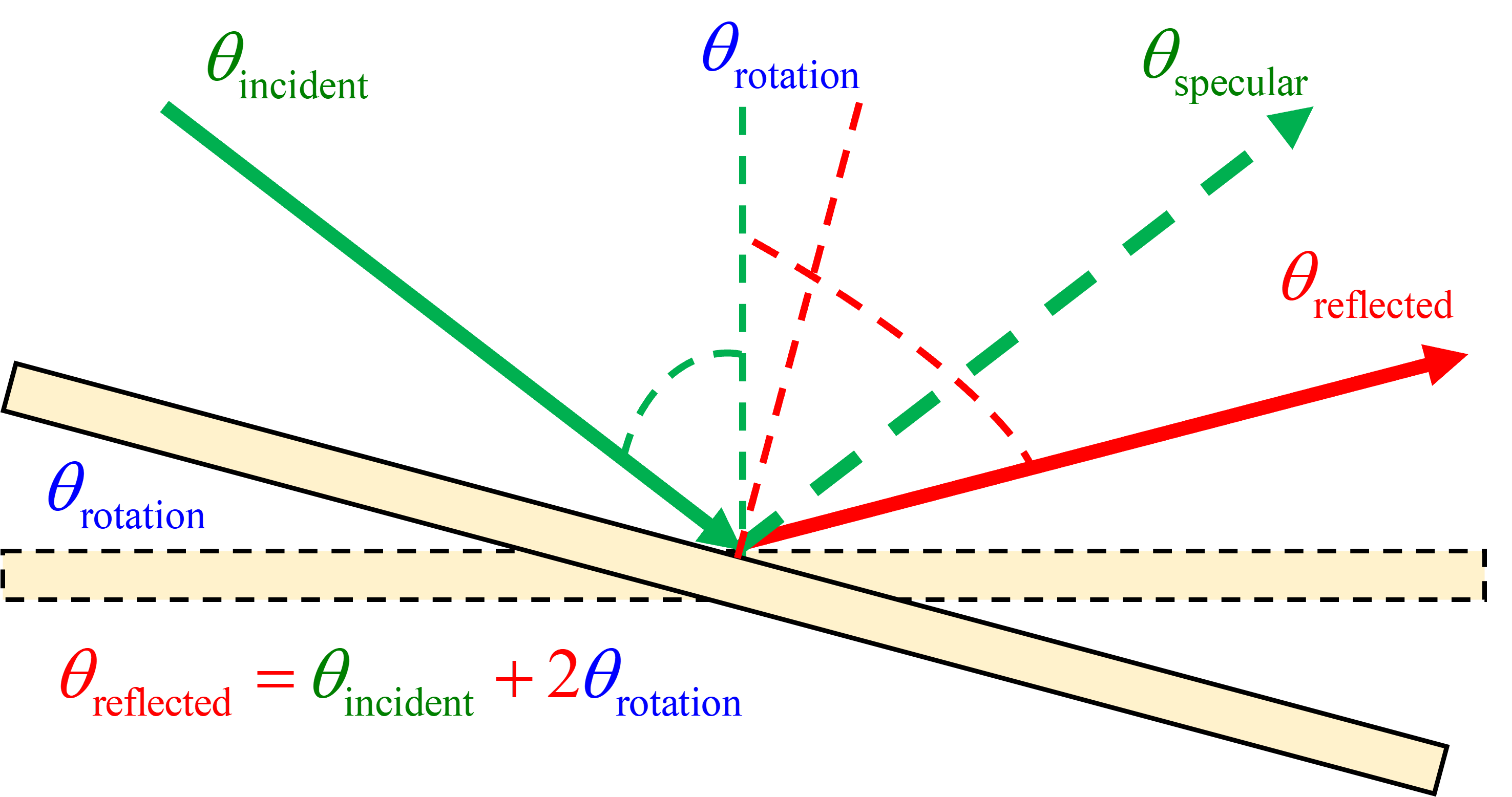}}
\caption{Titled (rotated) perfect electric conductor mimicking a reflecting engineered surface (i.e., an anomalous reflector).}\label{fig:TiltedMirror}
\end{figure}
{\subsection{Reflecting Engineered Surfaces vs. Tilted Homogeneous Surfaces}}
{An RIS is capable of realizing wave transformations that non-engineered surfaces cannot realize. In some cases, however, homogeneous surfaces can be utilized to mimic the behavior of an RIS if some mechanical operations are applied to them. This is the notable case of an anomalous reflector, whose function can be realized by appropriately tilting a homogeneous reflecting surface, i.e., a perfect electric conducting surface. Therefore, it is instructive to elaborate on the advantages of realizing an anomalous reflector through engineered and reconfigurable surfaces, i.e., an RIS, with respect to an appropriately tilted (rotated) perfect electric conductor. Besides the apparent benefit of avoiding the mechanical rotation for realizing anomalous reflection, which cannot be always possible in practice, there are more fundamental benefits in using an RIS instead of a rotated perfect electric conductor. As illustrated in Fig. \ref{fig:TiltedMirror}, a perfect electric conductor can steer an electromagnetic wave that impinges upon it from the direction $\theta_{\rm{incident}}$ towards the direction of reflection $\theta_{\rm{reflected}}$ if the homogeneous surface is rotated by an angle $\theta_{\rm{rotation}}$. More precisely, the angle of reflection is $\theta_{\rm{reflected}} = \theta_{\rm{incident}} + 2\theta_{\rm{rotation}}$, if $\theta_{\rm{incident}}$ and $\theta_{\rm{reflected}}$ are computed with the respect to the normal to the non-rotated perfect electric conductor. In principle, therefore, we can realize anomalous reflection by simply rotating a perfect electric conductor by $\theta_{\rm{rotation}}$, which is chosen to steer the desired reflected electromagnetic wave towards $\theta_{\rm{reflected}}$ for a given $\theta_{\rm{incident}}$, similar to a reflecting RIS. A rotated perfect electric conductor and a reflecting RIS are, however, not equivalent from the power efficiency point of view. The power that impinges upon a rotated perfect electric conductor is equal to ${P_{{\rm{incident}}}} = {P_0}\cos \left( {{\theta _{{\rm{incident}}}}} \right)$, where $P_0$ is the incident power in the absence of rotation. Since a perfect electric conductor is an electrically large homogeneous surface of non-resonant size and is not an engineered surface, it is capable of reflecting an amount of power that is at most equal to the incident power, i.e., ${P_{{\rm{reflected}}}}= {P_{{\rm{incident}}}} = {P_0}\cos \left( {{\theta _{{\rm{incident}}}}} \right)$. The larger the rotation, therefore, the smaller the reflected power. An appropriately designed RIS is, on the other hand, capable of overcoming this limitation and, in theory, it is capable of reflecting an amount of power equal to $P_0$, i.e., ${P_{{\rm{reflected}}}}=P_0$, regardless of the angle of incidence and hence with 100\% of efficiency. This is possible by appropriately modulating the amplitude and the phase of the incident electromagnetic wave. A practical realization of this perfect anomalous reflector is available in \cite{Sergei_RealizedSurface}. In conclusion, an efficient RIS cannot be replaced, in general, by a titled homogeneous surface. In addition, a perfect electric conductor cannot operate as a focusing lens, which provides optimal performance in the near-field.}
\vspace{0.5cm}

\section{RIS as an Inhomogeneous Impedance Boundary: Electromagnetically Consistent Modeling and Optimization}
In this section, we concentrate on RISs that are modeled as an impenetrable inhomogeneous sheet of surface impedance, i.e., an inhomogeneous impedance boundary. Thus, an incident electromagnetic wave is only reflected by the surface, while no refracted  electromagnetic waves are possible. The objective of this section is to provide the readers with a detailed tutorial-style, step-by-step, example on how to formulate electromagnetically consistent analytical models for RISs and how to use them for evaluating the performance of RISs in wireless networks, as well as for optimizing their surface impedance in order to fulfill specific design requirements. To this end, this section is structured in three interlinked macro parts.
\begin{itemize}
    \item First, we introduce the system and signal models departing from Maxwell's equations, and elaborate on the conditions that need to be fulfilled for ensuring that the RIS model is electromagnetically consistent.
    \item Then, we overview the concepts of local and global designs by departing from the considered electromagnetically consistent model RISs. We discuss the corresponding design criteria in terms of surface impedance and the implications associated with the practical implementation of specified surface impedances.
    \item Finally, we formulate optimization problems for designing the surface impedance of an RIS based on local and global designs, and we discuss the associated design constraints that determine the surface power efficiency, the reradiated power flux at some specified locations, the amount of reradiated power towards spurious directions, and the realization of RISs made of purely reactive surface boundaries that require no local power amplification.
\end{itemize}

\subsection{Electromagnetically Consistent Modeling of RISs}
Throughout the present paper, we adopt the notation in \cite{Orfanidis_Book}. In particular, we assume the universal time dependency $e^{j\omega t}$, where $j$ is the imaginary unit and $\omega$ is the angular frequency. Also, ${\nabla _{\boldsymbol{\xi}}}$ denotes the gradient computed with respect to the vector ${\boldsymbol{\xi}} = a{\bf{\hat x}} + b{\bf{\hat y}} + c{\bf{\hat z}}$, and $\cdot$ and $\times$ denote the scalar and vector products between vectors, respectively, $\Re(\cdot)$ and $\Im(\cdot)$ denote the real and imaginary parts of a complex function, respectively, and $\left\| {{{\boldsymbol{\xi }}_1} - {{\boldsymbol{\xi }}_2}} \right\|$ denotes the Euclidean distance between the points ${{{\boldsymbol{\xi }}_1}}$ and ${{{\boldsymbol{\xi }}_2}}$.

\subsubsection{System Model}
We consider a system model in a three-dimensional space, which includes a transmitter (Tx), a receiver (Rx), and an RIS (a flat surface $\SS$) that is modeled as an inhomogeneous boundary of surface impedance with negligible thickness with respect to the  wavelength of the electromagnetic waves. A sketched representation of the considered system model is given in Fig. \ref{fig:SystemModel}.

The RIS $\SS$ is a rectangle that lies in the $xy$-plane (i.e., $z=0$) whose center is located at the origin. The sides of $\SS$ are parallel to the $x$-axis and $y$-axis, and they have lengths $2L_x$ and $2L_y$, respectively. Hence, $\SS$ is defined as follows:
	\begin{equation} \label{SurfaceDefinition}
		\SS = \left\{\vect{s} = x\hat{\vect{x}} + y\hat{\vect{y}}:|x|\leq L_x,|y|\leq L_y\right\}
	\end{equation}

We consider a reflecting RIS, i.e., Tx and Rx are located in the same side of $\SS$. The transmitter, Tx, emits an electromagnetic wave that propagates through the vacuum whose permittivity and permeability are $\epsilon_0$ and $\mu_0$, respectively. The electromagnetic wave emitted by Tx travels at the speed of light $c = 1/\sqrt{\epsilon_0\mu_0}$ and the free-space impedance is defined as $\eta_0 = \sqrt{\mu_0/\epsilon_0}$. The carrier frequency, the wavelength, and the wavenumber of the electromagnetic wave are denoted by $f$, $\lambda=c/f$, $k = 2\pi/\lambda$, respectively, and $\omega = 2\pi f$.

We denote the center location of Tx and Rx as $\vect{r}_{\textup{Tx}} = x_{\textup{Tx}}\hat{\vect{x}} + y_{\textup{Tx}}\hat{\vect{y}} + z_{\textup{Tx}}\hat{\vect{z}}$ and $\vect{r}_{\textup{Rx}} = x_{\textup{Rx}}\hat{\vect{x}} + y_{\textup{Rx}}\hat{\vect{y}} + z_{\textup{Rx}}\hat{\vect{z}}$, respectively. The transmitter is characterized by a charge density $\rho(\cdot)$ and a current density $\vect{J}(\cdot)$ that fulfill the continuity equation \cite{Orfanidis_Book} and that are non-zero only in a small volume $V_{\textup{Tx}}$ centered at $\vect{r}_{\textup{Tx}}$. For ease of exposition, but without loss of generality, we assume that the transmitter Tx is in the far field region of the RIS and the receiver Rx. Thus, we assume that the incident signal is a plane wave, while the reflected signal is not necessarily a plane wave. We consider only the signal reflected by the RIS and assume that the Tx-Rx direct signal can be ignored due to the presence of blocking objects.

\subsubsection{Electric and Magnetic Fields}
More precisely, we assume that the incident plane wave emitted by the transmitter Tx impinges upon the RIS at an elevation angle $\theta_i(\vect{r}_{\textup{Tx}})$ and at an azimuth angle $\varphi_i(\vect{r}_{\textup{Tx}})$, and that it is reflected (reradiated), not necessarily as a plane wave, towards the receiver Rx, whose location is specified by the elevation angle $\theta_r(\vect{r}_{\textup{Rx}})$ and the azimuth angle $\varphi_r(\vect{r}_{\textup{Rx}})$. For ease of analysis, we assume $\varphi_i(\vect{r}_{\textup{Tx}})  = \varphi_r(\vect{r}_{\textup{Rx}})  = \pi/2$, i.e., the incident and reflected fields propagate in the $yz$-plane.
	
Let $\vect{k}_i(\vect{r}_{\textup{Tx}})$ and $\vect{k}_r(\vect{r}_{\textup{Rx}})$ be the wavevectors of the incident and reflected electromagnetic waves, respectively, whose amplitudes are equal to the wavenumber $k$. In particular, $\vect{k}_r(\vect{r}_{\textup{Rx}})$ identifies the desired direction of reflection, i.e., the direction towards which the RIS is intended to maximize the reradiated power. The wavevectors are defined as follows:
	\begin{align}
		\vect{k}_i(\vect{r}_{\textup{Tx}})
		&= k \sin \theta_i(\vect{r}_{\textup{Tx}}) \cos \varphi_i(\vect{r}_{\textup{Tx}}) \hat{\vect{x}} \nonumber\\ &+ k \sin \theta_i(\vect{r}_{\textup{Tx}}) \sin \varphi_i(\vect{r}_{\textup{Tx}}) \hat{\vect{y}} - k \cos \theta_i(\vect{r}_{\textup{Tx}}) \hat{\vect{z}}\nonumber\\
		&= k \left(\sin \theta_i(\vect{r}_{\textup{Tx}}) \hat{\vect{y}} - \cos \theta_i(\vect{r}_{\textup{Tx}}) \hat{\vect{z}}\right)
		\label{eq:wavevector-incident}
		\\
		\vect{k}_r(\vect{r}_{\textup{Rx}})
		&= k \sin \theta_r(\vect{r}_{\textup{Rx}}) \cos \varphi_r(\vect{r}_{\textup{Rx}}) \hat{\vect{x}} \nonumber\\ &+ k \sin \theta_r(\vect{r}_{\textup{Rx}}) \sin \varphi_r(\vect{r}_{\textup{Rx}}) \hat{\vect{y}} + k \cos \theta_r(\vect{r}_{\textup{Rx}}) \hat{\vect{z}}\nonumber\\
		&= k \left(\sin \theta_r (\vect{r}_{\textup{Rx}}) \hat{\vect{y}} + \cos \theta_r (\vect{r}_{\textup{Rx}}) \hat{\vect{z}}\right)
		\label{eq:wavevector-reflected}
	\end{align}
	
The electric field transmitted by the transmitter Tx is assumed to be $x$-polarized and to propagate in free-space. To formulate the incident and reflected fields in the close proximity of the surface $\SS$ located at $z=0$, we introduce the general vector $\vect{r} = x\hat{\vect{x}} + y\hat{\vect{y}} + z\hat{\vect{z}}$ in the reflection half plane (i.e., $z \ge 0$) with $\tau_m < z  < \tau_M$, where $\tau_m$ and $\tau_M$ are positive constant values that are very small but $\tau_m$ is sufficiently large for the impedance boundary model to be applicable, as elaborated in Section II-C. The condition $z  < \tau_M$ is necessary to validate the electromagnetic consistency of the reflected electric and magnetic fields, as detailed next. The reradiated field for $z  > \tau_M$, which encompasses the near-field and far-field regions of the RIS, is defined and detailed next by utilizing the radiation integrals. With this notation, the incident electric field is formulated as follows:
	\begin{align}
		\vect{E}_{\textup{inc}}(\vect{r},\vect{r}_{\textup{Tx}})
		&= {E}^i_{x,0} e^{-j\vect{k}_i(\vect{r}_{\textup{Tx}})\cdot {\vect{r}}} \hat{\vect{x}} \nonumber\\
		&= {E}^i_{x,0}e^{-jk(\sin\theta_i(\vect{r}_{\textup{Tx}})y - \cos\theta_i(\vect{r}_{\textup{Tx}})z)} \hat{\vect{x}} \label{eq:incident-electric-field-plane-wave}
	\end{align}
where ${E}^i_{x,0}$ is a constant (independent of $\vect{r}$) complex amplitude as dictated by the plane wave assumption.

The corresponding incident magnetic field evaluated at $\vect{r}$ is obtained from Maxwell's equations, as follows:
	\begin{align}\label{eq:Hinc-plane-wave-general}
		\vect{H}_{\textup{inc}}(\vect{r},\vect{r}_{\textup{Tx}})
		&= -\frac{1}{j\omega\mu_0} \nabla_{\vect{r}} \times \vect{E}_{\textup{inc}}(\vect{r},\vect{r}_{\textup{Tx}}) \nonumber \\
		&= -\frac{1}{\eta_0} {E}^i_{x,0}e^{-jk(\sin\theta_i(\vect{r}_{\textup{Tx}})y - \cos\theta_i(\vect{r}_{\textup{Tx}})z)} \nonumber \\ &*\bigg[\cos\theta_i(\vect{r}_{\textup{Tx}}) \hat{\vect{y}} + \sin\theta_i(\vect{r}_{\textup{Tx}}) \hat{\vect{z}}\bigg]
	\end{align}
where the identity $k/{(\omega\mu_0)} = 1/\eta_0$ is used.

In order to model the field reradiated by the RIS in the close proximity of $\SS$, we utilize the physical optics approximation method \cite{Osipov_Book}, \cite{Fadil_TCOM}. Accordingly, the field reflected by the RIS is assumed to be radiated by secondary currents that are induced on the surface $\SS$ by the incident electromagnetic wave. These currents determine, in turn, the surface (at $z=0^+$) electric and magnetic fields. Since the surface electric and magnetic fields are not known in advance, the physical optics approximation substitutes them with their geometrical optics approximation \cite{Osipov_Book}. In particular, it is assumed that the secondary currents (and the corresponding surface fields) that are induced on the finite-size surface $\SS$ are the same as those that would be induced on an infinite-size (i.e., $L_x \to \infty$ and $L_y \to \infty$) RIS. This implies that the perturbations of the induced currents near the edges of the surface $\SS$ are neglected. Even though approximated, the physical optics method is a suitable approach to guide the design and optimization of RISs and to shed light on their achievable performance, without using numerically-intensive techniques, such as the method of moments. In this section, we concentrate on the reradiated electric and magnetic fields in the close proximity of the RIS, i.e., evaluated at $\vect{r}$ for $\tau_m < z < \tau_M$.

Based on the physical optics method and assuming that the RIS does not change the polarization of the incident field, the reflected electric field at $\vect{r}$ for $\tau_m < z < \tau_M$ can be formulated as:
	\begin{align}\label{eq:Einc-Eref-relationship-general}
		&\vect{E}_{\textup{ref}}(\vect{r}_{\textup{Rx}},\vect{r},\vect{r}_{\textup{Tx}})
		\approx
		\Gamma_{\textup{ref}}(\vect{r}_{\textup{Rx}},\vect{r},\vect{r}_{\textup{Tx}}) \vect{E}_{\textup{inc}}(\vect{r},\vect{r}_{\textup{Tx}})
	\end{align}
where $\Gamma_{\textup{ref}}(\vect{r}_{\textup{Rx}},\vect{s},\vect{r}_{\textup{Tx}})$ is a complex-valued function, which is usually referred to as the field reflection coefficient.

In general, $\Gamma_{\textup{ref}}(\vect{r}_{\textup{Rx}},\vect{r},\vect{r}_{\textup{Tx}})$ is an arbitrary function provided that the reflected field in \eqref{eq:Einc-Eref-relationship-general} fulfills Maxwell's equations. For ease of analysis, but without loss of generality, we assume that $\Gamma_{\textup{ref}}(\vect{r}_{\textup{Rx}},\vect{r},\vect{r}_{\textup{Tx}})$ is formulated as follows:
\begin{align}\label{eq:reflection-coefficient-explicit-at-RIS-complete}
		\Gamma_{\textup{ref}}(\vect{r}_{\textup{Rx}},\vect{r},\vect{r}_{\textup{Tx}})
		&= \Gamma(\vect{r}_{\textup{Rx}},\vect{s},\vect{r}_{\textup{Tx}})e^{j \Phi(\vect{r}_{\textup{Rx}},\vect{r},\vect{r}_{\textup{Tx}})}
	\end{align}
where the following definitions hold:
	\begin{align}\label{eq:phase-geometric-optics-explicit}
		\Phi(\vect{r}_{\textup{Rx}},\vect{r},\vect{r}_{\textup{Tx}})
		= &-k \left(\sin\theta_r(\vect{r}_{\textup{Rx}}) - \sin\theta_i(\vect{r}_{\textup{Tx}})\right)y \nonumber \\ & -k \left(\cos\theta_r(\vect{r}_{\textup{Rx}}) + \cos\theta_i(\vect{r}_{\textup{Tx}})\right)z
	\end{align}
\begin{align}\label{eq:reflection-coefficient-complex-amplitude}
		\Gamma(\vect{r}_{\textup{Rx}},\vect{s},\vect{r}_{\textup{Tx}})
		= R(\vect{r}_{\textup{Rx}},\vect{s},\vect{r}_{\textup{Tx}})e^{j\phi(\vect{r}_{\textup{Rx}},\vect{s},\vect{r}_{\textup{Tx}})}
	\end{align}
where $R(\vect{r}_{\textup{Rx}},\vect{s},\vect{r}_{\textup{Tx}}) = \left|\Gamma(\vect{r}_{\textup{Rx}},\vect{s},\vect{r}_{\textup{Tx}})\right|$ and $\phi(\vect{r}_{\textup{Rx}},\vect{s},\vect{r}_{\textup{Tx}}) = \angle \Gamma(\vect{r}_{\textup{Rx}},\vect{s},\vect{r}_{\textup{Tx}})$ denote the amplitude and the phase of $\Gamma(\vect{r}_{\textup{Rx}},\vect{s},\vect{r}_{\textup{Tx}})$, respectively.

In the considered case study, the incident electric and magnetic fields in \eqref{eq:incident-electric-field-plane-wave} and \eqref{eq:Hinc-plane-wave-general}, respectively, and the corresponding reradiated electric and magnetic fields are assumed to be independent of $x$. Hence, we obtain the simplified expressions $\Phi(\vect{r}_{\textup{Rx}},\vect{r},\vect{r}_{\textup{Tx}}) = \Phi(\vect{r}_{\textup{Rx}},(y,z),\vect{r}_{\textup{Tx}})$, $\Gamma(\vect{r}_{\textup{Rx}},\vect{s},\vect{r}_{\textup{Tx}}) = \Gamma(\vect{r}_{\textup{Rx}},y,\vect{r}_{\textup{Tx}})$, $R(\vect{r}_{\textup{Rx}},\vect{s},\vect{r}_{\textup{Tx}}) = R(\vect{r}_{\textup{Rx}},y,\vect{r}_{\textup{Tx}})$, and $\phi(\vect{r}_{\textup{Rx}},\vect{s},\vect{r}_{\textup{Tx}}) = \phi(\vect{r}_{\textup{Rx}},y,\vect{r}_{\textup{Tx}})$. If more general formulations for the incident and reradiated electromagnetic fields are considered, the approach described in this paper applies \textit{mutatis mutandis}.

The function $\Gamma_{\textup{ref}}(\vect{r}_{\textup{Rx}},\vect{r}=\vect{s},\vect{r}_{\textup{Tx}})$ for $\vect{s} = x \hat{\vect{x}} + y \hat{\vect{y}} \in \SS$ is the field reflection coefficient evaluated on the surface $\SS$ (i.e., at $z=0^+$), which is usually referred to as the surface reflection coefficient and characterizes the reradiation properties of the RIS. With a slight abuse of terminology, we refer to $\Gamma(\vect{r}_{\textup{Rx}},\vect{s},\vect{r}_{\textup{Tx}})$ as the surface reflection coefficient as well, since $\Gamma_{\textup{ref}}(\vect{r}_{\textup{Rx}},\vect{r}=\vect{s},\vect{r}_{\textup{Tx}})$ and  $\Gamma(\vect{r}_{\textup{Rx}},\vect{s},\vect{r}_{\textup{Tx}})$ differ by a linear phase shift, i.e., $\Phi(\vect{r}_{\textup{Rx}},(y,z=0^+),\vect{r}_{\textup{Tx}})$.

The analytical formulation in \eqref{eq:reflection-coefficient-explicit-at-RIS-complete} is not only analytically convenient, as it will be apparent next, but it has a useful physical interpretation. The function $\Phi(\vect{r}_{\textup{Rx}},\vect{r},\vect{r}_{\textup{Tx}})$ corresponds to the phase shift that needs to be applied by the RIS, based on the geometrical optics approximation, in order to steer (reradiate) an electromagnetic wave the impinges upon $\SS$ from the direction identified by $\vect{k}_i(\vect{r}_{\textup{Tx}})$ towards the direction identified by $\vect{k}_r(\vect{r}_{\textup{Rx}})$ \cite{Capasso_Science}. The complex-valued function $\Gamma(\vect{r}_{\textup{Rx}},\vect{s},\vect{r}_{\textup{Tx}})$ models the amplitude ($R(\vect{r}_{\textup{Rx}},\vect{s},\vect{r}_{\textup{Tx}})$) and the phase ($\phi(\vect{r}_{\textup{Rx}},\vect{s},\vect{r}_{\textup{Tx}})$) correction terms that are necessary for optimizing an RIS based on more advanced criteria than the canonical geometrical optics method. By setting $\Gamma(\vect{r}_{\textup{Rx}},\vect{s},\vect{r}_{\textup{Tx}})=1$, the solution based on the geometrical optics approximation is retrieved. The function $R(\vect{r}_{\textup{Rx}},\vect{s},\vect{r}_{\textup{Tx}})$ accounts for designs of RISs that may need a non-uniform amplitude control along the surface $\SS$. In general, as elaborated next, the amplitude correction function $R(\vect{r}_{\textup{Rx}},\vect{s},\vect{r}_{\textup{Tx}})$ and the phase correction function $\phi(\vect{r}_{\textup{Rx}},\vect{s},\vect{r}_{\textup{Tx}})$ may not be independent of one another for some optimization criteria for the RIS.

As mentioned, $\Gamma_{\textup{ref}}(\vect{r}_{\textup{Rx}},\vect{r},\vect{r}_{\textup{Tx}})$ is an arbitrary function provided that the reflected field in \eqref{eq:Einc-Eref-relationship-general} fulfills Maxwell's equations. This implies that $\vect{E}_{\textup{ref}}(\vect{r}_{\textup{Rx}},\vect{r},\vect{r}_{\textup{Tx}})$ in \eqref{eq:Einc-Eref-relationship-general} needs to satisfy Helmholtz's equation in the source-free region \cite{Orfanidis_Book}. Since the analytical formulation of the reflected field in \eqref{eq:Einc-Eref-relationship-general} is an approximation, it is customary to ensure that Helmholtz's equation is fulfilled approximately as well. By defining ${E}_{\textup{ref},x}(\vect{r}_{\textup{Rx}},\vect{r},\vect{r}_{\textup{Tx}}) = {\bf{\hat x}} \cdot \vect{E}_{\textup{ref}}(\vect{r}_{\textup{Rx}},\vect{r},\vect{r}_{\textup{Tx}})$, Helmholtz's equation on the surface $\SS$ (i.e., at $z=0^+$) reads as follows:
\begin{align}\label{eq:Helmholtz}
		&{\left. {{\nabla _{\bf{r}}} \cdot {\nabla _{\bf{r}}}\left( {{E_{{\rm{ref}},x}}\left( {{{\vect{r}}_{{\textup{Rx}}}},{\vect{r}},{{\vect{r}}_{{\textup{Tx}}}}} \right)} \right)} \right|_{z = 0^+}} \nonumber \\ & \hspace{3cm}\approx  - {k^2}{E_{{\rm{ref}},x}}\left( {{{\vect{r}}_{{\textup{Rx}}}},{\vect{s}},{{\vect{r}}_{{\textup{Tx}}}}} \right)
	\end{align}

By inserting \eqref{eq:Einc-Eref-relationship-general} in \eqref{eq:Helmholtz} and by taking into account that $\Gamma(\vect{r}_{\textup{Rx}},\vect{s},\vect{r}_{\textup{Tx}}) = \Gamma(\vect{r}_{\textup{Rx}},y,\vect{r}_{\textup{Tx}})$, i.e., it depends only on $y$, we obtain the following condition:
\begin{align}\label{eq:Helmholtz-reflected-electric-field}
			& \left|\frac{d^2}{d y^2} \Gamma(\vect{r}_{\textup{Rx}},y,\vect{r}_{\textup{Tx}}) - 2jk \sin\theta_r(\vect{r}_{\textup{Rx}})\frac{d}{d y}\Gamma(\vect{r}_{\textup{Rx}},y,\vect{r}_{\textup{Tx}}) \right| \nonumber \\ & \hspace{3cm} \ll k^2 \left|\Gamma(\vect{r}_{\textup{Rx}},y,\vect{r}_{\textup{Tx}})\right|
		\end{align}

To obtain an electromagnetically consistent design for the RIS, it is necessary to add the condition in \eqref{eq:Helmholtz-reflected-electric-field} as a constraint when optimizing $\Gamma_{\textup{ref}}(\vect{r}_{\textup{Rx}},\vect{s},\vect{r}_{\textup{Tx}})$. This is elaborated next. It is worth mentioning that the Helmholtz equation in \eqref{eq:Helmholtz-reflected-electric-field} satisfied with equality if $\Gamma(\vect{r}_{\textup{Rx}},y,\vect{r}_{\textup{Tx}})$ is independent of $y$, i.e., it is a constant function along the entire surface $\SS$. {This is expected because the reflected electric field in \eqref{eq:Einc-Eref-relationship-general} would be a perfect plane wave in this case.}

In addition to \eqref{eq:Helmholtz}, it is necessary that the reflected electric field in \eqref{eq:Einc-Eref-relationship-general} fulfills the zero-divergence condition in the source-free region, i.e., ${\nabla _{\vect{r}}} \cdot {{\bf{E}}_{{\rm{ref}}}}\left({{{\vect{r}}_{{\textup{Rx}}}},{\vect{r}},{{\vect{r}}_{{\textup{Tx}}}}} \right) = 0$, for it to be a consistent electric field. This imposes an additional constraint on the design and optimization of $\Gamma_{\textup{ref}}(\vect{r}_{\textup{Rx}},\vect{r},\vect{r}_{\textup{Tx}})$. The zero-divergence condition is always fulfilled, as shown as follows:
\begin{align}\label{eq:Divergence}
& {\nabla _{\vect{r}}} \cdot {{\bf{E}}_{{\rm{ref}}}}\left( {{{\vect{r}}_{{\textup{Rx}}}},{\vect{r}},{{\vect{r}}_{{\textup{Tx}}}}} \right) \\ & \quad \approx {\nabla _{\vect{r}}} \cdot \left( {{\Gamma _{{\rm{ref}}}}\left( {{{\vect{r}}_{{\textup{Rx}}}},{\vect{r}},{{\vect{r}}_{{\textup{Tx}}}}} \right){{\bf{E}}_{{\rm{inc}}}}\left( {{\vect{r}},{{\vect{r}}_{{\textup{Tx}}}}} \right)} \right) \nonumber \\
 & \quad = \left( {{\nabla _{\vect{r}}}{\Gamma _{{\rm{ref}}}}\left( {{{\vect{r}}_{{\textup{Rx}}}},{\vect{r}},{{\vect{r}}_{{\textup{Tx}}}}} \right)} \right) \cdot {{\bf{E}}_{{\rm{inc}}}}\left( {{\vect{r}},{{\vect{r}}_{{\textup{Tx}}}}} \right) \nonumber \\ & \quad + {\Gamma _{{\rm{ref}}}}\left( {{{\vect{r}}_{{\textup{Rx}}}},{\vect{r}},{{\vect{r}}_{{\textup{Tx}}}}} \right){\nabla _{\bf{r}}} \cdot {{\bf{E}}_{{\rm{inc}}}}\left( {{\vect{r}},{{\vect{r}}_{{\textup{Tx}}}}} \right) \nonumber \\
 & \quad = {E_{{\rm{inc}},x}}\left( {\left( {y,z} \right),{{\vect{r}}_{{\textup{Tx}}}}} \right)\left( {\frac{\partial }{{\partial y}}{\Gamma _{{\rm{ref}}}}\left( {{{\vect{r}}_{{\textup{Rx}}}},\left( {y,z} \right),{{\vect{r}}_{{\textup{Tx}}}}} \right)} \right){\bf{\hat y}} \cdot {\bf{\hat x}} \nonumber
		\end{align}
    \begin{align}
 & \quad + {E_{{\rm{inc}},x}}\left( {\left( {y,z} \right),{{\bf{r}}_{{\rm{Tx}}}}} \right)\left( {\frac{\partial }{{\partial z}}{\Gamma _{{\rm{ref}}}}\left( {{{\bf{r}}_{{\rm{Rx}}}},\left( {y,z} \right),{{\bf{r}}_{{\rm{Tx}}}}} \right)} \right){\bf{\hat z}} \cdot {\bf{\hat x}} \nonumber \\
 & \quad + {\Gamma _{{\rm{ref}}}}\left( {{{\bf{r}}_{{\rm{Rx}}}},{\bf{r}},{{\bf{r}}_{{\rm{Tx}}}}} \right)\left( {\frac{\partial }{{\partial x}}{E_{{\rm{inc}},x}}\left( {\left( {y,z} \right),{{\bf{r}}_{{\rm{Tx}}}}} \right)} \right) \nonumber \\
 & \quad = 0 \nonumber
		\end{align}
where ${E}_{\textup{inc},x}(\vect{r},\vect{r}_{\textup{Tx}}) = {\bf{\hat x}} \cdot \vect{E}_{\textup{inc}}(\vect{r},\vect{r}_{\textup{Tx}})$, and the identities are obtained by taking into account that neither the incident field nor the field reflection coefficient depend on $x$.

Given the reflected electric field in \eqref{eq:Einc-Eref-relationship-general} that fulfills \eqref{eq:Helmholtz-reflected-electric-field}, the final step for ensuring that the reradiated field in the close proximity of the RIS is electromagnetically consistent lies in calculating the reflected magnetic field. Similar to \eqref{eq:Hinc-plane-wave-general}, $\vect{H}_{\textup{ref}}(\vect{r}_{\textup{Rx}},\vect{r},\vect{r}_{\textup{Tx}})$ is obtained from Maxwell's equations:
\begin{align}\label{eq:Href-at-RIS-explicit}
		\vect{H}_{\textup{ref}}(\vect{r}_{\textup{Rx}},\vect{r},\vect{r}_{\textup{Tx}})
		&= -\frac{1}{j\omega\mu_0} \nabla_{\vect{r}} \times \vect{E}_{\textup{ref}}(\vect{r}_{\textup{Rx}},\vect{r},\vect{r}_{\textup{Tx}})  \\
		& = -\frac{{E}^i_{x,0}}{j\omega\mu_0} \nabla_{\vect{r}} \times \bigg[\Gamma(\vect{r}_{\textup{Rx}},\vect{r},\vect{r}_{\textup{Tx}}) g(y,z) \hat{\vect{x}}\bigg] \nonumber
	\end{align}
	\begin{align}
		 &\mathop  = \limits^{\left( a \right)} \frac{-{E}^i_{x,0}}{j\omega\mu_0}
		 \bigg[\left(-\frac{d}{d y}\Gamma(\vect{r}_{\textup{Rx}},y,\vect{r}_{\textup{Tx}}) \right) g(y,z)  \hat{\vect{z}} \nonumber\\
		 &\quad -jk \cos\theta_r(\vect{r}_{\textup{Rx}})
		 \Gamma(\vect{r}_{\textup{Rx}},y,\vect{r}_{\textup{Tx}}) g(y,z) \hat{\vect{y}} \nonumber \\
		 &\quad + jk \sin\theta_r(\vect{r}_{\textup{Rx}}) \Gamma(\vect{r}_{\textup{Rx}},y,\vect{r}_{\textup{Tx}}) g(y,z) \hat{\vect{z}} \bigg] \nonumber
	\end{align}
where $g(y,z)=e^{-jk(\sin\theta_r(\vect{r}_{\textup{Rx}})y + \cos\theta_r(\vect{r}_{\textup{Rx}})z)}$, $(a)$ follows from ${\nabla _{\bf{r}}}\left( {{\Gamma}\left( {{{\bf{r}}_{{\rm{Rx}}}},{\bf{r}},{{\bf{r}}_{{\rm{Tx}}}}} \right)} \right) = {\bf{\hat y}}\left( {{{d{\Gamma }\left({{{\bf{r}}_{{\rm{Rx}}}},y,{{\bf{r}}_{{\rm{Tx}}}}} \right)} \mathord{\left/ {\vphantom {{d{\Gamma _{{\rm{ref}},x}}\left( {{{\bf{r}}_{{\rm{Rx}}}},y,{{\bf{r}}_{{\rm{Tx}}}}} \right)} {dy}}} \right. \kern-\nulldelimiterspace} {dy}}} \right)$, and ${\bf{\hat y}} \times {\bf{\hat x}} =  - {\bf{\hat z}}$.

The electromagnetic fields in \eqref{eq:incident-electric-field-plane-wave}, \eqref{eq:Hinc-plane-wave-general}, \eqref{eq:Einc-Eref-relationship-general}, and \eqref{eq:Href-at-RIS-explicit} with the constraint \eqref{eq:Helmholtz-reflected-electric-field} constitute a set of electromagnetically consistent equations for analyzing and optimizing RIS-assisted communications under the assumptions of the physical optics approximation.

\textbf{Plane wave spectrum and Floquet's expansion.} Before proceeding, it is instructive to note that the electric field reradiated from an RIS may be formulated in different manners. For example, (i) according to the plane-wave representation of an electromagnetic field \cite{Orfanidis_Book}, the reflected electric field may be represented as the sum of plane waves with different wavevectors; and (ii) according to Floquet's theorem for periodic structures \cite[Section 7.1]{Ishimaru_Book}, the reflected electric field may be represented as the sum of multiple diffracted modes with different wavevectors, similar to a diffraction grating. {Floquet's theorem, specifically, stipulates the following: If we illuminate an infinite-size and periodic surface with a plane wave, the reflected electromagnetic field can, in general, be formulated as the superposition of plane waves that propagate in different directions. The intensity and the direction of each plane wave depend on how the surface is realized. Further details about Floquet's theorem are given next.} In general terms, the following expression for the reradiated electric field (in the close proximity of the surface $\SS$) can be considered:
    \begin{align} \label{eq:GeneralReflection}
			{{\bf{E}}_{{\rm{ref}}}}\left( {{{\bf{r}}_{{\rm{Rx}}}},{\bf{r}},{{\bf{r}}_{{\rm{Tx}}}}} \right) \approx \sum\limits_n {{r_n}\left( {{{\bf{k}}_n}\left( {{{\bf{r}}_{{\rm{Tx}}}}} \right)} \right){E_{0,n}}{e^{ - j{{\bf{k}}_n}\left( {{{\bf{r}}_{{\rm{Tx}}}}} \right) \cdot {\bf{r}}}}{\bf{\hat x}}}
	\end{align}
where ${r_n}\left( {{{\bf{k}}_n}\left( {{{\bf{r}}_{{\rm{Tx}}}}} \right)} \right) = {\Gamma _{{\rm{ref}},n}}\left( {{{\bf{r}}_{{\rm{Rx}}}},{\bf{r}},{{\bf{r}}_{{\rm{Tx}}}};{{\bf{k}}_n}\left( {{{\bf{r}}_{{\rm{Tx}}}}} \right)} \right)$ denotes the field reflection coefficient of the $n$th reradiated mode for a given incident electromagnetic wave \cite{Sergei_MacroscopicModel}.

In general, ${r_n}\left( {{{\bf{k}}_n}\left( {{{\bf{r}}_{{\rm{Tx}}}}} \right)} \right)$ depends on the direction of the incident electromagnetic wave, i.e., it depends on ${{{\bf{k}}_n}\left( {{{\bf{r}}_{{\rm{Tx}}}}} \right)}$, which is determined by the location of the transmitter. Each term of Floquet's expansion in \eqref{eq:GeneralReflection} is a plane wave that propagates towards a specific direction. The corresponding magnetic fields can be calculated by using wave impedances that depend on the propagation angles of each plane wave component, which can be obtained from \eqref{eq:Href-at-RIS-explicit}. Based on Floquet's model, the Helmholtz equation is satisfied with equality and there is no need of imposing \eqref{eq:Helmholtz-reflected-electric-field}, which ensures that the surface parameters (i.e., the surface reflection coefficient) are slowly varying.

In the present paper, we consider the approximated formulation of the reflected electric field in \eqref{eq:Einc-Eref-relationship-general} subject to the constraint in \eqref{eq:Helmholtz-reflected-electric-field}, since it is the most widely used definition for the reflected electric field in wireless communications. The definition and methods described in the present paper apply, \textit{mutatis mutandis}, to \eqref{eq:GeneralReflection}. Before proceeding, however, it is instructive to study whether the analytical formulation in \eqref{eq:GeneralReflection} can be retrieved from \eqref{eq:Einc-Eref-relationship-general}. To this end, we can leverage Floquet's theorem under the assumption that the RIS is a periodic structure. This is elaborated in further text, after introducing the notion of surface impedance.

\subsubsection{Surface Impedance}
In the previous sub-section, the electromagnetic fields are formulated in terms of the field reflection coefficient $\Gamma_{\textup{ref}}(\vect{r}_{\textup{Rx}},\vect{r},\vect{r}_{\textup{Tx}})$. We have adopted this formulation because it is widely used in wireless communications. In order to get insights for system design, however, it is more convenient to design an RIS as a function of the surface impedance, which depends on the tangential components of the incident and reradiated electric and magnetic fields on the surface $\SS$, i.e., at $z=0^+$. By direct inspection of the surface impedance, it is relatively simple to identify the structural properties of the RIS and the corresponding options for its practical implementation. For example, it is possible to understand whether an RIS is lossless and whether it can be realized without using active components. This is elaborated next.

Before introducing the definition of surface impedance, we summarize the tangential components (evaluated at $z=0^+$) of the incident and reflected electric and magnetic fields. Let ${{\bf{F}}\left( {{{\bf{r}}_{{\rm{Rx}}}},{\bf{r}},{{\bf{r}}_{{\rm{Tx}}}}} \right)}$ be a generic vector field. The component of ${{\bf{F}}\left( {{{\bf{r}}_{{\rm{Rx}}}},{\bf{r}},{{\bf{r}}_{{\rm{Tx}}}}} \right)}$ that is tangential to the plane occupied by $\SS$ and that is evaluated at $z=0^+$ is defined as follows:
    \begin{align}
			{{\bf{F}}^t}\left( {{{\bf{r}}_{{\rm{Rx}}}},\left( {x,y} \right),{{\bf{r}}_{{\rm{Tx}}}}} \right) = {\left. {\left( {{\bf{\hat z}} \times {\bf{F}}\left( {{{\bf{r}}_{{\rm{Rx}}}},{\bf{r}},{{\bf{r}}_{{\rm{Tx}}}}} \right)} \right) \times {\bf{\hat z}}} \right|_{z = 0^+}}
	\end{align}

From \eqref{eq:incident-electric-field-plane-wave}, \eqref{eq:Hinc-plane-wave-general}, \eqref{eq:Einc-Eref-relationship-general}, and \eqref{eq:Href-at-RIS-explicit}, the tangential components of the incident and reflected electric and magnetic fields are:
    \begin{align}
			&\vect{E}^t_{\textup{inc}}(y,\vect{r}_{\textup{Tx}})
			= {{\mathcal{E}}^i}\left( {y,{{\bf{r}}_{{\rm{Tx}}}}} \right) \hat{\vect{x}}
			\label{eq:Einc-tangential-planewave-at-RIS}
			\\
			&\vect{H}^t_{\textup{inc}}(y,\vect{r}_{\textup{Tx}})
			= -\frac{\cos\theta_i(\vect{r}_{\textup{Tx}})}{\eta_0} {{\mathcal{E}}^i}\left( {y,{{\bf{r}}_{{\rm{Tx}}}}} \right) \hat{\vect{y}} \label{eq:Hinc-tangential-planewave-at-RIS}
			\\
			&\vect{E}^t_{\textup{ref}}(\vect{r}_{\textup{Rx}},y,\vect{r}_{\textup{Tx}})
			= \Gamma_{\SS}(\vect{r}_{\textup{Rx}},y,\vect{r}_{\textup{Tx}}) {{\mathcal{E}}^i}\left( {y,{{\bf{r}}_{{\rm{Tx}}}}} \right)\hat{\vect{x}} \label{eq:Eref-tangential-at-RIS}\\
			&\vect{H}^t_{\textup{ref}}(\vect{r}_{\textup{Rx}},y,\vect{r}_{\textup{Tx}})
			= \frac{\cos\theta_r(\vect{r}_{\textup{Rx}})}{\eta_0} \Gamma_{\SS}(\vect{r}_{\textup{Rx}},y,\vect{r}_{\textup{Tx}}) {{\mathcal{E}}^i}\left( {y,{{\bf{r}}_{{\rm{Tx}}}}} \right) \hat{\vect{y}} \label{eq:Href-tangential-planewave-at-RIS}
		\end{align}
where $\Gamma_{\SS}(\vect{r}_{\textup{Rx}},y,\vect{r}_{\textup{Tx}}) = \Gamma_{\textup{ref}}(\vect{r}_{\textup{Rx}},(y,z=0^+),\vect{r}_{\textup{Tx}})$ and ${{\mathcal{E}}^i}\left( {y,{{\bf{r}}_{{\rm{Tx}}}}} \right) =
{E}^i_{x,0}e^{-jk\sin\theta_i(\vect{r}_{\textup{Tx}})y}$. Similarly, we use the notation $\Phi_{\SS}(\vect{r}_{\textup{Rx}},y,\vect{r}_{\textup{Tx}}) = \Phi(\vect{r}_{\textup{Rx}},(y,z=0^+),\vect{r}_{\textup{Tx}})$. The electromagnetic fields in \eqref{eq:Einc-tangential-planewave-at-RIS}-\eqref{eq:Href-tangential-planewave-at-RIS} are referred to as surface electromagnetic fields, and $\Gamma_{\SS}(\vect{r}_{\textup{Rx}},y,\vect{r}_{\textup{Tx}})$ is the surface reflection coefficient.

As mentioned, \eqref{eq:Eref-tangential-at-RIS} and \eqref{eq:Href-tangential-planewave-at-RIS} are applicable under the assumption that the inequality constraint in \eqref{eq:Helmholtz-reflected-electric-field} is satisfied, i.e., the surface reflection coefficient is slowly varying, at the wavelength scale, along the surface $\SS$.  We note, in addition, that the reflected electric field and the reflected magnetic field are closely related to those of a plane wave that propagates towards the direction $\theta_r(\vect{r}_{\textup{Rx}})$. This is due to the assumption of plane wave for the incident electromagnetic wave and to the definition of the field reflection coefficient in \eqref{eq:reflection-coefficient-explicit-at-RIS-complete}, which results in the term $\cos\theta_r(\vect{r}_{\textup{Rx}})$ in \eqref{eq:Href-tangential-planewave-at-RIS}. In general, this latter term depends on the specific expression of the function $g(y,z)$ in \eqref{eq:Href-at-RIS-explicit}. {More precisely, the reflected electric field in \eqref{eq:Eref-tangential-at-RIS} can be explicitly written as $\vect{E}^t_{\textup{ref}}(\vect{r}_{\textup{Rx}},y,\vect{r}_{\textup{Tx}}) = {E}^i_{x,0} R(\vect{r}_{\textup{Rx}},y,\vect{r}_{\textup{Tx}})e^{j\phi(\vect{r}_{\textup{Rx}},y,\vect{r}_{\textup{Tx}})} e^{-jk\sin\theta_r(\vect{r}_{\textup{Rx}})y}$, which is an electromagnetic wave that propagates towards the direction $\theta_r(\vect{r}_{\textup{Rx}})$. If $R(\vect{r}_{\textup{Rx}},y,\vect{r}_{\textup{Tx}})$ is independent of $y$, $\vect{E}^t_{\textup{ref}}(\vect{r}_{\textup{Rx}},y,\vect{r}_{\textup{Tx}})$ is a plane wave \cite[Eq. (13)]{MDR_JSAC}. Locally (i.e., at any point $y$ or, in general, $\vect{s}$), in addition, the reflected electric and magnetic fields in \eqref{eq:Eref-tangential-at-RIS} and \eqref{eq:Href-tangential-planewave-at-RIS} are in agreement with the local non-specular reflection model illustrated in Fig. \ref{fig:SheetModel}. Provided that the inequality constraint in \eqref{eq:Helmholtz-reflected-electric-field} is fulfilled, therefore, the analytical formulation in \eqref{eq:Einc-tangential-planewave-at-RIS}-\eqref{eq:Href-tangential-planewave-at-RIS} subsumes and generalizes the typical analytical formulation in terms of plane waves in, e.g., \cite[Eq. (13)]{MDR_JSAC}, thanks to the amplitude and phase terms $R(\vect{r}_{\textup{Rx}},y,\vect{r}_{\textup{Tx}})$ and $\phi(\vect{r}_{\textup{Rx}},y,\vect{r}_{\textup{Tx}})$, respectively. The role played by $R(\vect{r}_{\textup{Rx}},y,\vect{r}_{\textup{Tx}})$ and $\phi(\vect{r}_{\textup{Rx}},y,\vect{r}_{\textup{Tx}})$ is discussed next, when describing the global design for aperiodic and finite-size RISs.}

By assuming that the RIS does not alter the polarization of the incident electromagnetic wave, the surface impedance, $Z(\vect{r}_{\textup{Rx}},y,\vect{r}_{\textup{Tx}})$, is defined as follows \cite[Sec. 2.4.3]{SEM_Book}:
	\begin{align}\label{eq:surface-impedance-definition}	\vect{E}^t_{\textup{tot}}(\vect{r}_{\textup{Rx}},y,\vect{r}_{\textup{Tx}}) = Z(\vect{r}_{\textup{Rx}},y,\vect{r}_{\textup{Tx}}) \bigg(\hat{\vect{z}} \times \vect{H}^t_{\textup{tot}}(\vect{r}_{\textup{Rx}},y,\vect{r}_{\textup{Tx}})\bigg)
	\end{align}
where the following total surface fields are introduced:
    \begin{align}
			&\vect{E}^t_{\textup{tot}}(\vect{r}_{\textup{Rx}},y,\vect{r}_{\textup{Tx}}) = \vect{E}^t_{\textup{inc}}(y,\vect{r}_{\textup{Tx}}) + \vect{E}^t_{\textup{ref}}(\vect{r}_{\textup{Rx}},y,\vect{r}_{\textup{Tx}}) \label{eq:surface-field-total-E}
			\\
			&\vect{H}^t_{\textup{tot}}(\vect{r}_{\textup{Rx}},y,\vect{r}_{\textup{Tx}}) = \vect{H}^t_{\textup{inc}}(y,\vect{r}_{\textup{Tx}}) + \vect{H}^t_{\textup{ref}}(\vect{r}_{\textup{Rx}},y,\vect{r}_{\textup{Tx}}) \label{eq:surface-field-total-H}
		\end{align}

Therefore, $Z(\vect{r}_{\textup{Rx}},y,\vect{r}_{\textup{Tx}})$ can be formulated as follows:
	\begin{align}\label{eq:surface-impedance-explicit}
		Z(\vect{r}_{\textup{Rx}},y,\vect{r}_{\textup{Tx}}) = \eta_0 \frac{1 + \Gamma_{\SS}(\vect{r}_{\textup{Rx}},y,\vect{r}_{\textup{Tx}})}{\cos \theta_i(\vect{r}_{\textup{Tx}}) - \Gamma_{\SS}(\vect{r}_{\textup{Rx}},y,\vect{r}_{\textup{Tx}}) \cos \theta_r (\vect{r}_{\textup{Rx}})}
	\end{align}

The analytical formulation of the surface impedance $Z(\vect{r}_{\textup{Rx}},y,\vect{r}_{\textup{Tx}})$ in \eqref{eq:surface-impedance-explicit} is consistent with the general definition of an RIS as an inhomogeneous sheet given in \eqref{GSTC_E} and \eqref{GSTC_H}, which fulfills the generalized sheet transition conditions. The only difference is that the transmitted fields are assumed to be equal to zero in the considered example, i.e., the RIS is assumed to be impenetrable and is a purely reflecting surface. In this case, the RIS can be described only through an electric surface impedance boundary (i.e., \eqref{eq:surface-impedance-explicit}) and the magnetic surface admittance is redundant.
	
By direct inspection and analysis of $Z(\vect{r}_{\textup{Rx}},y,\vect{r}_{\textup{Tx}})$, it is possible to draw important conclusions on the inherent features of an RIS and on the associated implementation requirements for realizing specified wave transformations. For example:
\begin{itemize}
    \item An RIS is locally passive if $\Re(Z(\vect{r}_{\textup{Rx}},y,\vect{r}_{\textup{Tx}})) > 0$;
    \item An RIS is locally active if $\Re(Z(\vect{r}_{\textup{Rx}},y,\vect{r}_{\textup{Tx}})) < 0$;
    \item An RIS is locally capacitive if $\Im(Z(\vect{r}_{\textup{Rx}},y,\vect{r}_{\textup{Tx}})) < 0$;
    \item An RIS is locally inductive if $\Im(Z(\vect{r}_{\textup{Rx}},y,\vect{r}_{\textup{Tx}})) > 0$.
\end{itemize}

In the next sub-sections, we formally introduce the criteria for the locally-optimum and the globally-optimum designs of an RIS as a function of \eqref{eq:surface-impedance-explicit}, and formulate the corresponding optimization problems in order to obtain the optimal surface impedance. Therefore, it is convenient to reformulate the Helmholtz constraint in \eqref{eq:Helmholtz-reflected-electric-field} in terms of the surface impedance $Z(\vect{r}_{\textup{Rx}},y,\vect{r}_{\textup{Tx}})$. To this end, we first express $Z(\vect{r}_{\textup{Rx}},y,\vect{r}_{\textup{Tx}})$ in terms of $\Gamma_{\SS}(\vect{r}_{\textup{Rx}},y,\vect{r}_{\textup{Tx}})$. By inverting \eqref{eq:surface-impedance-explicit}, we obtain:
	\begin{align}\label{eq:reflection-coefficient-in-surface-impedance}
		\Gamma_{\SS}(\vect{r}_{\textup{Rx}},y,\vect{r}_{\textup{Tx}})
		= \frac{Z(\vect{r}_{\textup{Rx}},y,\vect{r}_{\textup{Tx}}) \cos\theta_i(\vect{r}_{\textup{Tx}}) - \eta_0}{Z(\vect{r}_{\textup{Rx}},y,\vect{r}_{\textup{Tx}}) \cos\theta_r(\vect{r}_{\textup{Rx}}) + \eta_0}
	\end{align}
	
From \eqref{eq:reflection-coefficient-explicit-at-RIS-complete}, in addition, we have:
\begin{align}\label{eq:GammaS_Gamma}
		\Gamma(\vect{r}_{\textup{Rx}},y,\vect{r}_{\textup{Tx}})= \Gamma_{\SS}(\vect{r}_{\textup{Rx}},y,\vect{r}_{\textup{Tx}}) e^{jk\left(\sin\theta_r(\vect{r}_{\textup{Rx}}) - \sin(\theta_i(\vect{r}_{\textup{Tx}}))\right)y}
	\end{align}

By first inserting \eqref{eq:reflection-coefficient-in-surface-impedance} in \eqref{eq:GammaS_Gamma} and then plugging the resulting analytical expression in \eqref{eq:Helmholtz-reflected-electric-field}, the Helmholtz constraint can be equivalently rewritten, as a function of the surface impedance $Z(\vect{r}_{\textup{Rx}},y,\vect{r}_{\textup{Tx}})$, as follows:
\begin{align}\label{eq:Helmholtz-condition-reflection-approximate-3}
& \bigg|\frac{d^2}{d y^2} \bigg(\frac{Z^-(y)}{Z^+(y)}e^{jk\left(\sin\theta_r(\vect{r}_{\textup{Rx}}) - \sin(\theta_i(\vect{r}_{\textup{Tx}}))\right)y}\bigg) \nonumber\\
&- 2jk \sin\theta_r(\vect{r}_{\textup{Rx}})\frac{d}{d y}\bigg(\frac{Z^-(y)}{Z^+(y)}e^{jk\left(\sin\theta_r(\vect{r}_{\textup{Rx}}) - \sin(\theta_i(\vect{r}_{\textup{Tx}}))\right)y}\bigg) \bigg| \nonumber\\
&\ll k^2 \left|\frac{Z^-(y)}{Z^+(y)}\right|
		\end{align}
where the following auxiliary functions are defined:
   \begin{align}
			& Z^-(y) = Z(\vect{r}_{\textup{Rx}},y,\vect{r}_{\textup{Tx}})\cos\theta_i(\vect{r}_{\textup{Rx}}) - \eta_0 \nonumber \\
			& Z^+(y) = Z(\vect{r}_{\textup{Rx}},y,\vect{r}_{\textup{Tx}})\cos\theta_r(\vect{r}_{\textup{Rx}}) + \eta_0
		\end{align}

Similar to \eqref{eq:Helmholtz-reflected-electric-field}, the inequality in \eqref{eq:Helmholtz-condition-reflection-approximate-3} implies that the surface impedance is slowly varying, at the wavelength scale, along the surface $\SS$.

\textbf{Field vs. load reflection coefficient.} Before proceeding, it is instructive to analyze the difference between the surface reflection coefficient $\Gamma_{\SS}(\vect{r}_{\textup{Rx}},y,\vect{r}_{\textup{Tx}})$ and the so-called load reflection coefficient, $\Gamma_{\textup{load}}(\vect{r}_{\textup{Rx}},y,\vect{r}_{\textup{Tx}})$, that it is often considered when designing an RIS \cite{MDR_JSAC}, \cite{Sergei_MacroscopicModel}. The load reflection coefficient is defined as follows:
\begin{align}\label{eq:load-reflection-coefficient}
		\Gamma_{\textup{load}}(\vect{r}_{\textup{Rx}},y,\vect{r}_{\textup{Tx}})
		= \frac{Z(\vect{r}_{\textup{Rx}},y,\vect{r}_{\textup{Tx}}) \cos\theta_i(\vect{r}_{\textup{Tx}}) - \eta_0}{Z(\vect{r}_{\textup{Rx}},y,\vect{r}_{\textup{Tx}}) \cos\theta_i(\vect{r}_{\textup{Rx}}) + \eta_0}
	\end{align}

The two reflection coefficients $\Gamma_{\SS}(\vect{r}_{\textup{Rx}},y,\vect{r}_{\textup{Tx}})$ in \eqref{eq:reflection-coefficient-in-surface-impedance} and $\Gamma_{\textup{load}}(\vect{r}_{\textup{Rx}},y,\vect{r}_{\textup{Tx}})$ in \eqref{eq:load-reflection-coefficient} are not independent of each other, but they are not exactly the same. {From the physical point of view, the difference between $\Gamma_{\textup{load}}(\vect{r}_{\textup{Rx}},y,\vect{r}_{\textup{Tx}})$ and $\Gamma_{\SS}(\vect{r}_{\textup{Rx}},y,\vect{r}_{\textup{Tx}})$ is illustrated in Fig. \ref{fig:LocalModel} and Fig. \ref{fig:SheetModel}, respectively. Specifically, the load reflection coefficient $\Gamma_{\textup{load}}(\vect{r}_{\textup{Rx}},y,\vect{r}_{\textup{Tx}})$ is defined by assuming that the RIS operates locally, i.e., at every point $y$, as a local specular reflector. The surface reflection coefficient $\Gamma_{\SS}(\vect{r}_{\textup{Rx}},y,\vect{r}_{\textup{Tx}})$ is, on the other hand, defined by assuming that the RIS operates locally as a local non-specular reflector.} In mathematical terms, the difference is apparent by comparing the denominators of $\Gamma_{\textup{load}}(\vect{r}_{\textup{Rx}},y,\vect{r}_{\textup{Tx}})$ and  $\Gamma_{\SS}(\vect{r}_{\textup{Rx}},y,\vect{r}_{\textup{Tx}})$, and by noting that $\cos\theta_i(\vect{r}_{\textup{Rx}})$ is present in \eqref{eq:reflection-coefficient-in-surface-impedance} and $\cos\theta_r(\vect{r}_{\textup{Rx}})$ is present in \eqref{eq:load-reflection-coefficient}. Notably, the field and load reflection coefficients are identical if specular reflection, i.e., ${{\theta_r(\vect{r}_{\textup{Rx}})}} = \theta _i(\vect{r}_{\textup{Rx}})$, is considered. The minor difference in the denominators of \eqref{eq:reflection-coefficient-in-surface-impedance} and \eqref{eq:load-reflection-coefficient} results in major differences in the properties of the two reflection coefficients. Let us assume, for example, that the real part of the surface impedance is positive, i.e., $\Re(Z(\vect{r}_{\textup{Rx}},y,\vect{r}_{\textup{Tx}})) > 0$. In this case, $\left| {\Gamma_{\textup{load}}(\vect{r}_{\textup{Rx}},y,\vect{r}_{\textup{Tx}})} \right| < 1$. In particular, $\left| {\Gamma_{\textup{load}}(\vect{r}_{\textup{Rx}},y,\vect{r}_{\textup{Tx}})} \right| = 1$ if and only if $\Re(Z(\vect{r}_{\textup{Rx}},y,\vect{r}_{\textup{Tx}})) = 0$. In other words, the load reflection coefficient has an amplitude that is locally equal to one if and only if the surface impedance is locally reactive, i.e., the RIS is lossless. This is, however, not necessarily true for $\Gamma_{\SS}(\vect{r}_{\textup{Rx}},y,\vect{r}_{\textup{Tx}})$, since its amplitude depends on the relation between the angles of incidence and reflection.

By direct inspection of \eqref{eq:surface-impedance-explicit}, specifically, we obtain that $\left|\Gamma_{\SS}(\vect{r}_{\textup{Rx}},y,\vect{r}_{\textup{Tx}})\right| \leq 1$ if and only if the following condition is fulfilled:
\begin{align}\label{eq:abs-reflection-coefficicient-less-than-1-condition-rewritten}
		& \hspace{-0.35cm} \left|\Gamma_{\SS}(\vect{r}_{\textup{Rx}},y,\vect{r}_{\textup{Tx}})\right| \leq 1
		\iff \nonumber \\
		& \hspace{0.5cm} \frac{\cos\theta_i(\vect{r}_{\textup{Tx}}) - \cos\theta_r(\vect{r}_{\textup{Rx}})}{2\eta_0} \leq \frac{\Re\left(Z(\vect{r}_{\textup{Rx}},y,\vect{r}_{\textup{Tx}})\right)}{\left|Z(\vect{r}_{\textup{Rx}},y,\vect{r}_{\textup{Tx}})\right|^2}
	\end{align}
	
From \eqref{eq:abs-reflection-coefficicient-less-than-1-condition-rewritten}, therefore, we evince that the condition $\Re(Z(\vect{r}_{\textup{Rx}},y,\vect{r}_{\textup{Tx}})) > 0$ is not sufficient for ensuring $\left|\Gamma_{\SS}(\vect{r}_{\textup{Rx}},y,\vect{r}_{\textup{Tx}})\right| \leq 1$. The angle of incidence and the desired angle of reflection cannot be ignored. If the angle of incidence is zero, i.e., $\theta_i(\vect{r}_{\textup{Tx}})=0$, as is often assumed, the condition $\Re(Z(\vect{r}_{\textup{Rx}},y,\vect{r}_{\textup{Tx}})) > 0$ is sufficient for ensuring $\left|\Gamma_{\SS}(\vect{r}_{\textup{Rx}},y,\vect{r}_{\textup{Tx}})\right| \leq 1$.

More specifically, let us analyze the design constraints to be imposed on the surface impedance $Z(\vect{r}_{\textup{Rx}},y,\vect{r}_{\textup{Tx}})$ for ensuring $\left|\Gamma_{\SS}(\vect{r}_{\textup{Rx}},y,\vect{r}_{\textup{Tx}})\right| = 1$. In this case, the inequalities in \eqref{eq:abs-reflection-coefficicient-less-than-1-condition-rewritten} are replaced by equalities. Therefore, by direct inspection of \eqref{eq:abs-reflection-coefficicient-less-than-1-condition-rewritten}, the following design guidelines are drawn:
\begin{itemize}
    \item If $\theta_i(\vect{r}_{\textup{Tx}}) = \theta_r(\vect{r}_{\textup{Rx}})$, we obtain $\Re\left(Z(\vect{r}_{\textup{Rx}},y,\vect{r}_{\textup{Tx}})\right) = 0$. This implies that the surface impedance is purely reactive and that the RIS is lossless.
    \item If $\theta_i(\vect{r}_{\textup{Tx}}) > \theta_r(\vect{r}_{\textup{Rx}})$, we obtain $\Re\left(Z(\vect{r}_{\textup{Rx}},y,\vect{r}_{\textup{Tx}})\right) < 0$. This implies that the RIS is locally active.
    \item If $\theta_i(\vect{r}_{\textup{Tx}}) < \theta_r(\vect{r}_{\textup{Rx}})$, we obtain $\Re\left(Z(\vect{r}_{\textup{Rx}},y,\vect{r}_{\textup{Tx}})\right) > 0$. This implies that the RIS is locally passive.
\end{itemize}

In order to have a field reflection coefficient with unit amplitude at a given point $y$ on the surface $\SS$, the obtained findings allow us to conclude that the real part of the surface impedance of the RIS may be negative or positive depending on the angles of incidence and reflection. This implies that the RIS needs to introduce local power amplifications (a negative resistance is equivalent to an amplification) or local power losses along the surface $\SS$, depending on the desired angle of reflection for a given angle of incidence. By assuming, e.g., normal incidence (i.e., $\theta_i(\vect{r}_{\textup{Tx}})=0$), we evince that a reflection coefficient with unit amplitude corresponds to an engineered surface $\SS$ with a positive surface impedance, which implies that no power amplifiers or other sophisticated methods for creating virtual power amplifications through the use of, e.g., surface waves \cite{Sergei_RealizedSurface}, \cite{Do-Hoon_2021}, \cite{Eleftheriades_Arbitrary2021}, are needed to realize the RIS. The design constraints to be imposed on $Z(\vect{r}_{\textup{Rx}},y,\vect{r}_{\textup{Tx}}))$ for ensuring that the RIS has a high power efficiency for any pair $(\theta_i(\vect{r}_{\textup{Tx}}), \theta_r(\vect{r}_{\textup{Rx}}))$ are discussed in the next sub-sections.

In the rest of this sub-section, we discuss the structural properties of an RIS and optimize its operation as a function of the surface impedance $Z(\vect{r}_{\textup{Rx}},y,\vect{r}_{\textup{Tx}})$. Whenever necessary to gain engineering insights onto the design of RISs, we will refer to $\Gamma_{\SS}(\vect{r}_{\textup{Rx}},y,\vect{r}_{\textup{Tx}})$ in \eqref{eq:reflection-coefficient-in-surface-impedance}, i.e., the surface reflection coefficient, as well, since it is directly related to the electric and magnetic fields defined in \eqref{eq:Einc-Eref-relationship-general} and, therefore, to Maxwell's equations. Also, $\Gamma_{\SS}(\vect{r}_{\textup{Rx}},y,\vect{r}_{\textup{Tx}})$ is the most commonly utilized representation for an RIS in wireless communications \cite{MDR_JSAC}, \cite{Rui_Tutorial}.

\textbf{Surface reflection coefficient and Floquet’s theorem}
As mentioned, it is instructive to analyze whether and how the analytical formulation in \eqref{eq:GeneralReflection} can be retrieved from \eqref{eq:Einc-Eref-relationship-general}. We illustrate whether this is possible with an example. Let us assume that the RIS is a periodic structure of infinite size. The periodicity is usually determined by the surface reflection coefficient $\Gamma_{\SS}(\vect{r}_{\textup{Rx}},y,\vect{r}_{\textup{Tx}})$ and the surface impedance $Z(\vect{r}_{\textup{Rx}},y,\vect{r}_{\textup{Tx}})$. Specifically, an RIS is periodic with period $\mathcal{P}$ if $\Gamma_{\SS}(\vect{r}_{\textup{Rx}},y,\vect{r}_{\textup{Tx}}) = \Gamma_{\SS}(\vect{r}_{\textup{Rx}},y+\mathcal{P},\vect{r}_{\textup{Tx}})$ $\forall y$  or, equivalently, $Z(\vect{r}_{\textup{Rx}},y,\vect{r}_{\textup{Tx}}) = Z(\vect{r}_{\textup{Rx}},y+\mathcal{P},\vect{r}_{\textup{Tx}})$ $\forall y$. For simplicity, we limit ourselves to analyze $\Gamma_{\SS}(\vect{r}_{\textup{Rx}},y,\vect{r}_{\textup{Tx}})$.

First of all, let us analyze whether the reflected electric field in \eqref{eq:Eref-tangential-at-RIS} is an electromagnetically consistent solution for an electric field in a generic periodic RIS structure. Floquet’s theorem states that any electromagnetic field evaluated at a point $y$ in an infinite periodic structure with period $\mathcal{P}$ differs from the field evaluated one period $\mathcal{P}$ away from it only by a complex constant \cite[Section 7.1]{Ishimaru_Book}. The reflected electric field in \eqref{eq:Eref-tangential-at-RIS} fulfills this condition if $\Gamma_{\SS}(\vect{r}_{\textup{Rx}},y,\vect{r}_{\textup{Tx}}) = \Gamma_{\SS}(\vect{r}_{\textup{Rx}},y+\mathcal{P},\vect{r}_{\textup{Tx}})$ $\forall y$. Specifically, we have the following:
\begin{align} \label{eq:Floquet_1}
&E_{{\rm{ref}},x}^t\left( {{{\vect{r}}_{{\textup{Rx}}}},y + {\mathcal{P}},{{\vect{r}}_{{\textup{Tx}}}}} \right) \\ & \hspace{0.25cm}= {\Gamma _{\mathcal{S}}}\left( {{{\vect{r}}_{{\textup{Rx}}}},y + {\mathcal{P}},{{\vect{r}}_{{\textup{Tx}}}}} \right)E_{x,0}^i{e^{ - jk\sin {\theta _i}\left( {{{\vect{r}}_{{\textup{Tx}}}}} \right)\left( {y + {\mathcal{P}}} \right)}} \nonumber \\
& \hspace{0.25cm} \mathop  = \limits^{\left( a \right)} {\Gamma _{\mathcal{S}}}\left( {{{\vect{r}}_{{\textup{Rx}}}},y,{{\vect{r}}_{{\textup{Tx}}}}} \right)E_{x,0}^i{e^{ - jk\sin {\theta _i}\left( {{{\vect{r}}_{{\textup{Tx}}}}} \right)y}}{e^{ - jk\sin {\theta _i}\left( {{{\vect{r}}_{{\textup{Tx}}}}} \right){\mathcal{P}}}} \nonumber \\
& \hspace{0.25cm} \mathop  = \limits^{\left( b \right)} E_{{\rm{ref}},x}^t\left( {{{\vect{r}}_{{\textup{Rx}}}},y,{{\vect{r}}_{{\textup{Tx}}}}} \right){e^{ - jk\sin {\theta _i}\left( {{{\vect{r}}_{{\textup{Tx}}}}} \right){\mathcal{P}}}} \nonumber
\end{align}
where $(a)$ follows from $\Gamma_{\SS}(\vect{r}_{\textup{Rx}},y,\vect{r}_{\textup{Tx}}) = \Gamma_{\SS}(\vect{r}_{\textup{Rx}},y+\mathcal{P},\vect{r}_{\textup{Tx}})$ $\forall y$ and $(b)$ follows from \eqref{eq:Einc-Eref-relationship-general}.

From \eqref{eq:Floquet_1}, it follows that the reflected electric field fulfills the condition imposed by Floquet’s theorem. In particular, the difference between the electric field evaluated at two points that differ of one period is equal to ${e^{ - jk\sin {\theta _i}\left( {{{\vect{r}}_{{\textup{Tx}}}}} \right){\mathcal{P}}}}$.

If $\Gamma_{\SS}(\vect{r}_{\textup{Rx}},y,\vect{r}_{\textup{Tx}})$ is a periodic function, it can be expressed in terms of Fourier series. Specifically, we can write:
\begin{align} \label{eq:Floquet_2}
{\Gamma _{\mathcal{S}}}\left( {{{\vect{r}}_{{\textup{Rx}}}},y,{{\vect{r}}_{{\textup{Tx}}}}} \right) = \sum\limits_n {{\mu _n}{e^{ - j\frac{{2\pi n }}{{\mathcal{P}}}y}}}
\end{align}
where $\mu_n$ for $n=\ldots, -2, -1, 0, 1, 2, \ldots$ are the coefficients of the Fourier series, which depend on the specified angle of incidence of the electromagnetic wave and on the desired angle of reflection, i.e., they depend on the specific wave transformation that the RIS needs to realize.

With the aid of \eqref{eq:Floquet_2}, the reflected electric field in \eqref{eq:Eref-tangential-at-RIS} can be rewritten as follows:
\begin{align} \label{eq:Floquet_3pre}
& E_{{\rm{ref}},x}^t\left( {{{\vect{r}}_{{\textup{Rx}}}},y,{{\vect{r}}_{{\textup{Tx}}}}} \right) \\ & \hspace{1.0cm} = {\Gamma _{\mathcal{S}}}\left( {{{\vect{r}}_{{\textup{Rx}}}},y,{{\vect{r}}_{{\textup{Tx}}}}} \right)E_{x,0}^i{e^{ - jk\sin {\theta _i}\left( {{{\vect{r}}_{{\textup{Tx}}}}} \right)y}} \nonumber\\
& \hspace{1.0cm} = E_{x,0}^i\sum\limits_n {{\mu _n}{e^{ - jk\left( {\sin {\theta _i}\left( {{{\vect{r}}_{{\textup{Tx}}}}} \right) + \frac{\lambda }{{\mathcal{P}}}n} \right)y}}} \nonumber
\end{align}
which unveils that the reflected electric field is expressed as a summation of plane waves, similar to \eqref{eq:GeneralReflection}. This confirms that the analytical formulation in \eqref{eq:Einc-Eref-relationship-general} allows us to retrieve results that are consistent with infinite periodic structures in agreement with Floquet’s theorem. {It is interesting to evaluate the special case when $\Gamma(\vect{r}_{\textup{Rx}},y,\vect{r}_{\textup{Tx}})$ is a constant independent of $y$, i.e., $\Gamma(\vect{r}_{\textup{Rx}},y,\vect{r}_{\textup{Tx}})=\Gamma_0$. In this case, we obtain $\Gamma_{\SS}(\vect{r}_{\textup{Rx}},y,\vect{r}_{\textup{Tx}}) = \Gamma_0 e^{-jk\left(\sin\theta_r(\vect{r}_{\textup{Rx}}) - \sin(\theta_i(\vect{r}_{\textup{Tx}}))\right)y}$. Therefore, the Fourier series coincides with $\Gamma_{\SS}(\vect{r}_{\textup{Rx}},y,\vect{r}_{\textup{Tx}})$, which is the harmonic obtained by setting $n=1$, while all the other harmonics are equal to zero. This is consistent with \cite[Eq. (13)]{MDR_JSAC} in which a single plane wave is reradiated from an RIS.}

It is worth mentioning that the analytical formulation in \eqref{eq:Floquet_3pre} holds true for an arbitrary angle of incidence, even though the RIS is designed and optimized for the specified angle of incidence ${{\theta _i}\left( {{{\vect{r}}_{{\textup{Tx}}}}} \right)}$, which depends on the location of the transmitter of interest and determines the period $\mathcal{P}$ of the RIS. In other words, \eqref{eq:Floquet_3pre} can be written, in its most general formulation, as follows:
\begin{align} \label{eq:Floquet_3}
& E_{{\rm{ref}},x}^t\left( {{{\vect{r}}_{{\textup{Rx}}}},y,{{\vect{r}}_{{\textup{Tx}}}}} \right) \\ & \hspace{1.0cm} = E_{x,0}^i\sum\limits_n {{\mu _n}{e^{ - jk\left( {\sin {\theta _i} + \frac{\lambda }{{{\mathcal{P}}\left( {{\theta _i}\left( {{{\vect{r}}_{{\textup{Tx}}}}} \right),{\theta _r}\left( {{{\vect{r}}_{{\textup{Rx}}}}} \right)} \right)}}n} \right)y}}} \nonumber
\end{align}
where $\theta _i$ denotes a generic angle of incidence and, for the avoidance of doubt, we have made explicit that the period of the RIS depends on the angles of design ${{\theta _i}\left( {{{\vect{r}}_{{\textup{Tx}}}}} \right)}$ (the specified angle of incidence) and ${{\theta _r}\left( {{{\vect{r}}_{{\textup{Rx}}}}} \right)}$ (the desired angle of reflection), i.e., $\mathcal{P} = {{\mathcal{P}}\left( {{\theta _i}\left( {{{\vect{r}}_{{\textup{Tx}}}}} \right),{\theta _r}\left( {{{\vect{r}}_{{\textup{Rx}}}}} \right)} \right)}$.

Departing from the surface electric field in \eqref{eq:Floquet_3}, it is possible to obtain, with similar analytical steps, the electric field for $\tau_m < z < \tau_M$ and, by using \eqref{eq:Href-at-RIS-explicit}, the corresponding magnetic field that fulfills Maxwell's equations. This provides a complete representation of the electric and magnetic fields reradiated by an RIS according to Floquet's theorem \cite{Sergei_MacroscopicModel}.

The general analytical formulation in \eqref{eq:Floquet_3} is useful to characterize the electromagnetic field reradiated by an RIS when it is illuminated by an electromagnetic wave that originates from the direction ${{\theta _i}\left( {{{\vect{r}}_{{\textup{Tx}}}}} \right)}$, as well as the electromagnetic field reradiated by the RIS when it is illuminated by an interfering electromagnetic wave (including the multipath generated by scatterers that are not digitally controllable) that originates from any direction $\theta_i \ne {{\theta _i}\left( {{{\vect{r}}_{{\textup{Tx}}}}} \right)}$. Therefore, \eqref{eq:Floquet_3} is a relatively general formula that can be applied to any RIS of infinite size and with a periodical structure.

In particular, \eqref{eq:Floquet_3} unveils that the reflected electric field is given by the summation of plane waves whose tangential wave numbers are:
\begin{align} \label{eq:Floquet_4}
{k_{y,n}} = k\left( {\sin {\theta _i} + \frac{\lambda }{{\mathcal{P}}}n} \right), \; n= \ldots, -2, -1, 0, 1, 2, \ldots
\end{align}

Among the infinite number of plane waves (diffracted modes) in \eqref{eq:Floquet_4}, only those that fulfill the condition $k \ge \left| {{k_{y,n}}} \right|$ are propagating modes, while the others are evanescent modes, i.e., they are not observed in the far field region of the RIS microstructure \cite[Fig. 29]{MDR_JSAC}. As far as the propagating modes are concerned, specifically, their corresponding direction of reradiation is:
\begin{align} \label{eq:Floquet_5}
{\theta _{r,n}} = {\rm{arctan}}\left( {\frac{{{k_{y,n}}}}{{\sqrt {{k^2} - k_{y,n}^2} }}} \right)
\end{align}

To obtain further insights, let us analyze the case study $\theta_i = {{\theta _i}\left( {{{\vect{r}}_{{\textup{Tx}}}}} \right)} = 0$, i.e., the RIS is designed for reflecting a normally incident electromagnetic wave and the actual incident electromagnetic wave illuminates the RIS from the normal direction as well. Based on \eqref{eq:Floquet_4}, the condition that the propagating modes needs to fulfill reduces to:
\begin{align} \label{eq:Floquet_6}
\left| n \right| \le {{\mathcal{P}} \mathord{\left/ {\vphantom {{\mathcal{P}} \lambda }} \right. \kern-\nulldelimiterspace} \lambda }
\end{align}

Therefore, two case studies are worth of analysis.
\begin{itemize}
\item $\mathcal{P} > \lambda$: In this case, the RIS is characterized by multiple reflected modes.
\item $\mathcal{P} < \lambda$: In this case, the RIS is characterized by one single reflected mode, which corresponds to $n=0$.
\end{itemize}

If $\lambda < \mathcal{P} < 2\lambda$, in addition, the RIS is characterized by three main reflected modes, which correspond to $n= -1, 0, 1$. This is a typical case study that has been observed through experiments and full-wave simulations in several research works, e.g., \cite{Sergei_RealizedSurface}, \cite{Vittorio_RayTracing}, \cite{Sergei_MacroscopicModel}. Further comments will be given in Section IV when illustrating the numerical results.

\begin{figure}[!t]
		\begin{center}
			\includegraphics[width=\linewidth]{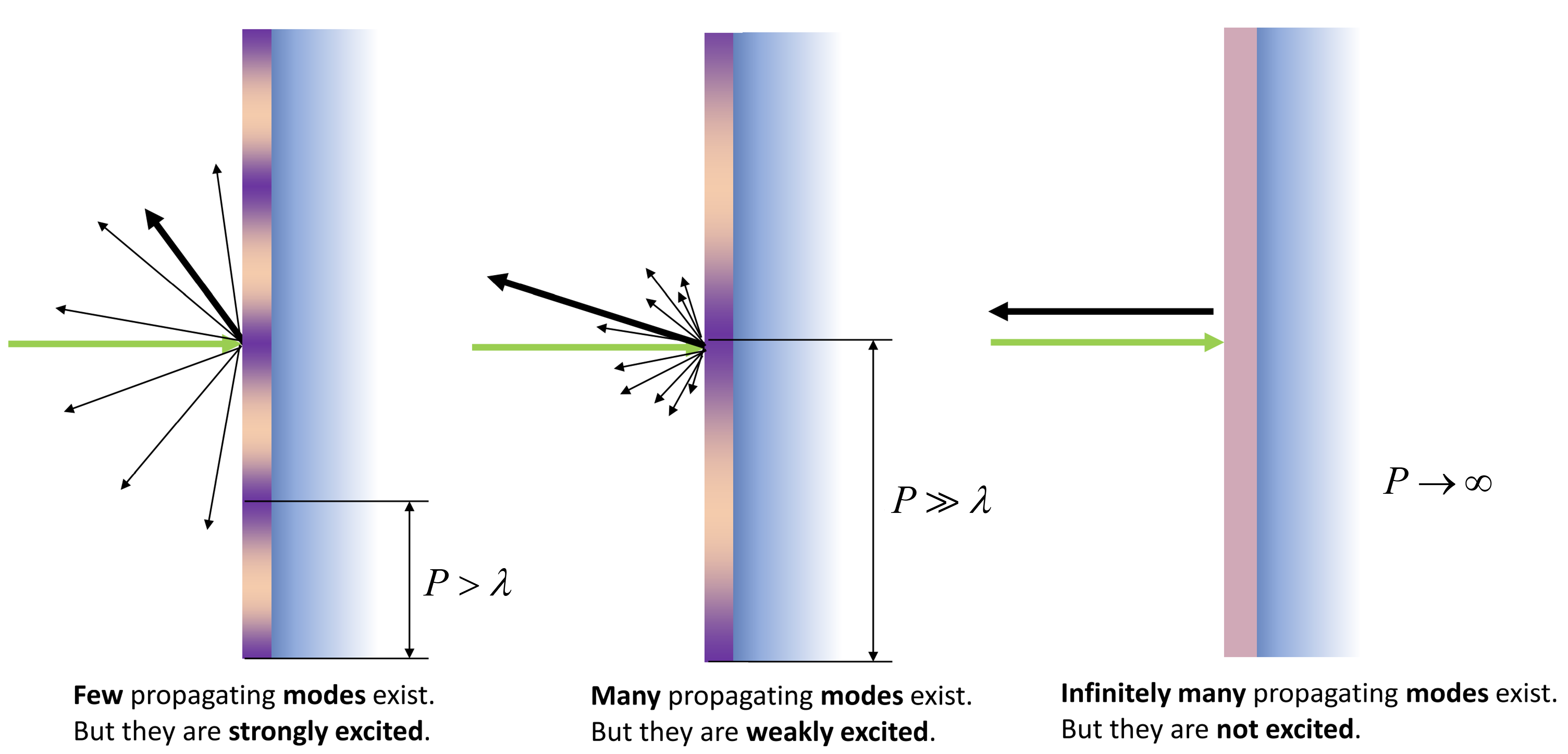}
			\caption{Illustration of the reradiated modes for periodic RISs based on Floquet’s theorem.}
			\label{fig:FloquetQualitative}
		\end{center}
	\end{figure}
For all the considered case studies, the amount of power that is reflected towards the direction of each propagating mode is determined by the complex coefficient $\mu_n$ of the Fourier series expansion of the surface reflection coefficient in \eqref{eq:Floquet_3}. By taking into account that the total reradiated power is distributed among the reradiated modes, a qualitative illustration of typical reradiation conditions for a periodic RIS is given in Fig. \ref{fig:FloquetQualitative}.  {Even though multiple reflected modes may exist based on \eqref{eq:Floquet_6}, it is important to emphasize that many of them may be weekly excited, i.e., the amount of power carried by them is negligible with respect to the amount of power carried by the other propagating modes. In other words, the reradiation properties of an RIS are jointly determined by the allowed propagating modes and by the amount of power that each of them carries, which is quantified by the Fourier coefficients $\left|\mu_n\right|$.}

{As a concrete case study, let us assume that the surface reflection coefficient is, according to \eqref{eq:reflection-coefficient-explicit-at-RIS-complete}, given by:
\begin{align} \label{eq:Floquet_7_1}
{\Gamma _{\mathcal{S}}}\left( {{{\vect{r}}_{{\textup{Rx}}}},y,{{\vect{r}}_{{\textup{Tx}}}}} \right) = \Gamma(\vect{r}_{\textup{Rx}},y,\vect{r}_{\textup{Tx}}) {e^{ - jk\left({\sin {\theta _r}\left( {{{\vect{r}}_{{\textup{Rx}}}}} \right)} - {\sin {\theta _i}\left( {{{\vect{r}}_{{\textup{Tx}}}}} \right)} \right)y}}
\end{align}
where $\Gamma(\vect{r}_{\textup{Rx}},y,\vect{r}_{\textup{Tx}})$ is assumed to be a periodic function with period $\mathcal{P}$, i.e., $\Gamma(\vect{r}_{\textup{Rx}},y,\vect{r}_{\textup{Tx}}) = \Gamma(\vect{r}_{\textup{Rx}},y+\mathcal{P},\vect{r}_{\textup{Tx}})$ $\forall y$. Specifically, we assume that the period $\mathcal{P}$ is equal to:
\begin{align} \label{eq:Floquet_7}
{\mathcal{P}} = \frac{\lambda }{{\left| {\sin {\theta _i}\left( {{{\vect{r}}_{{\textup{Tx}}}}} \right) - \sin {\theta _r}\left( {{{\vect{r}}_{{\textup{Rx}}}}} \right)} \right|}}
\end{align}}

{This implies that ${\Gamma _{\mathcal{S}}}\left( {{{\vect{r}}_{{\textup{Rx}}}},y,{{\vect{r}}_{{\textup{Tx}}}}} \right)$ is a periodic function with period $\mathcal{P}$ as well. For simplicity, let us assume again that ${{\theta _i}\left( {{{\vect{r}}_{{\textup{Tx}}}}} \right)} = 0$. Therefore, we obtain:
\begin{align} \label{eq:Floquet_7bis}
{\mathcal{P}} = \frac{\lambda }{{\sin {\theta _r}\left( {{{\vect{r}}_{{\textup{Rx}}}}} \right)}} \ge \lambda
\end{align}}

In this case, \eqref{eq:Floquet_6} simplifies to:
\begin{align} \label{eq:Floquet_8}
\left| n \right| \le \frac{1}{{\sin {\theta _r}\left( {{{\vect{r}}_{{\textup{Rx}}}}} \right)}}
\end{align}

From \eqref{eq:Floquet_8}, we evince that at most three modes (i.e., $n=-1, 0, 1$) may be reradiated from the RIS if ${{\theta _r}\left( {{{\vect{r}}_{{\textup{Rx}}}}} \right)}$ is close to $90^\circ$, which corresponds to the case study in which the desired angle of reradiation is very different from the angle of incidence assumed when designing the RIS. From \eqref{eq:Floquet_5}, in particular, the possible angles of reradiation are:
\begin{align} \label{eq:Floquet_9}
& {\theta _{r,n =  - 1}} =  - {\theta _r}\left( {{{\vect{r}}_{{\textup{Rx}}}}} \right) \nonumber\\
& {\theta _{r,n = 0}} = 0\\
& {\theta _{r,n =  + 1}} = {\theta _r}\left( {{{\vect{r}}_{{\textup{Rx}}}}} \right) \nonumber
\end{align}
{and their associated reradiated power is determined by the corresponding three coefficients of the Fourier series expansion of $\Gamma(\vect{r}_{\textup{Rx}},y,\vect{r}_{\textup{Tx}})$.}

From \eqref{eq:Floquet_9}, we conclude that, besides the desired angle of reradiation, we may observe two additional radiated modes towards the specular direction and towards the direction that is symmetric to the desired direction of reradiation. The amount of power that is radiated towards these three modes depends, however, on the specific design of the surface reflection coefficient and surface impedance. In fact, different functions may have the same period, but their corresponding Fourier series are usually different. This point is further clarified in Section IV with the aid of numerical results, where it is shown that surface impedances with almost identical periods result in different reradiated electromagnetic fields. {This is because they are characterized by different functions $\Gamma(\vect{r}_{\textup{Rx}},y,\vect{r}_{\textup{Tx}})$.}

From \eqref{eq:Floquet_4}, we observe, as mentioned, that the possible modes reradiated by an RIS depend on the actual angle of incidence of the electromagnetic waves, regardless of the angles of incidence and reflection for which the RIS is designed, which, however, determine the period $\mathcal{P}$ of the RIS. If the RIS operates in the presence of other (interfering) electromagnetic waves, this implies that these latter waves are reradiated towards directions that are specified by \eqref{eq:Floquet_5}. Therefore, the model for the reflected electric field considered in \eqref{eq:GeneralReflection} is not only consistent with Floquet’s theory if the RIS is periodic but it can be used to analyze both intended and interfering electromagnetic waves.

\subsection{Power Efficiency and Reradiated Power Flux}
In this sub-section, we introduce the concepts of power efficiency and reradiated power flux of an RIS as a function of the surface impedance $Z(\vect{r}_{\textup{Rx}},y,\vect{r}_{\textup{Tx}})$. For application to wireless communications and for completeness, we discuss the implications in terms of field reflection coefficient as well. For this analysis, we utilize the notion of Poynting vector \cite{Orfanidis_Book}. In particular, we consider two designs for an RIS that are referred to as (i) local design with unitary power efficiency according to a local power conservation principle, and (ii) global design with unitary power efficiency according to an average or global power conservation principle.

\subsubsection{Power Efficiency -- Surface Poynting Vector}
The power efficiency characterizes the amount of power that is reradiated towards the desired direction by an RIS, given the amount of incident power. The power efficiency can be deduced by analyzing the net power flow of an RIS in the close vicinity of the surface $\SS$ (i.e., at $z=0^+$). A typical design objective when optimizing an RIS is to engineer the surface impedance $Z(\vect{r}_{\textup{Rx}},y,\vect{r}_{\textup{Tx}})$ so that the power efficiency is unitary \cite{Sergei_PerfectReflector}. For a lossless RIS, this design criterion implies that the amount of power that is reradiated towards a specified direction of reflection coincides with the amount of incident power, i.e., the net power flow is equal to zero. In the present paper, we consider only lossless RISs. The net power flow of a lossless RIS can be defined either locally, i.e., for each individual point $\vect{s} \in \SS$, or globally, i.e., for the entire surface $\SS$. In this sub-section, we overview the local and global designs for lossless RISs, and discuss their properties, advantages, and limitations from the theoretical and implementation points of view.

The departing point to formulate local and global designs for a lossless RIS is the notion of surface Poynting vector. By definition, the normal (to the surface $\SS$) component of the surface Poynting vector evaluated at a point $\vect{s} \in \SS$ is defined as follows \cite{Orfanidis_Book}:
	\begin{align} \label{eq:Poynting-at-RIS}
		\vect{P}_{\SS}(\vect{r}_{\textup{Rx}},\vect{s},\vect{r}_{\textup{Tx}})
		= \frac{1}{2}\Re\left( \vect{E}_{\textup{tot}}^t(\vect{r}_{\textup{Rx}},\vect{s},\vect{r}_{\textup{Tx}}) \times \left( \vect{H}_{\textup{tot}}^t(\vect{r}_{\textup{Rx}},\vect{s},\vect{r}_{\textup{Tx}})\right)^*\right)
	\end{align}
where $(\cdot)^*$ denotes the complex conjugate operation. Similar to the preceding text, the total surface electric and magnetic fields in \eqref{eq:Poynting-at-RIS} are macroscopic electromagnetic fields that are averaged over the area of a unit cell.

The surface Poynting vector accounts for the interaction between the incident and the reradiated electromagnetic fields in the close vicinity of the surface $\SS$, since the total surface electric and magnetic fields defined in \eqref{eq:surface-field-total-E} and \eqref{eq:surface-field-total-H} are utilized in \eqref{eq:Poynting-at-RIS}. This is true regardless of whether the direct link is available or is blocked at the receiver Rx.

\paragraph{Local Design}
Based on $\vect{P}_{\SS}(\vect{r}_{\textup{Rx}},\vect{s},\vect{r}_{\textup{Tx}})$ in \eqref{eq:Poynting-at-RIS}, a lossless RIS is defined to be locally passive if the following condition holds true:
    \begin{align}\label{eq:locally-passive-condition-0}
			P_{\SS}(\vect{r}_{\textup{Rx}},\vect{s},\vect{r}_{\textup{Tx}}) =  \left| {\vect{P}_{\SS}(\vect{r}_{\textup{Rx}},\vect{s},\vect{r}_{\textup{Tx}})} \right| \leq 0 ~ \quad \forall \vect{s} \in \SS
	\end{align}

By inserting the total surface electric and magnetic fields in \eqref{eq:surface-field-total-E} and \eqref{eq:surface-field-total-H} into \eqref{eq:locally-passive-condition-0}, and by using \eqref{eq:Einc-tangential-planewave-at-RIS}-\eqref{eq:Href-tangential-planewave-at-RIS}, the surface Poynting vector evaluated at a point $y \in [-L_y,L_y]$ is formulated as follows:
		\begin{align}\label{eq:Poynting-at-RIS-amplitude-control}
			\vect{P}_{\SS}(\vect{r}_{\textup{Rx}},y,\vect{r}_{\textup{Tx}})
			&= \hat{\vect{z}}\frac{1}{2\eta_0}|{E}^i_{x,0}|^2  S(\vect{r}_{\textup{Rx}},y,\vect{r}_{\textup{Tx}})
		\end{align}
		where $S(\vect{r}_{\textup{Rx}},y,\vect{r}_{\textup{Tx}})$ for $y \in [-L_y,L_y]$ is defined as:
		\begin{align}
			S(\vect{r}_{\textup{Rx}},y,\vect{r}_{\textup{Tx}})
			&= \left|\Gamma_{\SS}(\vect{r}_{\textup{Rx}},y,\vect{r}_{\textup{Tx}})\right|^2 \cos \theta_r(\vect{r}_{\textup{Rx}}) - \cos \theta_i(\vect{r}_{\textup{Tx}}) \nonumber\\
			& \hspace{-1.5cm} + \Re\left(\Gamma_{\SS}(\vect{r}_{\textup{Rx}},y,\vect{r}_{\textup{Tx}})\right) (\cos \theta_r(\vect{r}_{\textup{Rx}}) - \cos \theta_i(\vect{r}_{\textup{Tx}})) \label{eq:powerflow-at-RIS-normalized-amplitude-control}
		\end{align}

Therefore, $P_{\SS}(\vect{r}_{\textup{Rx}},\vect{s},\vect{r}_{\textup{Tx}})$ in \eqref{eq:locally-passive-condition-0} can be written as follows:
\begin{align}\label{eq:Poynting-at-RIS-amplitude-control-scalar}
			P_{\SS}(\vect{r}_{\textup{Rx}},y,\vect{r}_{\textup{Tx}})
			= \frac{1}{2\eta_0}|{E}^i_{x,0}|^2  S(\vect{r}_{\textup{Rx}},y,\vect{r}_{\textup{Tx}})
		\end{align}

By using \eqref{eq:reflection-coefficient-in-surface-impedance}, $\vect{P}_{\SS}(\vect{r}_{\textup{Rx}},\vect{s},\vect{r}_{\textup{Tx}})$ can be (equivalently) formulated, in terms of the surface impedance $Z(\vect{r}_{\textup{Rx}},y,\vect{r}_{\textup{Tx}})$, as $\vect{P}_{\SS}(\vect{r}_{\textup{Rx}},y,\vect{r}_{\textup{Tx}}) = \hat{\vect{z}}{P}_{\SS}(\vect{r}_{\textup{Rx}},y,\vect{r}_{\textup{Tx}})$, where:
    \begin{align} \label{eq:powerflow-at-RIS-impedance}
			P_{\SS}(\vect{r}_{\textup{Rx}},y,\vect{r}_{\textup{Tx}})
			&= - \frac{\left|{E}^i_{x,0}\right|^2}{2} \left|\frac{ \cos\theta_i(\vect{r}_{\textup{Tx}}) + \cos\theta_r(\vect{r}_{\textup{Rx}})}{Z(\vect{r}_{\textup{Rx}},y,\vect{r}_{\textup{Tx}}) \cos\theta_r(\vect{r}_{\textup{Rx}}) + \eta_0}\right|^2 \nonumber \\
			&* \Re\left({Z(\vect{r}_{\textup{Rx}},y,\vect{r}_{\textup{Tx}})} \right)
		\end{align}

By definition of locally passive design according to \eqref{eq:locally-passive-condition-0}, we evince from \eqref{eq:powerflow-at-RIS-impedance} that a lossless RIS is locally passive if and only if $\Re\left({Z(\vect{r}_{\textup{Rx}},y,\vect{r}_{\textup{Tx}})}\right) \ge 0$ $\forall y \in [-L_y,L_y]$. This is consistent with the engineering insights that can be gained from the surface impedance introduced in the previous sub-section (see \eqref{eq:surface-impedance-explicit} and related comments). Notably, a lossless RIS has a locally unitary power efficiency if and only if $\Re\left({Z(\vect{r}_{\textup{Rx}},y,\vect{r}_{\textup{Tx}})}\right) = 0$ $\forall y \in [-L_y,L_y]$. Accordingly, the surface $\SS$ needs to be realized by using only reactive components, without using resistive elements. It is worth mentioning that the Helmholtz constraint in \eqref{eq:Helmholtz-condition-reflection-approximate-3} needs to be always fulfilled and then taken into consideration. {A simple example of local design with unit power efficiency that satisfies the Helmholtz constraint in \eqref{eq:Helmholtz-condition-reflection-approximate-3} with equality is when an RIS operates as a simple specular reflector by setting $\Gamma(\vect{r}_{\textup{Rx}},y,\vect{r}_{\textup{Tx}})=1$ $\forall y \in [-L_y,L_y]$ and $\cos \theta_r(\vect{r}_{\textup{Rx}}) = \cos \theta_i(\vect{r}_{\textup{Tx}})$.} In general, an infinite number of surface impedances whose real part is equal to zero and that fulfill the Helmholtz constraint in \eqref{eq:Helmholtz-condition-reflection-approximate-3} can be found, and each of them leads to a different observed power at the receiver Rx. This is further elaborated next.

To reap further engineering insights, especially in the context of wireless communication systems, it is instructive to analyze the design constraints that the surface reflection coefficient $\Gamma_{\SS}(\vect{r}_{\textup{Rx}},y,\vect{r}_{\textup{Tx}})$ needs to fulfill in order for an RIS to be locally passive, and, in particular, for a lossless RIS to have unitary power efficiency. By utilizing the definition of surface reflection coefficient in \eqref{eq:reflection-coefficient-complex-amplitude} as a function of the amplitude $R(\vect{r}_{\textup{Rx}},\vect{s},\vect{r}_{\textup{Tx}})$ and the phases $\phi(\vect{r}_{\textup{Rx}},\vect{s},\vect{r}_{\textup{Tx}})$ and $\Phi_{\SS}(\vect{r}_{\textup{Rx}},y,\vect{r}_{\textup{Tx}})$, $S(\vect{r}_{\textup{Rx}},y,\vect{r}_{\textup{Tx}})$ in \eqref{eq:powerflow-at-RIS-normalized-amplitude-control} can be reformulated as:
		\begin{align}\label{eq:S-with-local-amplitude-expanded}
			S(\vect{r}_{\textup{Rx}},y,\vect{r}_{\textup{Tx}})
			&= \left(R(\vect{r}_{\textup{Rx}},y,\vect{r}_{\textup{Tx}})\right)^2 \cos \theta_r(\vect{r}_{\textup{Rx}}) - \cos \theta_i(\vect{r}_{\textup{Tx}}) \nonumber \\
			& + R(\vect{r}_{\textup{Rx}},y,\vect{r}_{\textup{Tx}}) 	(\cos \theta_r(\vect{r}_{\textup{Rx}}) - \cos \theta_i(\vect{r}_{\textup{Tx}})) \nonumber \\
			&* \left(\cos(\phi(\vect{r}_{\textup{Rx}},y,\vect{r}_{\textup{Tx}}) + \Phi_{\SS}(\vect{r}_{\textup{Rx}},y,\vect{r}_{\textup{Tx}}))\right)
		\end{align}

Equation \eqref{eq:S-with-local-amplitude-expanded} indicates that $S(\vect{r}_{\textup{Rx}},y,\vect{r}_{\textup{Tx}})$ is a quadratic polynomial in the amplitude $R(\vect{r}_{\textup{Rx}},\vect{s},\vect{r}_{\textup{Tx}})$ of the field (and surface) reflection coefficient. Therefore, it can be readily analyzed. By definition, as mentioned, a locally lossless RIS has a locally  unitary power efficiency if and only if $S(\vect{r}_{\textup{Rx}},y,\vect{r}_{\textup{Tx}}) = 0$ $\forall y \in [-L_y,L_y]$. Based on \eqref{eq:S-with-local-amplitude-expanded}, and subject to Helmholtz's condition in \eqref{eq:Helmholtz-reflected-electric-field}, this leads to the design constraint:
    \begin{align}\label{eq:reflection-coefficient-amplitude-locally-passive-general}
R\left( {{{\bf{r}}_{{\rm{Rx}}}},y,{{\bf{r}}_{{\rm{Tx}}}}} \right) &= \frac{1}{2}\cos \left( {\Psi \left( y \right)} \right)\left( {{F_{i,r}} - 1} \right)\\
& + \frac{1}{2}\sqrt {{{\cos }^2}\left( {\Psi \left( y \right)} \right){{\left( {{F_{i,r}} - 1} \right)}^2} + 4{F_{i,r}}}  \nonumber
		\end{align}
where the following shorthand notation is introduced:
\begin{align}
& {F_{i,r}} = F\left( {{{\bf{r}}_{{\rm{Rx}}}},{{\bf{r}}_{{\rm{Tx}}}}} \right) = \frac{{\cos {\theta _i}\left( {{{\bf{r}}_{{\rm{Tx}}}}} \right)}}{{\cos {\theta _r}\left( {{{\bf{r}}_{{\rm{Rx}}}}} \right)}} \\
&\Psi \left( y \right) = \Psi \left( {{{\bf{r}}_{{\rm{Rx}}}},y,{{\bf{r}}_{{\rm{Tx}}}}} \right) = \phi \left( {{{\bf{r}}_{{\rm{Rx}}}},y,{{\bf{r}}_{{\rm{Tx}}}}} \right) + \Phi_{\SS} \left( {{{\bf{r}}_{{\rm{Rx}}}},y,{{\bf{r}}_{{\rm{Tx}}}}} \right)
		\end{align}

From \eqref{eq:reflection-coefficient-amplitude-locally-passive-general}, we conclude that the amplitude $R\left( {{{\bf{r}}_{{\rm{Rx}}}},y,{{\bf{r}}_{{\rm{Tx}}}}} \right)$ and the total phase $\Psi \left( {{{\bf{r}}_{{\rm{Rx}}}},y,{{\bf{r}}_{{\rm{Tx}}}}} \right)$ of the surface reflection coefficient are intertwined, and they are not, in general, independent of each other if a lossless RIS needs to have a locally unitary power efficiency. By definition, \eqref{eq:powerflow-at-RIS-impedance} and \eqref{eq:reflection-coefficient-amplitude-locally-passive-general} are equivalent. Therefore, imposing $\Re\left({Z(\vect{r}_{\textup{Rx}},y,\vect{r}_{\textup{Tx}})}\right) = 0$ $\forall y \in [-L_y,L_y]$ is equivalent to imposing a well-defined relation between the amplitude and the phase of the field reflection coefficient, as obtained in \eqref{eq:reflection-coefficient-amplitude-locally-passive-general}. From \eqref{eq:abs-reflection-coefficicient-less-than-1-condition-rewritten}, notably, we know that the condition $\Re\left({Z(\vect{r}_{\textup{Rx}},y,\vect{r}_{\textup{Tx}})}\right) = 0$ does not necessarily imply $\left|\Gamma_{\SS}(\vect{r}_{\textup{Rx}},y,\vect{r}_{\textup{Tx}})\right| = R(\vect{r}_{\textup{Rx}},y,\vect{r}_{\textup{Tx}}) \leq 1$. Indeed, whether the amplitude $R(\vect{r}_{\textup{Rx}},y,\vect{r}_{\textup{Tx}})$ of the reflection coefficient is smaller or greater than unity depends on the angles of incidence and reflection. If we consider the special case of specular reflection, i.e., ${\theta _i}\left( {{{\bf{r}}_{{\rm{Tx}}}}} \right) = {\theta _r}\left( {{{\bf{r}}_{{\rm{Rx}}}}} \right)$, then ${F_{i,r}}=1$ in \eqref{eq:reflection-coefficient-amplitude-locally-passive-general} and we obtain $R(\vect{r}_{\textup{Rx}},y,\vect{r}_{\textup{Tx}}) = 1$ $\forall y \in [-L_y,L_y]$. If $ {\theta _i}\left( {{{\bf{r}}_{{\rm{Tx}}}}} \right) \ne  {\theta _r}\left( {{{\bf{r}}_{{\rm{Rx}}}}} \right)$, on the other hand, the amplitude $R(\vect{r}_{\textup{Rx}},y,\vect{r}_{\textup{Tx}})$ of the reflection coefficient is, in general, not unitary.

In wireless communications, the constraint $R(\vect{r}_{\textup{Rx}},y,\vect{r}_{\textup{Tx}})= 1$ is typically assumed when optimizing an RIS \cite{Rui_Tutorial}. Also, the total phase $\Psi \left( {{{\bf{r}}_{{\rm{Rx}}}},y,{{\bf{r}}_{{\rm{Tx}}}}} \right)$ is often optimized by assuming that it is independent of $R(\vect{r}_{\textup{Rx}},y,\vect{r}_{\textup{Tx}})$. This is not in agreement with the the condition obtained in \eqref{eq:reflection-coefficient-amplitude-locally-passive-general} for designing lossless RISs with a locally unitary power efficiency. By letting $R(\vect{r}_{\textup{Rx}},y,\vect{r}_{\textup{Tx}})= \left|\Gamma_{\SS}(\vect{r}_{\textup{Rx}},y,\vect{r}_{\textup{Tx}})\right| =1$ in \eqref{eq:abs-reflection-coefficicient-less-than-1-condition-rewritten}, specifically, the real part of the surface impedance can only be equal to:
\begin{align}\label{eq:abs-reflection-coefficicient-less-than-1-condition-rewritten-unitary}
		\Re\left(Z(\vect{r}_{\textup{Rx}},y,\vect{r}_{\textup{Tx}})\right) & = \frac{\left|Z(\vect{r}_{\textup{Rx}},y,\vect{r}_{\textup{Tx}})\right|^2}{2\eta_0} \nonumber \\ &*(\cos\theta_i(\vect{r}_{\textup{Tx}}) - \cos\theta_r(\vect{r}_{\textup{Rx}}))
	\end{align}

Therefore, the condition $\Re\left(Z(\vect{r}_{\textup{Rx}},y,\vect{r}_{\textup{Tx}})\right) = 0$, i.e., locally unitary power efficiency, can be ensured only for specular reflection. If $ {\theta _i}\left( {{{\bf{r}}_{{\rm{Tx}}}}} \right) \ne  {\theta _r}\left( {{{\bf{r}}_{{\rm{Rx}}}}} \right)$, on the other hand, $\Re\left(Z(\vect{r}_{\textup{Rx}},y,\vect{r}_{\textup{Tx}})\right)$ can be either positive or negative, which results in a locally passive or locally active RIS, respectively, as dictated by \eqref{eq:S-with-local-amplitude-expanded}. Interested readers are invited to consult \cite{Sergei_PerfectReflector,Alu_PerfectReflector,Sergei_RealizedSurface} for examples and discussions.

Let us analyze in further detail the case study for which the constraint $R(\vect{r}_{\textup{Rx}},y,\vect{r}_{\textup{Tx}}) = 1$ $\forall y \in [-L_y,L_y]$ is imposed by design. In this case, $S(\vect{r}_{\textup{Rx}},y,\vect{r}_{\textup{Tx}})$ in \eqref{eq:powerflow-at-RIS-normalized-amplitude-control} simplifies to:
    \begin{align}\label{eq:powerflow-at-RIS-normalized-amplitude-contro-aux-1l}
S\left( {{{\bf{r}}_{{\rm{Rx}}}},y,{{\bf{r}}_{{\rm{Tx}}}}} \right) &= \left( {\cos {\theta _r}\left( {{{\bf{r}}_{{\rm{Rx}}}}} \right) - \cos {\theta _i}\left( {{{\bf{r}}_{{\rm{Tx}}}}} \right)} \right) \nonumber \\
& *\left( {1 + \cos \left( {\Psi \left( {{{\bf{r}}_{{\rm{Rx}}}},y,{{\bf{r}}_{{\rm{Tx}}}}} \right)} \right)} \right)
		\end{align}

From \eqref{eq:powerflow-at-RIS-normalized-amplitude-contro-aux-1l}, we retrieve, in agreement with \eqref{eq:abs-reflection-coefficicient-less-than-1-condition-rewritten-unitary}, that, for phase-gradient reflectors, a unitary power efficiency is obtained only for specular reflection (i.e., $ {\theta _i}\left( {{{\bf{r}}_{{\rm{Tx}}}}} \right) = {\theta _r}\left( {{{\bf{r}}_{{\rm{Rx}}}}} \right)$). If the angle of incidence is normal to the surface $\SS$ (i.e., ${\theta _i}\left( {{{\bf{r}}_{{\rm{Tx}}}}} \right) =0 $), as is often assumed or implied in the literature, we evince that the net power flow in \eqref{eq:powerflow-at-RIS-normalized-amplitude-contro-aux-1l} is negative, i.e., $S(\vect{r}_{\textup{Rx}},y,\vect{r}_{\textup{Tx}}) <0$, for any angle of reflection. This implies that it is possible to steer a normally incident wave towards any directions of reflection while ensuring $R(\vect{r}_{\textup{Rx}},y,\vect{r}_{\textup{Tx}}) = 1$ and without necessitating any local power amplification and active components. This is in agreement with \eqref{eq:abs-reflection-coefficicient-less-than-1-condition-rewritten}, which yields $\Re\left(Z(\vect{r}_{\textup{Rx}},y,\vect{r}_{\textup{Tx}})\right) > 0$ in this considered case study. The power efficiency, however, highly depends on the angle of reflection and it typically decreases as the angle of reflection increases. If the specified angle of reflection is smaller than the angle of incidence (i.e., $\cos {\theta _i}\left( {{{\bf{r}}_{{\rm{Tx}}}}} \right) < \cos {\theta _r}\left( {{{\bf{r}}_{{\rm{Rx}}}}} \right)$), \eqref{eq:powerflow-at-RIS-normalized-amplitude-contro-aux-1l} unveils that the net power flow is locally positive, i.e., $S(\vect{r}_{\textup{Rx}},y,\vect{r}_{\textup{Tx}}) >0$, which implies that the corresponding wave transformation, with the constraint $R(\vect{r}_{\textup{Rx}},y,\vect{r}_{\textup{Tx}}) = 1$, cannot be realized without local power amplification or active components. This is in agreement with \eqref{eq:abs-reflection-coefficicient-less-than-1-condition-rewritten}, i.e., the real part of the surface impedance needs to be negative, and, thus, the RIS needs to locally amplify the incident wave. In practice, this implies that, to realize this wave transformation without active components, the amplitude of the reflection coefficient cannot be equal to one, i.e., $R(\vect{r}_{\textup{Rx}},y,\vect{r}_{\textup{Tx}}) \ne 1$ and the total phase $\Psi \left( {{{\bf{r}}_{{\rm{Rx}}}},y,{{\bf{r}}_{{\rm{Tx}}}}} \right)$ of the surface reflection coefficient needs to be carefully engineered. Based on \eqref{eq:S-with-local-amplitude-expanded}, specifically, $S(\vect{r}_{\textup{Rx}},y,\vect{r}_{\textup{Tx}}) \le0$ if and only if:
 \begin{align}\label{eq:reflection-coefficient-amplitude-locally-passive-general-inequality}
0 \le R\left( {{{\bf{r}}_{{\rm{Rx}}}},y,{{\bf{r}}_{{\rm{Tx}}}}} \right) & \le \frac{1}{2}\cos \left( {\Psi \left( y \right)} \right)\left( {{F_{i,r}} - 1} \right)\\
& + \frac{1}{2}\sqrt {{{\cos }^2}\left( {\Psi \left( y \right)} \right){{\left( {{F_{i,r}} - 1} \right)}^2} + 4{F_{i,r}}}  \nonumber
		\end{align}

The condition $S(\vect{r}_{\textup{Rx}},y,\vect{r}_{\textup{Tx}}) \le0$ implies, however, that a smaller amount of power is reradiated towards the specified direction of reflection and, consequently, a smaller amount of power is available at the receiver Rx. This is elaborated next.

{The remarks that originate from \eqref{eq:abs-reflection-coefficicient-less-than-1-condition-rewritten} and \eqref{eq:reflection-coefficient-amplitude-locally-passive-general} allow us to conclude that the optimization problems typically formulated in wireless communications result in RISs whose power efficiency is not necessarily unitary, even though the reflection coefficient is unitary. This is because the condition $R(\vect{r}_{\textup{Rx}},y,\vect{r}_{\textup{Tx}})= \left|\Gamma_{\SS}(\vect{r}_{\textup{Rx}},y,\vect{r}_{\textup{Tx}})\right| =1$ results in $\Re\left(Z(\vect{r}_{\textup{Rx}},y,\vect{r}_{\textup{Tx}})\right) = 0$ only if $ {\theta _i}\left( {{{\bf{r}}_{{\rm{Tx}}}}} \right) =  {\theta _r}\left( {{{\bf{r}}_{{\rm{Rx}}}}} \right)$. This apparent inconsistency has a simple justification. In wireless communications, the definition of reflection coefficient that is typically utilized is ${\Gamma_{\textup{load}}(\vect{r}_{\textup{Rx}},y,\vect{r}_{\textup{Tx}})}$ in \eqref{eq:load-reflection-coefficient}. We have already noted that $\left| {\Gamma_{\textup{load}}(\vect{r}_{\textup{Rx}},y,\vect{r}_{\textup{Tx}})} \right| = 1$ if and only if $\Re(Z(\vect{r}_{\textup{Rx}},y,\vect{r}_{\textup{Tx}})) = 0$. With this definition, as a result, we can indeed conclude that a lossless RIS has a local unitary power efficiency if $\left| {\Gamma_{\textup{load}}(\vect{r}_{\textup{Rx}},y,\vect{r}_{\textup{Tx}})} \right| = 1$. Based on ${\Gamma_{\textup{load}}(\vect{r}_{\textup{Rx}},y,\vect{r}_{\textup{Tx}})}$, therefore, the conclusions drawn in the context of wireless communications are consistent. Specifically, the load reflection coefficient ${\Gamma_{\textup{load}}(\vect{r}_{\textup{Rx}},y,\vect{r}_{\textup{Tx}})}$ is consistent with the locally periodic discrete model in Section II-A and with the local specular reflection model illustrated in Fig. \eqref{fig:LocalModel}.}

\paragraph{Global Design}
According to the local design criterion, a lossless RIS has locally unitary unit power efficiency if and only if $S(\vect{r}_{\textup{Rx}},y,\vect{r}_{\textup{Tx}}) = 0$ $\forall y \in [-L_y,L_y]$. Based on this definition, the real part of the surface impedance needs to be identically equal to zero along the entire surface $\SS$, i.e., $\Re\left({Z(\vect{r}_{\textup{Rx}},y,\vect{r}_{\textup{Tx}})}\right) = 0$ $\forall y \in [-L_y,L_y]$. The main advantage of this design criterion is the relatively simple structure of a lossless RIS, which can be implemented without using resistive components while ensuring a unitary power efficiency. The main disadvantages are, on the other hand, the restricted feasible set of possible implementation options that the constraint $\Re\left({Z(\vect{r}_{\textup{Rx}},y,\vect{r}_{\textup{Tx}})}\right) = 0$ $\forall y \in [-L_y,L_y]$ offers, the possibly low reradiated power at the location of the receiver Rx, and the impossibility of realizing some wave transformations. To elucidate this latter point, let us analyze the canonical example of anomalous reflection \cite{Sergei_PerfectReflector}.

Let us consider the design of a lossless RIS for which $R(\vect{r}_{\textup{Rx}},y,\vect{r}_{\textup{Tx}}) = R_0$,  $\phi(\vect{r}_{\textup{Rx}},y,\vect{r}_{\textup{Tx}}) = 0$, and $\Phi(\vect{r}_{\textup{Rx}},y,\vect{r}_{\textup{Tx}})$ is given in \eqref{eq:phase-geometric-optics-explicit} $\forall y \in [-L_y,L_y]$. This corresponds to an anomalous reflector \cite{Sergei_PerfectReflector} for which Helmholtz's condition in \eqref{eq:Helmholtz-reflected-electric-field} is fulfilled with equality. If $R_0=1$ and $R_0 = \sqrt {{{\cos {\theta _i}\left( {{{\bf{r}}_{{\rm{Tx}}}}} \right)} \mathord{\left/
{\vphantom {{\cos {\theta _i}\left( {{{\bf{r}}_{{\rm{Tx}}}}} \right)} {\cos {\theta _r}\left( {{{\bf{r}}_{{\rm{Rx}}}}} \right)}}} \right. \kern-\nulldelimiterspace} {\cos {\theta _r}\left( {{{\bf{r}}_{{\rm{Rx}}}}} \right)}}}$, we retrieve the geometrical optics solution and the perfect anomalous reflector {under the assumption that the RIS is periodic and an integer number of periods is available in $[-L_y,L_y]$}, respectively, \cite{Sergei_PerfectReflector}, \cite{Alu_PerfectReflector}. In this case, the surface impedance $Z(\vect{r}_{\textup{Rx}},y,\vect{r}_{\textup{Tx}})$ in \eqref{eq:surface-impedance-explicit} simplifies as follows:
	\begin{align}\label{eq:Global-1}
Z\left( {{{\bf{r}}_{{\rm{Rx}}}},y,{{\bf{r}}_{{\rm{Tx}}}}} \right) = \frac{{{\eta _0}}}{{\cos {\theta _r}\left( {{{\bf{r}}_{{\rm{Rx}}}}} \right)}}\frac{{1 + {R_0}\mathcal{W}\left( y \right)}}{{{{\mathcal{C}}_0} - {R_0}\mathcal{W}\left( y \right)}}
	\end{align}
where the following notation is introduced:
\begin{align}\label{eq:Global-1a}
& \mathcal{W}\left( y \right) = {e^{j\Phi_{\SS} \left( {{{\bf{r}}_{{\rm{Rx}}}},y,{{\bf{r}}_{{\rm{Tx}}}}} \right)}} \\
& {{\mathcal{C}}_0} = {{\cos {\theta _i}\left( {{{\bf{r}}_{{\rm{Tx}}}}} \right)} \mathord{\left/
 {\vphantom {{\cos {\theta _i}\left( {{{\bf{r}}_{{\rm{Tx}}}}} \right)} {\cos {\theta _r}\left( {{{\bf{r}}_{{\rm{Rx}}}}} \right)}}} \right.
 \kern-\nulldelimiterspace} {\cos {\theta _r}\left( {{{\bf{r}}_{{\rm{Rx}}}}} \right)}}
	\end{align}

Then, the real part of $Z\left( {{{\bf{r}}_{{\rm{Rx}}}},y,{{\bf{r}}_{{\rm{Tx}}}}} \right)$ in \eqref{eq:Global-1} is:
	\begin{align}\label{eq:Global-2}
{{\Re}} \left( {Z\left( {{{\bf{r}}_{{\rm{Rx}}}},y,{{\bf{r}}_{{\rm{Tx}}}}} \right)} \right) = \frac{\eta_0}{{\cos {\theta _r}\left( {{{\bf{r}}_{{\rm{Rx}}}}} \right)}}\frac{{{{\mathcal{N}}_Z}\left( y \right)}}{{{{\mathcal{D}}_Z}\left( y \right)}}
	\end{align}
where the following shorthand notation is used:
\begin{align}
& {{\mathcal{N}}_Z}\left( y \right) = {\mathcal{C}}_0 - R_0^2 + \left( {{{\mathcal{C}}_0} - 1} \right){R_0}\cos \left( {\Phi_{\SS} \left( {{{\bf{r}}_{{\rm{Rx}}}},y,{{\bf{r}}_{{\rm{Tx}}}}} \right)} \right)\\
& {{\mathcal{D}}_Z}\left( y \right) = {{\mathcal{C}}_0^2 + R_0^2 - 2{{\mathcal{C}}_0}{R_0}\cos \left( {\Phi_{\SS} \left( {{{\bf{r}}_{{\rm{Rx}}}},y,{{\bf{r}}_{{\rm{Tx}}}}} \right)} \right)}
	\end{align}

By direct inspection of \eqref{eq:Global-2}, it follows that ${{\Re}} \left( {Z\left( {{{\bf{r}}_{{\rm{Rx}}}},y,{{\bf{r}}_{{\rm{Tx}}}}} \right)} \right)=0$ if and only if ${{\mathcal{N}}_Z}\left( y \right) = 0$ $\forall y \in [-L_y,L_y]$. However, this condition cannot be ensured $\forall y \in [-L_y,L_y]$ either for $R_0=1$ or for $R_0 = \sqrt {{{\cos {\theta _i}\left( {{{\bf{r}}_{{\rm{Tx}}}}} \right)} \mathord{\left/
{\vphantom {{\cos {\theta _i}\left( {{{\bf{r}}_{{\rm{Tx}}}}} \right)} {\cos {\theta _r}\left( {{{\bf{r}}_{{\rm{Rx}}}}} \right)}}} \right. \kern-\nulldelimiterspace} {\cos {\theta _r}\left( {{{\bf{r}}_{{\rm{Rx}}}}} \right)}}} = \sqrt{\mathcal{C}_0}$. In fact, ${{\Re}} \left( {Z\left( {{{\bf{r}}_{{\rm{Rx}}}},y,{{\bf{r}}_{{\rm{Tx}}}}} \right)} \right)$ is, in general, an oscillatory function in $y$, since ${{\mathcal{D}}_Z}\left( y \right) \ge 0$ $\forall y \in [-L_y,L_y]$ and ${{\mathcal{N}}_Z}\left( y \right)$ is positive or negative in $[-L_y,L_y]$. To gain further insights and for illustrative purposes, let us assume $\mathcal{C}_0 \ge 1$. This is always true if, e.g., ${\cos {\theta _i}\left( {{{\bf{r}}_{{\rm{Tx}}}}} \right)}=1$, which corresponds to normal incidence. Then, we have the following findings for $R_0=1$ and $R_0 = \sqrt{\mathcal{C}_0}$, respectively.
\begin{itemize}
    \item $R_0=1$: In this case, ${{\mathcal{D}}_Z}\left( y \right) \ge 0$ and ${{\mathcal{N}}_Z}\left( y \right) \ge 0$ $\forall y \in [-L_y,L_y]$. This implies that a lossless RIS can be implemented without any power amplification, and the surface power flow in \eqref{eq:powerflow-at-RIS-impedance} is negative and can be very different from zero.
    \item $R_0 = \sqrt{\mathcal{C}_0}$. In this case, ${{\mathcal{D}}_Z}\left( y \right) \ge 0$ $\forall y \in [-L_y,L_y]$ and ${{\mathcal{N}}_Z}\left( y \right) = {{\mathcal{C}}_0}\left( {{{\mathcal{C}}_0} - 1} \right)\cos \left( {\Phi_{\SS} \left( {{{\bf{r}}_{{\rm{Rx}}}},y,{{\bf{r}}_{{\rm{Tx}}}}} \right)} \right)$, which is an oscillating function in $[-L_y,L_y]$, i.e., it can take positive and negative values. This implies that this wave transformation cannot be realized without utilizing any sort of power amplification. The power amplification can be virtual, e.g., by using surface waves, or can be realized through actual power amplifiers \cite{Sergei_RealizedSurface}. The corresponding power flow in \eqref{eq:powerflow-at-RIS-impedance} is oscillatory in $[-L_y,L_y]$ as well.
\end{itemize}

This simple example that corresponds to the design of a canonical anomalous reflector allows us to understand that the constraint ${{\Re}} \left( {Z\left( {{{\bf{r}}_{{\rm{Rx}}}},y,{{\bf{r}}_{{\rm{Tx}}}}} \right)} \right) = 0$ may be too restrictive for enabling the realization of some wave transformations and other design criteria may be more appropriate to this end. The global design criterion falls in this category, since it allows us to relax the inherent constraints of the local design and to enlarge the feasible set of wave transformations that can be realized at a high power efficiency. The main essence of the global design is to relax the local power efficiency constraint, i.e., for each point of the surface $\SS$, with an average power efficiency constraint, i.e., by considering the entire surface $\SS$. {For completeness, it is worth mentioning that an RIS optimized based on the global design is different from a hybrid or an active RIS \cite{Hybrid_RIS}. In a hybrid RIS, some RIS elements have the capability of amplifying the incident electromagnetic waves so that the total reradiated power is greater than the total incident power. An RIS designed based on the global design does not increase the amount of incident power: The available power budget is just carefully redistributed along the surface \cite{Sergei_RealizedSurface}.}

Specifically, based on the definition of surface Poynting vector  $\vect{P}_{\SS}(\vect{r}_{\textup{Rx}},\vect{s},\vect{r}_{\textup{Tx}})$ in \eqref{eq:Poynting-at-RIS}, a lossless RIS is defined to be globally passive if the following condition holds true:
    \begin{align}\label{eq:globally-passive-condition-0}
			{P_{\mathcal{S}}}\left( {{{\bf{r}}_{{\rm{Rx}}}},{{\bf{r}}_{{\rm{Tx}}}}} \right) = \int_{{\bf{s}} \in {\mathcal{S}}} {\left| {{{\bf{P}}_{\mathcal{S}}}\left( {{{\bf{r}}_{{\rm{Rx}}}},{\bf{s}},{{\bf{r}}_{{\rm{Tx}}}}} \right)} \right|d{\bf{s}}}  \le 0
	\end{align}

By utilizing the same notation as for the local design, the average power flow ${P_{\mathcal{S}}}\left( {{{\bf{r}}_{{\rm{Rx}}}},{{\bf{r}}_{{\rm{Tx}}}}} \right)$ in \eqref{eq:globally-passive-condition-0} can be explicitly written as follows:
\begin{align}\label{eq:globally-passive-condition-0-explicit}
{P_{\mathcal{S}}}\left( {{{\bf{r}}_{{\rm{Rx}}}},{{\bf{r}}_{{\rm{Tx}}}}} \right) &= \int\nolimits_{ - {L_x}}^{ + {L_x}} {\int\nolimits_{ - {L_y}}^{ + {L_y}} {{P_{\mathcal{S}}}\left( {{{\bf{r}}_{{\rm{Rx}}}},y,{{\bf{r}}_{{\rm{Tx}}}}} \right)dydx} } \nonumber \\
& = 2{L_x}\int\nolimits_{ - {L_y}}^{ + {L_y}} {{P_{\mathcal{S}}}\left( {{{\bf{r}}_{{\rm{Rx}}}},y,{{\bf{r}}_{{\rm{Tx}}}}} \right)dy}
	\end{align}
where ${{P_{\mathcal{S}}}\left( {{{\bf{r}}_{{\rm{Rx}}}},y,{{\bf{r}}_{{\rm{Tx}}}}} \right)}$ is given in
\eqref{eq:powerflow-at-RIS-impedance}.

Based on \eqref{eq:globally-passive-condition-0}, therefore, a lossless RIS has, on average, unitary power efficiency if and only if ${P_{\mathcal{S}}}\left( {{{\bf{r}}_{{\rm{Rx}}}},{{\bf{r}}_{{\rm{Tx}}}}} \right) = 0$. In other words, the power flow does not need to be equal to zero for each point of the surface but only the power flow integrated along the surface $\SS$ needs to be equal to zero. This implies that the global design accounts for solutions that may correspond to practical implementations that require positive and negative values of the surface impedance ${Z\left( {{{\bf{r}}_{{\rm{Rx}}}},y,{{\bf{r}}_{{\rm{Tx}}}}} \right)}$. Then, the design constraint is milder: The integral of the local power flow along the surface is equal to zero. In a global design, in particular, the electromagnetic field reradiated by an RIS is not obtained as the local contribution of each RIS element but as the collective action of all the RIS elements as a whole.

Let us consider again the case study of perfect anomalous reflection already analyzed from the point of view of the local design. The wave transformation for which $R_0 = \sqrt {{{\cos {\theta _i}\left( {{{\bf{r}}_{{\rm{Tx}}}}} \right)} \mathord{\left/
{\vphantom {{\cos {\theta _i}\left( {{{\bf{r}}_{{\rm{Tx}}}}} \right)} {\cos {\theta _r}\left( {{{\bf{r}}_{{\rm{Rx}}}}} \right)}}} \right. \kern-\nulldelimiterspace} {\cos {\theta _r}\left( {{{\bf{r}}_{{\rm{Rx}}}}} \right)}}}$ corresponds to the globally optimum design with unitary power efficiency, i.e., $R_0$ is obtained by setting ${P_{\mathcal{S}}}\left( {{{\bf{r}}_{{\rm{Rx}}}},{{\bf{r}}_{{\rm{Tx}}}}} \right) = 0$ in \eqref{eq:globally-passive-condition-0-explicit} {under the assumption that the RIS is periodic and an integer number of periods is available in $[-L_y,L_y]$}. Therefore, a global design offers a greater flexibility than a local design.

\paragraph{Local Design vs. Global Design}
In general, a locally optimal solution with unitary power efficiency is a globally optimal solution with unitary power efficiency as well. The opposite is, however, not true in general. This implies that a global design enlarges the set of feasible solutions that can be found when considering a local design. However, there exist inherent performance and implementation tradeoffs between a local and global designs.

The solutions found according to a local design result in implementations of RISs whose real part of the surface impedance is identically equal to zero. This implies that the corresponding RISs can be realized without utilizing resistive components and by utilizing only capacitive and inductive elements. In general, this simplifies the implementation of the surface $\SS$. However, the set of feasible solutions that fulfill the local design criterion may be limited and some wave transformations may not be allowed by the design constraint.

A global design allows, on the other hand, solutions for which the real part of the surface impedance is not necessarily equal to zero. This enlarges the set of feasible solutions. Also, wave transformations that may require positive and negative values of the surface impedance are allowed. However, RISs with resistive elements and with active elements, which correspond to negative values of the surface impedance, are  difficult to design and to implement. The need of active elements can be avoided by realizing virtual power losses and virtual power gains along the surface $\SS$ through carefully engineered evanescent Floquet's harmonics (surface waves) that are excited at the metasurface \cite{Sergei_RealizedSurface,Wang_2020,Do-Hoon_2021,Eleftheriades_Arbitrary2021}.

Therefore, it would be convenient to identify designs for RIS that have a unitary power efficiency, that allow the same feasible set of solutions as the global design, and that have the same implementation simplicity as the local design, i.e., they can be realized by using purely reactive components. With this in mind, it is convenient to depart from the global design with unitary power efficiency as the starting point for designing an RIS, in order to have a large set of feasible solutions. Once the corresponding optimal solution is found, one can find an approximate solution that corresponds to an implementation of the RIS with a surface impedance whose real part is equal to zero. This would make the implementation of the RIS easier while ensuring a high power efficiency. Examples of similar design methods do exist in the recent scientific literature, e.g., \cite{Grbic_Homogenization}. In the next sub-section, we illustrate examples of this design paradigm with application to wireless communications and with focus on how approximated solutions can be found. We anticipate that the approximated solutions can be implemented by utilizing only reactive components, but they may result in slightly lower beam pattern gains and higher side lobes that need to be accurately controlled when formulating the problem. This is elaborated in the next sub-section.

\begin{figure}[!t]
		\begin{center}
			\includegraphics[width=\linewidth]{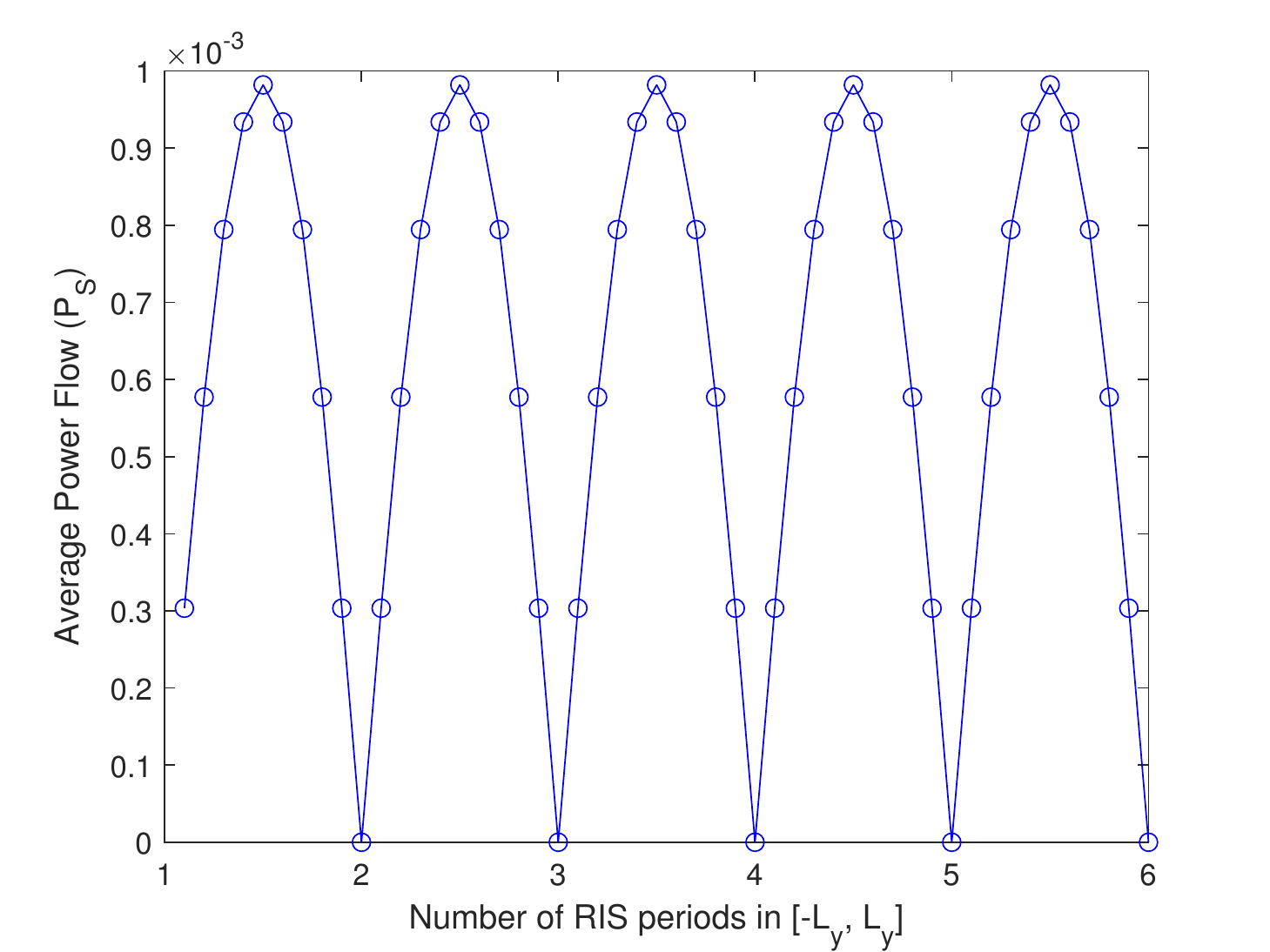}
			\caption{Average power flow ${P_{\mathcal{S}}}\left( {{{\bf{r}}_{{\rm{Rx}}}},{{\bf{r}}_{{\rm{Tx}}}}} \right)$ in \eqref{eq:globally-passive-condition-0} as a function of the length of the RIS expressed in number of RIS periods.}
			\label{fig:PeriodicVsAperiodic_IntegerPeriods}
		\end{center}
	\end{figure}
{\textbf{Periodic vs. aperiodic surfaces.}}
{In Section I, we have discussed the difference between periodic and aperiodic RISs with the help of the example illustrated in Fig. \ref{fig:RIS_PeriodicAperiodic}. It is instructive to elaborate further on the concept of periodicity of an engineered surface in the context of optimizing an RIS according to the global design criterion. Specifically, we are interested in the \textit{RIS period} that depends on the specified wave transformation that a given RIS needs to realize. The results available in the literature on the global design criterion are usually applicable only to periodic surfaces, e.g., \cite{MDR_JSAC}, \cite{Sergei_RealizedSurface}. In the previous text, in fact, we have emphasized that the solution $R_0 = \sqrt {{{\cos {\theta _i}\left( {{{\bf{r}}_{{\rm{Tx}}}}} \right)} \mathord{\left/
{\vphantom {{\cos {\theta _i}\left( {{{\bf{r}}_{{\rm{Tx}}}}} \right)} {\cos {\theta _r}\left( {{{\bf{r}}_{{\rm{Rx}}}}} \right)}}} \right. \kern-\nulldelimiterspace} {\cos {\theta _r}\left( {{{\bf{r}}_{{\rm{Rx}}}}} \right)}}}$ corresponds to the globally optimum design with unit power efficiency \textit{under the assumption that the RIS is periodic and that an integer number of periods is available in $[-L_y,L_y]$}. If either the RIS is not periodic or the size (length) of the RIS is not equal to an integer number of RIS periods, the solution $R_0 = \sqrt {{{\cos {\theta _i}\left( {{{\bf{r}}_{{\rm{Tx}}}}} \right)} \mathord{\left/
{\vphantom {{\cos {\theta _i}\left( {{{\bf{r}}_{{\rm{Tx}}}}} \right)} {\cos {\theta _r}\left( {{{\bf{r}}_{{\rm{Rx}}}}} \right)}}} \right. \kern-\nulldelimiterspace} {\cos {\theta _r}\left( {{{\bf{r}}_{{\rm{Rx}}}}} \right)}}}$ is not globally optimal with unit power efficiency anymore, i.e., ${P_{\mathcal{S}}}\left( {{{\bf{r}}_{{\rm{Rx}}}},{{\bf{r}}_{{\rm{Tx}}}}} \right) \ne 0$, in general.}

{An illustration of this case study is given in Fig. \ref{fig:PeriodicVsAperiodic_IntegerPeriods}. Specifically, we plot the average power flow ${P_{\mathcal{S}}}\left( {{{\bf{r}}_{{\rm{Rx}}}},{{\bf{r}}_{{\rm{Tx}}}}} \right)$ in \eqref{eq:globally-passive-condition-0} by using $S(\vect{r}_{\textup{Rx}},y,\vect{r}_{\textup{Tx}})$ in \eqref{eq:powerflow-at-RIS-normalized-amplitude-control}, and by setting $R(\vect{r}_{\textup{Rx}},y,\vect{r}_{\textup{Tx}}) = R_0 = \sqrt {1/{\cos {\theta _r}\left( {{{\bf{r}}_{{\rm{Rx}}}}} \right)}}$, $\phi(\vect{r}_{\textup{Rx}},y,\vect{r}_{\textup{Tx}}) = 0$, and $\Phi(\vect{r}_{\textup{Rx}},y,\vect{r}_{\textup{Tx}})$ as defined in \eqref{eq:phase-geometric-optics-explicit} $\forall y \in [-L_y,L_y]$. This corresponds to the canonical case study analyzed in the literature \cite{MDR_JSAC}, \cite{Sergei_RealizedSurface}, when ${ {\theta _i}\left( {{{\bf{r}}_{{\rm{Tx}}}}} \right)}=0$ and the RIS period is equal to $\mathcal{P} = \lambda/{\cos {\theta _r}\left( {{{\bf{r}}_{{\rm{Rx}}}}} \right)}$. We see that ${P_{\mathcal{S}}}\left( {{{\bf{r}}_{{\rm{Rx}}}},{{\bf{r}}_{{\rm{Tx}}}}} \right)$  is equal to zero only if the length $2L_y$ of the RIS  is equal to an integer number of RIS periods. This implies that the solution $R(\vect{r}_{\textup{Rx}},y,\vect{r}_{\textup{Tx}}) = R_0 = \sqrt {1/{\cos {\theta _r}\left( {{{\bf{r}}_{{\rm{Rx}}}}} \right)}}$ is, in general, not optimal if the RIS is either aperiodic or it has an arbitrary size that is not equal to an integer number of RIS periods. This is because, under these conditions, the integration of the addend in the second line of $S(\vect{r}_{\textup{Rx}},y,\vect{r}_{\textup{Tx}})$ in \eqref{eq:powerflow-at-RIS-normalized-amplitude-control} is not equal to zero. This is thoroughly discussed and formulated in \cite[Eq. (17), Eq. (18)]{MDR_JSAC}. As mentioned in Section I, however, the RISs studied in wireless communications are often aperiodic surfaces and their physical size is fixed and is not usually adapted to the RIS period, e.g., to the specified wave transformation to realize.}

{This simple example allows us to further motivate the signal model in \eqref{eq:Einc-tangential-planewave-at-RIS}-\eqref{eq:Href-tangential-planewave-at-RIS}, and specifically, the assumption that the surface reflection coefficient $\Gamma_{\SS}(\vect{r}_{\textup{Rx}},y,\vect{r}_{\textup{Tx}})$ may depend on $y$ (in general, it may depend on $(x,y) \in \SS$), provided that the Helmholtz constraint in \eqref{eq:Helmholtz} is fulfilled. By allowing $\Gamma_{\SS}(\vect{r}_{\textup{Rx}},y,\vect{r}_{\textup{Tx}})$ to depend on $y$, i.e., either the amplitude, the phase, or both may depend on $y$ subject to the constraint in \eqref{eq:Helmholtz}, it may be possible to design an RIS so that ${P_{\mathcal{S}}}\left( {{{\bf{r}}_{{\rm{Rx}}}},{{\bf{r}}_{{\rm{Tx}}}}} \right)$ is equal zero (or it is numerically very close to zero), even if the RIS is an aperiodic surface and its size is arbitrarily chosen, regardless of the RIS period associated to the specified wave transformation to realize. This offers a more general parametric model compared to those typically employed to design periodic surfaces.}

\subsubsection{Reradiated Power Flux -- Poynting Vector}
The surface power flow introduced in preceding text allows us to characterize the efficiency of an RIS as a device that realizes specific wave transformations. It does not offer, however, any information on the amount of power that is available at the receiver Rx, which ultimately determines the performance of a communication link. The surface power efficiency provides information only on the difference between the incident and the reradiated powers in the close vicinity of the surface $\SS$. If a lossless RIS has a unit power efficiency, the total reradiated power is equal to the total incident power. The power efficiency is, therefore, an important key performance indicator to characterize a communication link in the presence of an RIS. However, it is not sufficient. It is necessary to characterize the power observed at the location of the intended receiver Rx as well. This is possible by introducing the notion of reradiated power flux.

The reradiated power flux characterizes the amount of power that is reradiated by an RIS at an arbitrary point of observation, which can be located in the radiative (Fresnel) near-field and the (Fraunhofer) far-field regions of the surface $\SS$ \cite{MDR_JSAC}, \cite{Fadil_TCOM}. Therefore, the reradiated power flux is not defined only in the close proximity of the surface $\SS$, i.e., at $z=0^+$. The reradiated power flux provides information on the angular response of the RIS and, in particular, how the incident power is reradiated as a function of the angle of observation. In the far-field region of the RIS, the reradiated power flux is proportional to the radiation pattern (or array factor) of the RIS. The reradiated power flux allows us to characterize the amount of power that is reradiated towards the specified direction of design and towards undesired (spurious) directions of observation. Therefore, both the main lobe and the side lobes of the RIS are characterized. The reradiated power flux is an essential performance indicator in wireless communications, since it determines the amount of received power and, therefore, the signal-to-noise and the signal-to-interference ratios. {In addition, the recent study reported in \cite{DigitalRIS_Elsevier} highlights the importance of characterizing the reradiation pattern of an RIS for every allowed angle of observation, so as to ensure that the incident power is not directed towards unwanted directions.}

The reradiated power flux is defined from the Poynting vector evaluated at any observation point in a three-dimensional volume, $V$, of interest. Let $\vect{r}_{\textup{obs}} = x_{\textup{obs}} \hat{\vect{x}} + y_{\textup{obs}} \hat{\vect{y}} + z_{\textup{obs}} \hat{\vect{z}} \in V$ denote a generic observation point in $V$. For simplicity, we assume that $\vect{r}_{\textup{obs}}$ is located outside the volume $V_{\textup{Tx}}$ occupied by the transmitter Tx. Therefore, the canonical source-free scenario is considered. For generality, we assume that $\vect{r}_{\textup{obs}}$ is located at a distance $R_{\textup{obs}}$ from the center of the surface $\SS$ (the origin of the reference system) and that the elevation angle and the azimuth angle are equal to $\theta_o$ and $\varphi_o$, respectively, with respect to the origin. Thus, $\vect{r}_{\textup{obs}} = R_{\textup{obs}} (\sin\theta_o\cos\varphi_o \hat{\vect{x}} + \sin\theta_o\sin\varphi_o \hat{\vect{y}} + \cos\theta_o \hat{\vect{z}})$ with ${R_{{\rm{obs}}}} = \left\| {{{\bf{r}}_{{\rm{obs}}}}} \right\|$.

As mentioned in previous sections, we assume, for simplicity but without loss of generality, that the direct link is blocked at the observation point $\vect{r}_{\textup{obs}}$. This is known to be the most useful case study when deploying an RIS \cite{Stefan_TWC} and, in addition, the focus of the present paper is on the electromagnetic field reradiated by the RIS. The direct link can be taken into account as described in \cite{Fadil_TCOM}. By definition, the Poynting vector evaluated at $\vect{r}_{\textup{obs}} = x_{\textup{obs}} \hat{\vect{x}} + y_{\textup{obs}} \hat{\vect{y}} + z_{\textup{obs}} \hat{\vect{z}}$ is formulated as follows \cite{Orfanidis_Book}:
	\begin{align}\label{eq:Poynting-at-obs}
		& \vect{P}_{\textup{obs}}(\vect{r}_{\textup{Rx}},\vect{r}_{\textup{obs}},\vect{r}_{\textup{Tx}})
		\\ & \hspace{1.5cm} = \frac{1}{2} \Re\left(\vect{E}_{\textup{ref}}(\vect{r}_{\textup{Rx}},\vect{r}_{\textup{obs}},\vect{r}_{\textup{Tx}}) \times \vect{H}^*_{\textup{ref}}(\vect{r}_{\textup{Rx}},\vect{r}_{\textup{obs}},\vect{r}_{\textup{Tx}})\right) \nonumber
	\end{align}
where $\vect{E}_{\textup{ref}}(\vect{r}_{\textup{Rx}},\vect{r}_{\textup{obs}},\vect{r}_{\textup{Tx}})$ and $\vect{H}_{\textup{ref}}(\vect{r}_{\textup{Rx}},\vect{r}_{\textup{obs}},\vect{r}_{\textup{Tx}})$ denote the reradiated electric and magnetic fields evaluated at $\vect{r}_{\textup{obs}}$, respectively.

It is worth emphasizing that $\vect{E}_{\textup{ref}}(\vect{r}_{\textup{Rx}},\vect{r}_{\textup{obs}},\vect{r}_{\textup{Tx}})$ and $\vect{H}_{\textup{ref}}(\vect{r}_{\textup{Rx}},\vect{r}_{\textup{obs}},\vect{r}_{\textup{Tx}})$ in \eqref{eq:Poynting-at-obs} are different from the electric and magnetic fields in the close vicinity of the surface $\SS$, as defined in \eqref{eq:Einc-Eref-relationship-general}, since the latter fields are defined only for $\tau_m < z = z_{\textup{obs}} < \tau_M$. The electric and magnetic fields $\vect{E}_{\textup{ref}}(\vect{r}_{\textup{Rx}},\vect{r}_{\textup{obs}},\vect{r}_{\textup{Tx}})$ and $\vect{H}_{\textup{ref}}(\vect{r}_{\textup{Rx}},\vect{r}_{\textup{obs}},\vect{r}_{\textup{Tx}})$ in \eqref{eq:Poynting-at-obs} are, however, uniquely determined by the surface electromagnetic fields in \eqref{eq:Einc-tangential-planewave-at-RIS}-\eqref{eq:Href-tangential-planewave-at-RIS}. By using Franz's formula \cite[Eq. 18.10.11]{Orfanidis_Book}, specifically, $\vect{E}_{\textup{ref}}(\vect{r}_{\textup{Rx}},\vect{r}_{\textup{obs}},\vect{r}_{\textup{Tx}})$ and $\vect{H}_{\textup{ref}}(\vect{r}_{\textup{Rx}},\vect{r}_{\textup{obs}},\vect{r}_{\textup{Tx}})$ can be formulated as follows:
\begin{align}
			&\vect{E}_{\textup{ref}}(\vect{r}_{\textup{Rx}},\vect{r}_{\textup{obs}},\vect{r}_{\textup{Tx}}) \nonumber \\
			& \hspace{1cm} = \frac{1}{j\omega\epsilon_0\mu_0} \left(\nabla_{\vect{r}_{\textup{obs}}} \times \left(\nabla_{\vect{r}_{\textup{obs}}} \times \vect{A}_s(\vect{r}_{\textup{Rx}},\vect{r}_{\textup{obs}},\vect{r}_{\textup{Tx}})\right) \right) \nonumber\\
			& \hspace{1cm} - \frac{1}{\epsilon_0}\nabla_{\vect{r}_{\textup{obs}}} \times \vect{A}_{ms}(\vect{r}_{\textup{Rx}},\vect{r}_{\textup{obs}},\vect{r}_{\textup{Tx}}) \label{eq:Eref-at-obs}\\
			&\vect{H}_{\textup{ref}}(\vect{r}_{\textup{Rx}},\vect{r}_{\textup{obs}},\vect{r}_{\textup{Tx}})
			= -\frac{1}{j\omega\mu_0} \nabla_{\vect{r}_{\textup{obs}}} \times \vect{E}_{\textup{ref}}(\vect{r}_{\textup{Rx}},\vect{r}_{\textup{obs}},\vect{r}_{\textup{Tx}}) \label{eq:Href-at-obs}
		\end{align}
where the following shorthand notation is introduced:
\begin{align}
			&\vect{A}_s(\vect{r}_{\textup{Rx}},\vect{r}_{\textup{obs}},\vect{r}_{\textup{Tx}}) \label{eq:As-ref}  \\
			& \hspace{1cm}= \mu_0 \int_{\SS}  (\hat{\vect{n}}_{\textup{out}} \times \vect{H}_{\textup{ref}}^t(\vect{r}_{\textup{Rx}},y,\vect{r}_{\textup{Tx}}))G(\vect{r}_{\textup{obs}},\vect{s})d\vect{s}  \nonumber
		\end{align}
\begin{align}
			&\vect{A}_{ms}(\vect{r}_{\textup{Rx}},\vect{r}_{\textup{obs}},\vect{r}_{\textup{Tx}}) \label{eq:Ams-ref} \\
			& \hspace{1cm} = -\epsilon_0\int_{\SS}  (\hat{\vect{n}}_{\textup{out}} \times \vect{E}_{\textup{ref}}^t(\vect{r}_{\textup{Rx}},y,\vect{r}_{\textup{Tx}}))G(\vect{r}_{\textup{obs}},\vect{s})d\vect{s}  \nonumber
		\end{align}
and $\hat{\vect{n}}_{\textup{out}} = -\hat{\vect{z}}$ is the unit norm vector that is perpendicular to the RIS and that points towards the transmission side of $\SS$, $\vect{E}_{\textup{ref}}^t(\vect{r}_{\textup{Rx}},y,\vect{r}_{\textup{Tx}})$ and $\vect{H}_{\textup{ref}}^t(\vect{r}_{\textup{Rx}},y,\vect{r}_{\textup{Tx}})$ are the surface electric and magnetic fields defined in \eqref{eq:Eref-tangential-at-RIS} and \eqref{eq:Href-tangential-planewave-at-RIS}, respectively, and $G(\vect{r}_{\textup{obs}},\vect{s})$ is the scalar Green function that corresponds to a point source located at $\vect{s} = x\hat{\vect{x}} + y\hat{\vect{y}} \in \SS$ and that is observed at $\vect{r}_{\textup{obs}}$:
\begin{align}\label{eq:Green's-function-definition}
		G\left( {{{\bf{r}}_{{\rm{obs}}}},{\bf{s}}} \right) = \frac{{{e^{ - jk\left\| {{{\bf{r}}_{{\rm{obs}}}} - {\bf{s}}} \right\|}}}}{{4\pi \left\| {{{\bf{r}}_{{\rm{obs}}}} - {\bf{s}}} \right\|}}
	\end{align}
with $\left\| {{{\bf{r}}_{{\rm{obs}}}} - {\bf{s}}} \right\| = \sqrt {{{\left( {{x_{{\rm{obs}}}} - x} \right)}^2} + {{\left( {{y_{{\rm{obs}}}} - y} \right)}^2} + z_{{\rm{obs}}}^2}$.

Therefore, the reradiated electromagnetic field in the near field and far field regions of the RIS is uniquely determined by the surface fields in the close proximity of the surface $\SS$, as dictated by the principles of surface electromagnetics \cite{SEM_Book}. From the obtained integral expression of $\vect{E}_{\textup{ref}}(\vect{r}_{\textup{Rx}},\vect{r}_{\textup{obs}},\vect{r}_{\textup{Tx}})$ and $\vect{H}_{\textup{ref}}(\vect{r}_{\textup{Rx}},\vect{r}_{\textup{obs}},\vect{r}_{\textup{Tx}})$, the Poynting vector in \eqref{eq:Poynting-at-obs} can be readily computed for any observation point $\vect{r}_{\textup{obs}}$ in the far field of the array microstructure, i.e., a few wavelengths far away from $\SS$ or $ {{z_{{\rm{obs}}}}} \gg \lambda$ \cite[Fig. 29]{MDR_JSAC}. This constraint is usually fulfilled for typical wireless applications.

The main assumptions for the validity of \eqref{eq:Eref-at-obs} and \eqref{eq:Href-at-obs} lie in the approximations made for applying \eqref{eq:Einc-Eref-relationship-general}, i.e., the physical optics approximation \cite{Osipov_Book}. More precisely, ${\bf{J}}(y)=\hat{\vect{n}}_{\textup{out}} \times \vect{H}_{\textup{ref}}^t(\vect{r}_{\textup{Rx}},y,\vect{r}_{\textup{Tx}})$ and ${\bf{M}}(y)=-\hat{\vect{n}}_{\textup{out}} \times \vect{E}_{\textup{ref}}^t(\vect{r}_{\textup{Rx}},y,\vect{r}_{\textup{Tx}})$ can be interpreted as equivalent electric and magnetic surface currents, respectively, that produce the electromagnetic fields reradiated by the RIS. These equivalent surface currents are obtained under the assumption that the RIS has an infinite size and no edge effects are accounted for in the currents distribution along the surface $\SS$. Even though the physical optics method yields an approximated solution, it allows us to obtain analytical expressions that are suitable for performance evaluation, for optimizing RISs based on relevant key performance criteria, and to get engineering insights for network design. The limitations of the physical optics approximation method may be overcome by resorting to numerical methods, e.g., the method of moments, which, however, offer limited design insights and have limited applicability for network optimization in the context of wireless communications.

Under the physical optics approximation, the reradiation integrals in \eqref{eq:Eref-at-obs} and \eqref{eq:Href-at-obs} have general applicability. In the present paper, for ease of illustration, we focus our attention on networks setups in which $\vect{r}_{\textup{obs}}$ lie in the Fraunhofer far field region of the RIS, i.e., ${R_{{\rm{obs}}}} \ge {{z_{{\rm{obs}}}}} \ge {{8\left( {L_x^2 + L_y^2} \right)} \mathord{\left/ {\vphantom {{8\left( {L_x^2 + L_y^2} \right)} \lambda }} \right. \kern-\nulldelimiterspace} \lambda }$. In this case, the electric and magnetic fields in \eqref{eq:Eref-at-obs} and \eqref{eq:Href-at-obs}, respectively, can be simplified.

More precisely, for ease of writing, let us introduce the following shorthand notation:
\begin{align}
& E_{{\rm{ref,}}x}^t\left( {{{\bf{r}}_{{\rm{Rx}}}},y,{{\bf{r}}_{{\rm{Tx}}}}} \right) = {\bf{\hat x}} \cdot {\bf{E}}_{{\rm{ref}}}^t\left( {{{\bf{r}}_{{\rm{Rx}}}},y,{{\bf{r}}_{{\rm{Tx}}}}} \right)\\
& H_{{\rm{ref,}}y}^t\left( {{{\bf{r}}_{{\rm{Rx}}}},y,{{\bf{r}}_{{\rm{Tx}}}}} \right) = {\bf{\hat y}} \cdot {\bf{H}}_{{\rm{ref}}}^t\left( {{{\bf{r}}_{{\rm{Rx}}}},y,{{\bf{r}}_{{\rm{Tx}}}}} \right)
\end{align}
\begin{align}
&\hat{\vect{r}}_{\textup{obs}}(\vect{s}) = \frac{(x_{\textup{obs}}-x) \hat{\vect{x}} + (y_{\textup{obs}}-y) \hat{\vect{y}} + z_{\textup{obs}} \hat{\vect{z}}}{\norm{\vect{r}_{\textup{obs}} - \vect{s}}}
\end{align}
where $\hat{\vect{r}}_{\textup{obs}}(\vect{s})$ is a unit norm vector that is directed from $\vect{s}$ to $\vect{r}_{\textup{obs}}$. Also, $E_{{\rm{ref,}}x}^t\left( {{{\bf{r}}_{{\rm{Rx}}}},y,{{\bf{r}}_{{\rm{Tx}}}}} \right)$ and $H_{{\rm{ref,}}y}^t\left( {{{\bf{r}}_{{\rm{Rx}}}},y,{{\bf{r}}_{{\rm{Tx}}}}} \right)$ are, by definition, independent of $\vect{r}_{\textup{obs}}$.

Based on this notation and considerations, we evince that the operator $\nabla_{\vect{r}_{\textup{obs}}}$ can be moved inside the integrals in \eqref{eq:As-ref} and \eqref{eq:Ams-ref}, and that it operates only on the Green function. Under the mild assumption ${R_{{\rm{obs}}}} \ge {{z_{{\rm{obs}}}}} \gg \lambda$, the following approximations can be applied to the integrand functions in \eqref{eq:As-ref} and \eqref{eq:Ams-ref}:
\begin{align}
			&\nabla_{\vect{r}_{\textup{obs}}} \times \bigg(\nabla_{\vect{r}_{\textup{obs}}} \times \left(E_{{\rm{ref,}}x}^t\left( {{{\bf{r}}_{{\rm{Rx}}}},y,{{\bf{r}}_{{\rm{Tx}}}}} \right)G(\vect{r}_{\textup{obs}},\vect{s}) \hat{\vect{x}}\right) \bigg) \label{eq:triple-product-Green-at-robs-E} \\
			& \approx -k^2  E_{{\rm{ref,}}x}^t\left( {{{\bf{r}}_{{\rm{Rx}}}},y,{{\bf{r}}_{{\rm{Tx}}}}} \right) G(\vect{r}_{\textup{obs}},\vect{s}) (\hat{\vect{r}}_{\textup{obs}}(\vect{s}) \times (\hat{\vect{r}}_{\textup{obs}}(\vect{s}) \times \hat{\vect{x}})) \nonumber \\
			&\nabla_{\vect{r}_{\textup{obs}}} \times \bigg(\nabla_{\vect{r}_{\textup{obs}}} \times \left(H_{{\rm{ref,}}y}^t\left( {{{\bf{r}}_{{\rm{Rx}}}},y,{{\bf{r}}_{{\rm{Tx}}}}} \right)G(\vect{r}_{\textup{obs}},\vect{s}) \hat{\vect{y}}\right) \bigg) \label{eq:triple-product-Green-at-robs-H} \\
			& \approx -k^2 H_{{\rm{ref,}}y}^t\left( {{{\bf{r}}_{{\rm{Rx}}}},y,{{\bf{r}}_{{\rm{Tx}}}}} \right) G(\vect{r}_{\textup{obs}},\vect{s}) (\hat{\vect{r}}_{\textup{obs}}(\vect{s}) \times (\hat{\vect{r}}_{\textup{obs}}(\vect{s}) \times \hat{\vect{y}})) \nonumber
		\end{align}

The approximations in \eqref{eq:triple-product-Green-at-robs-E} and \eqref{eq:triple-product-Green-at-robs-H} avoid the explicit computation of the derivatives of the electric and magnetic surface fields and, therefore, they make the computation of the power flux relatively simple. In addition, these approximations are applicable to the Fresnel and Fraunhofer regions of the RIS. In the Fraunhofer region of the RIS, \eqref{eq:Eref-at-obs} and \eqref{eq:Href-at-obs} can be further simplified, and, for illustrative purposes, we focus our attention on this regime for the rest of the present paper.

If ${R_{{\rm{obs}}}} \ge {{z_{{\rm{obs}}}}} \ge {{8\left( {L_x^2 + L_y^2} \right)} \mathord{\left/ {\vphantom {{8\left( {L_x^2 + L_y^2} \right)} \lambda }} \right. \kern-\nulldelimiterspace} \lambda }$, specifically, the Poynting vector in \eqref{eq:Poynting-at-obs} can be formulated as follows:
		\begin{align}\label{eq:Poynting-at-obs-amplitude-control}
			\vect{P}_{\textup{obs}}(\vect{r}_{\textup{Rx}},\vect{r}_{\textup{obs}},\vect{r}_{\textup{Tx}})
			& \approx \frac{k^2}{2\eta_0} |{E}^i_{x,0}|^2 \Theta(\theta_r(\vect{r}_{\textup{Rx}}),\theta_o,\varphi_o) \nonumber \\
			&*\left|\mathcal{A}(\vect{r}_{\textup{Rx}},\vect{r}_{\textup{obs}},\vect{r}_{\textup{Tx}})\right|^2 \hat{\vect{r}}_{\textup{obs},0}
		\end{align}
where the following functions are defined:
		\begin{align} \label{eq:A-tilde-with-amplitude-control-electrically-small}
			\mathcal{A}(\vect{r}_{\textup{Rx}},\vect{r}_{\textup{obs}},\vect{r}_{\textup{Tx}})
			& = \frac{e^{-jkR_{\textup{obs}}} }{4\pi R_{\textup{obs}}} \\ & \hspace{-1.0cm} * \int_{\SS} \Gamma_{\SS}(\vect{r}_{\textup{Rx}},y,\vect{r}_{\textup{Tx}})e^{-jk\sin\theta_i(\vect{r}_{\textup{Tx}})y} w_o(x,y) dxdy \nonumber
		\end{align}
\begin{align}
			\Theta(\theta_r(\vect{r}_{\textup{Rx}}),\theta_o,\varphi_o)
			&=
			 \left(1 - \sin^2\theta_o \cos^2\varphi_o \right)\cos^2\theta_r(\vect{r}_{\textup{Rx}}) \nonumber \\
			& \hspace{-1.5cm} + 2\cos\theta_o\cos \theta_r(\vect{r}_{\textup{Rx}}) + 1 - \sin^2\theta_o \sin^2\varphi_o
			\label{eq:C-with-amplitude-control}
		\end{align}
\begin{align}\label{eq:unit-length-vector-of-observation-point}
			\hat{\vect{r}}_{\textup{obs},0} &= \hat{\vect{r}}_{\textup{obs}}(\vect{s}=(0,0))
			= \frac{\vect{r}_{\textup{obs}}}{\norm{\vect{r}_{\textup{obs}}}} \nonumber \\
		& = (\sin\theta_o\cos\varphi_o \hat{\vect{x}} + \sin\theta_o\sin\varphi_o \hat{\vect{y}} + \cos\theta_o \hat{\vect{z}})
		\end{align}
where $w_o(x,y) = e^{jk\left(x \sin\theta_o\cos\varphi_o + y \sin\theta_o\sin\varphi_o\right)}$.

The analytical expression of the power flux evaluated at ${\vect{r}}_{\textup{obs}}$ in \eqref{eq:Poynting-at-obs-amplitude-control} explicitly depends on the surface reflection coefficient $\Gamma_{\SS}(\vect{r}_{\textup{Rx}},y,\vect{r}_{\textup{Tx}})$, it is electromagnetically consistent provided that the Helmholtz constraint in \eqref{eq:Helmholtz} is fulfilled, and it is simple enough for computing the power reradiated from an RIS as a function of the angle of observation. In the next sub-section, \eqref{eq:Poynting-at-obs-amplitude-control} is utilized for evaluating the performance of an RIS as a function of the surface impedance.

\subsection{Optimization of the Surface Impedance}
In this sub-section, we overview mathematical formulations of canonical optimization problems for RISs that fulfill the design criteria introduced in the preceding sub-sections and we evaluate their performance in terms of power flux, i.e., ${P_{{\rm{obs}}}}\left( {{{\bf{r}}_{{\rm{Rx}}}},{{\bf{r}}_{{\rm{obs}}}},{{\bf{r}}_{{\rm{Tx}}}}} \right) = \left| {{{\bf{P}}_{{\rm{obs}}}}\left( {{{\bf{r}}_{{\rm{Rx}}}},{{\bf{r}}_{{\rm{obs}}}},{{\bf{r}}_{{\rm{Tx}}}}} \right)} \right|$, as a function of the observation point. This allows us to characterize the angular response (the reradiation pattern) of an RIS. {Notably, we analyze and compare solutions to the optimization problems that fulfill Helmholtz's condition against solutions that are not subject to Helmholtz's constraint}. As case studies for the optimization criteria, we consider the local design, the global design, and an approximated solution to the global design that is realized by utilizing purely reactive surface impedances. We illustrate how some designs of RISs may lead to a large amount of reradiated power towards directions that are different from the direction of design, i.e., where the receiver Rx is. This is shown to be agreement with Floquet's theorem applied to periodic structures. In light of the power conservation principle, these spurious reflections result in a lower amount of reradiated power towards the intended direction. We discuss how these spurious reflections can be kept under control at the design stage as well.

\begin{table*}[!t]
		\centering
		\caption{Summary of the functions utilized for optimization (${\theta _i} = {\theta _i}\left( {{{\bf{r}}_{{\rm{Tx}}}}} \right)$, ${\theta _r} = {\theta _r}\left( {{{\bf{r}}_{{\rm{Rx}}}}} \right)$).}
		\label{Table_Summary}
		\footnotesize
		\newcommand{\tabincell}[2]{\begin{tabular}{@{}#1@{}}#2\end{tabular}}
		\begin{tabular}{c||c} \hline
			
			Original Equation &  Discretization \\ \hline \hline
			
			Surface impedance (vector) in \eqref{eq:surface-impedance-definition} &  ${\bf{Z}} = \left[ {{Z_1},{Z_2}, \ldots ,{Z_N}} \right], \quad {Z_n} = Z({y_n}),\;n = 1,2, \ldots ,N$ \\ \hline
			
			Surface reflection coefficient (vector) in \eqref{eq:reflection-coefficient-in-surface-impedance} &  ${{\bf{\Gamma }}_{\mathcal{S}}}\left( {\bf{Z}} \right) = \left[ {{\Gamma _1},{\Gamma _2}, \ldots ,{\Gamma _N}} \right],\quad {\Gamma _n}\left( {{Z_n}} \right) = \frac{{{Z_n}\cos {\theta _i} - {\eta _0}}}{{{Z_n}\cos {\theta _r} + {\eta _0}}}$ \\ \hline
			
			$\mathcal{O}_{\SS}({\bf{Z }}) = {P_{\mathcal{S}}}\left( {{{\bf{r}}_{{\rm{Rx}}}},{{\bf{r}}_{{\rm{Tx}}}}} \right)$ in \eqref{eq:globally-passive-condition-0-explicit} &  ${{\mathcal{O}}_{\mathcal{S}}}\left( {\bf{Z}} \right) = \frac{{{{\left| {E_{x,0}^i} \right|}^2}{L_x}}}{{{\eta _0}}}\left( { - 2{L_y}\cos {\theta _i} + {\Delta _y}\sum\limits_{n = 1}^N {\left[ {{{\left| {{\Gamma _n}\left( {{Z_n}} \right)} \right|}^2}\cos {\theta _r} + {\mathop{\Re}\nolimits} \left( {{\Gamma _n}\left( {{Z_n}} \right)} \right)\left( {\cos {\theta _r} - \cos {\theta _i}} \right)} \right]} } \right)$ \\ \hline
			
			
			Helmholtz's condition in \eqref{eq:Helmholtz-condition-reflection-approximate-3} &  $\begin{array}{l} {{\mathcal{H}}_n}\left( {\bf{Z}} \right) = \frac{{\left| {f_n^{''} - 2jk\sin {\theta _i}f_n^{'}} \right|}}{{{k^2}\left| {Z_n^{-} /Z_n^{+} } \right|}}, \quad Z_n^{-}  = {Z^{-} }\left( {{y_n}} \right), \quad Z_n^{+}  = {Z^{+} }\left( {{y_n}} \right) \\ {f_n} = f\left( {{y_n}} \right) = \frac{{Z_n^{-} }}{{Z_n^{+} }}{e^{jk\left( {\sin {\theta _r} - \sin {\theta _i}} \right){y_n}}}, \quad f_n^{'} = \frac{{{f_{n + 1}} - {f_n}}}{{{\Delta _y}}}, \quad f_n^{''} = \frac{{f_{n + 1}^{'} - f_n^{'}}}{{{\Delta _y}}}  \end{array}$ \\ \hline

			Power flow at ${{\bf{r}}_{{\rm{obs}}}}$ ($\varphi_o  = \pi/2$) in \eqref{eq:Poynting-at-obs-amplitude-control} &  $\begin{array}{l}
{{\mathcal{P}}_{{\rm{obs}}}}\left( {\bf{Z}} \right) = \frac{{{k^2}}}{{{\eta _0}}}\frac{{{{\left| {E_{x,0}^i} \right|}^2}{L_x^2}}}{{8{\pi ^2}R_{{\rm{obs}}}^2}}{\left| {{\mathcal{\tilde A}}\left( {\bf{Z}} \right)} \right|^2}\left( {\cos^2 {\theta _r} + \cos^2 {\theta _o} + 2\cos {\theta _r}\cos {\theta _o}} \right)\\
{\mathcal{\tilde A}}\left( {\bf{Z}} \right) = {\Delta _y}\sum\limits_{n = 1}^N {{\Gamma _n}\left( {{Z_n}} \right){e^{ - jk\left( {\sin {\theta _i} - \sin {\theta _o}} \right)y_n}}}
\end{array}$ \\ \hline

		Power flow at ${{\bf{r}}_{{\rm{obs}}}} = {{\bf{r}}_{{\rm{Rx}}}}$ in \eqref{eq:Poynting-at-obs-amplitude-control} &  $\begin{array}{l}
{{\mathcal{P}}_{{\rm{Rx}}}}\left( {\bf{Z}} \right) = \frac{{{k^2}}}{{{\eta _0}}}\frac{{{{\left| {E_{x,0}^i} \right|}^2}{L_x^2}}}{{8{\pi ^2}{{\left| {{{\bf{r}}_{{\rm{Rx}}}}} \right|}^2}}}{\left| {{\mathcal{\tilde A}}\left( {\bf{Z}} \right)} \right|^2}4\cos^2 {\theta _r}\\
{\mathcal{\tilde A}}\left( {\bf{Z}} \right) = {\Delta _y}\sum\limits_{n = 1}^N {{\Gamma _n}\left( {{Z_n}} \right){e^{ - jk\left( {\sin {\theta _i} - \sin {\theta _r}} \right)y_n}}}
\end{array}$ \\ \hline
			
		\end{tabular}
	\end{table*}

For ease of writing and to facilitate the implementation of the numerical algorithms and the computation of the corresponding numerical solutions, we summarize in Table \ref{Table_Summary} the main functions utilized in this sub-section. The functions reported in Table \ref{Table_Summary} are, specifically, discretized into unit cells of length ${\Delta _y}$, which denotes the spatial resolution at which the amplitude and phase of the incident wave are controlled and shaped by the RIS. The variable of optimization is the surface impedance, as it provides direct information on how an RIS is implemented. Since the surface impedance has variations only along the $y$-axis in the considered case study, it is sampled at the center-point of each unit cell, i.e., ${y_n} =  - {L_y} - {{{\Delta _y}} \mathord{\left/ {\vphantom {{{\Delta _y}} 2}} \right. \kern-\nulldelimiterspace} 2} + n{\Delta _y}$ for $n=1,2, \ldots, N$ and ${\Delta _y} = {{2{L_y}} \mathord{\left/ {\vphantom {{2{L_y}} N}} \right. \kern-\nulldelimiterspace} N}$. The resulting $N$ samples are collected in a vector $\bf{Z}$ of size $N$, as defined in Table \ref{Table_Summary}. For simplicity, similar to the incident and reradiated electromagnetic waves, we assume $\varphi_o  = \pi/2$, i.e., the observation point lies in the $yz$-plane. The RIS is optimized based on specified locations of the transmitter Tx (${{{\bf{r}}_{{\rm{Tx}}}}}$) and the receiver Rx (${{{\bf{r}}_{{\rm{Rx}}}}}$), while ${{{\bf{r}}_{{\rm{obs}}}}}$ characterizes the location at which the power flux is observed. In general, ${{\bf{r}}_{{\rm{obs}}}} \ne {{\bf{r}}_{{\rm{Rx}}}} \ne {{\bf{r}}_{{\rm{Tx}}}}$. For simplicity, similar to Table \ref{Table_Summary}, we use the notation ${\theta _i} = {\theta _i}\left( {{{\bf{r}}_{{\rm{Tx}}}}} \right)$ and ${\theta _r} = {\theta _r}\left( {{{\bf{r}}_{{\rm{Rx}}}}} \right)$.

\subsubsection{Benchmark Solution -- Generalized Geometrical Optics}
As a benchmark solution for the considered system designs, we consider the canonical linear phase-gradient design, which we refer to as the generalized geometrical optics solution, since it leads to the so-called generalized law of reflection \cite{Capasso_Science}. The generalized geometrical optics solution is a typical local design, which, however, does not necessarily guarantee a locally unitary power efficiency. It corresponds to the following surface reflection coefficient and surface impedance:
\begin{align}
		&	{\Gamma _{\mathcal{S}}}\left( {{{\bf{r}}_{{\rm{Rx}}}},y,{{\bf{r}}_{{\rm{Tx}}}}} \right) = {\Gamma _{{\rm{GO}}}}\left( y \right) = {e^{ - jk\left( {\sin {\theta _r} - \sin {\theta _i}} \right)y}} \label{eq:GO-Gamma} \\
		&{Z_{\mathcal{S}}}\left( {{{\bf{r}}_{{\rm{Rx}}}},y,{{\bf{r}}_{{\rm{Tx}}}}} \right) = {Z_{{\rm{GO}}}}\left( y \right) = {\eta _0}\frac{{1 + {\Gamma _{{\rm{GO}}}}\left( y \right)}}{{\cos {\theta _i} - {\Gamma _{{\rm{GO}}}}\left( y \right)\cos {\theta _r}}} \label{eq:GO-Z}
		\end{align}
		
The difference between the generalized and the conventional geometrical optics solutions is that in the former case the reflected rays are assumed to propagate towards a direction that is different from that dictated by the conventional law of reflection. In conventional geometrical optics, on the other hand, the rays at every point of the surface $\SS$ are reflected specularly, i.e., the angle of reflection is equal to the angle of incidence, and the surface impedance depends only on the angle of incidence. In mathematical terms, the surface impedance that corresponds to the conventional geometrical optics solution is obtained by replacing $\theta_r\rightarrow \theta_i$ in the denominator of \eqref{eq:GO-Z}, while keeping unchanged the definition of the surface reflection coefficient in \eqref{eq:GO-Gamma}. {Under the assumption that ${\Gamma _{\mathcal{S}}}\left( {{{\bf{r}}_{{\rm{Rx}}}},y,{{\bf{r}}_{{\rm{Tx}}}}} \right)$ is set as in \eqref{eq:GO-Gamma}, the difference between the approximations assumed by the conventional and generalized geometrical optics solutions is the same as the local specular reflection illustrated in Fig. \ref{fig:LocalModel} and the local non-specular reflection illustrated in Fig. \ref{fig:SheetModel}, respectively.}

For completeness, it is instructive to review the analytical steps that allow us to retrieve the geometrical optics solution of the surface reflection coefficient in \eqref{eq:GO-Gamma}. Geometrical optics, or ray optics, is a model for describing the propagation of electromagnetic waves in terms of rays. In geometrical optics, a ray is an abstraction that is useful for approximating the paths along which the electromagnetic waves propagate under certain circumstances. Generally speaking, the definition of ray follows from Fermat's principle, which states that the trajectory between two points taken by a ray is the path that is traversed in the least time. As far as the present paper is concerned, the main properties of the rays that we need to consider are that they propagate along straight-line trajectories as they travel in a homogeneous medium and that they bend at the interface between two dissimilar media.

\begin{figure}[!t]
		\begin{center}
			\includegraphics[width=\linewidth]{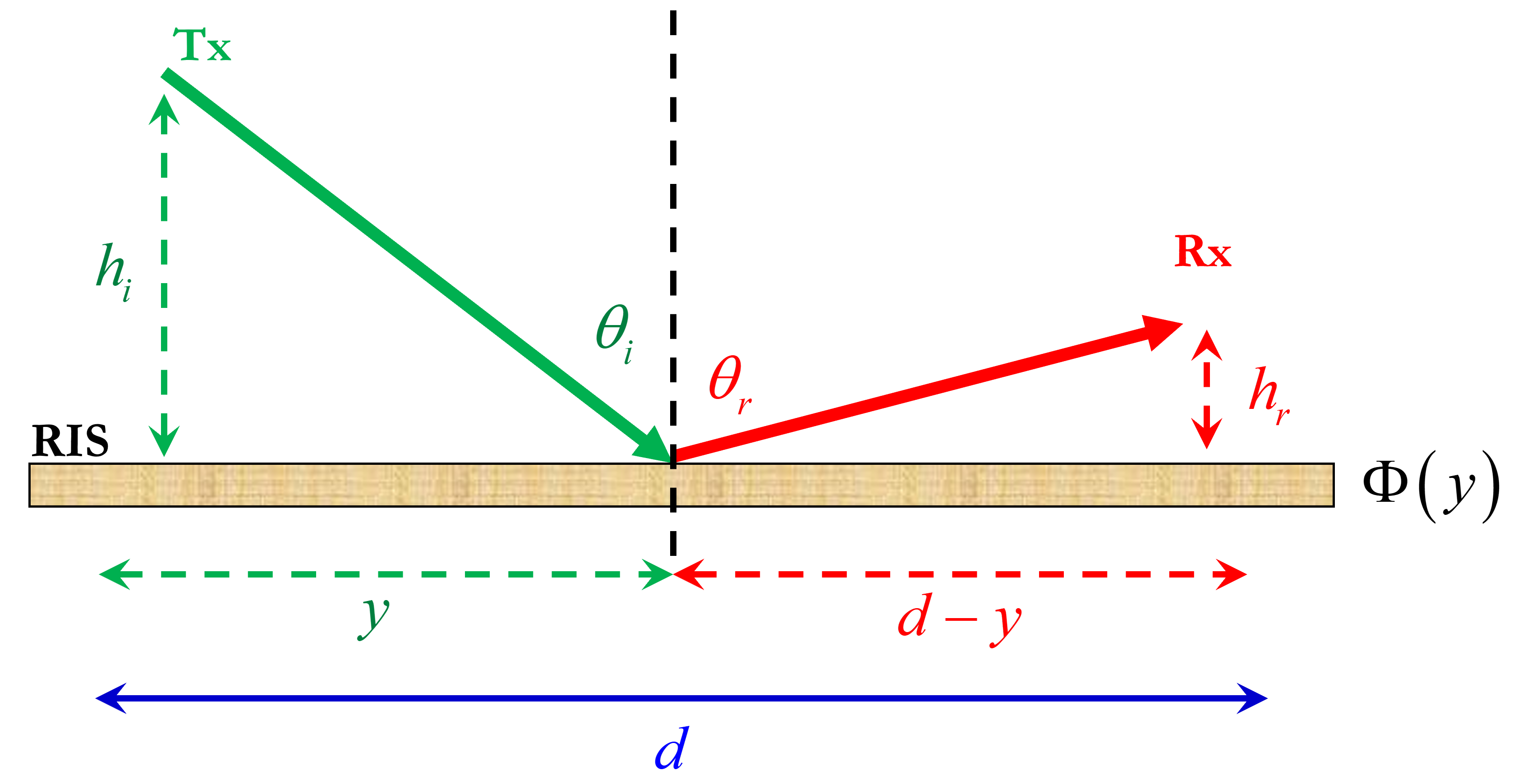}
			\caption{Generalized geometrical optics approximation.}
			\label{fig:GO}
		\end{center}
	\end{figure}
Based on this definition, let us consider the setup illustrated in Fig. \ref{fig:GO}. According to the geometrical optics approximation, an RIS is modeled as a device that is capable of introducing a phase modulation or phase shift, $\Phi \left( y \right)$, to the incident electromagnetic wave. A conventional surface is, on the other hand, characterized by a constant phase modulation along the surface, i.e., $\Phi \left( y \right) = \Phi_0$ $\forall y$. In geometrical optics, an RIS is assumed to be of infinite extend without edges. Also, the geometrical optics approximation does not allow us to model the power of the incident and reflected electromagnetic waves, therefore the reflection coefficient has a unit amplitude by definition. According to the geometrical optics approximation, the problem formulation consists of finding the phase modulation $\Phi \left( y \right)$ so that an electromagnetic wave that impinges upon an RIS from the direction ${\theta _{{i}}}$ is reflected towards the direction ${\theta _{{r}}}$, where the incident and reflected electromagnetic waves are modeled as the two rays illustrated in Fig. \ref{fig:GO}. This problem formulation amounts to identifying the position $y$ on the RIS according to Fermat's principle: The trajectory between two points that is taken by a ray is the path that is traversed in the least time. Since the time and the phase shift of a monochromatic electromagnetic wave with carrier frequency $f$ are proportional to each other, i.e., $\varphi \left( y \right) = 2\pi f\tau \left( y \right)$, where $\varphi \left( y \right)$ and $\tau \left( y \right)$ are the phase shift and the time, respectively, Fermat's principle can be equivalently stated as: The trajectory between two points that is taken by a ray is the path that is traversed by minimizing the phase shift.

Based on Fig. \ref{fig:GO}, the total accumulated phase of the ray that is emitted by the transmitter, that is bent by the RIS when it impinges upon it at $y$, and that reaches the receiver depends on $y$, and it can be formulated as follows:
\begin{align}
\varphi \left( y \right) & = 2\pi f\left( {\frac{{m\sqrt {{y^2} + h_{{i}}^2} }}{c}} \right) \nonumber\\
& + 2\pi f\left( {\frac{{m\sqrt {{{\left( {d - y} \right)}^2} + h_{{r}}^2} }}{c}} \right) \nonumber
 \\ &+ \Phi \left( y \right)
\end{align}
where $m$ denotes the index of refraction of the medium where the RIS is deployed and $c$ is the speed of light.

The trajectory of the ray according to Fermat's theorem is obtained by computing the first-order derivative of $\varphi \left( y \right)$ and equating it to zero, which yields:
\begin{align} \label{PhaseShift_GO}
mk\sin {{\theta _{{i}}}} - mk\sin {{\theta _{{r}}}}  + \frac{{d\Phi \left( y \right)}}{{dy}} = 0
\end{align}
where ${{\theta _{{i}}}}$ and ${{\theta _{{r}}}}$ are the angles of incidence and reflection, respectively. Notably, the expression in \eqref{PhaseShift_GO} is the main criterion for designing conventional reflectarray antennas. In the modern physics literature, \eqref{PhaseShift_GO} is often referred to as the \emph{generalized law of reflection} \cite{Capasso_Science}.

It is apparent from \eqref{PhaseShift_GO} that, given the angle of incidence, the angle of reflection can be appropriately configured by optimizing the first-order derivative of the phase modulation introduced by an RIS. This is the reason why an RIS is often referred to as, according to the (generalized) geometrical optics approximation, a phase-gradient metasurface. If, for example, we set $\Phi \left( y \right) = mk\left( {\sin {\theta _{{\rm{desired}}}} - \sin {\theta _i}} \right)y$, which corresponds to the phase modulation of the reflection coefficient in \eqref{eq:GO-Gamma}, we obtain, as desired, $\theta_r = {{\theta _{{\rm{desired}}}}}$. Therefore, the geometrical optics solution for the surface reflection coefficient and the corresponding surface impedance in \eqref{eq:GO-Z} is a direct consequence of Fermat's principle, under the assumption that an RIS applies a linear phase modulation to the incident electromagnetic wave.

The corresponding discretized versions of the surface reflection coefficient and surface impedance are ${\Gamma _{{\rm{GO}},n}} = {\Gamma _{{\rm{GO}}}}\left( {y_n} \right)$ and ${Z _{{\rm{GO}},n}} = {Z _{{\rm{GO}}}}\left( {y_n} \right)$ for $n=1, 2,\ldots,N$, respectively. In this case, no optimization problem needs to be solved, and the power flux evaluated at ${{\bf{r}}_{{\rm{obs}}}}$ is equal to ${{\mathcal{P}}_{{\rm{obs}}}}\left( {{{\bf{Z}}_{{\rm{GO}}}}} \right)$, as defined in Table \ref{Table_Summary}.

\subsubsection{Global Design -- Unit Power Efficiency}
As elaborated in the previous sub-section, an RIS designed based on a global design with unit power efficiency is a  solution of the following constrained optimization problem:
    \begin{subequations} \label{eq:Global-Opt-Formulation:main}
    \begin{align}
& \mathop {\min }\limits_{\bf{Z}} \left| {{{\mathcal{O}}_{\mathcal{S}}}\left( {\bf{Z}} \right)} \right| \\ 
{\rm{s}}{\rm{.t}}{\rm{.}}\quad \; & {{\mathcal{H}}_n}\left( {\bf{Z}} \right) \le \varepsilon \quad \forall n = 1,2, \ldots ,N-2 \label{eq:Global-Opt-Formulation:a}
		\end{align}
	\end{subequations}
where $\varepsilon$ is a small positive constant and the constraint in \eqref{eq:Global-Opt-Formulation:a} ensures that the obtained surface impedance fulfills Helmholtz's condition.

Given the problem formulation, the solution of the optimization problem in \eqref{eq:Global-Opt-Formulation:main} is not necessarily unique. In terms of system implementation, this can be considered as an advantage. In fact, additional optimization constraints may be added in order to find a solution that has some desired or desirable implementation features. The surface impedance that is solution to the optimization problem in \eqref{eq:Global-Opt-Formulation:main} is denoted by ${{\bf{Z}}_{{\rm{glo0}}}}$, since it is globally optimal and the surface power flow is zero by design (i.e., unit power efficiency). The corresponding power flux evaluated at ${{\bf{r}}_{{\rm{obs}}}}$ is ${{\mathcal{P}}_{{\rm{obs}}}}\left( {{{\bf{Z}}_{{\rm{glo0}}}}} \right)$, as defined in Table \ref{Table_Summary}.

\subsubsection{Approximated Global Design -- Purely Reactive Impedance Boundary}
The surface impedance ${{\bf{Z}}_{{\rm{glo0}}}}$ solution of the optimization problem in \eqref{eq:Global-Opt-Formulation:main} is usually characterized by a non-zero real part. As mentioned in the previous sub-section, this is not always a suitable design from the implementation point of view, since local power gains and local power losses are present along the surface $\SS$. A convenient solution from the implementation point of view is, on the other hand, to impose that the surface impedance is purely reactive, i.e., $\Re({Z_n})=0$ $\forall n=1, 2, \ldots, N$, which is, by definition, both locally and globally optimal and has a unit power efficiency. Inspired by \cite{Grbic_Homogenization}, a suitable approach in the context of wireless communications consists of finding a purely reactive surface impedance that provides almost the same power flux evaluated at ${{\bf{r}}_{{\rm{Rx}}}}$, which is the location of interest, as the power flux obtained with ${{\bf{Z}}_{{\rm{glo0}}}}$. By denoting with ${\bf{Z}}_{{\rm{glo0}}}^{{\rm{reactive}}}$ such a purely reactive surface impedance, the mentioned design criterion corresponds to the condition ${{\mathcal{P}}_{{\rm{Rx}}}}\left( {{\bf{Z}}_{{\rm{glo0}}}^{{\rm{reactive}}}} \right) \approx {{\mathcal{P}}_{{\rm{Rx}}}}\left( {{{\bf{Z}}_{{\rm{glo0}}}}} \right)$.

The corresponding optimization problem is, therefore, formulated as follows:
\begin{subequations} \label{eq:Pratical-Opt-Formulation:main}
\begin{align}
&\mathop {\min }\limits_{\bf{Z}} \left| {{{\mathcal{P}}_{{\rm{Rx}}}}\left( {\bf{Z}} \right) - {{\mathcal{P}}_{{\rm{Rx}}}}\left( {{{\bf{Z}}_{{\rm{glo0}}}}} \right)} \right| \\ 
{\rm{s}}{\rm{.t}}{\rm{.}}\quad \; & {{\mathcal{H}}_n}\left( {\bf{Z}} \right) \le \varepsilon \quad \forall n = 1,2, \ldots ,N - 2 \label{eq:Practical-Opt-Formulation:a} \\
\quad \;\quad & {\mathop{\Re}\nolimits} \left( {{Z_n}} \right) = 0\quad \forall n = 1,2, \ldots ,N \label{eq:Practical-Opt-Formulation:b}
\end{align}
\end{subequations}
where ${{{\mathcal{P}}_{{\rm{Rx}}}}\left( {{{\bf{Z}}_{{\rm{glo0}}}}} \right)}$ is the power flux evaluated at the receiver Rx by using the surface impedance that is solution of the optimization problem in \eqref{eq:Global-Opt-Formulation:main}, and the constraint in \eqref{eq:Practical-Opt-Formulation:b} ensures that the real part of the impedance is equal to zero. As mentioned, the corresponding power flux evaluated at ${{\bf{r}}_{{\rm{obs}}}}$ is equal to ${{\mathcal{P}}_{{\rm{obs}}}}\left( {{\bf{Z}}_{{\rm{glo0}}}^{{\rm{reactive}}}} \right)$.

\subsubsection{Optimization Constraints on Spurious Reflections}
By definition, the optimization problems formulated in \eqref{eq:Global-Opt-Formulation:main} and \eqref{eq:Pratical-Opt-Formulation:main} specify the power efficiency and the power flux only in correspondence of the location of the receiver Rx, while they do not explicitly account, in the problem formulation, for the power reradiated towards directions different from that of the receiver Rx. In general, this implies that a large amount of power may be reradiated towards undesired directions, i.e., directions are are different from the target direction of reflection (where the receiver Rx is). This is apparent by direct inspection of, e.g., \eqref{eq:Pratical-Opt-Formulation:main} and is consistent with Floquet's theory applied to periodic RISs, as discussed in the previous sub-sections. The problem formulation imposes that the power fluxes ${{\mathcal{P}}_{{\rm{Rx}}}}\left( {{\bf{Z}}_{{\rm{glo0}}}^{{\rm{reactive}}}} \right)$ and ${{\mathcal{P}}_{{\rm{Rx}}}}\left( {{{\bf{Z}}_{{\rm{glo0}}}}} \right)$ are approximately the same at the receiver Rx, but they may be different at locations ${{\bf{r}}_{{\rm{obs}}}} \ne {{\bf{r}}_{{\rm{Rx}}}}$. For example, strong reflections towards some directions may emerge since they are not controlled at the design stage. The same comment applies to \eqref{eq:Global-Opt-Formulation:main}, since a large amount of power may be reradiated towards directions different from ${{\bf{r}}_{{\rm{Rx}}}}$, as no specific constraint is added in the formulation of the optimization problem. Several research works have reported that spurious reflections are often and usually observed \cite{Vittorio_RayTracing}, \cite{Sergei_MacroscopicModel}. {The presence of undesired reflections towards directions that are different from the desired direction of reflection has recently been extensively discussed in \cite{DigitalRIS_Elsevier}, in the context of RISs that are optimized based on the locally periodic discrete model. In \cite{DigitalRIS_Elsevier}, it is shown that non-ideal RIS alphabets may result in a large amount of power that is directed towards undesired directions.}

In order to make sure that an optimized RIS does not produce undesired reflections \textit{by design}, the optimization problems in \eqref{eq:Global-Opt-Formulation:main} and \eqref{eq:Pratical-Opt-Formulation:main} need to be modified by adding specified constrains to the radiation pattern of the RIS. Specifically, the optimization problems in \eqref{eq:Global-Opt-Formulation:main} and \eqref{eq:Pratical-Opt-Formulation:main} can be reformulated, respectively, as follows:
    \begin{subequations} \label{eq:Mask-Global-Opt-Formulation:main}
    \begin{align}
& \mathop {\min }\limits_{\bf{Z}} \left| {{{\mathcal{O}}_{\mathcal{S}}}\left( {\bf{Z}} \right)} \right| \tag{\ref{eq:Mask-Global-Opt-Formulation:main}} \\
{\rm{s}}{\rm{.t}}{\rm{.}}\quad \; & {{\mathcal{H}}_n}\left( {\bf{Z}} \right) \le \varepsilon \quad \forall n = 1,2, \ldots ,N-2 \label{eq:Mask-Global-Opt-Formulation:a} \\
& {{\mathcal{P}}_{{\rm{obs}}}}\left( {\bf{Z}} \right) \le \delta \quad {\theta _o} \in \left[ {{\theta _{1l }},{\theta _{1u }}} \right],\;\left[ {{\theta _{2l }},{\theta _{2u }}} \right], \ldots \label{eq:Mask-Global-Opt-Formulation:b}
		\end{align}
	\end{subequations}
\begin{subequations} \label{eq:Mask-Pratical-Opt-Formulation:main}
\begin{align}
&\mathop {\min }\limits_{\bf{Z}} \left| {{{\mathcal{P}}_{{\rm{Rx}}}}\left( {\bf{Z}} \right) - {{\mathcal{P}}_{{\rm{Rx}}}}\left( {{{\bf{Z}}_{{\rm{glo0}}}}} \right)} \right| \\ 
{\rm{s}}{\rm{.t}}{\rm{.}}\quad \; & {{\mathcal{H}}_n}\left( {\bf{Z}} \right) \le \varepsilon \quad \forall n = 1,2, \ldots ,N - 2 \label{eq:Mask-Practical-Opt-Formulation:a} \\
\quad \;\quad & {\mathop{\Re}\nolimits} \left( {{Z_n}} \right) = 0\quad \forall n = 1,2, \ldots ,N \label{eq:Mark-Practical-Opt-Formulation:b} \\
& {{\mathcal{P}}_{{\rm{obs}}}}\left( {\bf{Z}} \right) \le \delta \quad {\theta _o} \in \left[ {{\theta _{1l }},{\theta _{1u }}} \right],\;\left[ {{\theta _{2l }},{\theta _{2u }}} \right], \ldots \label{eq:Mask-Global-Opt-Formulation:c}
\end{align}
\end{subequations}
where $\left[ {{\theta _{i,\min }},{\theta _{i,\max }}} \right]$ for $i=1, 2, \ldots$ are specified angular sectors where the reradiation of the RIS needs to be kept under some maximum power reradiation constraints and $\delta$ is a small positive constant that quantifies the reradiated power that is allowed towards the specified angular sectors.

\begin{figure}[!t]
		\begin{center}
			\includegraphics[width=\linewidth]{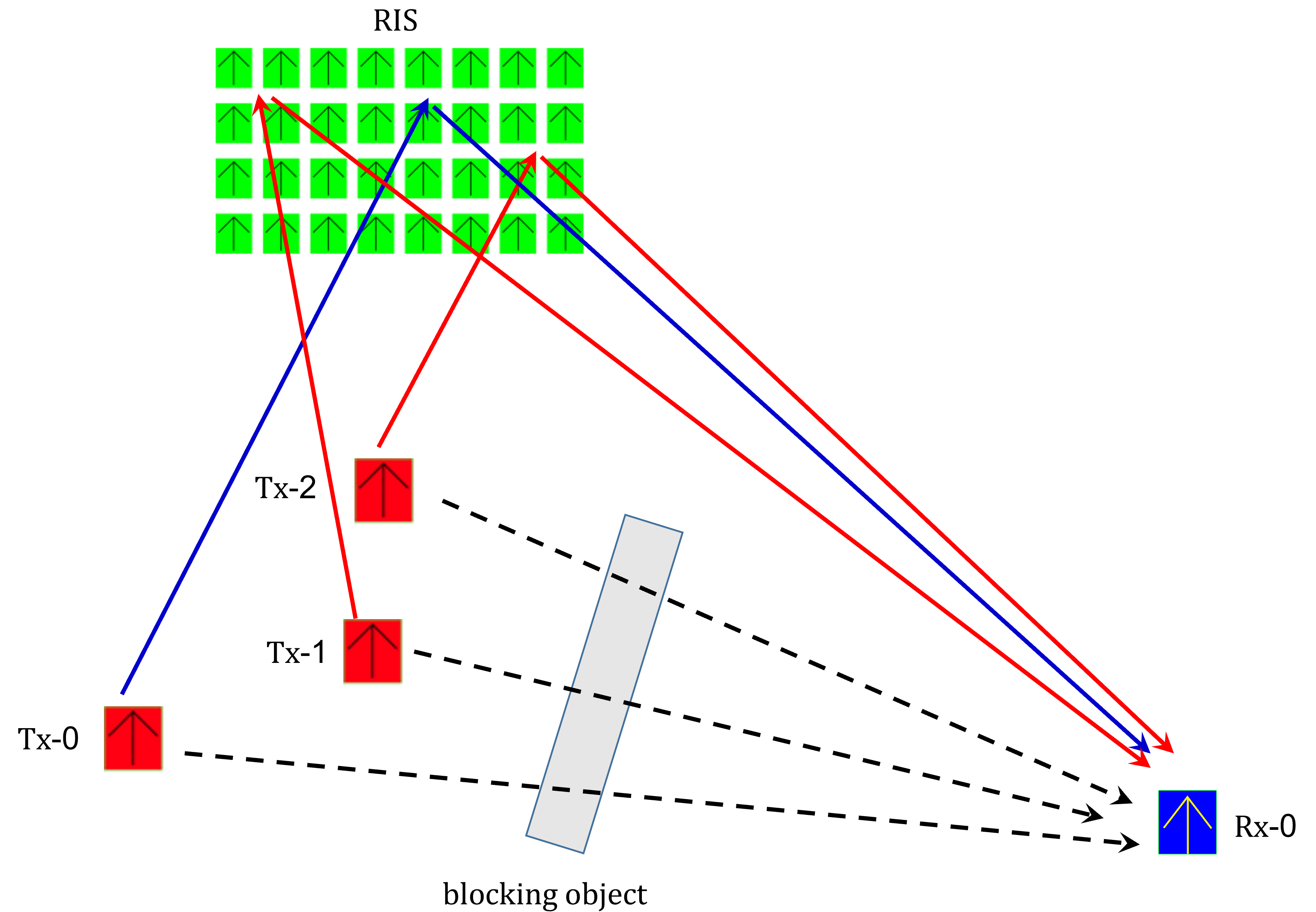}
			\caption{Typical RIS-aided wireless interference network with one intended transmitter-receiver pair and two interfering links.}
			\label{fig:WirelessApplication}
		\end{center}
	\end{figure}
{\textbf{How to use these models and methods in wireless communications?}}
{The formulated problems are focused on optimizing the received power at some locations of interest, i.e., where the intended receiver is. They constitute, therefore, the key ingredient for formulating more complex optimization problems that are of interest in wireless communications. As a case study, let us assume a wireless network with one intended link and $I$ interfering links. For simplicity, we analyze single antenna transmitters and receivers. The intended link is identified by the pair of transmitter ${\rm{Tx}}_0$ and receiver ${\rm{Rx}}_0$, respectively, and the $i$th interfering link is identified by the pair of transmitter ${\rm{Tx}}_i$ and receiver ${\rm{Rx}}_0$. We assume that the data transmission between ${\rm{Tx}}_0$ and ${\rm{Rx}}_0$ in the presence of the $I$ interfering transmitters ${\rm{Tx}}_i$ is assisted by an RIS. For simplicity, we assume no direct links between the transmitters and the receiver, i.e., only the RIS-aided links are available. An illustration of the considered network model is provided in Fig. \ref{fig:WirelessApplication}. The RIS is identified by the surface impedance $Z\left( {{{\bf{r}}_{{\rm{Rx}}_0}},y,{{\bf{r}}_{{\rm{Tx}}_0}}} \right)$ (or more, in general, $Z\left( {{{\bf{r}}_{{\rm{Rx}}_0}},\vect{s},{{\bf{r}}_{{\rm{Tx}}_0}}} \right)$). The intended link and interfering links are characterized by the received powers ${{\mathcal{P}}_{{\rm{Rx}},0}}\left( {\bf{Z}} \right)$ and ${{\mathcal{P}}_{{\rm{Rx}},i}}\left( {\bf{Z}} \right)$, respectively, where ${\bf{Z}}$ is the vector of the sampled surface impedance defined in Table \ref{Table_Summary}.}

{The performance of a wireless system is tightly linked to the statistical distribution of the signal-to-interference+noise-ratio (SINR) \cite{JeffFrancois_SG}, \cite{MDR_SG}, which is usually defined as follows \cite{Hamdi_SINR}:     
\begin{equation} \label{eq:SINR}
{\rm{SINR}}\left( {\bf{Z}} \right) = \frac{{{{\mathcal{P}}_{{\rm{Rx,0}}}}\left( {\bf{Z}} \right)}}{{\sigma _N^2{\rm{ + }}\sum\limits_{i = 1}^I {{{\mathcal{P}}_{{\rm{Rx,i}}}}\left( {\bf{Z}} \right)} }}
\end{equation}
where ${\sigma _N^2}$ is the noise power at the receiver ${\rm{Rx}}_0$.}

{The proposed optimization problems can be reformulated by replacing the received power with the SINR in \eqref{eq:SINR}, as the key performance indicator of interest. It is important to mention, however, that the problem formulations considered in the present paper are applicable to free-space communications, since the channel is modeled through the free-space Green's function. The proposed approach can, however, be generalized for application to fading channels, either by utilizing the stochastic Green's function \cite{StochasticGreenFunction_CONF}, \cite{StochasticGreenFunction_PAPER} or by resorting to the system-level method in \cite{MDR_StatisticalLocation}, \cite{MDR_MutualImpedancesMIMO}, where the RIS-aided links are assumed to be immersed in a statistical multipath channel in which the RIS is viewed as an additional digitally controllable scatterer. The generalization to multiple-antenna transmitters and receivers is possible as well, but the impact of the multiple antennas needs to be accounted for starting from \eqref{eq:Eref-at-obs} and \eqref{eq:Href-at-obs}. The proposed modeling and optimization framework can, therefore, serve as a starting point to analyze wireless networks and channels that include multiple-antenna transceivers and multipath propagation. Therefore, it has several applications in wireless communications.}

\begin{table}[!t]
		\centering
		\caption{Parameters setup.}
		\label{table:setup-optimization-with-null}
		\footnotesize
		\newcommand{\tabincell}[2]{\begin{tabular}{@{}#1@{}}#2\end{tabular}}
		\begin{tabular}{c|c} \hline
			Parameter &  Value \\ \hline
			$f$				& 28 GHz \\
			$\lambda = c/f$	& $10.7$ mm\\
			$\theta_i(\vect{r}_{\textup{Tx}})$ 	& $0^\circ$\\
			$\theta_r(\vect{r}_{\textup{Rx}})$ 	& {$\{30^\circ, 75^\circ\}$}\\
			$\norm{\vect{r}_{\textup{Rx}}} = R_{\textup{obs}}$		& $100$ m\\
			$\eta_0$							& $377~\Omega$\\
			$P_0$ & $1$ Watt/m$^2$\\
			$|E^i_{x,0}| = \sqrt{2P_0\eta_0}$		& $27.45$ V/m\\
			$L_x$						& $0.5$ m\\
			$L_y$						& $0.25$ m\\
			$\Delta_y$							& $\lambda/32$\\
			$\delta$ & $10^{-4}$\\
			$[\theta_{1l}, \theta_{1u}]$ & $\{0^\circ, 1^\circ\}$, step = $0.1^\circ$ \\
			\hline
		\end{tabular}
	\end{table}
\section{Numerical Examples}
In this section, we provide some numerical examples in order to compare and discuss the optimal designs for RISs that are obtained as solutions of the optimization problems formulated in \eqref{eq:Global-Opt-Formulation:main}, \eqref{eq:Pratical-Opt-Formulation:main}, \eqref{eq:Mask-Global-Opt-Formulation:main}, and \eqref{eq:Mask-Pratical-Opt-Formulation:main}. {The aim of this section is to showcase, with the aid of numerical results and illustrations, the properties of the obtained surface impedances, the impact of the Helmholtz constraint, and the reradiated power flux as a function of the angle of observation.}

The simulation setup is given in Table \ref{table:setup-optimization-with-null}. As examples, we assume that the incident electromagnetic wave impinges upon the RIS from the normal direction, i.e., ${{\theta _i}\left( {{{\bf{r}}_{{\rm{Tx}}}}} \right)}=0^\circ$. Two desired angles of reradiation are considered: ${{\theta _r}\left( {{{\bf{r}}_{{\rm{Rx}}}}} \right)} = 30^\circ$ and ${{\theta _r}\left( {{{\bf{r}}_{{\rm{Rx}}}}} \right)} = 75^\circ$. This choice is made in order to highlight the differences, in terms of surface impedance, between a relatively small and a relatively large angle of reflection with respect to the angle of incidence. As far as the nullification of possible spurious reflections is concerned, we focus our attention on the reradiation towards the specular direction, since this is one of the most important undesired reradiated modes in the considered case study. This is also in agreement with Floquet's theory that was reviewed in Section III. According to Floquet's theory, we may expect, in general, the existence of more than one spurious reflection, while parasitic specular reflection is always possible. Therefore, we focus our attention on the reradiation towards the specular direction as an example, and elaborate this point in further text. {As far as the Helmholtz constraint $\varepsilon$ is concerned, we consider different case studies in order to analyze the impact of fulfilling Helmholtz's condition on the surface impedance. Specifically, we analyze case studies for which $\varepsilon = 5 \cdot 10^{-2}$ and case studies for which $\varepsilon$ is large, which corresponds to solutions that are not subject to Helmholtz's constraint}.

The optimization problems are solved by using the \verb+fmincon+ function in Matlab, which is a gradient-based algorithm that is designed to work on problems where the objective function and the constraint functions are continuous and have continuous first-order derivatives. The \verb+fmincon+ function is designed to find the minimum of constrained nonlinear multivariable functions. As far the optimization problems in \eqref{eq:Global-Opt-Formulation:main} and \eqref{eq:Mask-Global-Opt-Formulation:main} are concerned, the \verb+fmincon+ function is initialized with the geometrical optics solution ${{\bf{Z}}_{\rm{0}}} = {{\bf{Z}}_{{\rm{GO}}}}$ in \eqref{eq:GO-Z}. As far the optimization problems in \eqref{eq:Pratical-Opt-Formulation:main} and \eqref{eq:Mask-Pratical-Opt-Formulation:main} are concerned, the \verb+fmincon+ function is initialized with the imaginary part of the solution of the optimization problems in \eqref{eq:Global-Opt-Formulation:main} and \eqref{eq:Mask-Global-Opt-Formulation:main}, i.e., ${{\bf{Z}}_{\rm{0}}} = j{\mathop{\Im}\nolimits} \left( {{{\bf{Z}}_{{\rm{glo0}}}}} \right)$.

\begin{figure*}[!t]
		\begin{center}
			\begin{subfigure}{0.66\columnwidth}
				{\includegraphics[width=\linewidth]{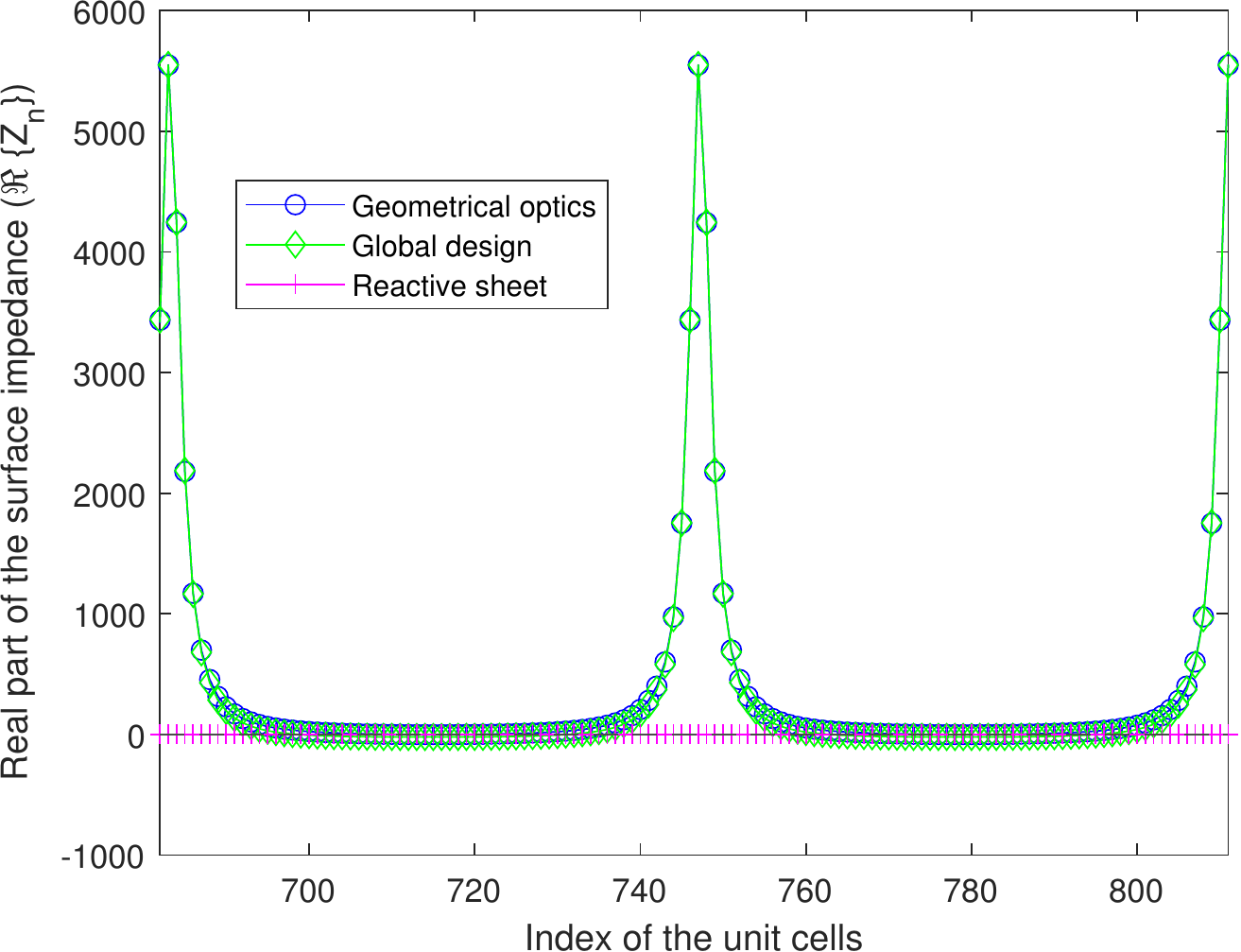}}
				\caption{Real part of surface impedance ${\mathop{\Re}\nolimits} \left( {{Z_n}} \right)$.}\label{fig:Z_real_no_null_thetaR_30_deg_FINAL}
			\end{subfigure}
			\begin{subfigure}{0.66\columnwidth}
				{\includegraphics[width=\linewidth]{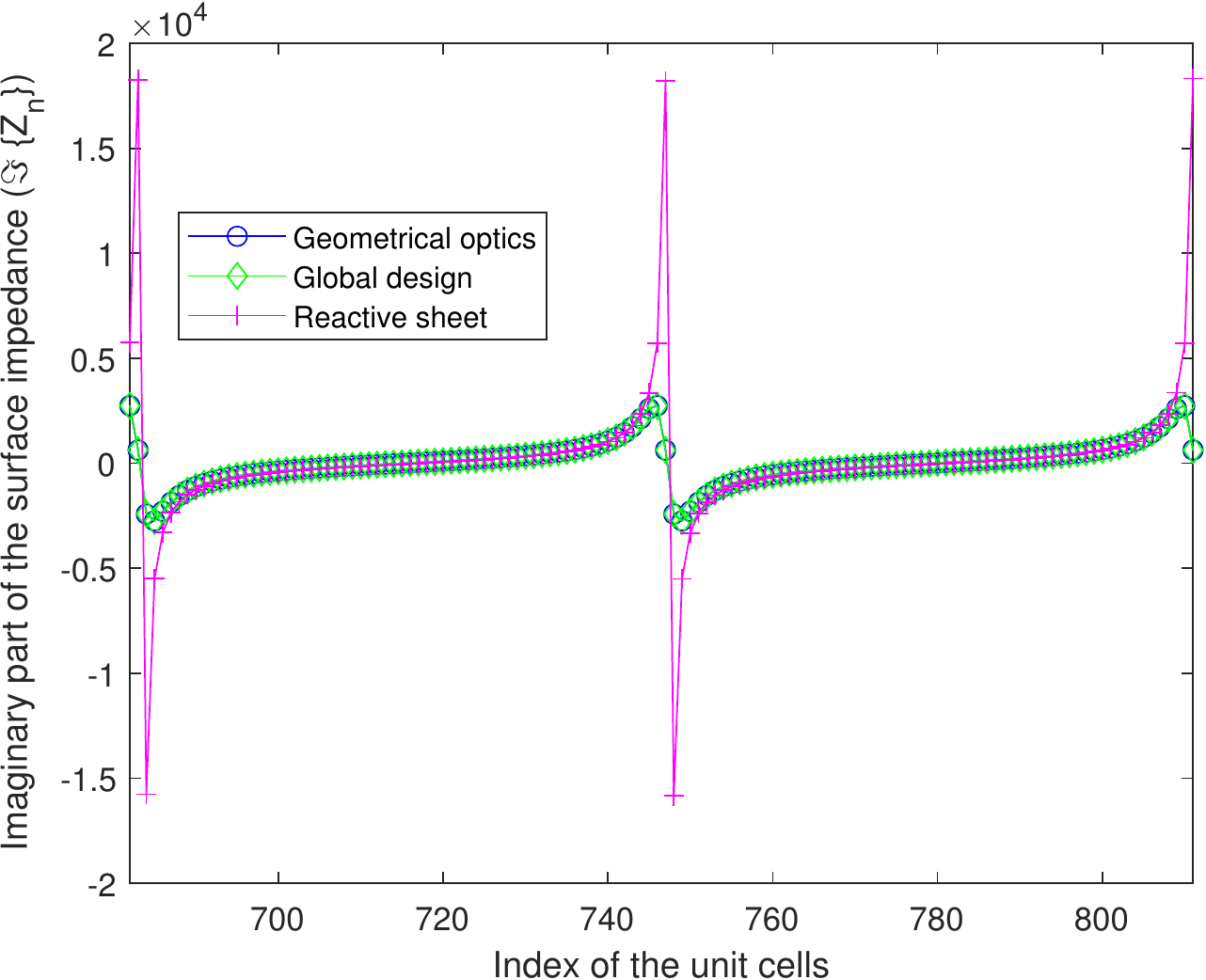}}
				\caption{Imaginary part of surface impedance ${\mathop{\Im}\nolimits} \left( {{Z_n}} \right)$.}\label{fig:Z_imag_no_null_thetaR_30_deg_FINAL}
			\end{subfigure}
			\begin{subfigure}{0.66\columnwidth}
				{\includegraphics[width=\linewidth]{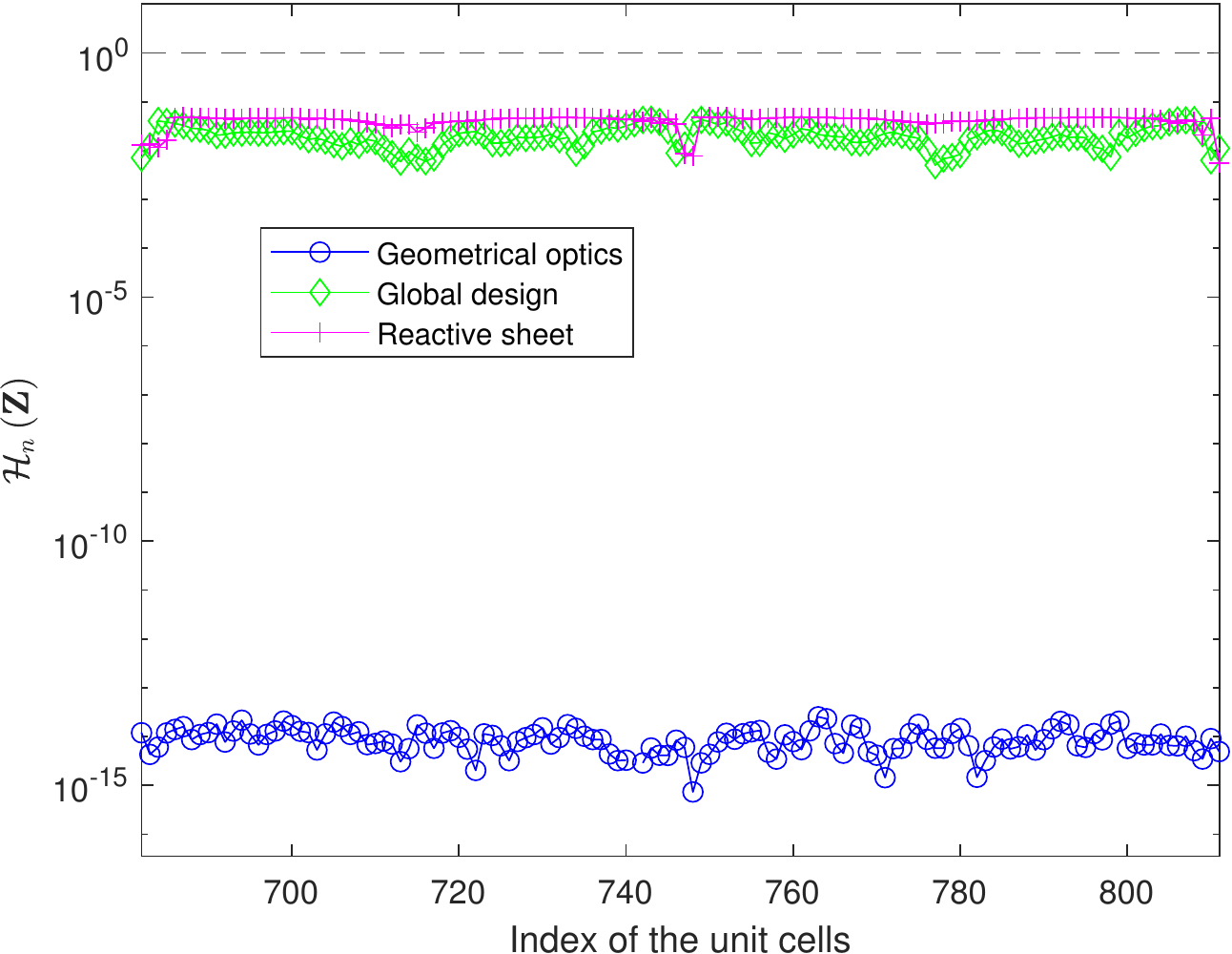}}
				\caption{Helmholtz's ratio ${{\mathcal{H}}_n}\left( {\bf{Z}}\right)$.}\label{fig:Helmholtz_Ratio_No_Null_thetaR_30_deg_FINAL}
			\end{subfigure}
			\begin{subfigure}{0.99\columnwidth}
				{\includegraphics[width=\linewidth]{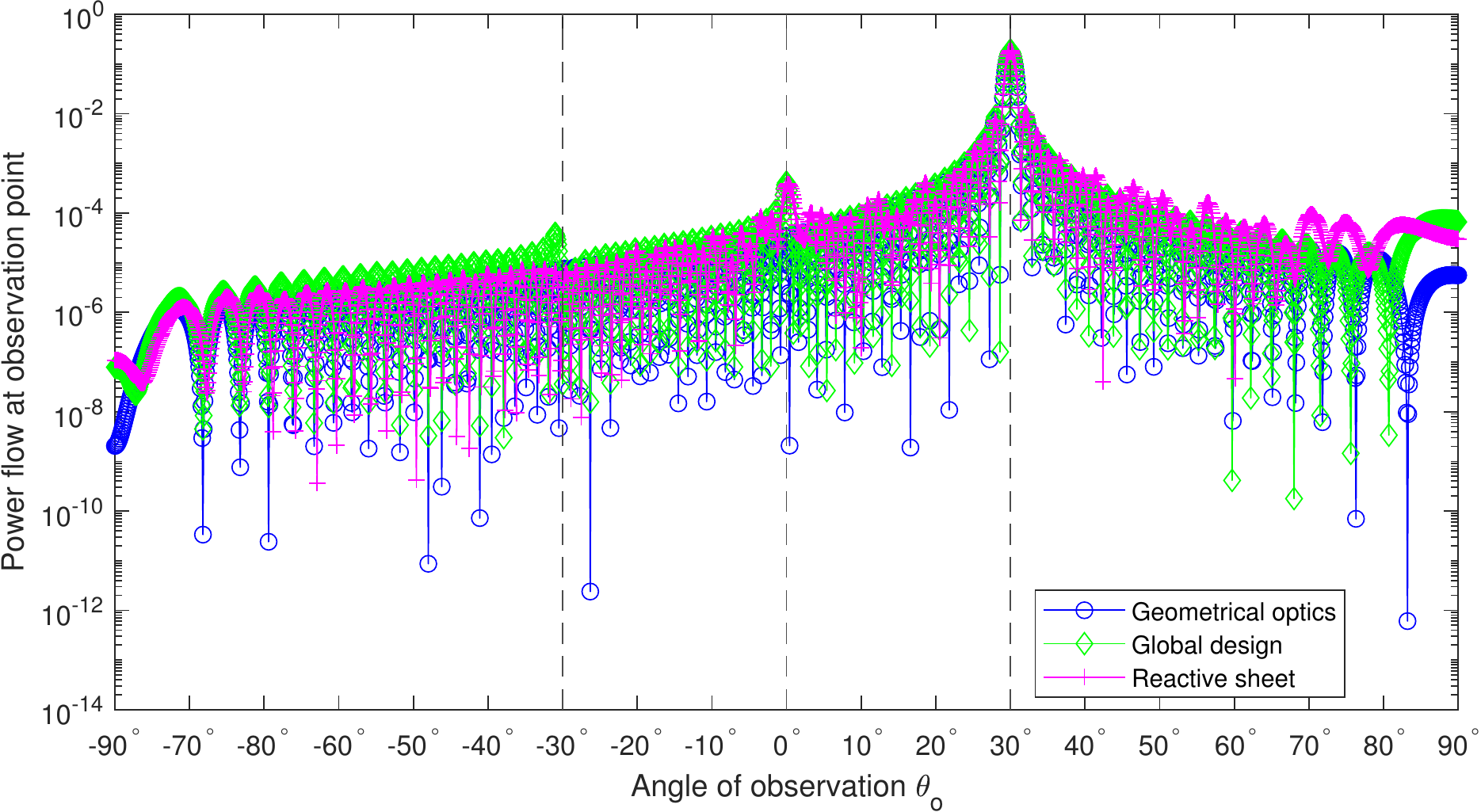}}
				\caption{Power flux vs. angle of observation ${{\mathcal{P}}_{{\rm{obs}}}}\left( {\bf{Z}} \right)$ (dB plot).}\label{fig:Poynting_Obs_No_Null_thetaR_30_deg_in_dB_FINAL}
			\end{subfigure}
			\begin{subfigure}{0.99\columnwidth}
				{\includegraphics[width=\linewidth]{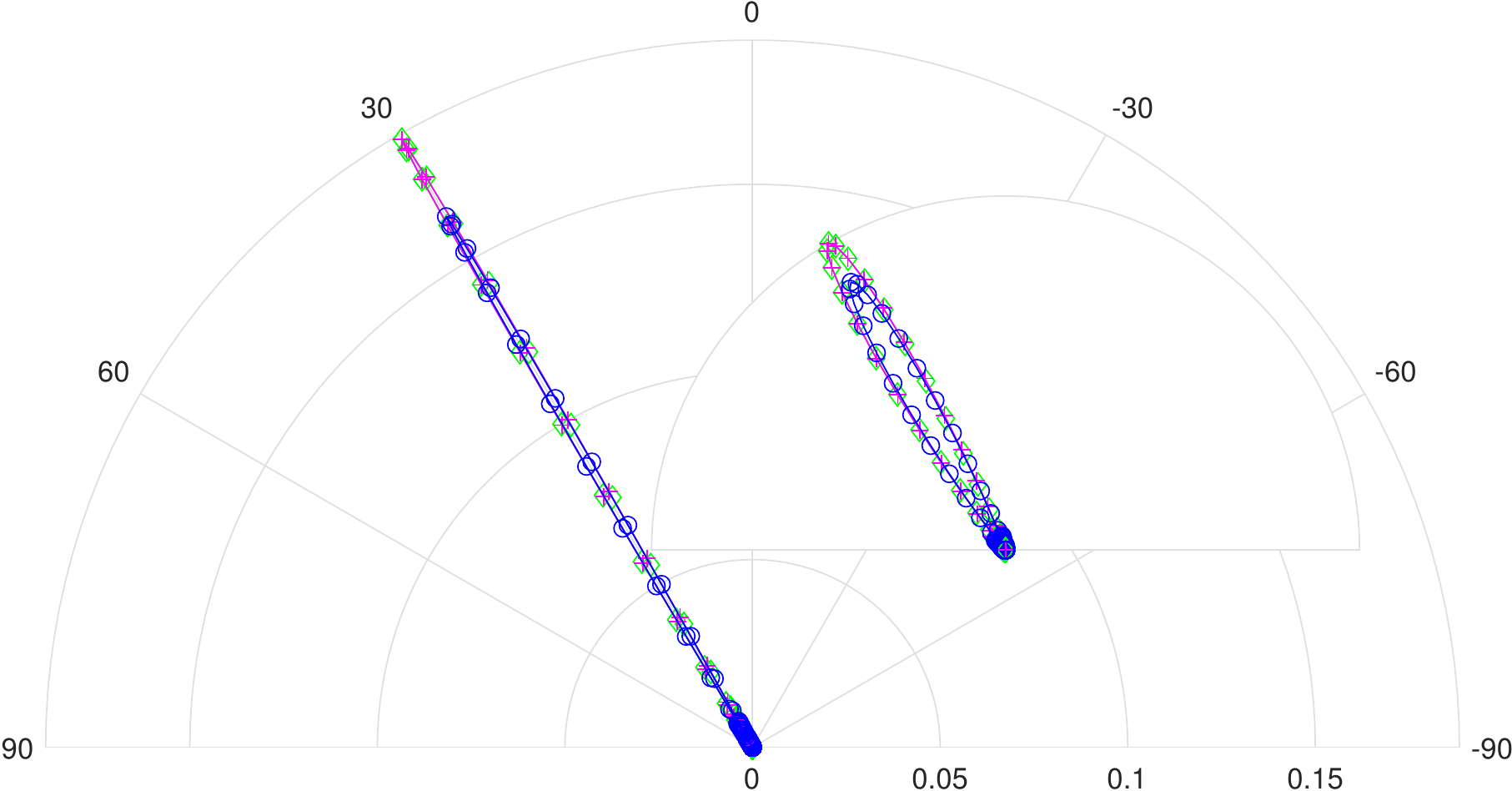}}
				\caption{Power flux vs. angle of observation ${{\mathcal{P}}_{{\rm{obs}}}}\left( {\bf{Z}} \right)$ (polar plot).}\label{fig:Poynting_Obs_No_Null_thetaR_30_deg_polar_FINAL}
			\end{subfigure}
		\end{center}
		\caption{Surface impedance in \eqref{eq:GO-Z} and solution of the optimization problems in \eqref{eq:Global-Opt-Formulation:main} and \eqref{eq:Pratical-Opt-Formulation:main} ($\theta_r(\vect{r}_{\textup{Rx}}) = 30^\circ$).}
		\label{fig:No_Null_thetaR_30_FINAL}
	\end{figure*}
	\begin{figure*}[!t]
		\begin{center}
			\begin{subfigure}{0.66\columnwidth}
				{\includegraphics[width=\linewidth]{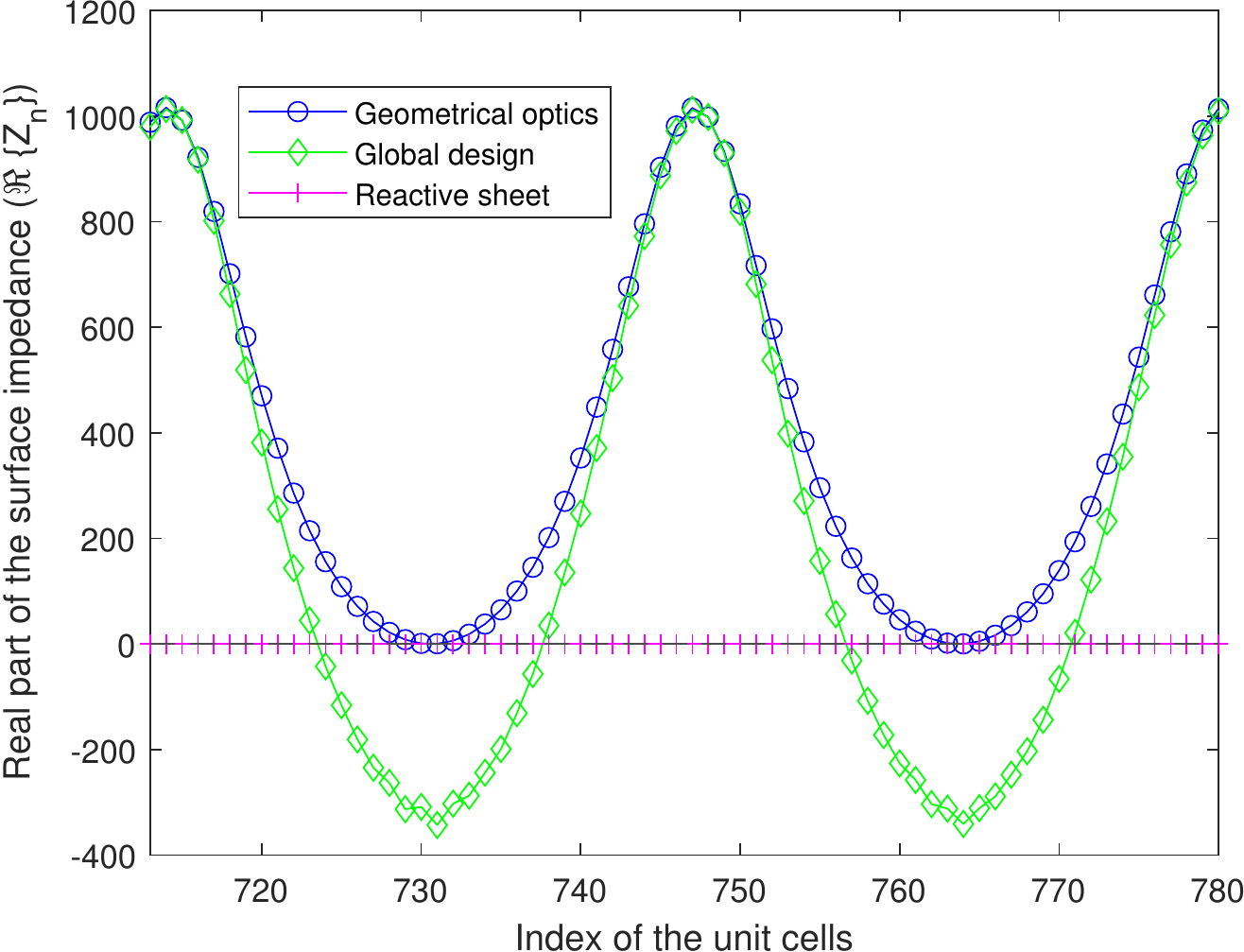}}
				\caption{Real part of surface impedance ${\mathop{\Re}\nolimits} \left( {{Z_n}} \right)$.}\label{fig:Z_real_no_null_thetaR_75_deg_FINAL}
			\end{subfigure}
			\begin{subfigure}{0.66\columnwidth}
				{\includegraphics[width=\linewidth]{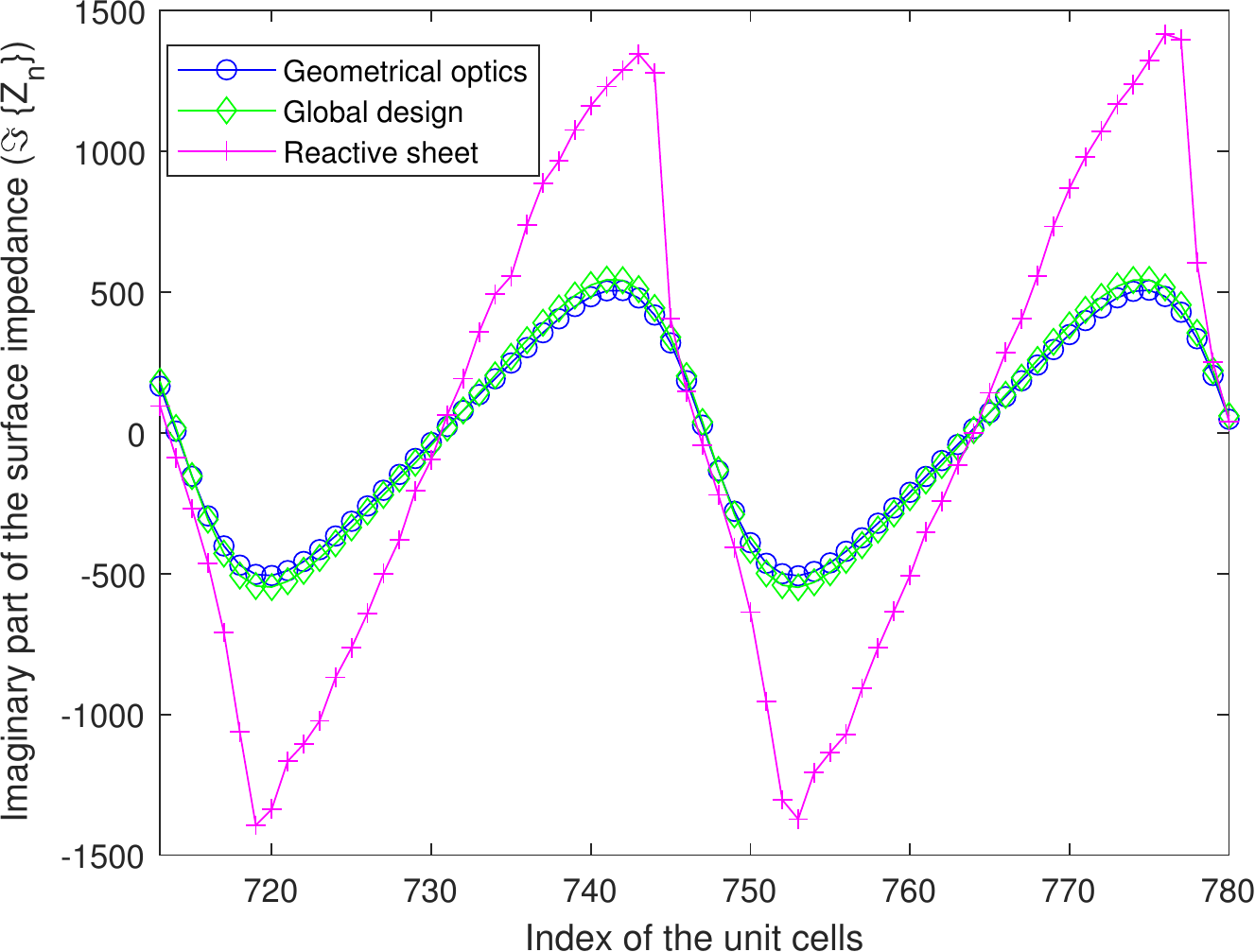}}
				\caption{Imaginary part of surface impedance ${\mathop{\Im}\nolimits} \left( {{Z_n}} \right)$.}\label{fig:Z_imag_no_null_thetaR_75_deg_FINAL}
			\end{subfigure}
			\begin{subfigure}{0.66\columnwidth}
				{\includegraphics[width=\linewidth]{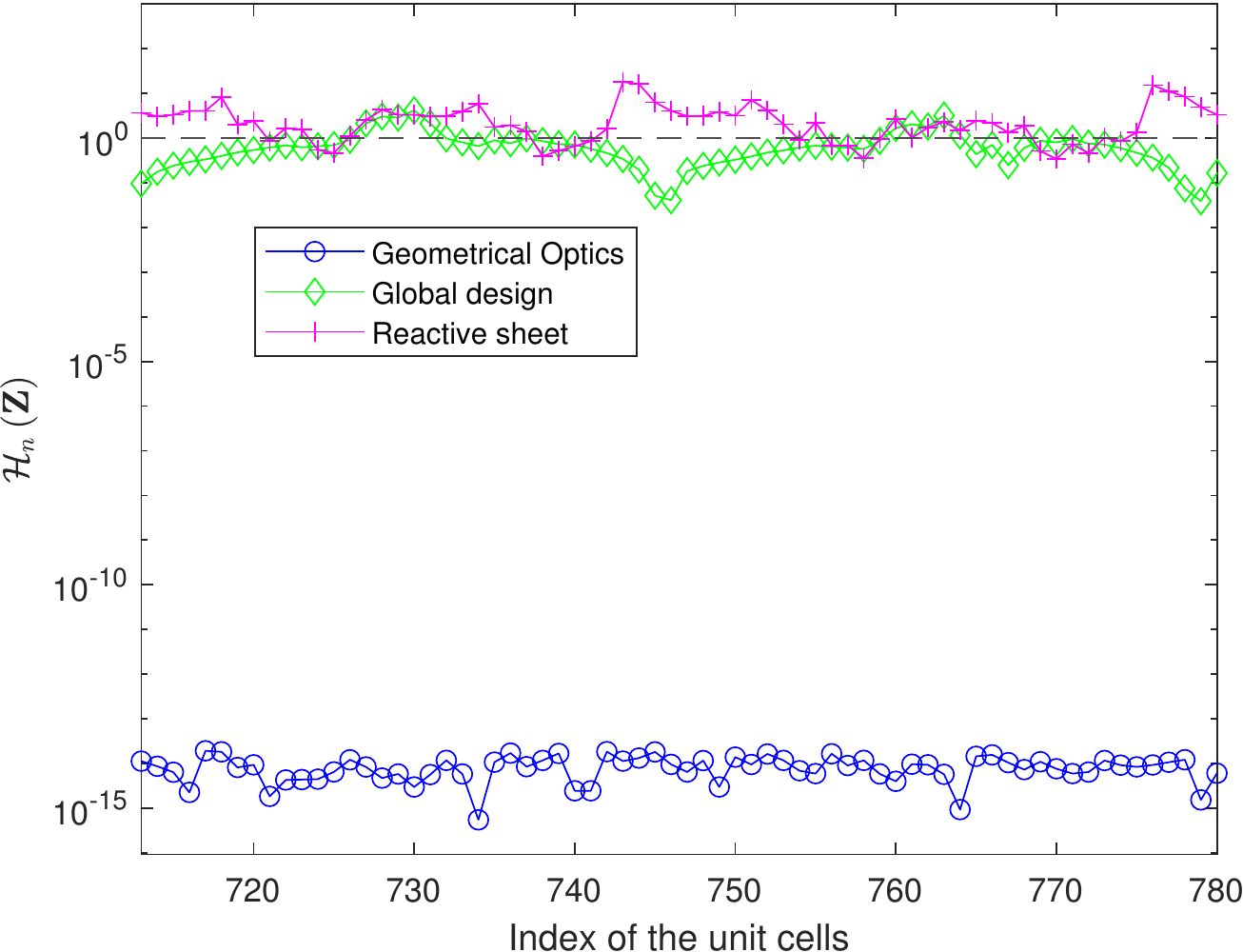}}
				\caption{Helmholtz's ratio ${{\mathcal{H}}_n}\left( {\bf{Z}} \right)$.}\label{fig:Helmholtz_Ratio_No_Null_thetaR_75_deg_FINAL}
			\end{subfigure}
			\begin{subfigure}{0.99\columnwidth}
				{\includegraphics[width=\linewidth]{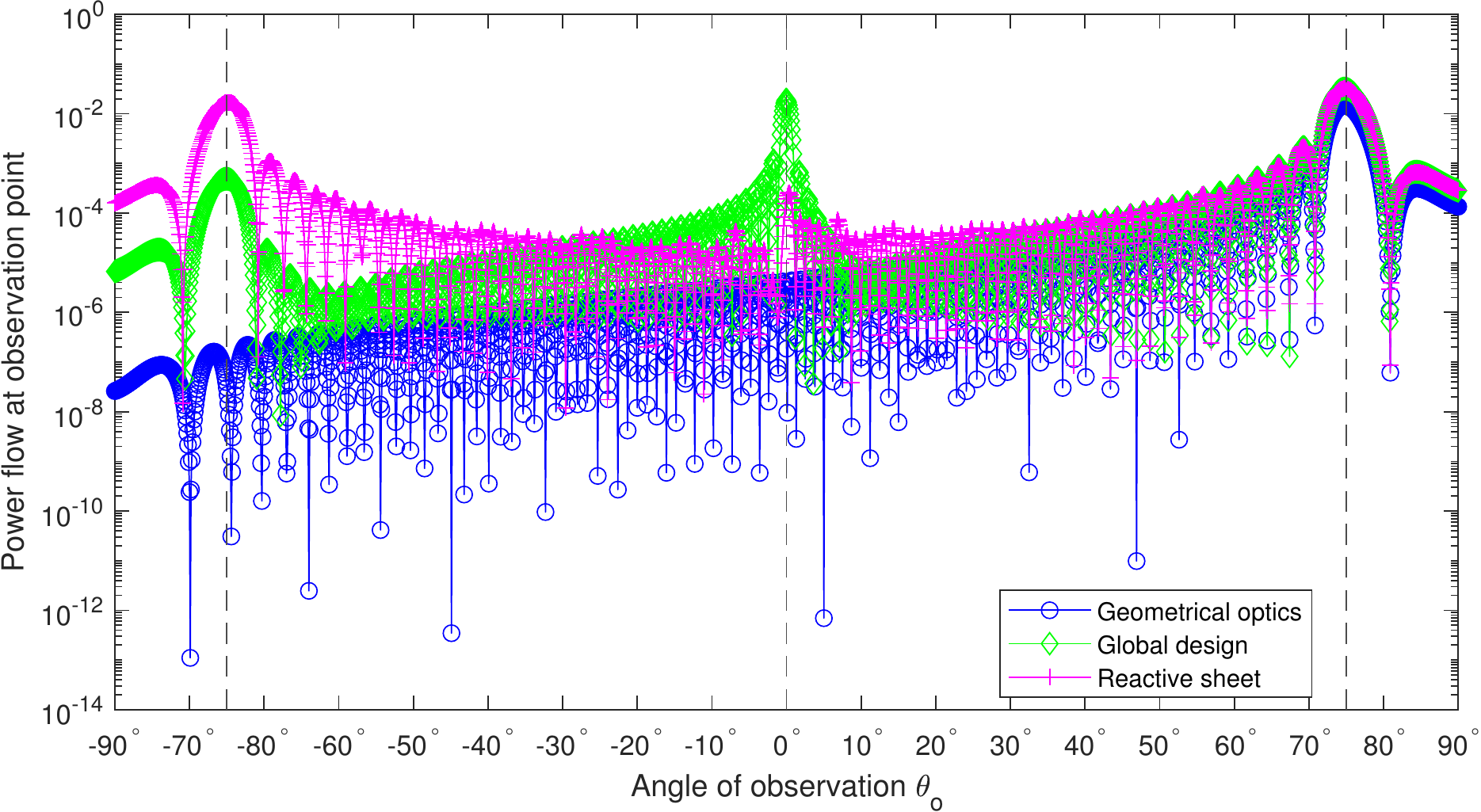}}
				\caption{Power flux vs. angle of observation ${{\mathcal{P}}_{{\rm{obs}}}}\left( {\bf{Z}} \right)$ (dB plot).}\label{fig:Poynting_Obs_No_Null_thetaR_75_deg_in_dB_FINAL}
			\end{subfigure}
			\begin{subfigure}{0.99\columnwidth}
				{\includegraphics[width=\linewidth]{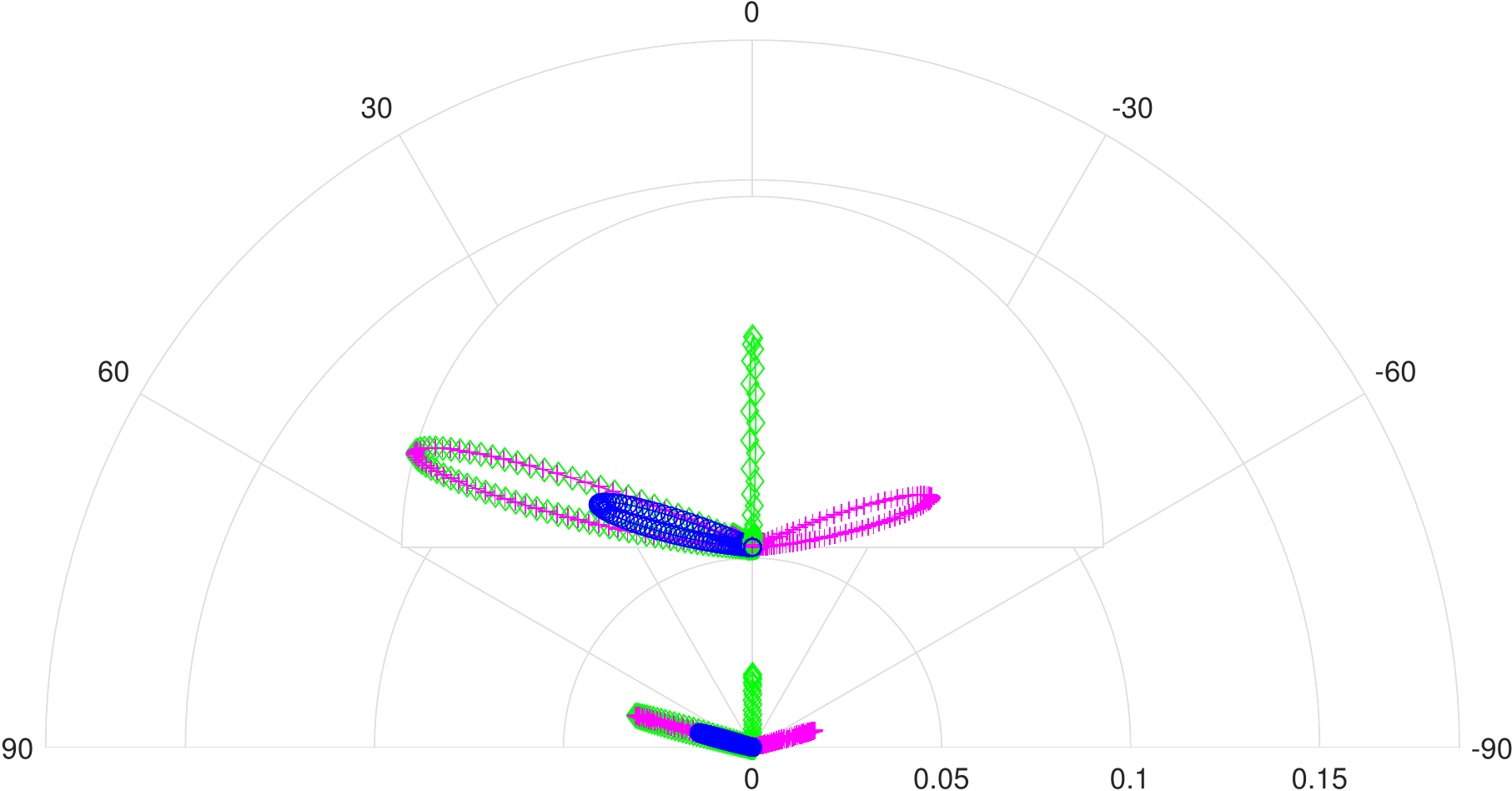}}
				\caption{Power flux vs. angle of observation ${{\mathcal{P}}_{{\rm{obs}}}}\left( {\bf{Z}} \right)$ (polar plot).}\label{fig:Poynting_Obs_No_Null_thetaR_75_deg_polar_FINAL}
			\end{subfigure}
		\end{center}
		\caption{Surface impedance in \eqref{eq:GO-Z} and solution of the optimization problems in \eqref{eq:Global-Opt-Formulation:main} and \eqref{eq:Pratical-Opt-Formulation:main} ($\theta_r(\vect{r}_{\textup{Rx}}) = 75^\circ$).}
		\label{fig:No_Null_thetaR_75_FINAL}
	\end{figure*}
In Figs. \ref{fig:No_Null_thetaR_30_FINAL} and \ref{fig:No_Null_thetaR_75_FINAL}, we illustrate the surface impedance and the reradiated power flux that correspond to the solution of the optimization problems in \eqref{eq:Global-Opt-Formulation:main} and \eqref{eq:Pratical-Opt-Formulation:main} for $\theta_r(\vect{r}_{\textup{Rx}}) = 30^\circ$ and $\theta_r(\vect{r}_{\textup{Rx}}) = 75^\circ$, respectively. {For illustrative purposes, the optimization problems for $\theta_r(\vect{r}_{\textup{Rx}}) = 30^\circ$ are solved by setting  $\varepsilon = 5 \cdot 10^{-2}$ (i.e., Helmholtz’s constraint is active), and the optimization problems for $\theta_r(\vect{r}_{\textup{Rx}}) = 75^\circ$ are solved by setting $\varepsilon$ to a large value (i.e., Helmholtz’s constraint is not active). In this latter case, the impact of Helmholtz’s constraint on the surface impedance is analyzed in Fig. \ref{fig:No_Null_thetaR_75_FINAL_correct}, where we set $\varepsilon = 5 \cdot 10^{-2}$ (i.e., Helmholtz’s constraint is active).} For ease of visualization, the sub-figures that show the surface impedance and the Helmholtz constraint illustrate a single period of the corresponding function. The numerical results confirm the general considerations made in previous sub-sections. In both figures, we observe that the main lobe of the reradiation pattern of the RIS is steered towards the desired direction of reflection. The surface impedance obtained as solution of the optimization problem in \eqref{eq:Pratical-Opt-Formulation:main} is, as desired, purely reactive. By comparing the reradiated power flux (in dB scale and in the polar representation), the performance vs. implementation tradeoff between the two surface impedances obtained as solutions of the optimization problems in \eqref{eq:Global-Opt-Formulation:main} and \eqref{eq:Pratical-Opt-Formulation:main} is apparent, especially for $\theta_r(\vect{r}_{\textup{Rx}}) = 75^\circ$. The surface impedance that is solution of the optimization problem in \eqref{eq:Pratical-Opt-Formulation:main} offers a good approximation of the main lobe of the reradiation pattern at the cost of slightly higher side lobes. The difference between the main lobe and the side lobes is, however, large and can be controlled by adding additional optimization constraints to the optimization problems, as discussed next for suppressing the specular reflection.  As for the case study $\theta_r(\vect{r}_{\textup{Rx}}) = 75^\circ$ in Fig. \ref{fig:No_Null_thetaR_75_FINAL}, we observe that the real part of the surface impedance obtained by solving the optimization problem in \eqref{eq:Global-Opt-Formulation:main} varies along the surface and can take positive and negative values. Finally, the geometrical optics solution results in worse reradiation performance as compared with the optimized solutions obtained from the optimization problems in \eqref{eq:Global-Opt-Formulation:main} and \eqref{eq:Pratical-Opt-Formulation:main}. {By direct inspection of Fig. \ref{fig:Helmholtz_Ratio_No_Null_thetaR_30_deg_FINAL}, we see that the Helmholtz condition is fulfilled according to the imposed optimization constraint, i.e., ${{\mathcal{H}}_n}\left( {\bf{Z}}\right) \le \varepsilon = 5 \cdot 10^{-2}$. By direct inspection of Fig. \ref{fig:Helmholtz_Ratio_No_Null_thetaR_75_deg_FINAL}, we see, on the other hand, that the obtained surface impedance results in a Helmholtz condition that is not significantly smaller than one, since no constraint is added to the optimization problem. The impact of Helmholtz's constraint is analyzed in Fig. \ref{fig:No_Null_thetaR_75_FINAL_correct}. As far as the geometrical optics solution is concerned, the numerical results confirm that the Helmholtz constraint is fulfilled by definition, as discussed in previous sections.}

	\begin{figure*}[!t]
		\begin{center}
			\begin{subfigure}{0.66\columnwidth}
				{\includegraphics[width=\linewidth]{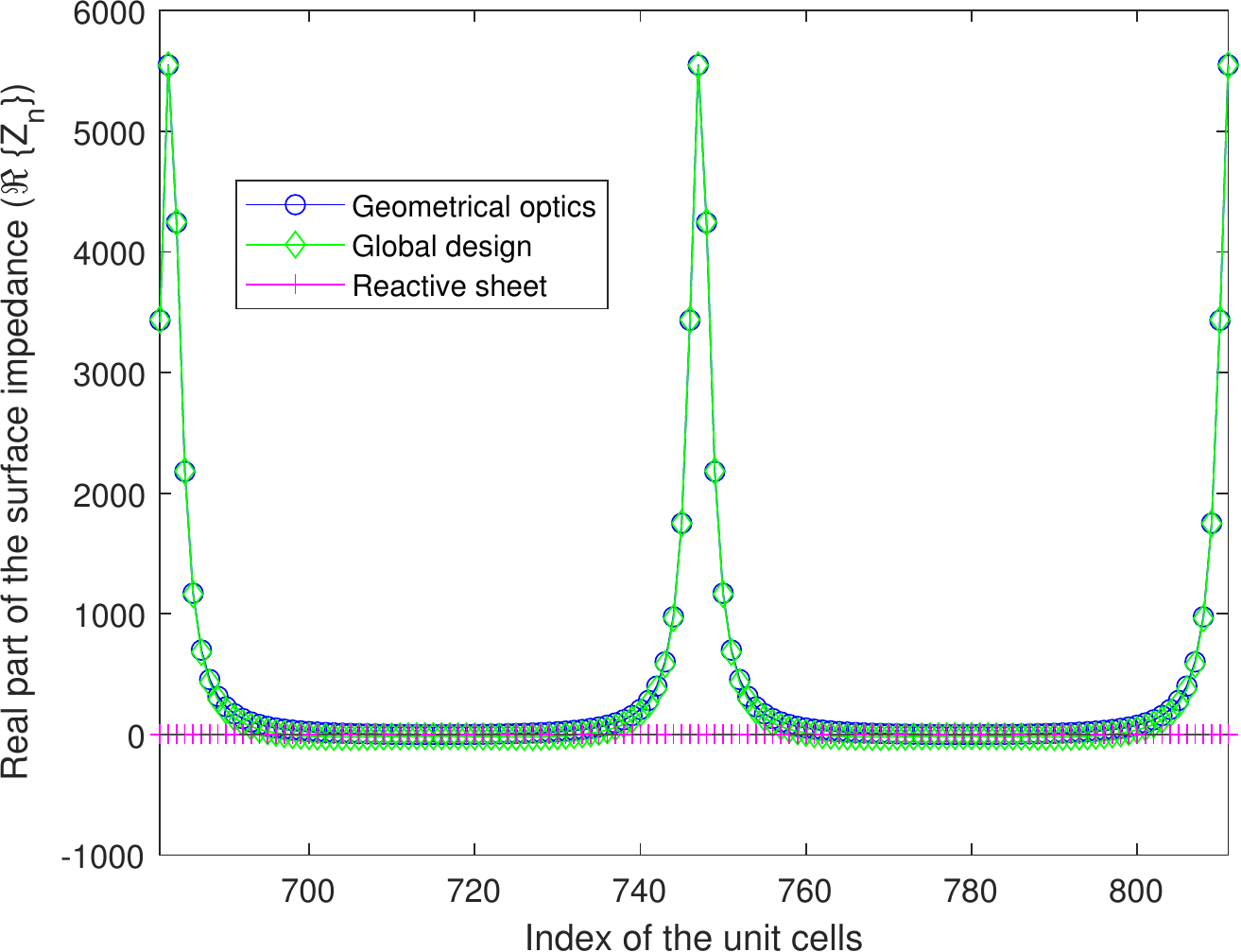}}
				\caption{Real part of surface impedance ${\mathop{\Re}\nolimits} \left( {{Z_n}} \right)$.}\label{fig:Z_real_with_null_thetaR_30_deg_FINAL}
			\end{subfigure}
			\begin{subfigure}{0.66\columnwidth}
				{\includegraphics[width=\linewidth]{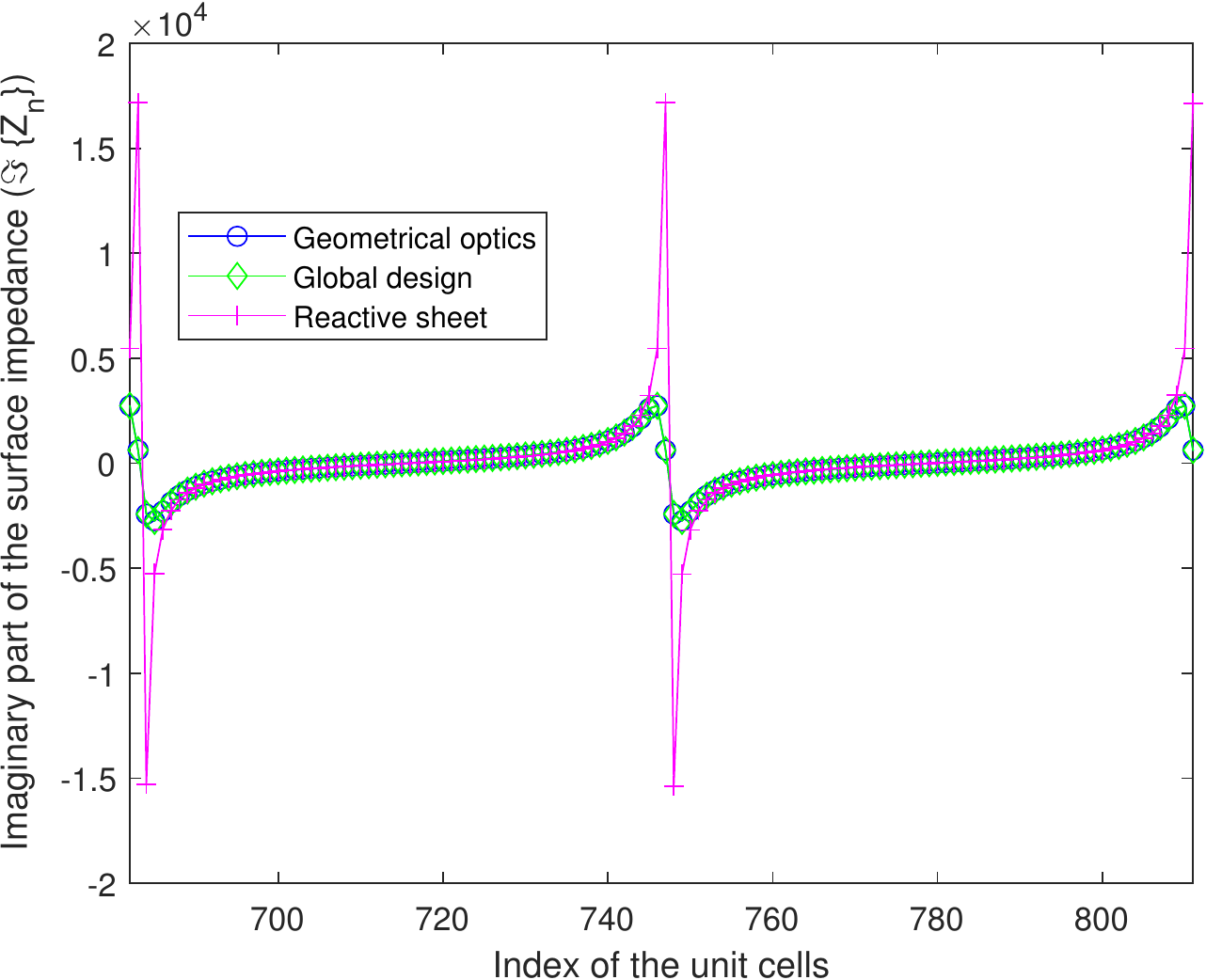}}
				\caption{Imaginary part of surface impedance ${\mathop{\Im}\nolimits} \left( {{Z_n}} \right)$.}\label{fig:Z_imag_with_null_thetaR_30_deg_FINAL}
			\end{subfigure}
			\begin{subfigure}{0.66\columnwidth}
				{\includegraphics[width=\linewidth]{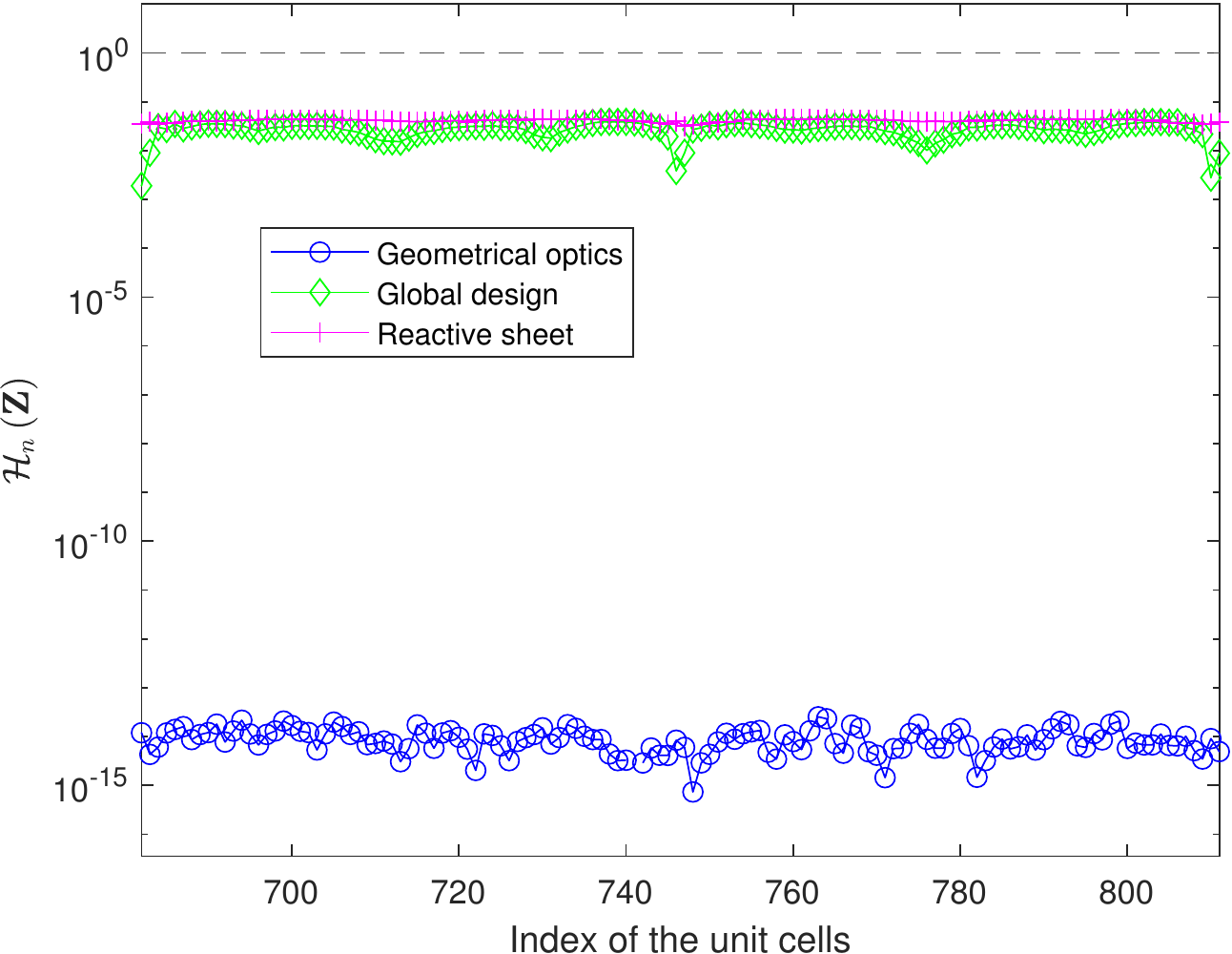}}
				\caption{Helmholtz's ratio ${{\mathcal{H}}_n}\left( {\bf{Z}} \right)$.}\label{fig:Helmholtz_Ratio_With_Null_thetaR_30_deg_FINAL}
			\end{subfigure}
			\begin{subfigure}{0.99\columnwidth}
				{\includegraphics[width=\linewidth]{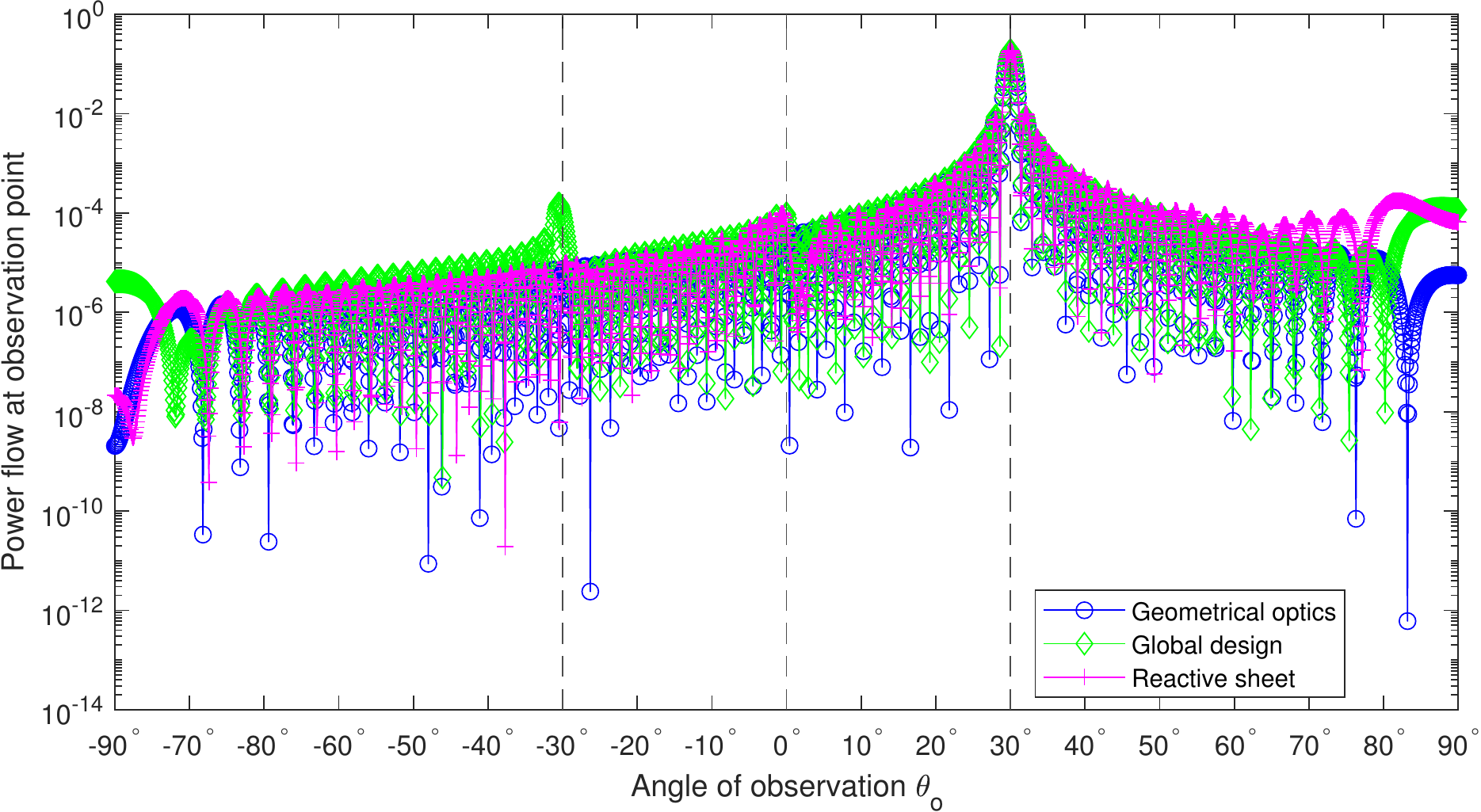}}
				\caption{Power flux vs. angle of observation ${{\mathcal{P}}_{{\rm{obs}}}}\left( {\bf{Z}} \right)$ (dB plot).}\label{fig:Poynting_Obs_With_Null_thetaR_30_deg_in_dB_FINAL}
			\end{subfigure}
			\begin{subfigure}{0.99\columnwidth}
				{\includegraphics[width=\linewidth]{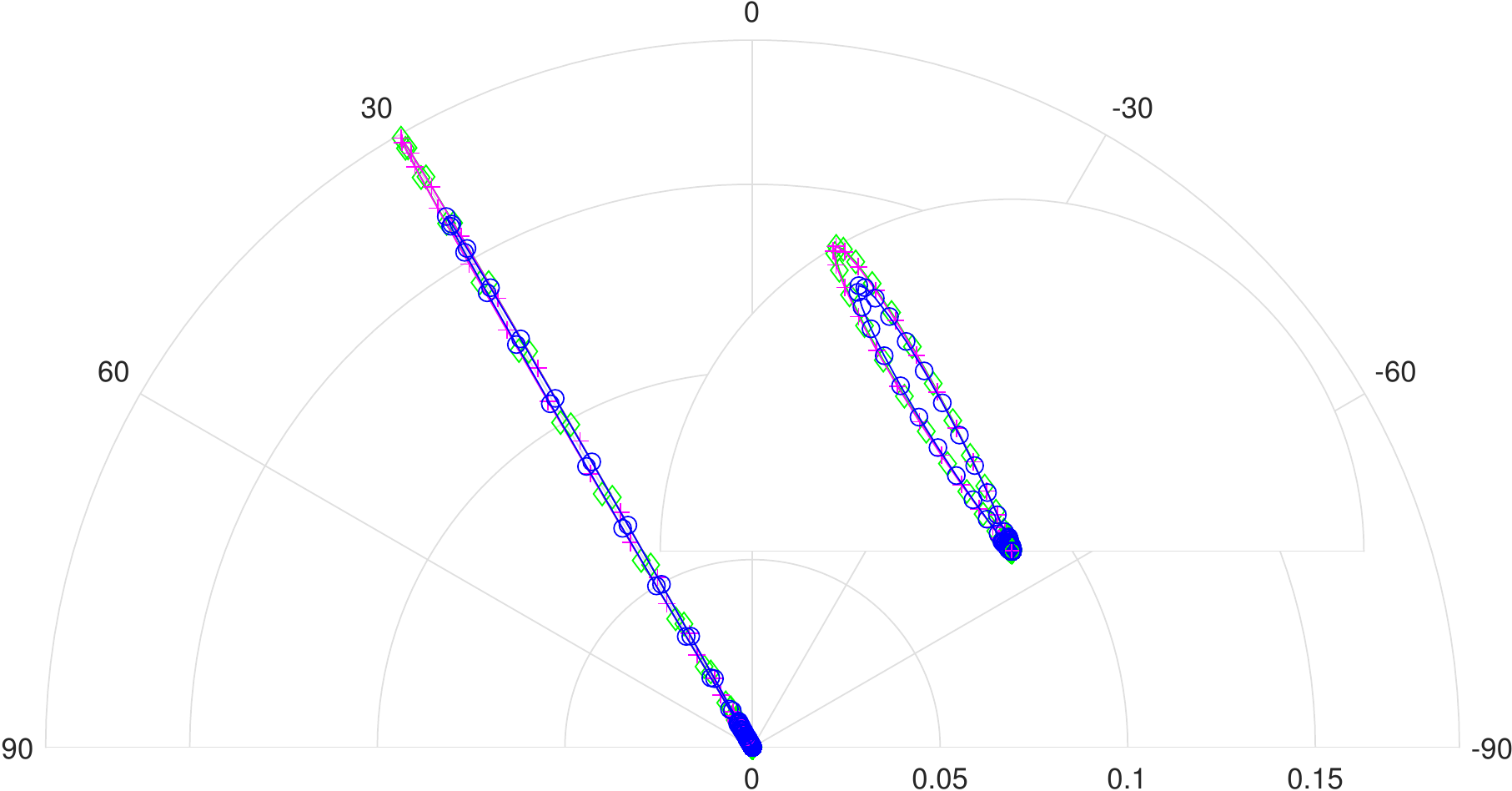}}
				\caption{Power flux vs. angle of observation ${{\mathcal{P}}_{{\rm{obs}}}}\left( {\bf{Z}} \right)$ (polar plot).}\label{fig:Poynting_Obs_With_Null_thetaR_30_deg_polar_FINAL}
			\end{subfigure}
		\end{center}
		\caption{Surface impedance in \eqref{eq:GO-Z} and solution of the optimization problems in \eqref{eq:Mask-Global-Opt-Formulation:main} and \eqref{eq:Mask-Pratical-Opt-Formulation:main} ($\theta_r(\vect{r}_{\textup{Rx}}) = 30^\circ$).}
		\label{fig:With_Null_thetaR_30_FINAL}
	\end{figure*}

	\begin{figure*}[!t]
		\begin{center}
			\begin{subfigure}{0.66\columnwidth}
				{\includegraphics[width=\linewidth]{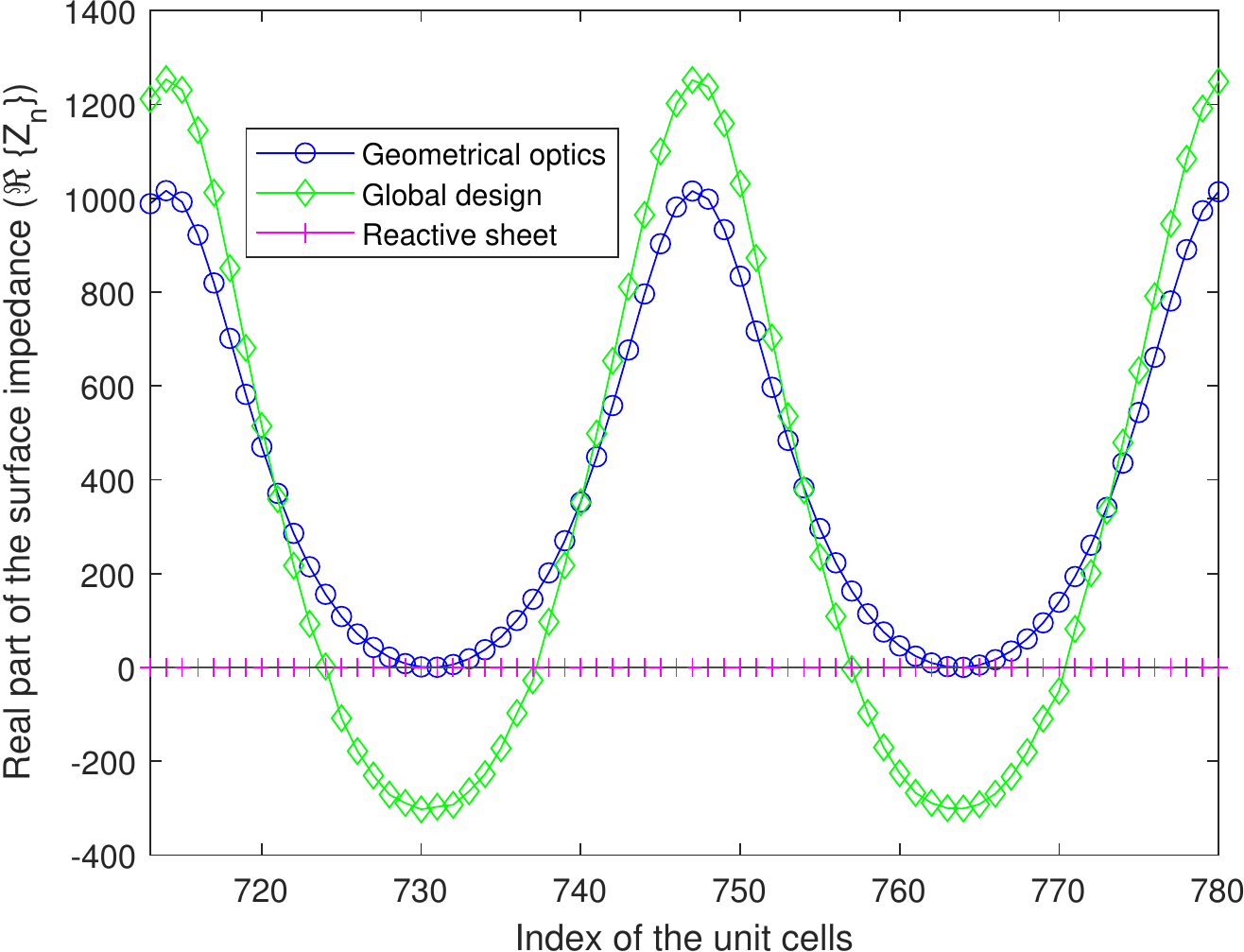}}
				\caption{Real part of surface impedance ${\mathop{\Re}\nolimits} \left( {{Z_n}} \right)$.}\label{fig:Z_real_with_null_thetaR_75_deg_FINAL}
			\end{subfigure}
			\begin{subfigure}{0.66\columnwidth}
				{\includegraphics[width=\linewidth]{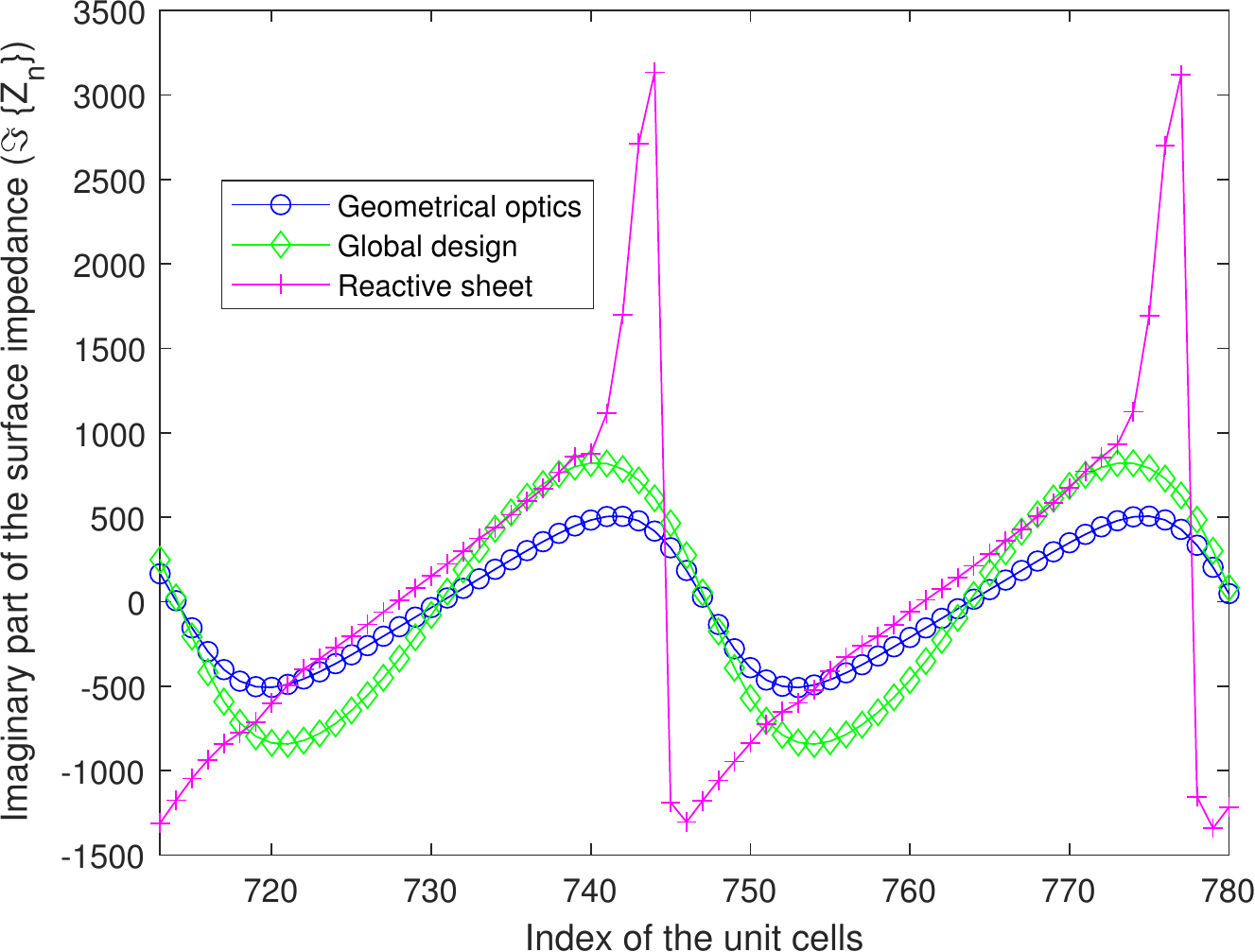}}
				\caption{Imaginary part of surface impedance ${\mathop{\Im}\nolimits} \left( {{Z_n}} \right)$.}\label{fig:Z_imag_with_null_thetaR_75_deg_FINAL}
			\end{subfigure}
			\begin{subfigure}{0.66\columnwidth}
				{\includegraphics[width=\linewidth]{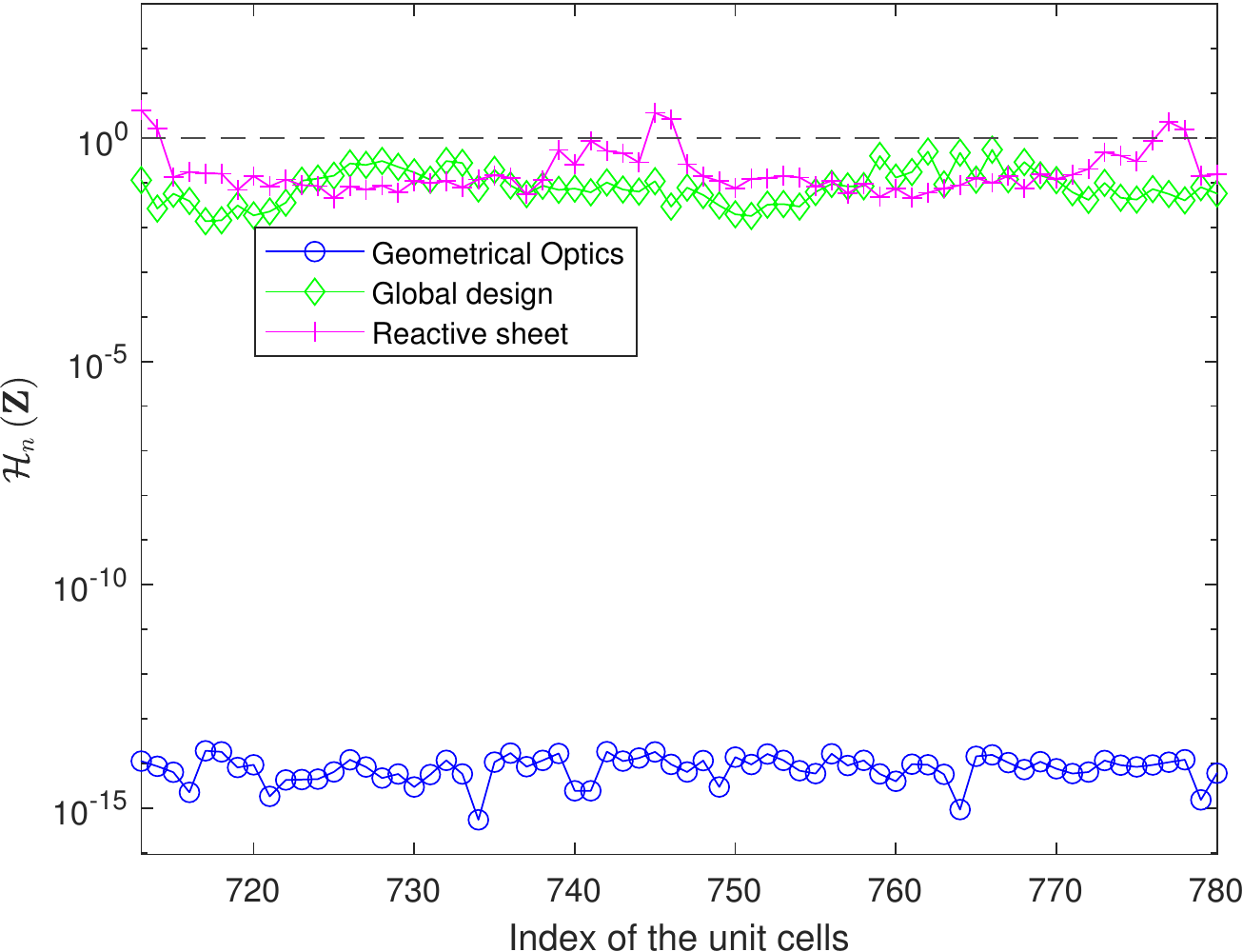}}
				\caption{Helmholtz's ratio ${{\mathcal{H}}_n}\left( {\bf{Z}} \right)$.}\label{fig:Helmholtz_Ratio_With_Null_thetaR_75_deg_FINAL}
			\end{subfigure}
			\begin{subfigure}{0.99\columnwidth}
				{\includegraphics[width=\linewidth]{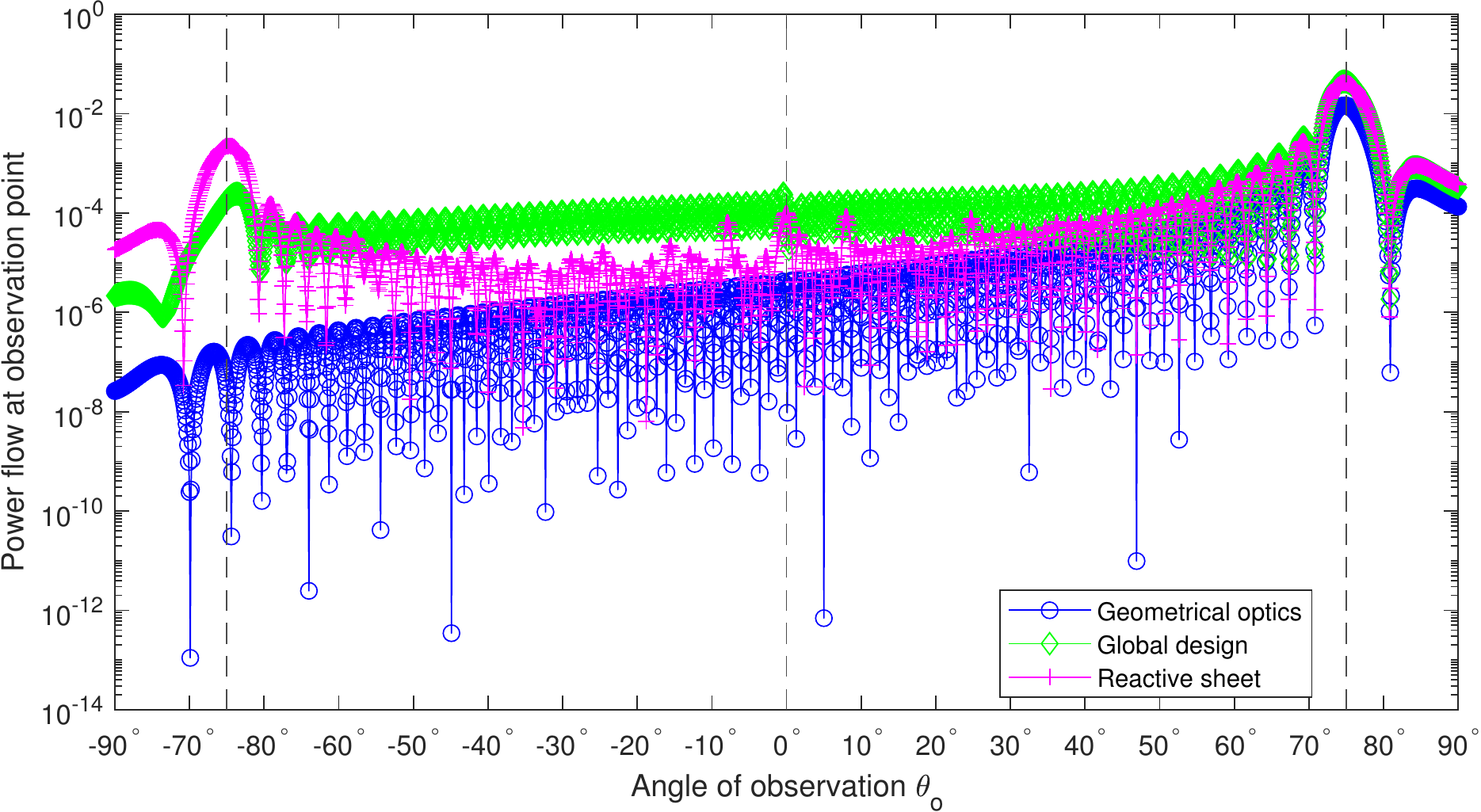}}
				\caption{Power flux vs. angle of observation ${{\mathcal{P}}_{{\rm{obs}}}}\left( {\bf{Z}} \right)$ (dB plot).}\label{fig:Poynting_Obs_With_Null_thetaR_75_deg_in_dB_FINAL}
			\end{subfigure}
			\begin{subfigure}{0.99\columnwidth}
				{\includegraphics[width=\linewidth]{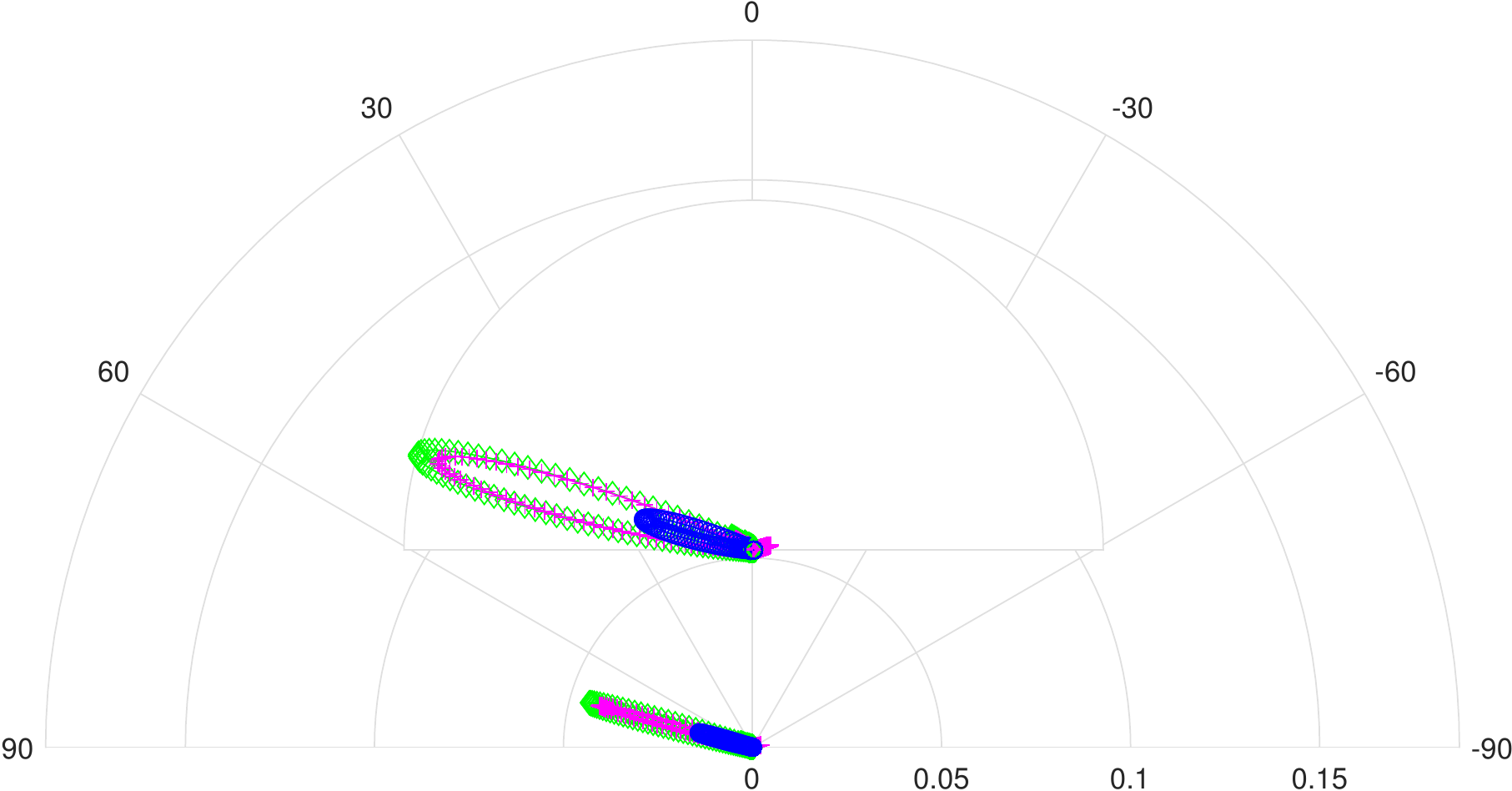}}
				\caption{Power flux vs. angle of observation ${{\mathcal{P}}_{{\rm{obs}}}}\left( {\bf{Z}} \right)$ (polar plot).}\label{fig:Poynting_Obs_With_Null_thetaR_75_deg_polar_FINAL}
			\end{subfigure}
		\end{center}
		\caption{Surface impedance in \eqref{eq:GO-Z} and solution of the optimization problems in \eqref{eq:Mask-Global-Opt-Formulation:main} and \eqref{eq:Mask-Pratical-Opt-Formulation:main} ($\theta_r(\vect{r}_{\textup{Rx}}) = 75^\circ$).}
		\label{fig:With_Null_thetaR_75_FINAL}
	\end{figure*}
In Figs. \ref{fig:No_Null_thetaR_30_FINAL} and \ref{fig:No_Null_thetaR_75_FINAL}, we observe the presence of unwanted reflections towards the specular direction, i.e., $\theta_r = 0$, and towards the direction that is symmetric with respect to the desired direction of reradiation, i.e., $\theta_r = -\theta_r(\vect{r}_{\textup{Rx}})$. This is in agreement with Floquet's theory in \eqref{eq:Floquet_6}. In fact, the figures illustrate that the surface impedance is a quasi-periodic function and the period $\mathcal{P}$ is slightly larger than the wavelength $\lambda$. Therefore, three main propagating reradiation modes towards the desired direction of reflection, the specular direction, and the direction that is symmetric with respect to desired direction of reflection may be present, whose intensity depends on the specific shape of the obtained surface impedance. This is an undesired effect and is particularly pronounced in Fig. \ref{fig:No_Null_thetaR_75_FINAL}, in which the difference between the angle of incidence and the angle of reflection is larger and no constraint on the Helmholtz condition is imposed.

The presence of spurious reflections has two negative consequences: (i) some power that could be directed towards the desired direction of reflection is steered towards other directions and (ii) the surface generates interference towards uncontrolled directions and this may increase the interference towards other devices. Therefore, it is necessary to keep under control these possible spurious reflections by design. The corresponding results are illustrated in Figs. \ref{fig:With_Null_thetaR_30_FINAL} and \ref{fig:With_Null_thetaR_75_FINAL}, which are obtained by solving the optimization problems formulated in \eqref{eq:Mask-Global-Opt-Formulation:main}, and \eqref{eq:Mask-Pratical-Opt-Formulation:main} for $\theta_r(\vect{r}_{\textup{Rx}}) = 30^\circ$ and $\theta_r(\vect{r}_{\textup{Rx}}) = 75^\circ$, respectively, under the assumption that only the reradiation mode towards the specular direction is minimized by design, while no optimization constraint is added to the second spurious reradiation mode. In this case, we observe that the specular reflection is below the predefined maximum level and, overall, the side lobes of the reradiation pattern are well below the main lobe. In spite of the additional design constraint that is added in the optimization problems formulated in \eqref{eq:Mask-Global-Opt-Formulation:main}, and \eqref{eq:Mask-Pratical-Opt-Formulation:main}, a purely reactive solution for the surface impedance exists and can be computed. {As far as the Helmholtz constraint is concerned, similar conclusions as for Fig. \ref{fig:Helmholtz_Ratio_No_Null_thetaR_30_deg_FINAL} and Fig. \ref{fig:Helmholtz_Ratio_No_Null_thetaR_75_deg_FINAL} can be drawn. Also in this case, in fact, $\varepsilon = 5 \cdot 10^{-2}$ is imposed in Fig. \ref{fig:Helmholtz_Ratio_With_Null_thetaR_30_deg_FINAL} and a large value of $\varepsilon$ is imposed in Fig.  \ref{fig:Helmholtz_Ratio_With_Null_thetaR_75_deg_FINAL}.}

\begin{figure*}[!t]
		\begin{center}
			\begin{subfigure}{0.66\columnwidth}
				{\includegraphics[width=\linewidth]{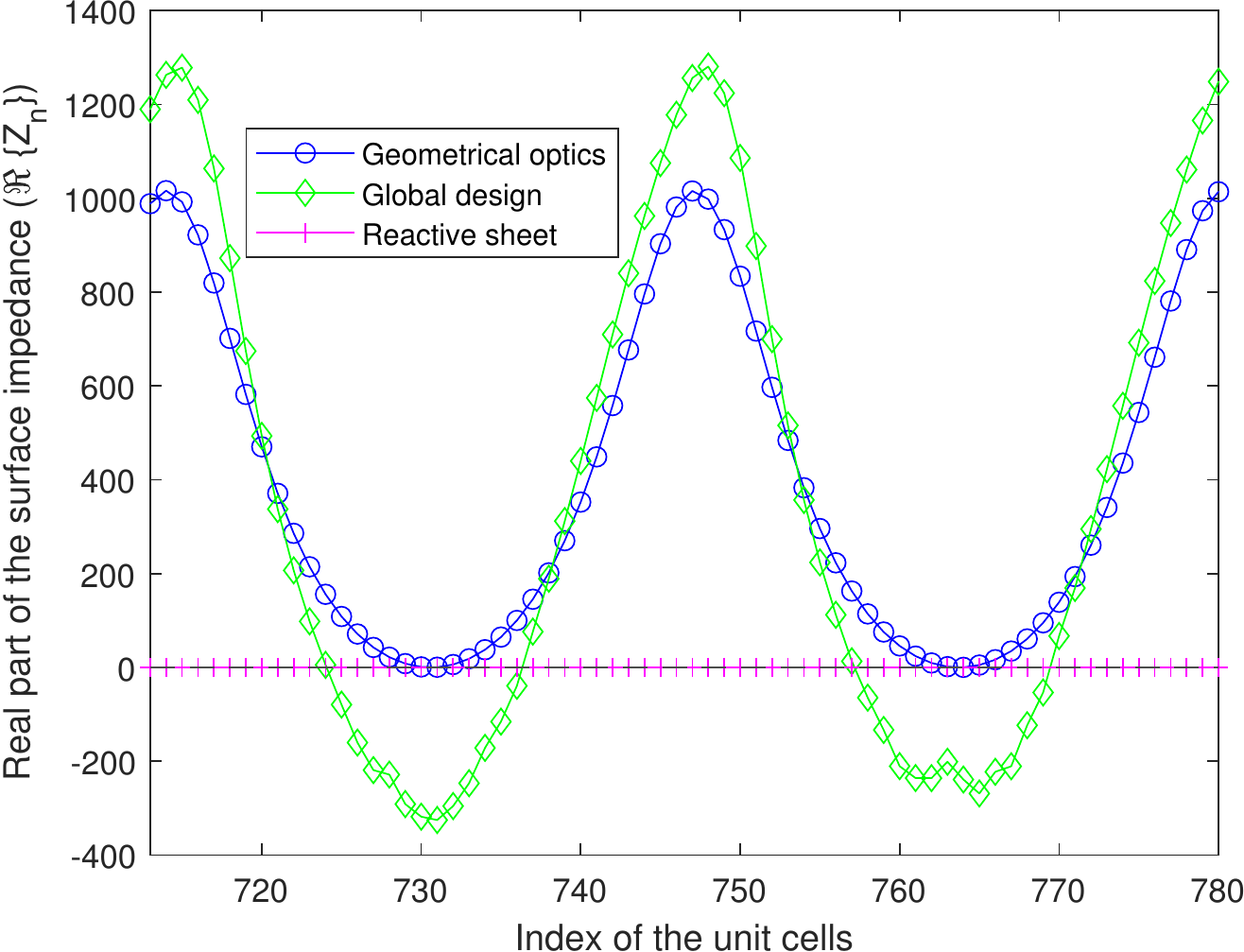}}
				\caption{Real part of surface impedance ${\mathop{\Re}\nolimits} \left( {{Z_n}} \right)$.}\label{fig:Z_real_with_DOUBLEnull_thetaR_75_deg_FINAL}
			\end{subfigure}
			\begin{subfigure}{0.66\columnwidth}
				{\includegraphics[width=\linewidth]{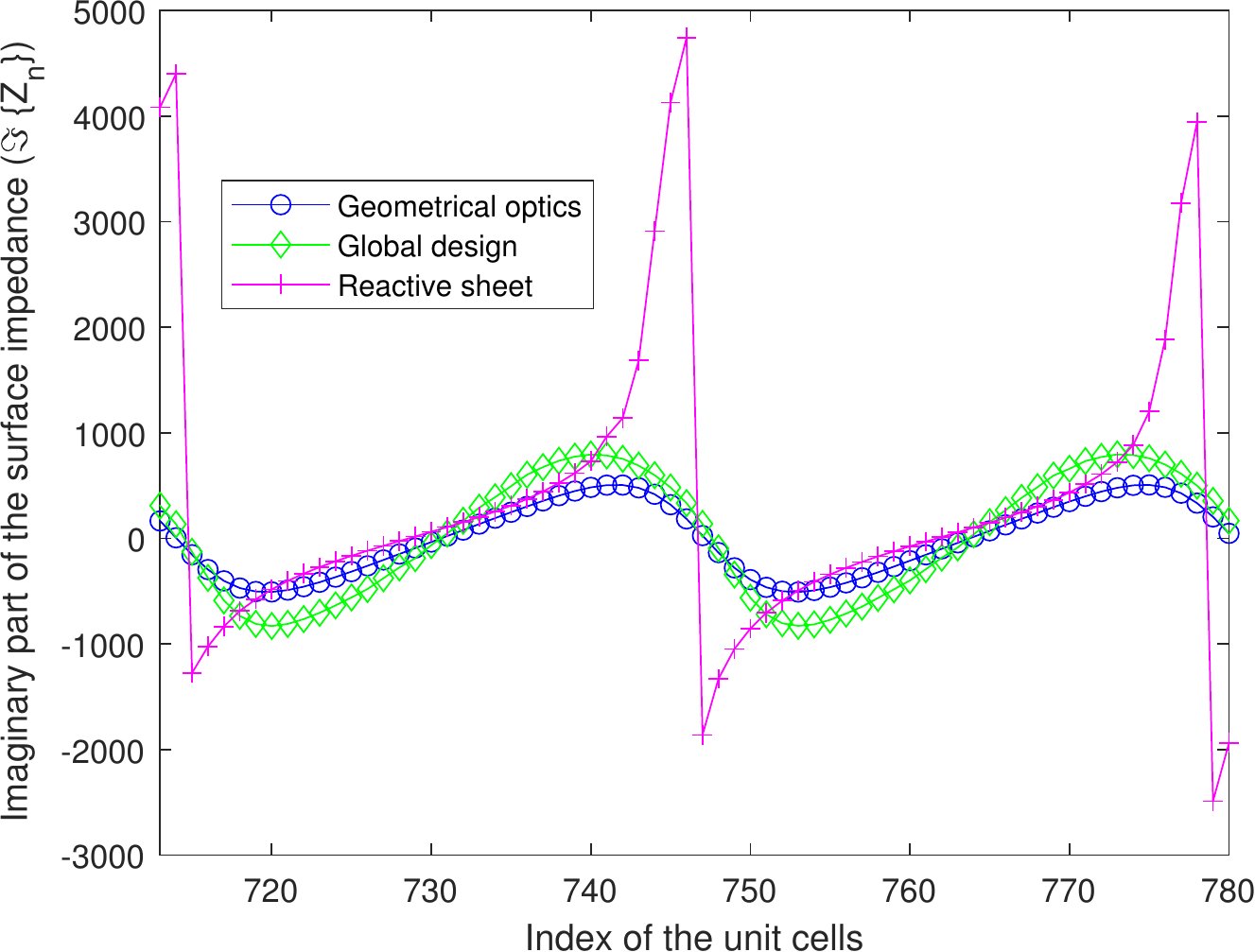}}
				\caption{Imaginary part of surface impedance ${\mathop{\Im}\nolimits} \left( {{Z_n}} \right)$.}\label{fig:Z_imag_with_DOUBLEnull_thetaR_75_deg_FINAL}
			\end{subfigure}
			\begin{subfigure}{0.66\columnwidth}
				{\includegraphics[width=\linewidth]{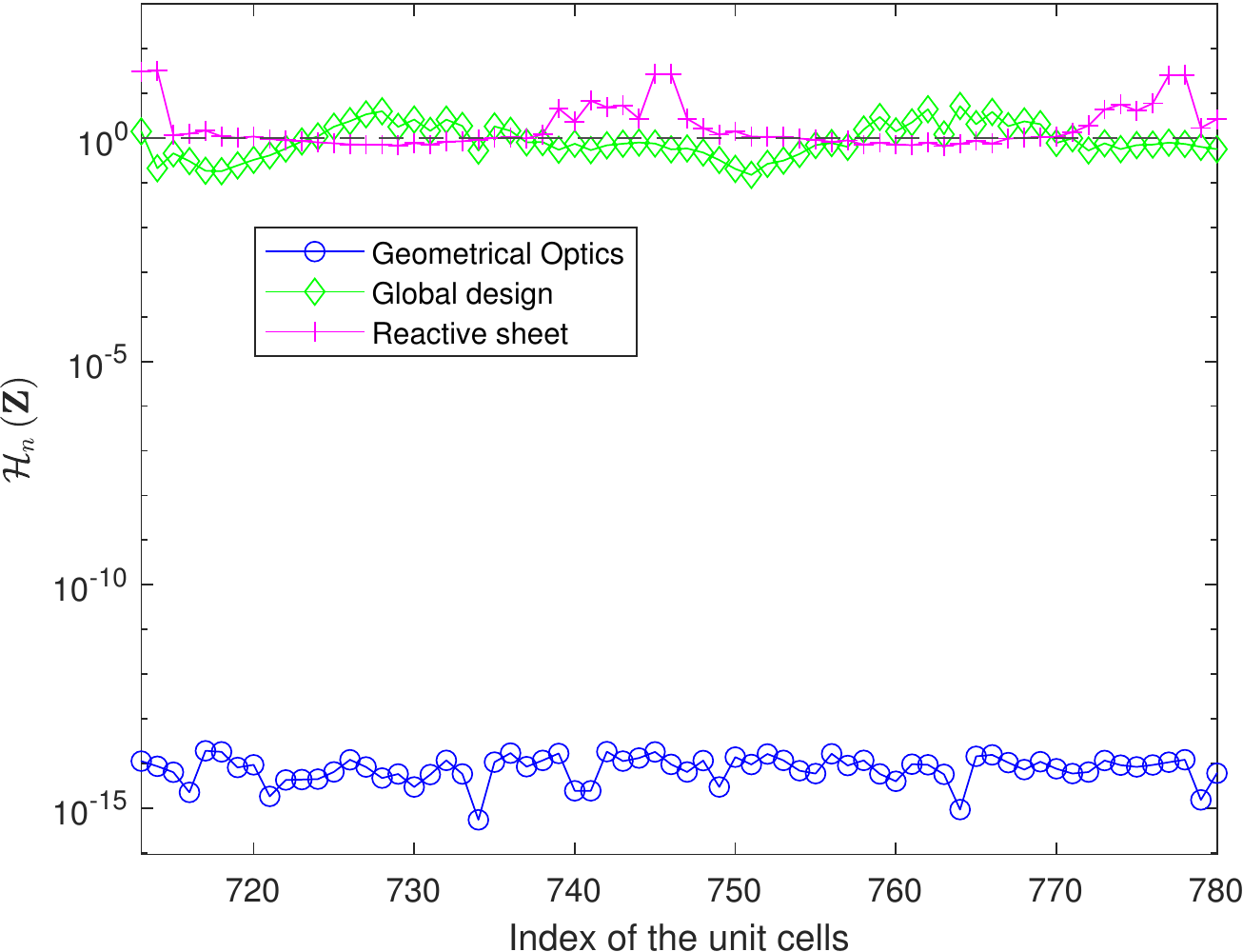}}
				\caption{Helmholtz's ratio ${{\mathcal{H}}_n}\left( {\bf{Z}} \right)$.}\label{fig:Helmholtz_Ratio_With_DOUBLEnull_thetaR_75_deg_FINAL}
			\end{subfigure}
			\begin{subfigure}{0.99\columnwidth}
				{\includegraphics[width=\linewidth]{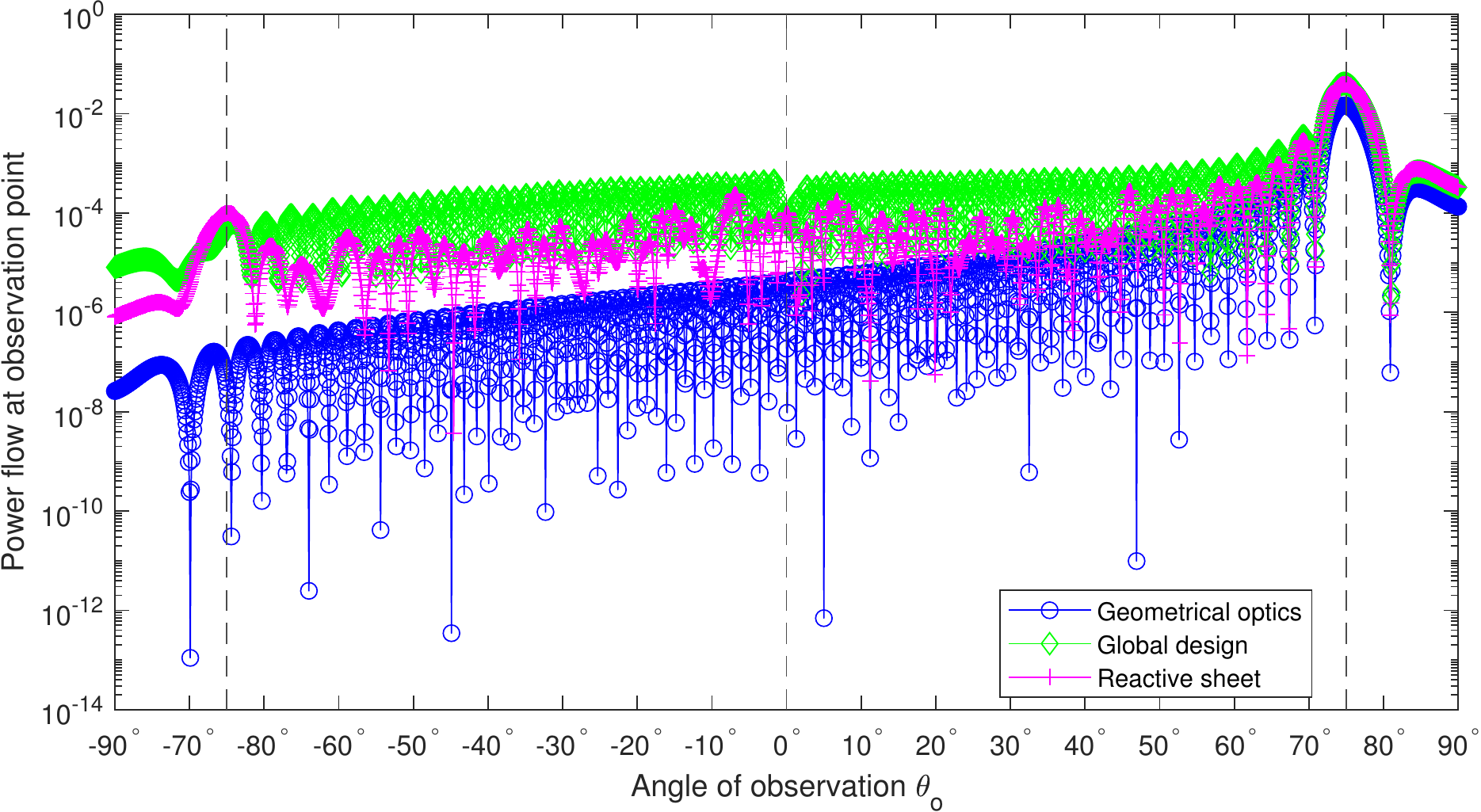}}
				\caption{Power flux vs. angle of observation ${{\mathcal{P}}_{{\rm{obs}}}}\left( {\bf{Z}} \right)$ (dB plot).}\label{fig:Poynting_Obs_With_DOUBLEnull_thetaR_75_deg_in_dB_FINAL}
			\end{subfigure}
			\begin{subfigure}{0.99\columnwidth}
				{\includegraphics[width=\linewidth]{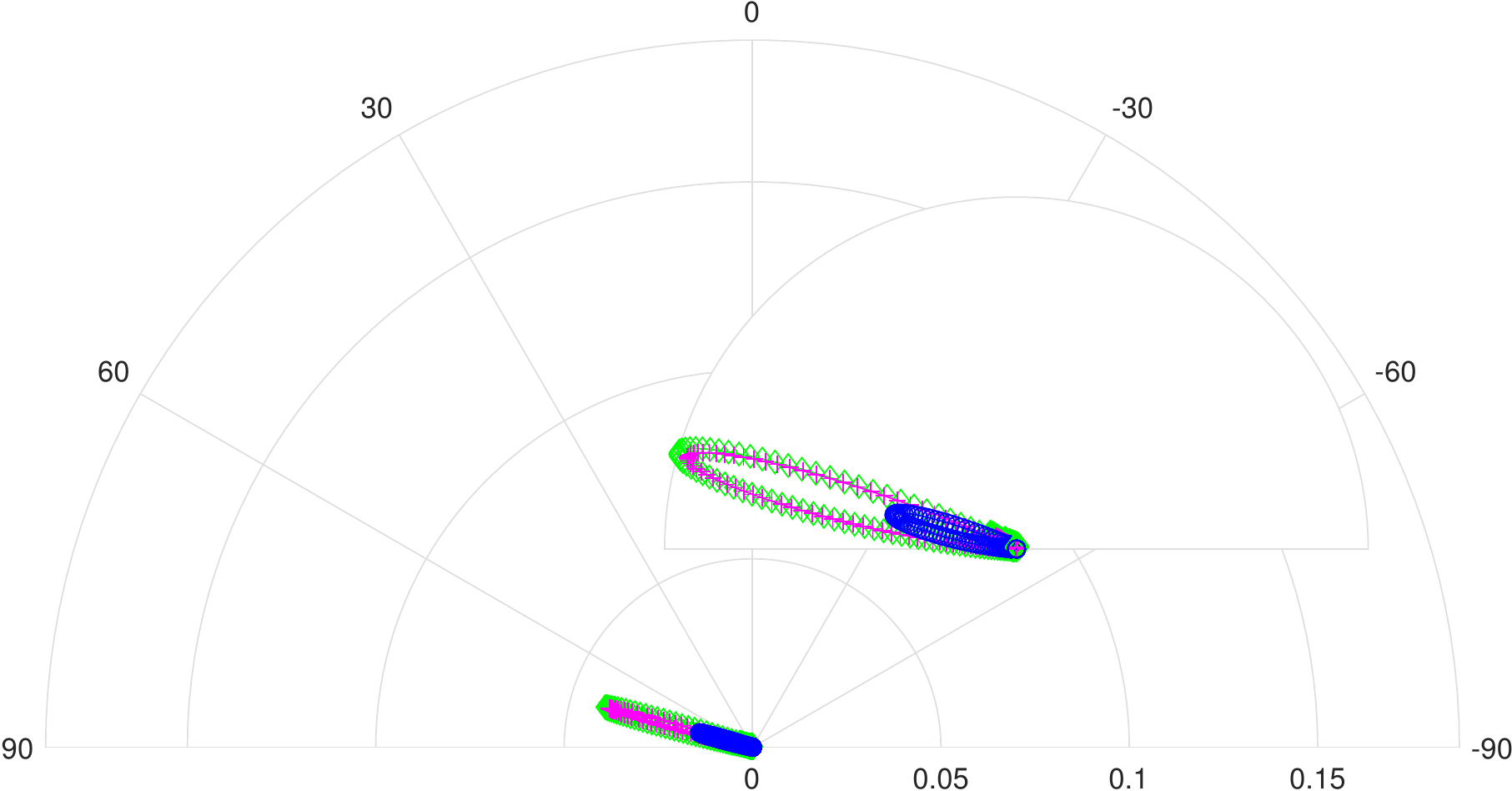}}
				\caption{Power flux vs. angle of observation ${{\mathcal{P}}_{{\rm{obs}}}}\left( {\bf{Z}} \right)$ (polar plot).}\label{fig:Poynting_Obs_With_DOUBLENull_thetaR_75_deg_polar_FINAL}
			\end{subfigure}
		\end{center}
		\caption{Surface impedance in \eqref{eq:GO-Z} and solution of the optimization problems in \eqref{eq:Mask-Global-Opt-Formulation:main} and \eqref{eq:Mask-Pratical-Opt-Formulation:main} ($\theta_r(\vect{r}_{\textup{Rx}}) = 75^\circ$). An additional constraint on the spurious reflection towards $\theta_r = - 75^\circ$ is added.}
		\label{fig:With_DOUBLENull_thetaR_75_FINAL}
	\end{figure*}
The reradiation efficiency of the considered RIS may be further enhanced by adding a design constraint that accounts for the spurious reradiation towards $\theta_r = -\theta_r(\vect{r}_{\textup{Rx}})$, besides the reradiation constraint towards $\theta_r = 0^\circ$. An illustrative example is reported in Fig. \ref{fig:With_DOUBLENull_thetaR_75_FINAL}, which corresponds to the same setup and optimization problem as for Fig. \ref{fig:With_Null_thetaR_75_FINAL} with a large value of $\varepsilon$, with the only addition that the intensity of the reradiated (undesired) mode towards $\theta_r = -75^\circ$ needs to be smaller than $\delta = 10^{-4}$. We observe that a feasible solution for the surface impedance exists, and that the intensity of the two spurious directions of reradiation can be made smaller than the predefined intensity threshold (i.e., $\delta$). This example shows that, by adding appropriate optimization constraints to the problem formulation, the reradiation towards the desired direction and the dominant undesired directions can be appropriately engineered through purely reactive surface impedances.

\begin{figure*}[!t]
	\begin{center}
		\begin{subfigure}{0.66\columnwidth}
			{\includegraphics[width=\linewidth]{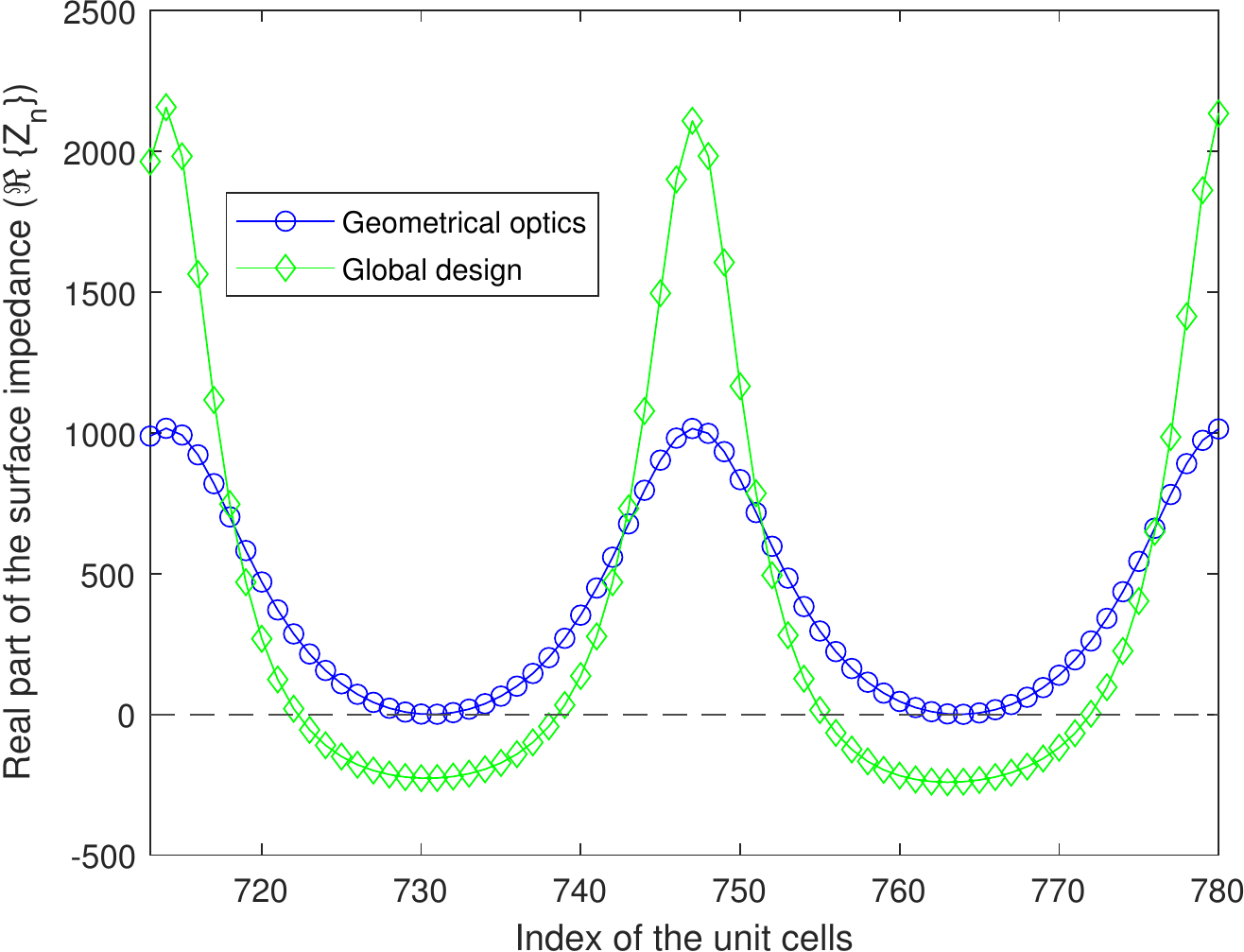}}
			\caption{Real part of surface impedance ${\mathop{\Re}\nolimits} \left( {{Z_n}} \right)$.}\label{fig:Z_real_no_null_thetaR_75_deg_FINAL_correct}
		\end{subfigure}
		\begin{subfigure}{0.66\columnwidth}
			{\includegraphics[width=\linewidth]{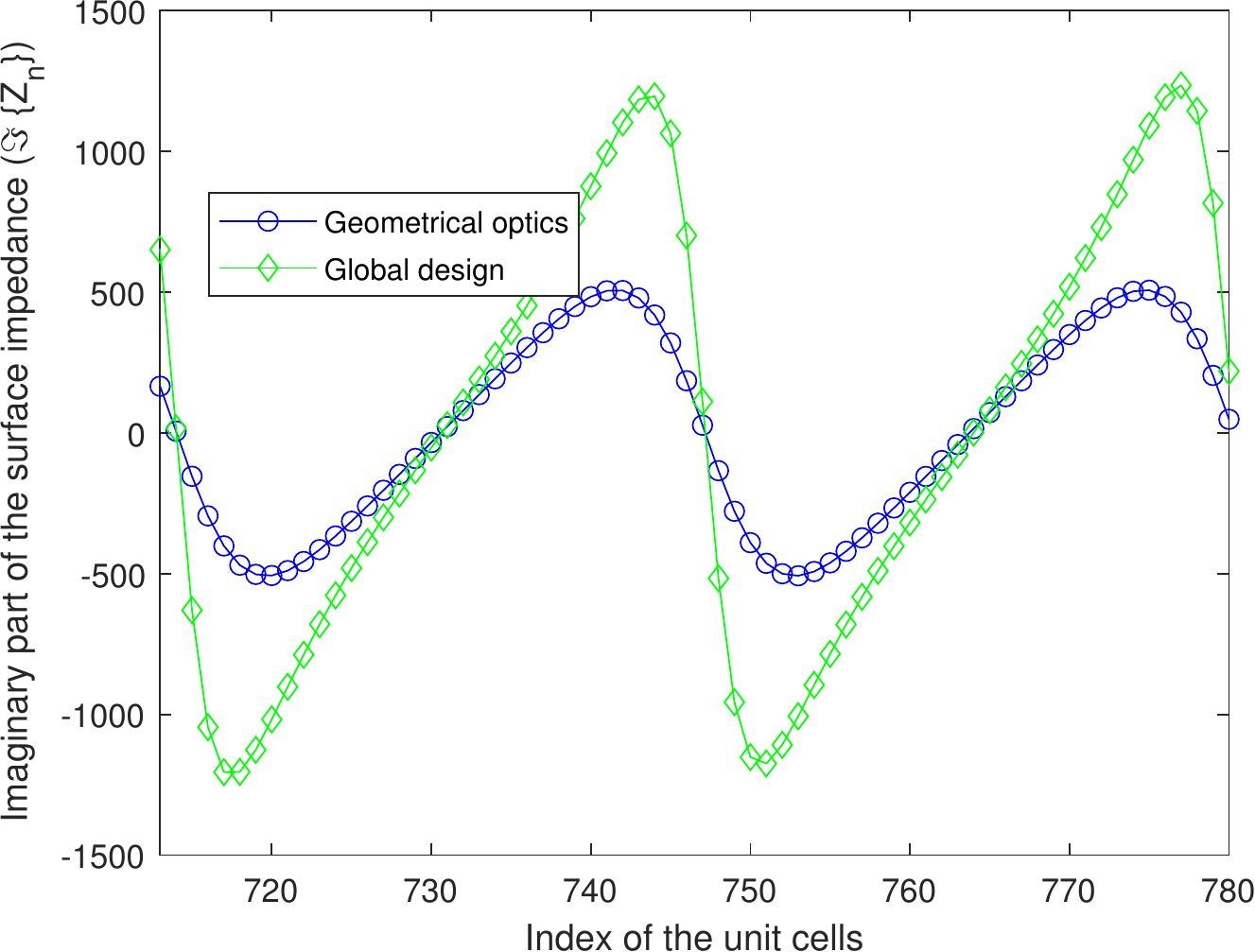}}
			\caption{Imaginary part of surface impedance ${\mathop{\Im}\nolimits} \left( {{Z_n}} \right)$.}\label{fig:Z_imag_no_null_thetaR_75_deg_FINAL_correct}
		\end{subfigure}
		\begin{subfigure}{0.66\columnwidth}
			{\includegraphics[width=\linewidth]{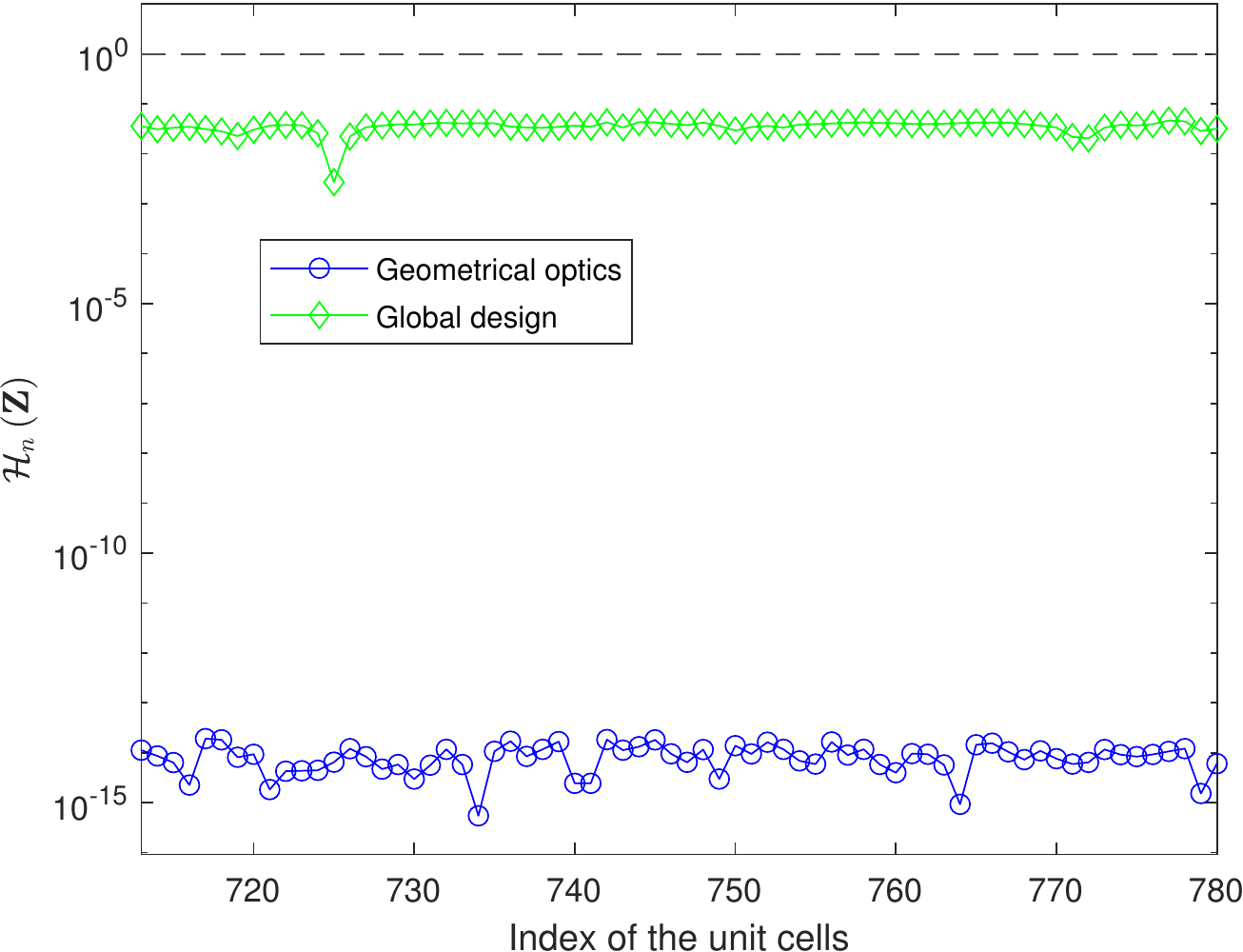}}
			\caption{Helmholtz's ratio ${{\mathcal{H}}_n}\left( {\bf{Z}} \right)$.}\label{fig:Helmholtz_Ratio_No_Null_thetaR_75_deg_FINAL_correct}
		\end{subfigure}
		\begin{subfigure}{0.99\columnwidth}
			{\includegraphics[width=\linewidth]{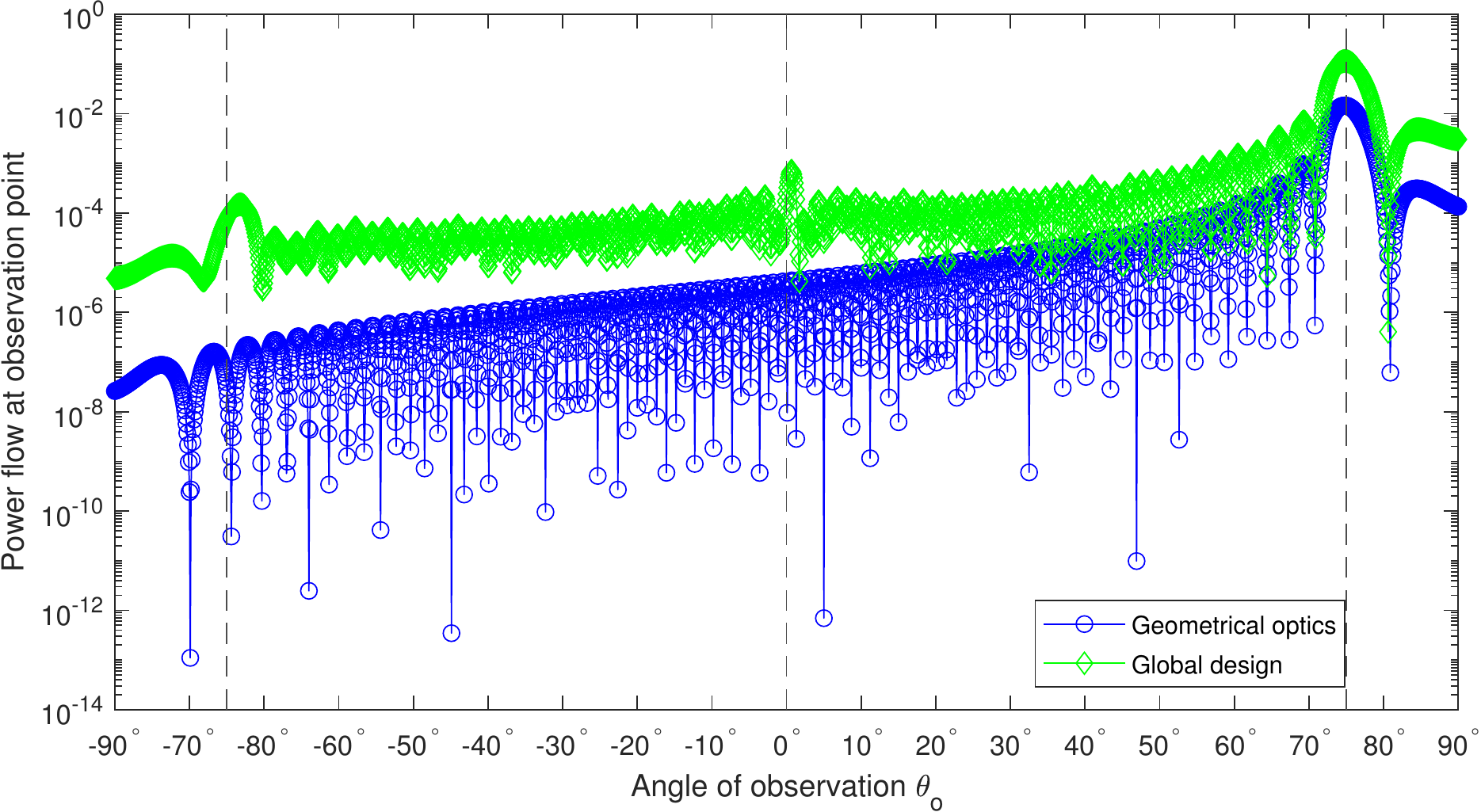}}
			\caption{Power flux vs. angle of observation ${{\mathcal{P}}_{{\rm{obs}}}}\left( {\bf{Z}} \right)$ (dB plot).}\label{fig:Poynting_Obs_No_Null_thetaR_75_deg_in_dB_FINAL_correct}
		\end{subfigure}
		\begin{subfigure}{0.99\columnwidth}
			{\includegraphics[width=\linewidth]{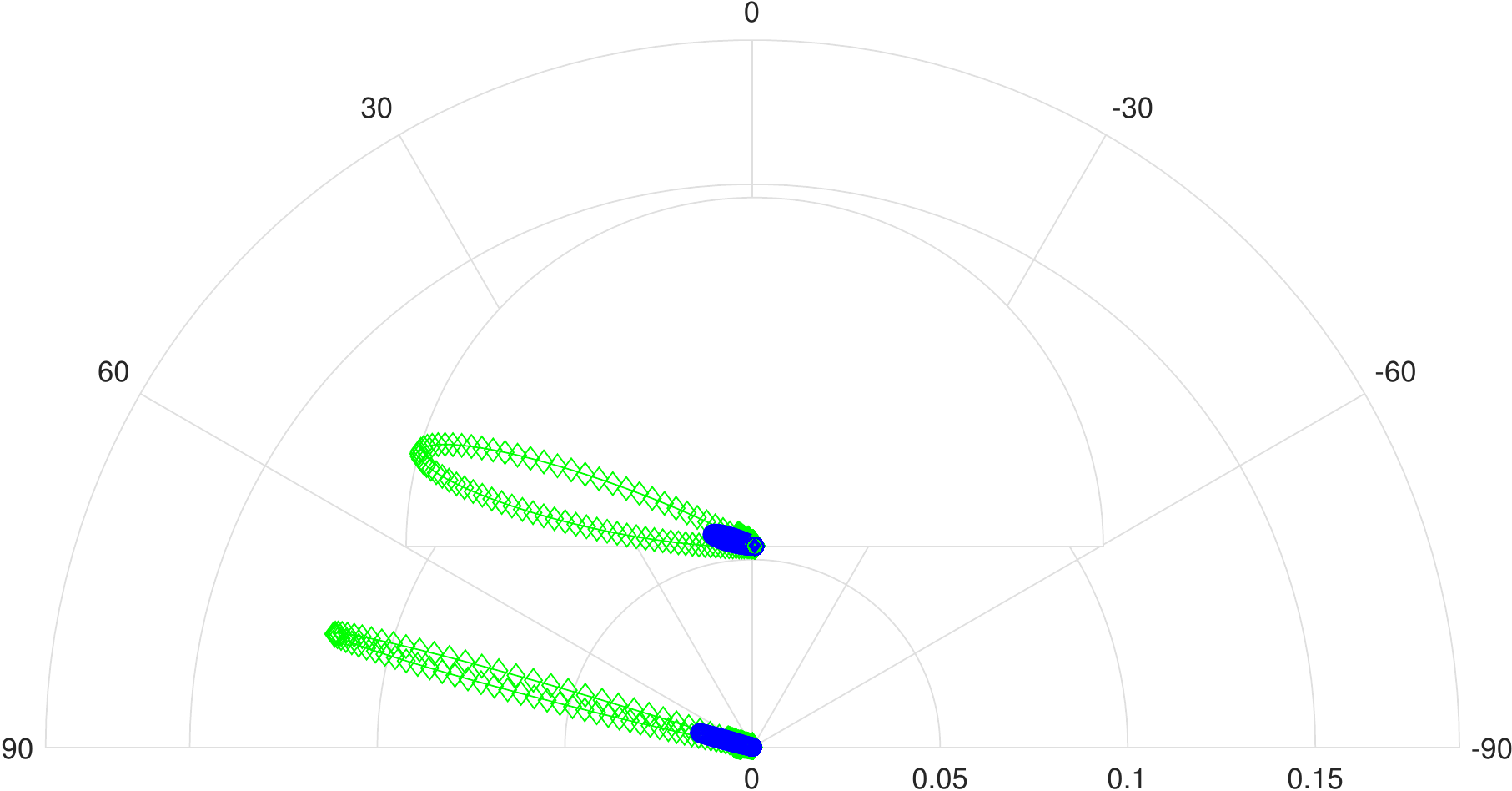}}
			\caption{Power flux vs. angle of observation ${{\mathcal{P}}_{{\rm{obs}}}}\left( {\bf{Z}} \right)$ (polar plot).}\label{fig:Poynting_Obs_No_Null_thetaR_75_deg_polar_FINAL_correct}
		\end{subfigure}
	\end{center}
	\caption{Surface impedance in \eqref{eq:GO-Z} and solution of the optimization problem in \eqref{eq:Global-Opt-Formulation:main} ($\theta_r(\vect{r}_{\textup{Rx}}) = 75^\circ$) -- The Helmholtz constraint is active.}
	\label{fig:No_Null_thetaR_75_FINAL_correct}
\end{figure*}
\begin{figure*}[!t]
	\begin{center}
		\begin{subfigure}{0.66\columnwidth}
			{\includegraphics[width=\linewidth]{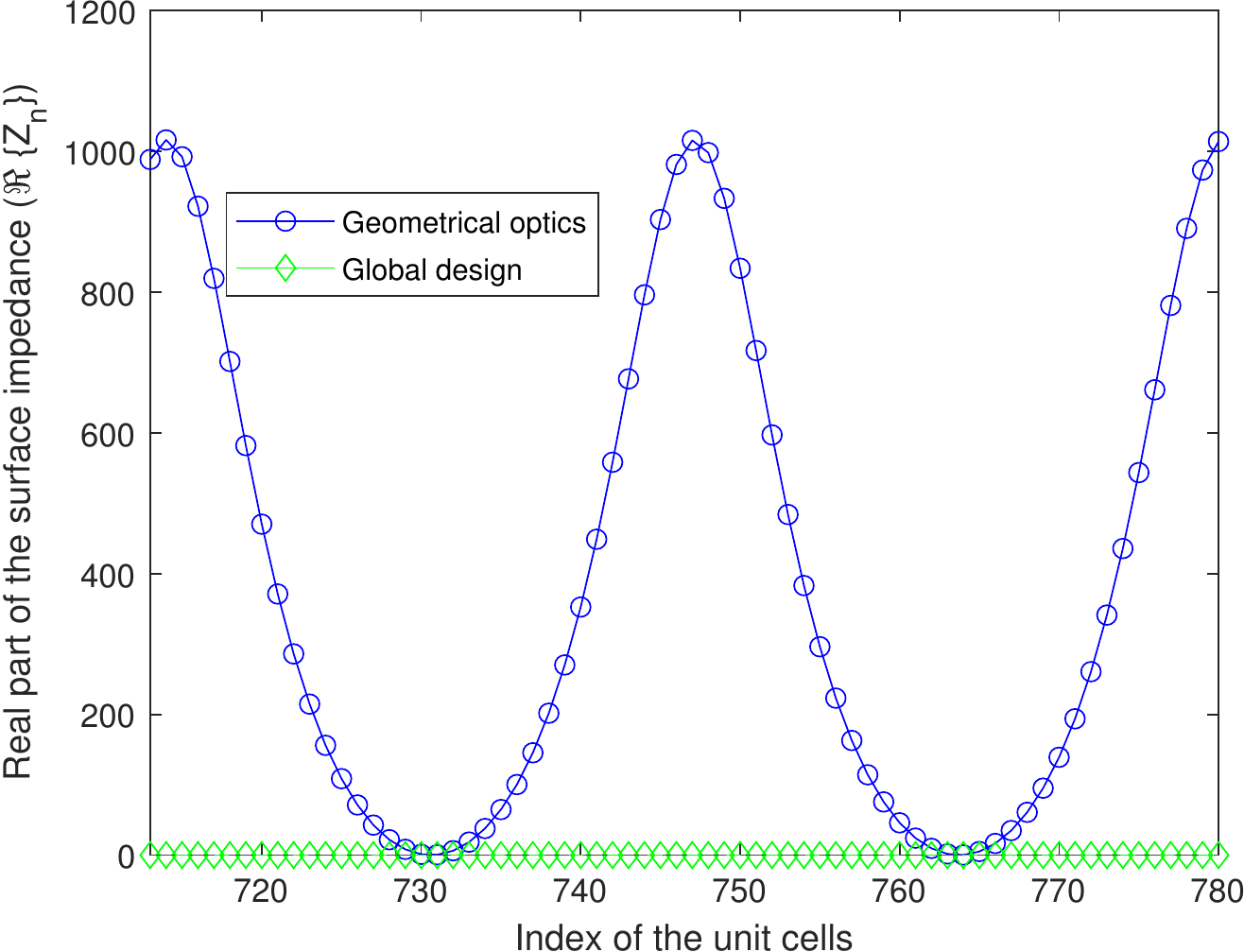}}
			\caption{Real part of surface impedance ${\mathop{\Re}\nolimits} \left( {{Z_n}} \right)$.}\label{fig:Z_real_no_null_thetaR_75_deg_FINAL_correct_imag_only}
		\end{subfigure}
		\begin{subfigure}{0.66\columnwidth}
			{\includegraphics[width=\linewidth]{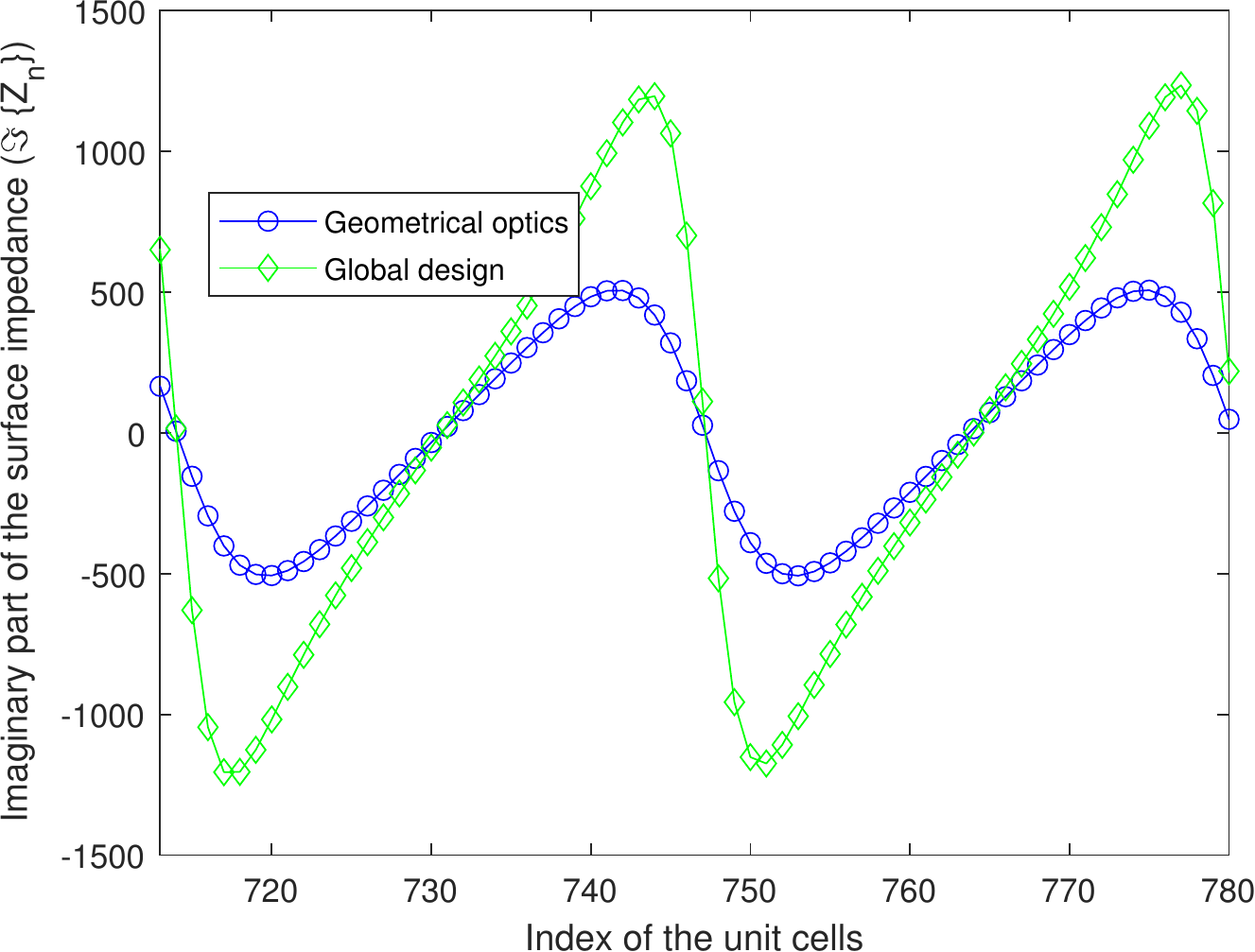}}
			\caption{Imaginary part of surface impedance ${\mathop{\Im}\nolimits} \left( {{Z_n}} \right)$.}\label{fig:Z_imag_no_null_thetaR_75_deg_FINAL_correct_imag_only}
		\end{subfigure}
		\begin{subfigure}{0.66\columnwidth}
			{\includegraphics[width=\linewidth]{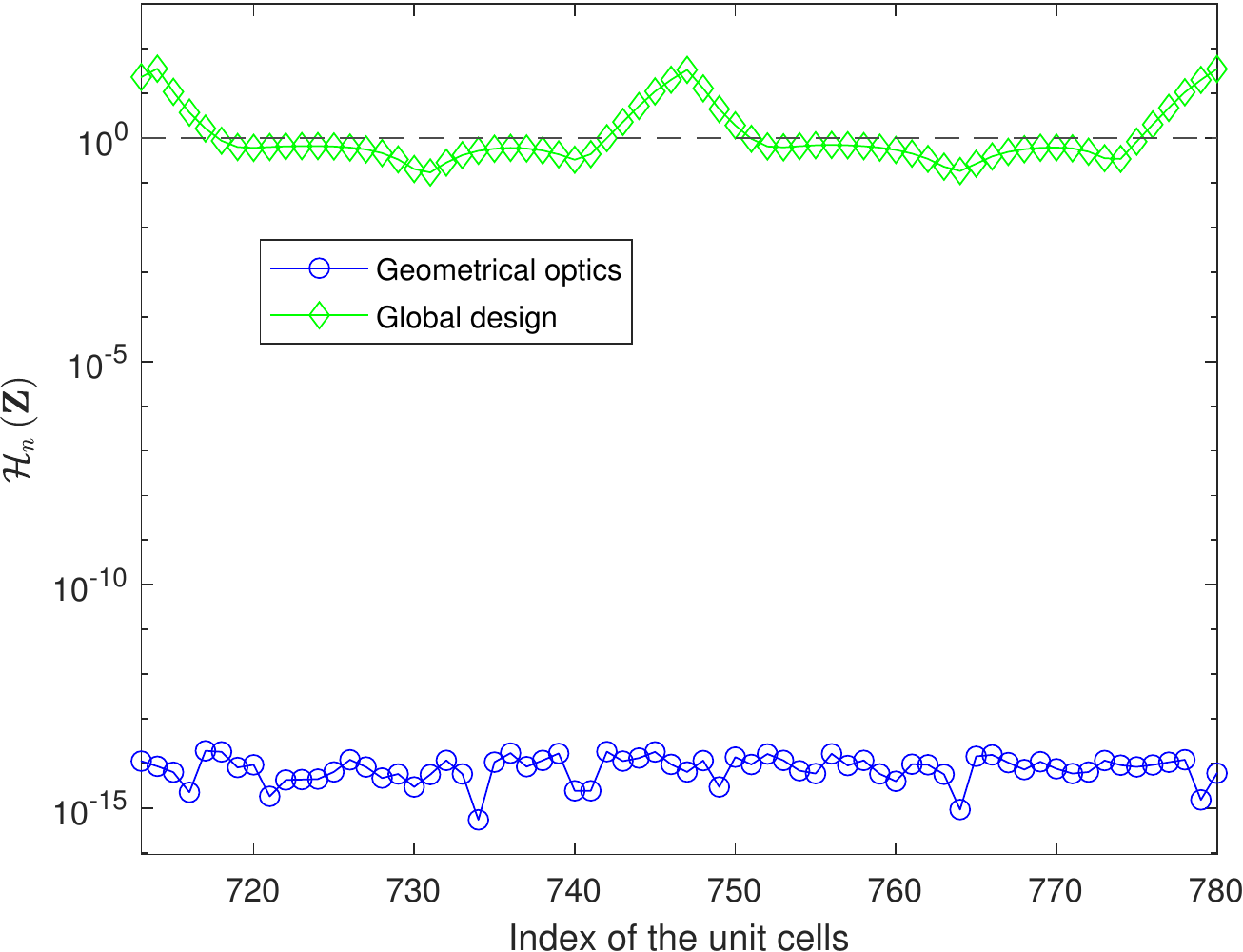}}
			\caption{Helmholtz's ratio ${{\mathcal{H}}_n}\left( {\bf{Z}} \right)$.}\label{fig:Helmholtz_Ratio_No_Null_thetaR_75_deg_FINAL_correct_imag_only}
		\end{subfigure}
		\begin{subfigure}{0.99\columnwidth}
			{\includegraphics[width=\linewidth]{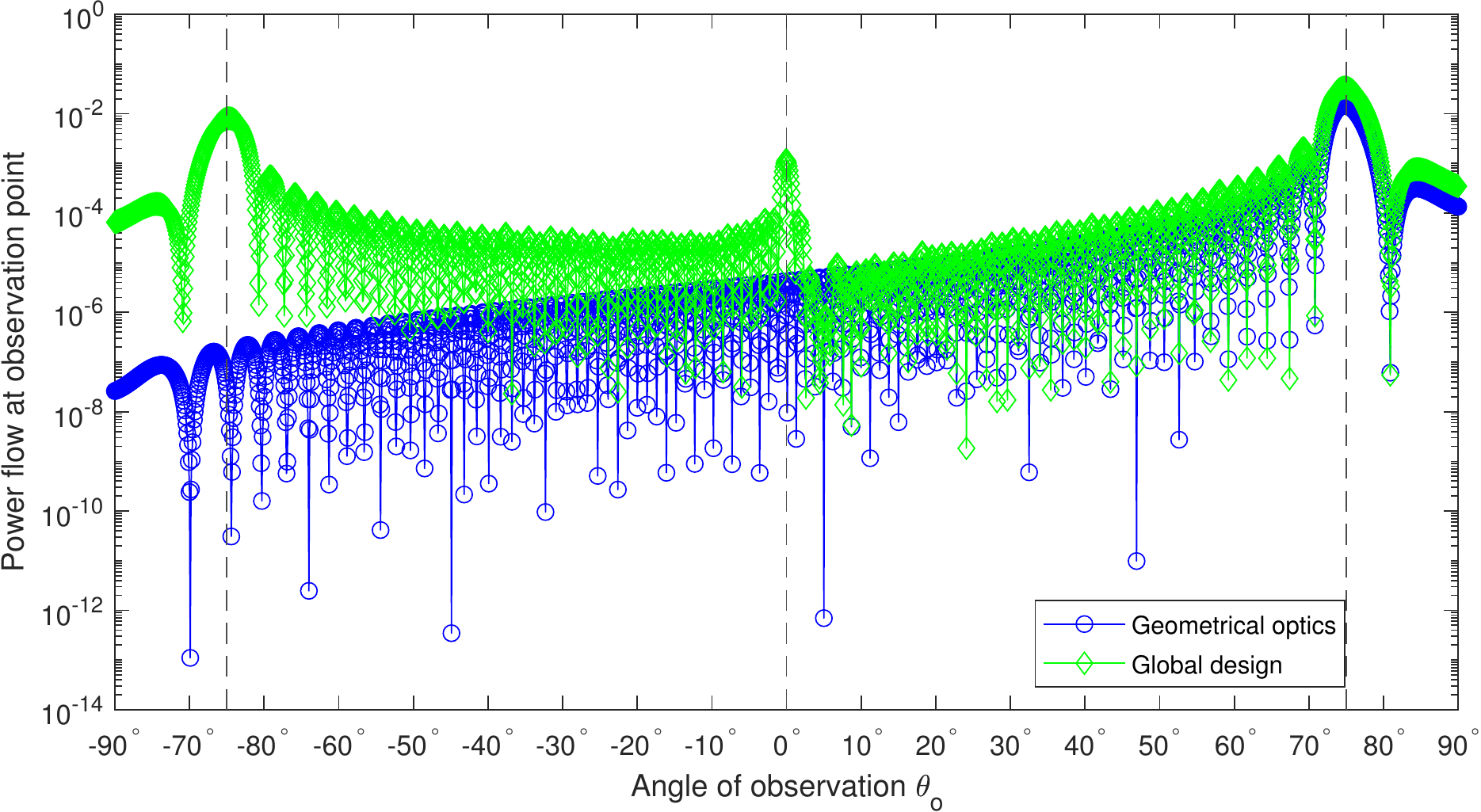}}
			\caption{Power flux vs. angle of observation ${{\mathcal{P}}_{{\rm{obs}}}}\left( {\bf{Z}} \right)$ (dB plot).}\label{fig:Poynting_Obs_No_Null_thetaR_75_deg_in_dB_FINAL_correct_imag_only}
		\end{subfigure}
		\begin{subfigure}{0.99\columnwidth}
			{\includegraphics[width=\linewidth]{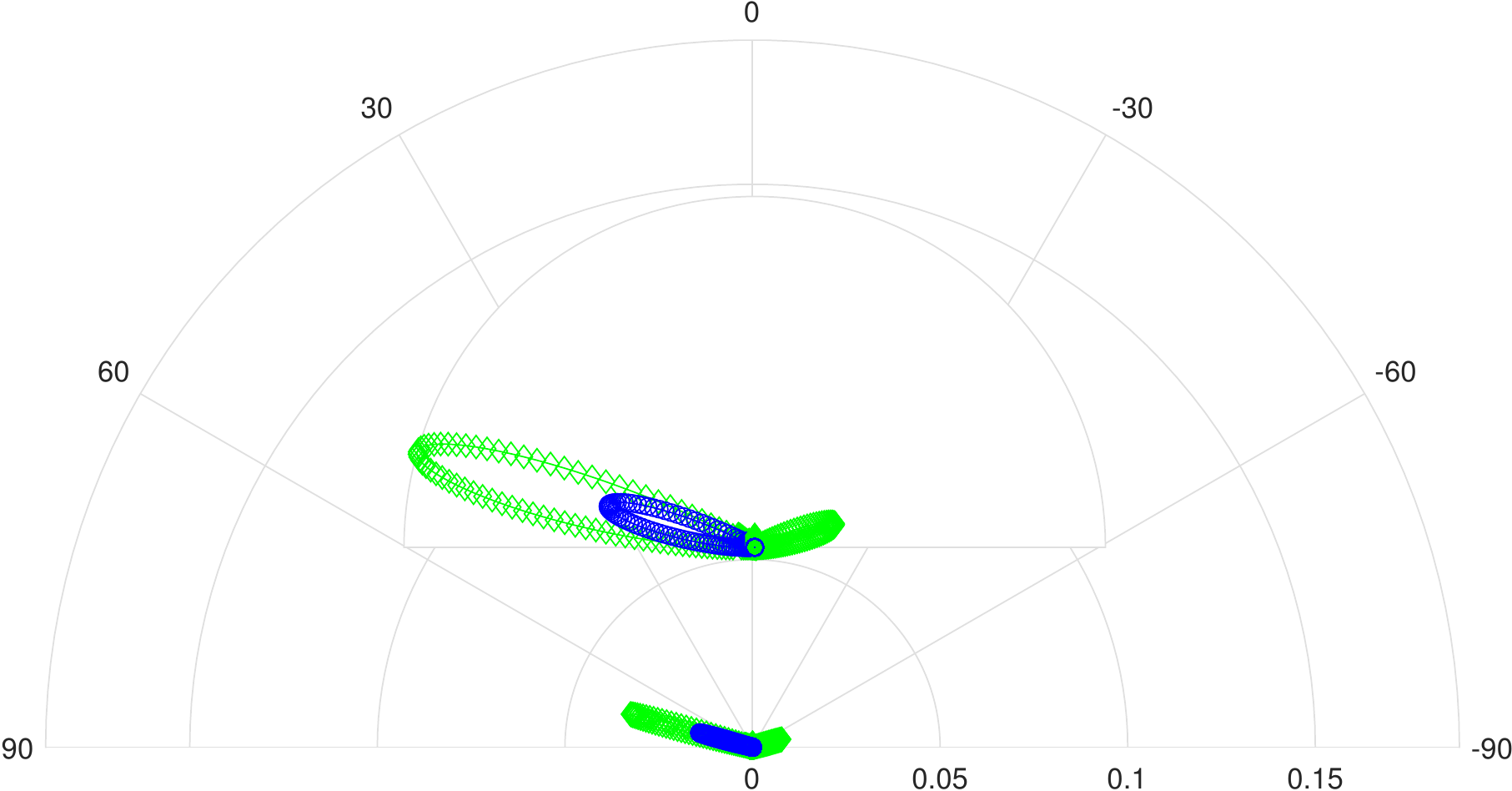}}
			\caption{Power flux vs. angle of observation ${{\mathcal{P}}_{{\rm{obs}}}}\left( {\bf{Z}} \right)$ (polar plot).}\label{fig:Poynting_Obs_No_Null_thetaR_75_deg_polar_FINAL_correct_imag_only}
		\end{subfigure}
	\end{center}
	\caption{Surface impedance in \eqref{eq:GO-Z} and solution of the optimization problem in \eqref{eq:Global-Opt-Formulation:main} ($\theta_r(\vect{r}_{\textup{Rx}}) = 75^\circ$). After obtaining the surface impedance, which is the same as in Fig. \ref{fig:No_Null_thetaR_75_FINAL_correct}, its real part is set equal to zero and the Poynting vector at the receiver is recomputed.}
	\label{fig:No_Null_thetaR_75_FINAL_correct_imag_only}
\end{figure*}
%
%
%
In the obtained numerical results, the surface impedances for the case study $\theta_r(\vect{r}_{\textup{Rx}}) = 75^\circ$ 
are obtained without imposing the fulfillment of Helmholtz's condition, i.e., $\varepsilon$ is large. It is interesting to analyze the impact of this assumption on the shape of the Poynting vector as a function of the angle of observation. This is illustrated in Fig \ref{fig:No_Null_thetaR_75_FINAL_correct} by imposing $\varepsilon = 5 \cdot 10^{-2}$. Due to the slow converge speed of the \verb+fmincon+ function, as an illustrative example, we focus our attention only on the optimization of the problem in \eqref{eq:Global-Opt-Formulation:main} without imposing any constraint on the nullification of the undesired directions of reradiation. By comparing Fig. \ref{fig:No_Null_thetaR_75_FINAL_correct} against Fig. \ref{fig:No_Null_thetaR_75_FINAL}, we note that adding the Helmholtz constraint in the optimization problem results in a better Poynting vector as a function of the angle of observation. Specifically, we see that a smaller power is steered towards the undesired directions of reradiation, even though no constraint is imposed to this end. In addition, we see that the real and imaginary parts of the surface impedance in  Fig. \ref{fig:No_Null_thetaR_75_FINAL_correct} are closer to those reported in Fig. \ref{fig:With_Null_thetaR_75_FINAL}, where a constraint for the reradiation towards the specular direction is imposed. Even though the reradiation constraint $\delta$ is not fulfilled in Fig. \ref{fig:With_Null_thetaR_75_FINAL}, we clearly see that imposing the Helmholtz constraint by design may inherently help in finding a better solution for the surface impedance. These observations that originate from the specific example analyzed in the present paper require further investigation, along with the development of efficient optimization algorithms for solving the considered optimization algorithms for any system setups and optimization constraints.

{Besides the impact of the Helmholtz constraint on the surface impedance obtained as solution of the considered optimization problems, it is interesting to analyze the reradiation properties of an RIS that is modeled as a purely reactive impedance boundary. More specifically, one relevant question is ``\textit{Given the surface impedance solution of the optimization problem in \eqref{eq:Global-Opt-Formulation:main}, what would the Poynting vector as a function of the angle of observation be if we set the surface impedance equal to the imaginary part of the surface impedance solution of the optimization problem in \eqref{eq:Global-Opt-Formulation:main}?} This approach would result in a surface impedance with a real part equal to zero, in agreement with the optimization problem in \ref{eq:Pratical-Opt-Formulation:main}. The answer to this question is illustrated in Fig.      \ref{fig:No_Null_thetaR_75_FINAL_correct_imag_only}. We see that the resulting surface impedance provides a Poynting vector with strong undesired reflections  and that the Helmholtz constraint is not fulfilled anymore. This case study highlights that RIS designs based on reactive boundary sheets cannot be obtained by simply setting equal to zero the real part of the surface impedance that is obtained by solving \eqref{eq:Global-Opt-Formulation:main}. Therefore, major research efforts need to be put in efficiently solving the optimization problems in \eqref{eq:Global-Opt-Formulation:main} and \eqref{eq:Pratical-Opt-Formulation:main}, in order to find solutions that offer good reradiation properties, that are electromagnetically consistent, and that are simple to implement (i.e., the surface impedance is purely reactive).}

\begin{table*}[!t]
		\centering
		\caption{Comparison of the power flux at the location of the receiver Rx.}
		\label{table:optimal_vs_initial_and_practical_at_desired}
		\newcommand{\tabincell}[2]{\begin{tabular}{@{}#1@{}}#2\end{tabular}}
		\begin{tabular}{l|c|c} \hline
			Without nullification of the specular reflection & $\theta_r(\vect{r}_{\textup{Rx}}) = 30^\circ$ (Fig. \ref{fig:No_Null_thetaR_30_FINAL}) & $\theta_r(\vect{r}_{\textup{Rx}}) = 75^\circ$ (Fig. \ref{fig:No_Null_thetaR_75_FINAL}) \\ \hline
			${{\mathcal{P}}_{{\rm{Rx}}}}\left( {{{\bf{Z}}_{{\rm{glo0}}}}} \right)/{{\mathcal{P}}_{{\rm{Rx}}}}\left( {{{\bf{Z}}_{{\rm{GO}}}}} \right)$ & $0.588$ dB & $3.392$ dB \\
			${{\mathcal{P}}_{{\rm{Rx}}}}\left( {{{\bf{Z}}_{{\rm{glo0}}}}} \right)/{{\mathcal{P}}_{{\rm{Rx}}}}\left( {{\bf{Z}}_{{\rm{glo0}}}^{{\rm{reactive}}}} \right)$ & $ \sim 10^{-6}$ dB & $\sim 10^{-11}$ dB \\ \hline \hline
			With nullification of the specular reflection & $\theta_r(\vect{r}_{\textup{Rx}}) = 30^\circ$ (Fig. \ref{fig:With_Null_thetaR_30_FINAL}) & $\theta_r(\vect{r}_{\textup{Rx}}) = 75^\circ$ (Fig. \ref{fig:With_Null_thetaR_75_FINAL}) \\ \hline
			${{\mathcal{P}}_{{\rm{Rx}}}}\left( {{{\bf{Z}}_{{\rm{glo0}}}}} \right)/{{\mathcal{P}}_{{\rm{Rx}}}}\left( {{{\bf{Z}}_{{\rm{GO}}}}} \right)$ & $0.599$ dB  & $4.822$ dB \\
			${{\mathcal{P}}_{{\rm{Rx}}}}\left( {{{\bf{Z}}_{{\rm{glo0}}}}} \right)/{{\mathcal{P}}_{{\rm{Rx}}}}\left( {{\bf{Z}}_{{\rm{glo0}}}^{{\rm{reactive}}}} \right)$ & $\sim 10^{-2}$ dB & $0.269$ dB \\ \hline
		\end{tabular}
	\end{table*}
	\begin{table*}[!t]
		\centering
		\caption{Comparison of the power flux at the location of the receiver Rx and the peak of the radiation pattern.}
		\label{table:peak_vs_desired}
		\newcommand{\tabincell}[2]{\begin{tabular}{@{}#1@{}}#2\end{tabular}}
		\begin{tabular}{l|c|c|c} \hline
		$\hspace{3.0cm}{ {\theta _r}\left( {{{\bf{r}}_{{\rm{Rx}}}}} \right)} = 30^\circ$ (Fig. \ref{fig:No_Null_thetaR_30_FINAL} and Fig. \ref{fig:With_Null_thetaR_30_FINAL}) & ${{\mathcal{P}}_{{\rm{max}}}}(\bf{Z}) = \mathop {\max }\limits_{{\theta _o}} \left( {{{\mathcal{P}}_{{\rm{obs}}}}\left( {\bf{Z}} \right)} \right)$ & ${{\mathcal{P}}_{{\rm{max}}}}\left( {\bf{Z}} \right)/{{\mathcal{P}}_{{\rm{Rx}}}}\left( {\bf{Z}} \right)$ \\ \hline
			\eqref{eq:GO-Z} -- Geometrical optics & $30^\circ$ & $0$ dB\\
			\eqref{eq:Global-Opt-Formulation:main} -- Global design & $30^\circ$ & $0$ dB \\
			\eqref{eq:Pratical-Opt-Formulation:main} -- Reactive boundary & $30^\circ$ & $0$ dB \\
			\eqref{eq:Mask-Global-Opt-Formulation:main} -- Global design with nullification of the specular reflection & $30^\circ$ & $0$ dB \\
			\eqref{eq:Mask-Pratical-Opt-Formulation:main} -- Reactive boundary with nullification of the specular reflection & $30^\circ$ & $0$ dB \\ \hline \hline
			$\hspace{3.0cm}{ {\theta _r}\left( {{{\bf{r}}_{{\rm{Rx}}}}} \right)} = 75^\circ$ (Fig. \ref{fig:No_Null_thetaR_75_FINAL} and Fig. \ref{fig:With_Null_thetaR_75_FINAL}) & ${{\mathcal{P}}_{{\rm{max}}}}(\bf{Z}) = \mathop {\max }\limits_{{\theta _o}} \left( {{{\mathcal{P}}_{{\rm{obs}}}}\left( {\bf{Z}} \right)} \right)$ & ${{\mathcal{P}}_{{\rm{max}}}}\left( {\bf{Z}} \right)/{{\mathcal{P}}_{{\rm{Rx}}}}\left( {\bf{Z}} \right)$ \\ \hline
			\eqref{eq:GO-Z} -- Geometrical optics &  $74.8^\circ$ & $0.0306$ dB\\
			\eqref{eq:Global-Opt-Formulation:main} -- Global design & $74.8^\circ$ & $0.0325$ dB\\
			\eqref{eq:Pratical-Opt-Formulation:main} -- Reactive boundary & $74.8^\circ$ & $0.0331$ dB\\
			\eqref{eq:Mask-Global-Opt-Formulation:main} -- Global design with nullification of the specular reflection & $74.8^\circ$ & $0.0508$ dB\\
			\eqref{eq:Mask-Pratical-Opt-Formulation:main} -- Reactive boundary with nullification of the specular reflection & $74.8^\circ$ & $0.0142$ dB\\
			\hline
		\end{tabular}
	\end{table*}
	\begin{table*}[!t]
		\centering
		\caption{Comparison of the power flux towards the direction of the receiver Rx and the direction of specular reflection.}
		\label{table:desired_vs_specular}
		\newcommand{\tabincell}[2]{\begin{tabular}{@{}#1@{}}#2\end{tabular}}
		\begin{tabular}{l|c|c|c} \hline
			$\hspace{3.0cm}{ {\theta _r}\left( {{{\bf{r}}_{{\rm{Rx}}}}} \right)} = 30^\circ$ (Fig. \ref{fig:No_Null_thetaR_30_FINAL} and Fig. \ref{fig:With_Null_thetaR_30_FINAL}) & ${{\mathcal{P}}_{{\rm{Rx}}}}\left( {\bf{Z}} \right)$ & ${{\mathcal{P}}_{{\rm{specular}}}}\left( {\bf{Z}} \right)$ & ${{\mathcal{P}}_{{\rm{Rx}}}}\left( {\bf{Z}} \right)/{{\mathcal{P}}_{{\rm{specular}}}}\left( {\bf{Z}} \right)$ \\ \hline
			\eqref{eq:GO-Z} -- Geometrical optics & $-7.871$ dB & $-45.854$ dB& $+37.983$ dB\\
			\eqref{eq:Global-Opt-Formulation:main} -- Global design & $-7.283$ dB & $-33.814$ dB & $+26.532$ dB\\
			\eqref{eq:Pratical-Opt-Formulation:main} -- Reactive boundary & $-7.283$ dB & $-34.397$ dB & $+27.114$ dB\\
			\eqref{eq:Mask-Global-Opt-Formulation:main} -- Global design with nullification of the specular reflection & $-7.271$ dB & $-40.089$ dB & $+32.817$ dB\\
			\eqref{eq:Mask-Pratical-Opt-Formulation:main} -- Reactive boundary with nullification of the specular reflection & $-7.249$ dB & $-40.343$ dB & $+33.094$ dB\\
			\hline \hline
			$\hspace{3.0cm} {{\theta _r}\left( {{{\bf{r}}_{{\rm{Rx}}}}} \right)} = 75^\circ$ (Fig. \ref{fig:No_Null_thetaR_75_FINAL} and Fig. \ref{fig:With_Null_thetaR_75_FINAL}) & ${{\mathcal{P}}_{{\rm{Rx}}}}\left( {\bf{Z}} \right)$ & ${{\mathcal{P}}_{{\rm{specular}}}}\left( {\bf{Z}} \right)$ & ${{\mathcal{P}}_{{\rm{Rx}}}}\left( {\bf{Z}} \right)/{{\mathcal{P}}_{{\rm{specular}}}}\left( {\bf{Z}} \right)$ \\ \hline
			\eqref{eq:GO-Z} -- Geometrical optics & $-18.362$ dB & $-67.317$ dB & $+48.955$ dB\\
			\eqref{eq:Global-Opt-Formulation:main} -- Global design & $-14.970$ dB & $-17.151$ dB & $+2.181$ dB\\
			\eqref{eq:Mask-Global-Opt-Formulation:main} -- Reactive boundary & $-14.970$ dB & $-37.186$ dB & $+22.216$ dB \\
			\eqref{eq:Mask-Global-Opt-Formulation:main} -- Global design with nullification of the specular reflection & $-13.540$ dB & $-40.001$ dB & $+26.462$ dB\\
			\eqref{eq:Mask-Pratical-Opt-Formulation:main} -- Reactive boundary with nullification of the specular reflection & $-13.809$ dB & $-40.000$ dB & $+26.191$ dB \\ \hline
		\end{tabular}
	\end{table*}
In Tables \ref{table:optimal_vs_initial_and_practical_at_desired},
\ref{table:peak_vs_desired}, and \ref{table:desired_vs_specular}, we analyze, in a more quantitative manner, the solutions obtained by solving the optimization problems in \eqref{eq:Global-Opt-Formulation:main}, \eqref{eq:Pratical-Opt-Formulation:main}, \eqref{eq:Mask-Global-Opt-Formulation:main}, and \eqref{eq:Mask-Pratical-Opt-Formulation:main}. The data reported in these tables are obtained from  Figs. \ref{fig:No_Null_thetaR_75_FINAL}-Fig. \ref{fig:With_Null_thetaR_75_FINAL}.

In Table \ref{table:optimal_vs_initial_and_practical_at_desired}, we compare the power flux at the location of the receiver Rx. We evince that the geometrical optics solution usually results in a power loss of a few decibels, as compared with the globally optimum design with unitary power efficiency. In the considered case study, the power difference can be of the order of 4.8 dB for ${{\theta _r}\left( {{{\bf{r}}_{{\rm{Rx}}}}} \right)} = 75^\circ$. On the other hand, we see that an RIS realized with a purely reactive impedance boundary results in a much smaller power loss, as compared with the globally optimum design with unitary power efficiency. In the considered case study, the power loss is about 0.27 dB for ${{\theta _r}\left( {{{\bf{r}}_{{\rm{Rx}}}}} \right)} = 75^\circ$. The results reported in Table \ref{table:optimal_vs_initial_and_practical_at_desired} provide a quantitative assessment of the implementation complexity versus the achievable performance of RISs that are implemented with purely reactive components.

In Table \ref{table:peak_vs_desired}, we analyze the steering accuracy of the considered designs for the surface impedance. More precisely, we compare the peak value of the power flux as a function of the angle of observation with the power flux evaluated at the location of the receiver Rx. If ${{\theta _r}\left( {{{\bf{r}}_{{\rm{Rx}}}}} \right)} = 30^\circ$, the considered designs result in perfect beamsteering capabilities. In fact, the maximum of the radiation pattern coincides with the desired direction of reradiation. If ${{\theta _r}\left( {{{\bf{r}}_{{\rm{Rx}}}}} \right)} = 75^\circ$, on the other hand, we observe some pointing errors. In the considered case study, the pointing error is relatively small and is about $0.2^\circ$, which results in a power loss of only a fraction of decibels.

In Table \ref{table:desired_vs_specular}, we analyze the reradiation properties of the considered designs for RISs in terms of power flux towards the direction of the desired receiver Rx and towards the direction of specular reflection. In general, the received power decreases with the desired angle of reflection. This is due to the so-called obliquity factor in the Poynting vector, i.e., the ${\cos {\theta _r}\left( {{{\bf{r}}_{{\rm{Rx}}}}} \right)}$ multiplication factor in the power flux in correspondence of the location of the receiver Rx. This is apparent from ${{\mathcal{P}}_{{\rm{Rx}}}}\left( {\bf{Z}} \right)$ in Table \ref{table:desired_vs_specular}. Due to the obliquity factor, for example, the power loss between ${{\theta _r}\left( {{{\bf{r}}_{{\rm{Rx}}}}} \right)} = 30^\circ$ and ${{\theta _r}\left( {{{\bf{r}}_{{\rm{Rx}}}}} \right)} = 75^\circ$ is about 6--11 dB in the considered case studies. In Table \ref{table:desired_vs_specular}, in addition, we see that, by taking into account the nullification of the specular reflection at the design stage, we can, at the same time, reduce the amount of power lost towards the specular direction and increase the amount of power steered towards the location of the receiver Rx. This is due to the total power conservation principle.

\section{Concluding Remarks}
In this paper, we have overviewed three communication models for RISs that are widely utilized in the context of performance evaluation and optimization of wireless communication systems and networks. We have focused our attention on models for RISs based on inhomogeneous surface impedance boundaries, in light of their ease of integration into Maxwell's equations and their inherent electromagnetic consistency under typical and practically relevant approximation regimes, e.g., physical optics. With the aid of several numerical examples, we have illustrated design criteria for RISs that are based on local and global optimality criteria, as well as an approximated design criterion that results in purely reactive impedance boundaries. We have discussed their inherent advantages and limitations, with the aid of mathematical analysis and numerical simulations.

As far as the considered optimization problems are concerned, we have focused our attention on problem formulations in which the objective function is given by the surface power efficiency, since it characterizes locally and globally optimal designs, and the power flux is utilized as a key performance indicator to showcase the steering capabilities of RISs. Similar optimization problems can be formulated by considering the power flux as the objective function and the surface power efficiency as a design constraint.

The considered optimization problems and methods are applicable to RISs whose surface impedance is slowly varying at the wavelength scale and the physical optics approximation is applicable. More advanced designs, which can lead to RISs that realize theoretically perfect anomalous reflections with complete suppression of parasitic scattering, require either an accurate control of the fast-varying surface modes excited at the surface or a careful design of the scattering from diffraction gratings, i.e., metagratings.

{Finally, we have discussed the applications of the proposed approach in wireless communications. An important extension of the proposed approach lies in developing efficient numerical algorithms for solving the proposed electromagnetically consistent optimization problems.}

\bibliographystyle{IEEEtran}
\bibliography{IEEEabrv,Ref}

\end{document}